\documentclass[aps,twocolumn,a4paper,showkeys,footinbib,longbibliography,nofootinbib]{revtex4-1}
\usepackage[utf8]{inputenc}
\usepackage{filecontents}
\usepackage{natbib}
\usepackage{amsmath,amssymb,bm,amsthm}
\usepackage{subfigure}
\usepackage{graphicx}
\usepackage{dcolumn}
\usepackage{bm}
\usepackage[mathlines]{lineno}
\usepackage{booktabs}
\usepackage{color}
\newcounter{one}
\usepackage{url}

\usepackage{textcase}

\usepackage{colortbl}
\usepackage{tabularx}
\usepackage{verbatim}
\usepackage{multirow}

\usepackage[T1]{fontenc} 
\usepackage{lmodern}
\usepackage{bbm}
\usepackage[utf8]{inputenc}
\usepackage{amsfonts}
\usepackage{array}

\usepackage{bbm}
\usepackage{enumitem}
\usepackage{umoline}
\usepackage[usenames,svgnames]{xcolor}
\usepackage{natbib}
\usepackage[hyperindex,breaklinks]{hyperref}
\hypersetup{
     colorlinks=true,       		
     linkcolor=Navy,          	
     citecolor=Navy,            
     filecolor=Navy,      		
     urlcolor=Navy,           	
    runcolor=cyan,
 }
\usepackage{graphicx}
\usepackage{amsfonts}
\usepackage{amssymb}
\usepackage{amsmath}
\usepackage{bbm}
\usepackage{enumitem}

\newcommand{\bra}[1]{\langle #1 |}
\newcommand{\ket}[1]{| #1 \rangle}

\newcommand{\tr}[0]{ {\rm tr}}

\newcommand{\half}[1]{{ \rm h}}
\newcommand{\Oorderof}{\mathcal{O}}
\newcommand{\orderof}[1]{\Oorderof(#1)} 

\newcommand{\for}[0]{\quad \textrm{for} \quad}
\newcommand{\with}[0]{\quad \textrm{with} \quad}

\newcommand{\dist}{d}

\newcommand{\co}{{\rm c}}

\newcommand{\diam}{{\rm diam}}

\newcommand{\Or}{\quad {\rm or} \quad}

\def\beq{\begin{equation}}
\def\eeq{\end{equation}}
\def\nbeq{\begin{equation*}}
\def\neeq{\end{equation*}}
\def\<{\langle}
\def\>{\rangle}

\def\tr{{\rm tr}}

\newtheorem{theorem}{Theorem}

\newtheorem{lemma}{Lemma}

\newtheorem{assump}{Assumption} 
\newtheorem{definition}{Definition}  
\newtheorem{prop}[theorem]{Proposition} 

\newcommand{\bal}[2]{#1[#2]} 

\newcommand{\fset}[1]{\mathcal{#1}}
\newcommand{\lsc}{\ell}
\newcommand{\abb}{\bigl |}
\newcommand{\vH}{\bar{v}}

\usepackage{mathtools}
\def\multiset#1#2{\ensuremath{\left(\kern-.3em\left(\genfrac{}{}{0pt}{}{#1}{#2}\right)\kern-.3em\right)}}

\begin{document}
\title{Strictly linear light cones in long-range interacting systems of arbitrary dimensions}

\author{Tomotaka Kuwahara$^{1,2}$}
\email{tomotaka.kuwahara@riken.jp}
\affiliation{$^{1}$
Mathematical Science Team, RIKEN Center for Advanced Intelligence Project (AIP),1-4-1 Nihonbashi, Chuo-ku, Tokyo 103-0027, Japan
}
\affiliation{$^{2}$Interdisciplinary Theoretical \& Mathematical Sciences Program (iTHEMS) RIKEN 2-1, Hirosawa, Wako, Saitama 351-0198, Japan}

\author{Keiji Saito$^{3}$}
\email{saitoh@rk.phys.keio.ac.jp}
\affiliation{$^{3}$Department of Physics, Keio University, Yokohama 223-8522, Japan}

\begin{abstract}


In locally interacting quantum many-body systems, the velocity of information propagation is finitely bounded and a linear light cone can be defined.
Outside the light cone, the amount of information rapidly decays with distance.
When systems have long-range interactions, it is highly nontrivial whether such a linear light cone exists.
Herein, we consider generic long-range interacting systems with decaying interactions, such as $R^{-\alpha}$ with distance $R$. 
We prove the existence of the linear light cone for  $\alpha>2D+1$ ($D$: the spatial dimension), where we obtain the Lieb--Robinson bound as $\|[O_i(t),O_j]\|\lesssim{t}^{2D+1}(R-\bar{v}t)^{-\alpha}$ with $\bar{v}=\mathcal{O}(1)$ for two arbitrary operators $O_i$ and $O_j$ separated by a distance $R$. 
Moreover, we provide an explicit quantum-state transfer protocol that achieves the above bound up to a constant coefficient and violates the linear light cone for $\alpha<2D+1$.
In the regime of $\alpha>2D+1$, our result characterizes the best general constraints on the information spreading.

%
%

\end{abstract}

\maketitle



\section{Introduction}

In deep understanding of many-body physics, we necessarily encounter the question on how fast information propagates in the dynamics.
In this context, the most fundamental principle is the causality; that is,
in relativistic systems, information propagation is completely prohibited outside the light cone.
On the other hand, in non-relativistic quantum many-body systems, a rigorous light cone is not defined.
Lieb and Robinson proved in 1972~\cite{ref:LR-bound72} that an effective light cone can be defined, outside which information propagation exponentially decreases with distance.
In their study, the effective light cone is characterized by the so-called Lieb--Robinson velocity.

The Lieb-Robinson bound imposes one of the most fundamental restrictions to the dynamics~\cite{
PhysRevLett.97.050401,
PhysRevLett.97.150404,
PhysRevA.78.010306,
PhysRevLett.102.017204,
PhysRevLett.113.187203,
lauchli2008spreading,
PhysRevA.80.052319,
PhysRevLett.111.230404,
PhysRevLett.120.020401,
PhysRevLett.113.127202,
PhysRevLett.102.240501,
PhysRevA.81.040102} 
and has been improved in various ways~~\cite{
ref:Nachtergaele2006-LR,
ref:Hastings2006-ExpDec,
PhysRevLett.99.167201,
nachtergaele2010lieb,
PhysRevLett.104.190401,
PhysRevA.84.032309,
PhysRevA.81.062107,
PhysRevLett.111.230404}.
Moreover, after Hastings' work on the Lieb-Schultz-Mattis theorem~\cite{PhysRevB.69.104431}, 
the Lieb-Robinson bound has been recognized as a crucial ingredient for analyzing the universal physics in many-body systems, such as
quasi-adiabatic continuation~\cite{PhysRevB.72.045141}, 
the area law of entanglement~\cite{
ref:Hastings-AL07,
PhysRevA.80.052104,
doi:10.1063/1.4932612,
PhysRevLett.111.170501}, 
thermalization~\cite{
ref:Mueller2013-Thermal,
Aolita_2015,
Gogolin_2016,
PhysRevLett.119.100601}, 
quantization of Hall conductance~\cite{
hastings2015quantization,
hastings2010quasi}, 
stability of the topological order~\cite{
bravyi2010topological,
bravyi2011short,
michalakis2013stability}, 
clustering theorem for correlation functions~\cite{
ref:Hastings2004-Markov,
ref:Hastings2006-ExpDec,
ref:Nachtergaele2006-LR,
PhysRevLett.119.110601,
PhysRevLett.93.126402}, 
effective Hamiltonian theory~\cite{
KUWAHARA201696,PhysRevX.10.011043}, 
classical simulation of many-body systems~\cite{
0034-4885-75-2-022001,
PhysRevLett.97.157202,
ref:Osborne2007-Adiabatic,
PhysRevLett.101.070503,
PhysRevB.76.201102,
Lightconematrixproduct,
PhysRevB.77.144302}.
More recently, the Lieb--Robinson bound has been further applied to the digital quantum simulation of many-body systems~\cite{8555119,PhysRevX.9.031006,PhysRevA.99.052332} and 
quantum information scrambling~\cite{
PhysRevLett.117.091602,
Maldacena2016,
Gu2017,
PhysRevB.100.064305,
PhysRevB.96.060301,
PhysRevX.7.031016,
PhysRevB.98.144304,
PhysRevB.99.014303,
PhysRevB.96.020406,
PhysRevX.7.031011,
PhysRevB.98.134303,
PhysRevX.9.041017,
zhou2019operator}, 
where the Lieb--Robinson velocity gives an upper bound of butterfly speed~\cite{PhysRevLett.117.091602}.
In addition, experimental advancement enables direct observation of the Lieb--Robinson bounds~\cite{
cheneau2012light,
langen2013local,
Meinert1259,
PhysRevLett.113.147205,
richerme2014non,
jurcevic2014quasiparticle,
PhysRevX.8.021070}.

In the case of short-range interacting spin systems, an effective light cone is characterized by a finite velocity, and information propagation is restricted inside the ``linear light cone.''
However, when we consider long-range interacting systems, the existence of a linear light cone is quite subtle because long-range interactions enable immediate communication between two arbitrarily distant parties.
Here, long-range interaction implies that the interaction strength between separated sites shows a power-law decay of $R^{-\alpha}$ with the distance $R$.
Depending on the exponent $\alpha$, both the linear and nonlinear light cones can appear.
Recent experiments have realized long-range interacting systems with various values of $\alpha$~\cite{
bendkowsky2009observation,
RevModPhys.80.885,
RevModPhys.82.2313,
yan2013observation,
PhysRevLett.108.210401,
britton2012engineered,
Islam583,
bernien2017probing,
zhang2017observation,
Neyenhuise1700672}, 
and hence, exploring the universal aspects of long-range interacting systems is attracting increasing attention~\cite{
PhysRevLett.113.156402,
PhysRevLett.109.267203,
PhysRevLett.119.023001,
Kuwahara_2017,
Kuwahara_2016_asymptotic,
kuwahara2019gaussian,
kuwahara2019area}.
From these backgrounds, one of the most important and intriguing open problems is the so-called \textit{linear light cone problem}, which clarifies whether linear right cones can exist in long-range interacting systems, and what is the general criterion for it.


So far, various studies have clarified dynamical properties in the long-range interacting systems~\cite{PhysRevLett.111.260401,
PhysRevLett.113.030602,
PhysRevLett.112.210601,
PhysRevLett.111.207202,
PhysRevB.90.174204,
PhysRevB.90.205438,
PhysRevLett.119.170503,
Cevolani_2016,
PhysRevLett.116.250402,
PhysRevB.94.144206,
Lepori_2017,
PhysRevB.98.024302,
PhysRevB.95.094205,
PhysRevLett.120.110602,
PhysRevA.99.010105,
PhysRevA.99.010105,
PhysRevA.99.032114,
PhysRevE.101.042118}.
As one of the generic aspects of the Lieb--Robinson bound, Refs.~\cite{ref:Hastings2006-ExpDec,nachtergaele2010lieb,Kuwahara_2016_njp,Sweke_2019,guo2019signaling} 
showed that the amount of information propagation is suppressed at least outside the effective light cone exponentially growing in time, irrespective of $\alpha$. 
Later, a more detailed universal upper bound was provided by Foss--Feig et al.~\cite{PhysRevLett.114.157201}. 
They proved that the effective light cone is at most polynomial with respect to time; in more detail, the shape of the light-cone was given by $t^{(\alpha-D+1)/(\alpha-2D)}$ ($\alpha>2D$) with $D$ as the spatial dimensions. 
However, despite significant efforts~\cite{PhysRevX.9.031006,Matsuta2017,PhysRevA.101.022333,tran2019locality}, 
the critical value of  $\alpha$ to obtain a linear light cone is still unclear even in the numerical level.

\begin{figure}[]
\centering
{
\includegraphics[clip, scale=0.38]{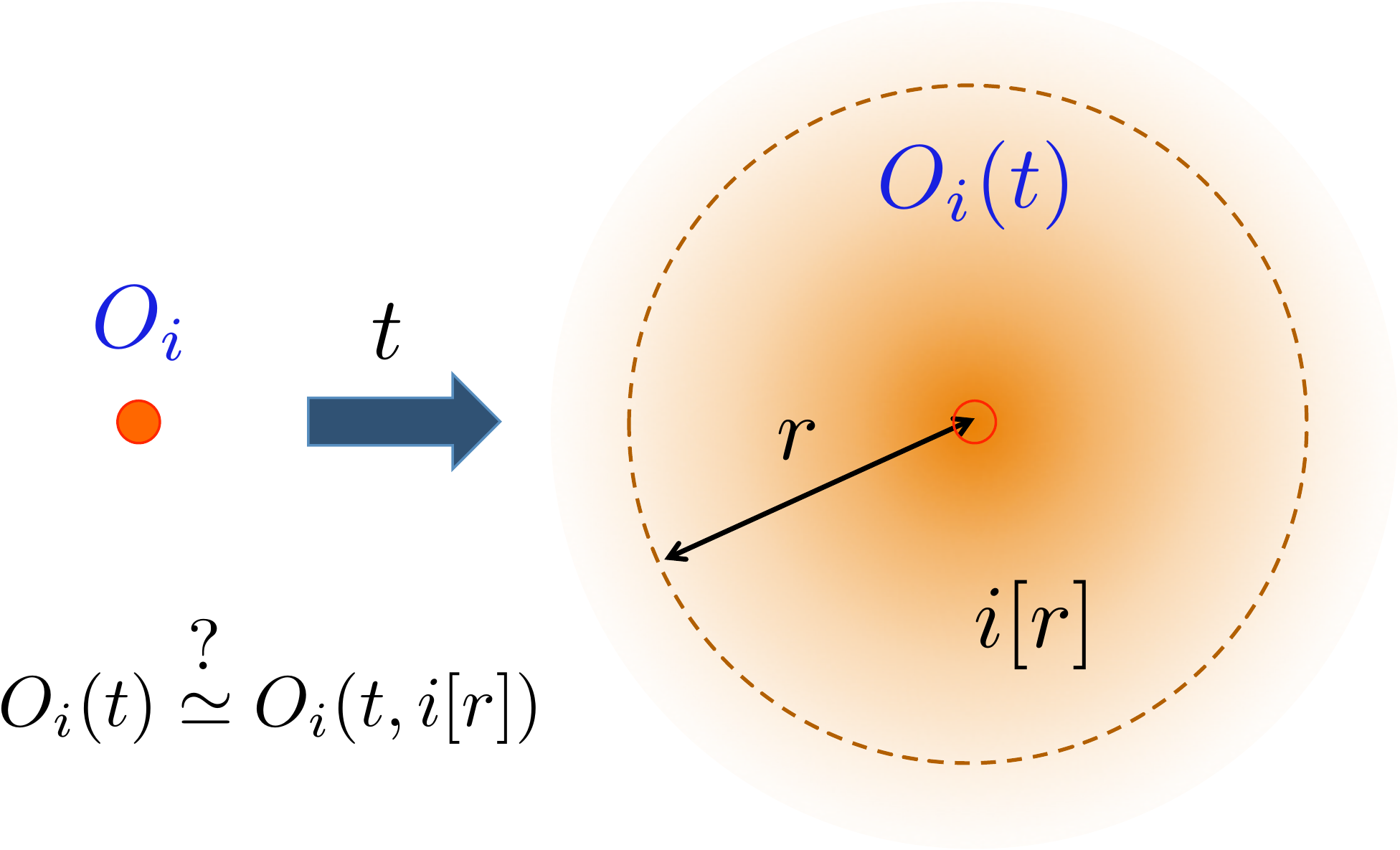}
}
\caption{When we consider time-evolution $O_i(t)$ of a local operator $O_i$, the quasi-locality of the interaction ensures that $O_i(t)$ is well-approximated by an operator defined on a ball region $\bal{i}{r}$ having the maximum distance of $r$ from $i$. The dynamics define the linear light cone if we can achieve an arbitrary approximation error for $r=\orderof{t}$ as in \eqref{def:linear_light_cone}.
}
\label{fig_local_approx}
\end{figure}

In this work, we rigorously prove that a linear light cone is obtained in generic long-range interacting systems under the condition of $\alpha>2D+1$.
As related work, in one-dimensional two-body interacting systems, 
the long-range Lieb--Robinson bound has been proved very recently in the form of $\|[O_i(t), O_j]\| \lesssim t/R$ for $\alpha>3$~\cite{chen2019finite}, which gives a non-trivial upper bound up to the time $t=\orderof{R}$.  
In our analyses, the Hamiltonian is not restricted to few-body interactions, and is applicable to arbitrary spatial dimensions.  
Our Lieb--Robinson bound is given in the stronger form of $\|[O_i(t), O_j]\| \lesssim t^{2D+1} (R-\bar{v} t)^{-\alpha}$. 
However, only the above commutation relation is not sufficient to upper-bound the entire information propagation outside the light cone. 
To obtain a linear light cone in a strict sense (see \eqref{def:linear_light_cone} and Fig.~\ref{fig_local_approx}), we also prove that the error of the local approximation of $O_i(t)$ decays as $t^{D+1}R^{-\alpha+D}$ outside the light cone (see \eqref{main_thm_ineq2_main} below).
Our result can improve various existing analyses that depend on a polynomial light cone~\cite{PhysRevLett.114.157201,PhysRevX.9.031006}.

We also discuss whether $\alpha=2D+1$ is the critical value to ensure a linear light cone in general setups.
In other words, we investigate the achievability of our Lieb--Robinson bound and the possibility to violate the linear-light cone for $\alpha<2D+1$, by which we show that our bound is the best general upper bound.
We consider a quantum state-transfer protocol through the dynamics on a spin network and give an explicit example that achieves our Lieb--Robinson bound for $\alpha>2D+1$ and violates the linear light cone for $\alpha<2D+1$. 
Our protocol is applied to (1/2)-spin systems and comprises only the Ising-type long-range interactions and simple short-range interactions that generate the controlled NOT gate operation.
This example ensures the optimality of our results.

The rest of this paper is organized as follows. 
In Sec. II, we formulate the precise setting and state the main
results and their implications.
In Sec. III, we sketch the intuitive explanation for the condition that linear light cone appears.
We also show the brief strategy for the proof.
In Sec. IV, we discuss the optimality of the present Lieb--Robinson bound.
In more detail, we explicitly show a quantum state transfer protocol which achieves our theoretical upper bound up to a coefficient.
Finally, in Sec. V, 
we summarize the paper making a brief discussion.

\section{Main results}

We consider a quantum many-body system with $n$ sites, where each site is located on a $D$-dimensional lattice with the total set $\Lambda$ ($|\Lambda|=n$).
We assume that a finite-dimensional Hilbert space is assigned to each of the sites.

For simplicity, we restrict ourselves to two-body interactions, but our results are extended to a more general setup, as shown in Appendix~\ref{appendixA}.
We focus on the following Hamiltonian $H$ with power-law decaying interactions:
\begin{align}
H=\sum_{i,j \in \Lambda} h_{i,j} + \sum_{i=1}^n h_i \with \| h_{i,j} \| \le \frac{g_0}{\dist_{i,j}^\alpha} ,\label{Ham:general_power}
\end{align} 
where $\dist_{i,j}$ is the distance between the sites $i$ and $j$, namely the minimum path length from the site $i$ to $j$,
$\{h_{i,j}\}_{i<j}$ are the bi-partite interaction operators, $\{h_i\}_{i=1}^n$ are the local potentials, $g_0$ is a positive constant of $\orderof{1}$, and $\|\cdots\|$ is the operator norm.
Although in some literatures~\cite{dauxois2002dynamics,CAMPA200957} the long-range interaction often implies the power-law decaying interactions with $\alpha\le D$, 
we refer to systems with \textit{arbitrary power-law decaying interactions} as long-range interacting systems in distinction from the finite-range (or, exponentially decaying) interactions.

Our analysis can be also applied to a time-dependent Hamiltonian, but for simplicity, we present the analysis only for the time-independent case.
One typical example is the following one-dimensional long-range Ising model:
\begin{align}
H= \sum_{i<j} \frac{g_0}{\dist_{i,j}^\alpha} \sigma_i^x  \sigma_j^x + B \sum_{i}  \sigma_i^z  \quad (D=1)  ,
\label{exp_long_range_trans_Ising}
\end{align}
where $\{\sigma^x,\sigma^y,\sigma^z\}$ are the Pauli matrices.
This class of Hamiltonians is experimentally realized for the power-law exponent in $\alpha\le 3$~\cite{
jurcevic2014quasiparticle,
zhang2017observation}.

We are now interested in the time-evolution using the Hamiltonian $H$.
For simplicity, we consider an operator $O_i$ that is locally defined on the site $i$ and analyze 
\begin{align}
O_i(t):= e^{iHt } O_i e^{-iHt }. \notag 
\end{align}
Mainly, we focus on the following two quantities:
\begin{align}
\| [O_i(t),O_j]\| \for  \forall j\in \Lambda,\label{commutator_approx}
\end{align}
 and
 \begin{align}
&\| [O_i(t)-O_i(t,\bal{i}{r}) ]\| \quad   {\rm with} \label{local_approx}\\
&O_i(t,\bal{i}{r}):= \frac{1}{\tr_{\bal{i}{r}^\co}(\hat{1})} \tr_{\bal{i}{r}^\co} \left[O_i(t)\right] \otimes \hat{1}_{\bal{i}{r}^\co},\notag 
\end{align}
where $\bal{i}{r}$ denotes the set of sites having a maximum distance of $r$ from the site $i$ and $\bal{i}{r}^\co$ is its complement set.
The quantity~\eqref{local_approx} characterizes the error of the local approximation for the time-evolved operator $O_i(t)$ into the region $\bal{i}{r}$ (see Fig.~\ref{fig_local_approx}).
We note that the decay of \eqref{local_approx} is not necessarily derived only from the decay of \eqref{commutator_approx}.
  
 Here, we define the linear light cone in the following sense.
We say that the Hamiltonian dynamics $e^{-iHt}$ has a linear light cone if the following inequality holds for an arbitrary error $\delta \in \mathbb{R}$ and $t$:
 \begin{align}
&\| [O_i(t)-O_i(t,\bal{i}{r}) ]\|\le \delta \for r \ge v_{t,\delta} |t|, \label{def:linear_light_cone}
\end{align}
where $v_{t,\delta}$ decreases in time and eventually converges to a finite value (i.e., $v_{\infty,\delta} = {\rm const.}$).
 From the definition, the amount of information propagation is smaller than $\delta$ outside the region separated by the distance $v_{t,\delta}|t|$.

 Here, we show our main results. 
For $\alpha>2D+1$, the Hamiltonian in Eq.~\eqref{Ham:general_power} satisfies the Lieb--Robinson bound for \eqref{commutator_approx} as 
\begin{align}
\| [O_i(t),O_j]\| \le \mathcal{C}_H  |t|^{2D+1} (R-\vH  |t| )^{-\alpha} 
\label{main_thm_ineq1_main}
\end{align}  
with $R=\dist_{i,j}$. In addition, the Lieb--Robinson bound for \eqref{local_approx} is given by
\begin{align}
&\| [O_i(t)-O_i(t,\bal{i}{R}) ]\|  \le \mathcal{C}'_H  |t|^{D+1}(R-\vH  |t| )^{-\alpha+D} , 
\label{main_thm_ineq2_main}
\end{align}  
where $R> \vH  |t| $ is considered and $\mathcal{C}_H$, $\mathcal{C}'_H$, and $\vH$ are constants that depend only on the parameters $\{D, g_0, \alpha\}$ and a geometric constant defined by the lattice structure.
We emphasize that the same upper bound is obtained for generic operators $O_X$ and $O_Y$ (see Appendix~\ref{appendixA}).

As long as we consider the commutator for local observables, the first inequality~\eqref{main_thm_ineq1_main} is stronger than the second one~\eqref{main_thm_ineq2_main} in the sense that asymptotic decay is as small as $\orderof{R^{-\alpha}}$.
From the inequality~\eqref{main_thm_ineq2_main}, we can calculate the Lieb--Robinson velocity $v_{\delta,t}$ defined in Eq.~\eqref{def:linear_light_cone}:
$$
v_{\delta,t} = \bar{v} + c\delta^{-\frac{1}{\alpha-D}}  |t|^{\frac{-\alpha+2D+1}{\alpha-D}} \xrightarrow{t\to \infty} \bar{v} 
$$
where $c$ is a constant of $\orderof{1}$, and where we use the condition $\alpha>2D+1$.

\begin{figure}[]
\centering
{
\includegraphics[clip, scale=0.35]{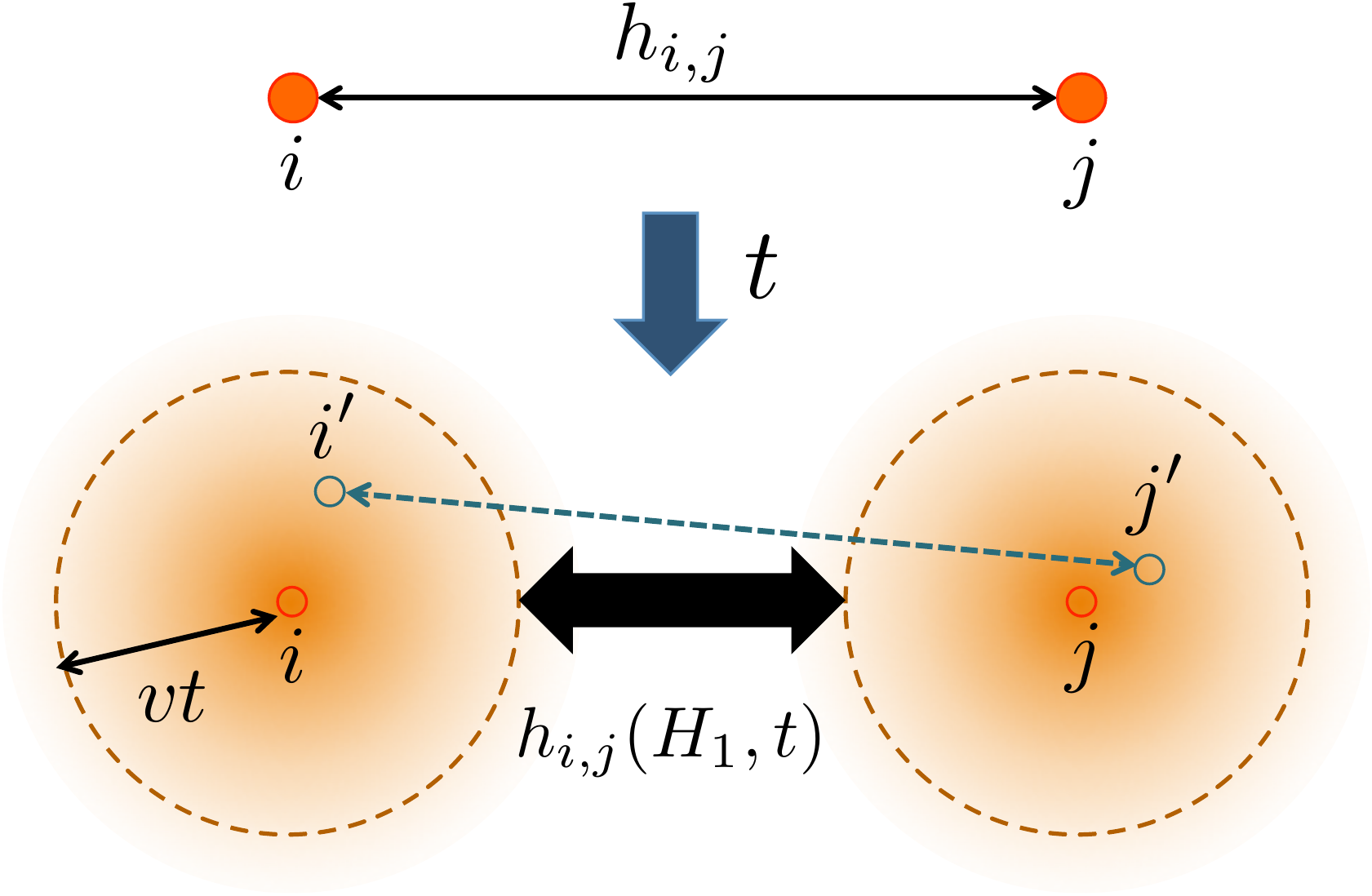}
}
\caption{Effect of the operator spreading. 
When a bipartite interaction $h_{i,j}$ evolves using the Hamiltonian $H_1$, 
 interactions between the sites $i'$ and $j'$ such that $\dist_{i,i'}\lesssim vt$ and $\dist_{j,j'}\lesssim vt$ are effectively induced. 
The operator $h_{i,j}(H_1,t)$ now acts on a subsystem with roughly $\orderof{t^D}$ sites.
Hence, one-site energy of $H_2 (H_1,t)$ defined as in Eq.~\eqref{one_site_energy} becomes $\orderof{t^D}$ times the original one. 
}
\label{fig:time_int}
\end{figure}

\section{Intuitive explanation of the condition $\alpha>2D+1$}

We here show an intuitive explanation of why the condition $\alpha>2D+1$ appears.
The point to obtain the linear light cone is that the contribution to the Lieb--Robinson velocity from the long-range interaction with very large distance becomes asymptotically negligible.
To get better insights into this condition, we consider the simplest setup as follows.

Let us consider the Hamiltonian given by $H= H_1+H_2$, where $H_1$ has only nearest-neighbor interactions,
while $H_2$ consists of the long-range interactions with the length scales from $\ell$ to $2\ell$ as $H_2= \sum_{\ell \le \dist_{i,j}\le 2\ell} h_{i,j}.
$ Note that the condition~\eqref{Ham:general_power} implies $\|h_{i,j}\| \le g_0 \dist_{i,j}^{-\alpha}$.
We eventually take the large $\ell$ limit to consider asymptotic behavior arising from the interactions of the large distance.
Note that the Hamiltonian $H_1$ consists of short-range interactions; hence, the unitary operator $e^{-iH_1 t}$ satisfies the standard Lieb--Robinson bound~\cite{ref:LR-bound72,ref:Hastings2006-ExpDec,ref:Nachtergaele2006-LR} giving a finite Lieb--Robinson velocity. Here, we denote it by $v_1$.
We focus on the time range of $t\lesssim \ell/v_1$ and then consider the condition for which 
the Lieb-Robinson velocity for the Hamiltonian $H$ is given by a finite velocity related to $v_1$ in the large $\ell$ limit.

As a simple exercise, we first consider the product of the unitary operators $e^{-iH_1 t}e^{-i H_2 t}$.
Then, to estimate the contribution from the long-range interactions, we only have to consider the Lieb--Robinson bound for $e^{-i H_2 t}$.
The Lieb--Robinson bound for $H_2$ is given by $e^{-c (x/\ell - v_\ell t)}$, and $v_\ell$ is proportional to the one-site energy:
$$g= \max_{i\in \Lambda} \sum_{j \in \Lambda} \|h_{i,j}\|  = \orderof{ \ell^{-\alpha+D}},$$
where $\sum_{j \in \Lambda} \|h_{i,j}\| \le g_0\sum_{j: \ell \le \dist_{i,j} \le 2\ell} \dist_{i,j}^{-\alpha}$ is a summation of all the interaction terms that act on the site $i$.
Hence, the Lieb--Robinson velocity is proportional to $\ell^{-\alpha+D+1}$,
which vanishes in the limit of $\ell \to \infty$ for $\alpha>D+1$.
Therefore, the unitary operator $e^{-iH_1 t}e^{-i H_2 t}$ has finite Lieb--Robinson velocity $v_1$ as long as $\alpha>D+1$.

Now, let us discuss the unitary operator $e^{-i (H_1+ H_2 )t}$.  
For this unitary operator, we use the following representation to decompose the contributions from $H_1$ and $H_2$:
\begin{align}
e^{-i (H_1+ H_2 )t} = e^{-iH_1 t} \mathcal{T} e^{-i\int_0^t H_2(H_1,\tau)d\tau},\label{unitary_decomp_exp}
\end{align}
where $H_2(H_1,\tau):= e^{iH_1\tau}H_2 e^{-iH_1\tau}$ and $\mathcal{T}$ denotes the time-ordering operator. 
Because the one-site operator spreads up to a distance $\orderof{\tau}$ owing to the time-evolution $e^{-iH_1\tau}$ (Fig.~\ref{fig:time_int}), the one-site energy is now given by
\begin{align}
g(\tau)= \max_{i\in \Lambda} \sum_{Z: Z\ni i} \|h_{\tau,Z}\| =\tau^D\orderof{ \ell^{-\alpha+D}},\label{one_site_energy}
\end{align}
where $h_{\tau,Z}$ is an interaction term on the subset $Z$ that constitutes $H_2(H_1,\tau)$, i.e., $H_2(H_1,\tau)=\sum_{Z} h_{\tau,Z}$. 
Therefore, the time-evolution $\mathcal{T} e^{-i\int_0^t H_2(H_1,\tau)d\tau}$ gives the Lieb--Robinson bound as 
$e^{-c (x/\ell - v_{t,\ell} t)}$ with the velocity $v_{t,\ell}\propto g(t)=t^D\orderof{ \ell^{-\alpha+D}}$.
For $t\lesssim \ell/v_1$, this estimation provides the Lieb--Robinson velocity as $t^{D}l^{-\alpha+D+1}\lesssim\orderof{\ell^{-\alpha+2D+1}}$.
Hence, the contribution to the Lieb--Robinson velocity from $H_2$ vanishes in the limit of $\ell \to \infty$ for $\alpha>2D+1$.
This leads to the finite Lieb--Robinson velocity $v_1$ for the unitary operator $e^{-i(H_1 +H_2) t}$.

In summary, the spread of the operator changes the effective one-site energy by $t^D$ times (see Eq.~\eqref{one_site_energy}), 
which yields the condition of $\alpha>2D+1$ for the linear light cone.
In our proof for the general Hamiltonian~\eqref{Ham:general_power}, we decompose the total length scale  into pieces and consider the multi-unitary decomposition by generalizing Eq.~\eqref{unitary_decomp_exp} (see also Appendix~\ref{appendixB}). 
We then obtain the Lieb--Robinson bound for each of the decomposed unitary operators and connect them into a single Lieb--Robinson bound.
The technical difficulties lie in that we need to connect infinitely many Lieb--Robinson bounds; in the step-by-step connections, 
a simple estimation makes the Lieb--Robinson velocity diverge rapidly, and hence, highly refined analyses are required to obtain a finite velocity.

\begin{figure}[t]
\centering
{
\includegraphics[clip, scale=0.31]{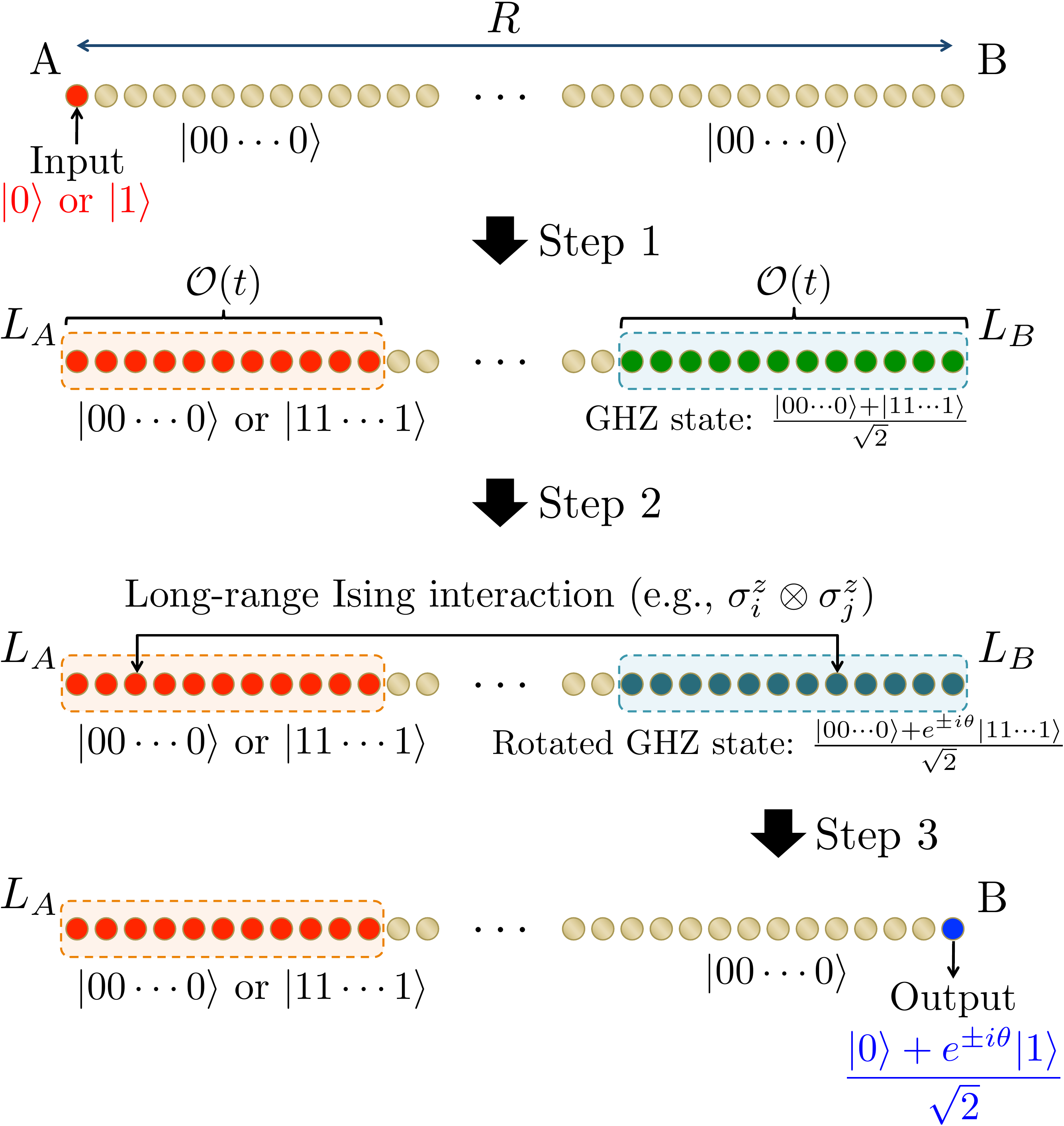}
}
\caption{
Schematic of our state-transfer protocol.
We show many copies of the input state here and transform the receiver into the GHZ state. 
The protocol effectively amplifies the long-range interaction strength between the sender and the receiver by $t^{2D}$ times ($D=1$). 
We can see that this simple protocol gives an example that realizes the intuitive picture in Fig.~\ref{fig:time_int}.
We here consider 1D systems, but the generalization to high-dimensional systems is straightforward.
}
\label{fig:Protocol}
\end{figure}



\section{Optimality of the present Lieb--Robinson bound}
We have proved that the condition $\alpha>2D+1$ is a sufficient condition for arbitrary Hamiltonians to have the linear-light cone. 
We here discuss whether this condition can be further improved. 
It has been conjectured in previous studies~\cite{PhysRevLett.119.170503,PhysRevX.9.031006} that the best general condition may be given by $\alpha > D+1$ from numerical and theoretical analyses of specific models.   
Against the conventional expectation, we show that any improvement from $\alpha > 2D+1$ is impossible as long as we consider the general long-range interacting systems~\eqref{Ham:general_power}.
In the following, we explicitly provide a quantum-state transfer protocol that achieves a nonlinear light cone for $\alpha < 2D+1$.

We follow a similar setup as in Ref.~\cite{PhysRevLett.97.050401} and consider the quantum-state transfer between two separated spins $A$ and $B$ through a spin network, where we define $R$ as the distance between the spins $A$ and $B$ (Fig.~\ref{fig:Protocol}). 
We start from the initial state $\ket{\psi}$ such that all spins are given by $\ket{0}$, namely $\ket{\psi}:=\ket{0}^{\otimes n}$.
We then apply the unitary operation $U_0$ or $U_1$ to the spin $A$, where $U_0$ is the identity operator and $U_1$ is the spin flip operator (i.e., $U_1=\ket{0}\bra{1}+\ket{1}\bra{0}$).
After applying $U_0$ or $U_1$, the initial quantum state is transformed into $\ket{\psi_s}=\ket{s}\otimes \ket{0}^{\otimes n-1}$ with $s=0,1$.
After time $t$, the quantum state evolves according to the unitary time operation $U(t)=\mathcal{T}[e^{-i\int_0^t H(\tau) d\tau}]$. 
We then define the output state for spin $B$ as 
\begin{align}
\rho_B^{(s)} := \tr_{B^\co} \left[ U(t) \ket{\psi_s}\bra{\psi_s}U(t)^\dagger\right].
\end{align}
If we can distinguish between the state $\rho_B^{(0)}$ from $\rho_B^{(1)}$ with the probability $1$, we can achieve the perfect quantum state transfer. 

By utilizing the Lieb--Robinson bound~\eqref{main_thm_ineq1_main}, we have~\cite{PhysRevLett.97.050401} 
\begin{align}
\| \rho_B^{(1)} - \rho_B^{(0)} \|_1 \lesssim |t|^{2D+1} R^{-\alpha} , \label{Lieb-Robinson_bound_1_0_thm}
\end{align}
where $\|\cdot\|_1$ is the trace norm. When $\| \rho_B^{(1)} - \rho_B^{(0)} \|_1=2$, the two states $\rho_B^{(0)}$ and $\rho_B^{(1)}$ are orthogonal to each other and completely distinguishable.
We prove that the upper bound of \eqref{Lieb-Robinson_bound_1_0_thm} is achievable by combining controlled-NOT-type short-range interactions and the Ising-type long-range interactions.

We here consider a one-dimensional (1/2)-spin system and decompose the time-evolution into three steps (see Fig.~\ref{fig:Protocol}).
In each of the steps, we take a time of $\orderof{t}$ (e.g., $t/3$).
In the first step, we copy the state of the spin A using the controlled NOT (CNOT) gate operation, which is generated by a simple bipartite interaction~\cite{PhysRevLett.119.170503}.
We define the subset $L_A$ as the spins that are the same state as that of spin $A$ (see Fig.~\ref{fig:Protocol}).  
We consider the dynamics by short-range interactions, and the number of spins in the subset $L_A$ is thus upper-bounded by $\orderof{t}$. 
At the same time, we prepare the Greenberger-Horne-Zeilinger (GHZ)  state that includes the spin $B$. 
The GHZ state is also generated by combining the rotation of the spin $B$ and the CNOT gate operation.
We define the subset $L_B$ as the spins that are involved in the GHZ state.
Owing to the short-rangeness of the interactions, the number of spins in $L_B$ is also of $\orderof{t}$.

In the second step, we apply long-range Ising interactions by the Hamiltonian
\begin{align}
H_{\rm Ising}=  \frac{g}{R^\alpha}  \sum_{i\in L_A} \sum_{\ j\in L_B} \sigma_i^z  \otimes \sigma_j^z ,
\end{align}
with $g\le g_0$.
Because the states of spins $L_A$ are given by $\ket{0\cdots 0}$ or $\ket{1\cdots 1}$, the unitary time evolution by $H_{\rm Ising}$ only changes the phase factor of the GHZ state of spins $L_B$; that is, the GHZ state $(\ket{00\cdots 0} + \ket{11\cdots 1})/\sqrt{2}$ is rotated as $(\ket{00\cdots 0} +e^{\pm 2 i \theta}  \ket{11\cdots 1})/\sqrt{2}$ with $\pm$ depending on the states of spins $L_A$.
Simple and straightforward calculations can give the phase shift $\theta$ as 
$
\theta=g \orderof{t} |L_A| \cdot |L_B| R^{-\alpha}.
$
Then, if $\theta$ is taken as $\theta=\pi/4$, the rotated GHZ states $(\ket{00\cdots 0} +e^{\pm 2 i \theta}  \ket{11\cdots 1})/\sqrt{2}$ are mutually orthogonal.

In the final step, we untangle the rotated GHZ state and concentrate the phase term on the spin $B$, which transforms this state to  $\ket{00\cdots 0} \otimes (\ket{0} + e^{\pm 2 i \theta} \ket{1})/\sqrt{2}$.
This procedure is also performed by the CNOT gate operation; hence, we only need the short-range interactions.
In this protocol, we obtain the lower bound of 
\begin{align}
\| \rho_B^{(1)} - \rho_B^{(0)} \|_1 =2\sin( 2\theta) \gtrsim t |L_A| \cdot |L_B| R^{-\alpha}. \label{Lieb-Robinson_bound_1_0}
\end{align}
Considering that $|L_A|=|L_B| = \orderof{t}$, we can achieve the theoretical upper bound of \eqref{main_thm_ineq1_main} with $D=1$.
Thus, as long as $\alpha < 3$, the information can reach a distance of $R=\orderof{t^{3/\alpha}}$. 
This protocol can be generalized to high-dimensional setups, and we obtain the same lower bound as \eqref{Lieb-Robinson_bound_1_0}, where we have $|L_A|=|L_B| = \orderof{t^D}$. 
Then, the shape of the light cone becomes $t^{(2D+1)/\alpha}$. 

This simple quantum model (i.e., two-body interaction and (1/2)-spin systems) already saturates the Lieb--Robinson bound of~\eqref{main_thm_ineq1_main}; hence, our condition 
$\alpha>2D+1$ for the linear light cone cannot be improved unless we consider a special class of Hamiltonians.
At the same time of our submission, a similar protocol to achieve the Lieb-Robinson bound~\eqref{main_thm_ineq1_main} was given~\cite{tran2020hierarchy} with a more explicit lower bound on the commutator $\| [O_i(t),O_j]\| $.


\section{Summary and discussion}

In this study, we proved the existence of the linear light cone (see \eqref{def:linear_light_cone} for the definition) in general long-range interacting systems, where the interaction decays as according to the relation $R^{-\alpha}$ ($\alpha>2D+1$) with respect to the distance $R$.
Our Lieb--Robinson bound in \eqref{main_thm_ineq1_main} provided an approximate commutation relation as $\|[O_i(t), O_j]\| \lesssim t^{2D+1} (R-\bar{v} t)^{-\alpha}$, with rapid decay beyond $r\gtrsim \bar{v}t$.
Moreover, the error of the local approximation for $O_i(t)$ was estimated as in \eqref{main_thm_ineq2_main}, namely $\|O_i(t)- O_i(t,i[R])\| \lesssim t^{D+1} (R-\bar{v} t)^{-\alpha+D}$.
Our result was obtained for Hamiltonians with two-body interactions but can be extended to a more general setup (see Eq.~\eqref{def:Ham_main} with \eqref{alternative_basic_assump_power_main} in Appendix~\ref{appendixA}), where even the few-body interactions are not assumed.
We also show an explicit example that our Lieb--Robinson bound is saturated for $\alpha>2D+1$, and the linearity of the light cone deteriorates for $\alpha<2D+1$.
Therefore, our condition for the linear light cone is optimal as long as we consider the general class of Hamiltonians.
Although we consider Hamiltonian dynamics throughout this work, we expect that the same analysis can be applied to generic Markovian dynamics using the procedures in Refs.~\cite{PhysRevLett.104.190401,PhysRevLett.108.230504}. 

%

We finally present an open question. 
In the present work, although we provided the optimal Lieb--Robinson bound for $\alpha>2D +1$, it is still unclear what can be obtained in the $\alpha\le 2D+1$ regimes.
The most important problem is to identify the regime of the exponent $\alpha$ that ensures the polynomial (or superlinear) light cone.
A state-of-the-art analysis~\cite{PhysRevX.9.031006} has defined the polynomial light cone in the form $r=t^{(\alpha-D)/(\alpha-2D)}$ for $\alpha>2D$.
In contrast, the super-polynomial light cone has been explicitly shown for $\alpha\le D$~\cite{PhysRevLett.119.170503}.
In tackling this problem, the simplest case with only two length scales as in Eq.~\eqref{unitary_decomp_exp} may be a good starting point. 
We expect that our present analysis will provide a better polynomial light cone for high-dimensional systems.

%
%
%
%
%
%
%
%
%
%
%
%
%
%

\section*{acknowledgments}

{~}\\
T.K. was supported by the RIKEN Center for AIP and JSPS KAKENHI grant no. 18K13475.
TK gives thanks to God for his wisdom. 
K.S. was supported by JSPS Grants-in-Aid for Scientific Research (grant nos. JP16H02211 and  JP19H05603).

\section{Formal expression of the theorem} \label{appendixA}
We demonstrate our theorem here in a general manner. 
First, some necessary notations are provided;
for arbitrary subsystems $X, Y \subset \Lambda$, we define $\dist_{X,Y}$ as the shortest path length on the lattice that connects $X$ and $Y$.
 If $X\cap Y \neq \emptyset$, $\dist_{X,Y}=0$.
 For a subset  $X\subseteq \Lambda$, we define $\diam(X) := \max_{i,j\in X} (\dist_{i,j})+1$, the cardinality $|X|$ as the number of vertices contained in $X$, and 
the complementary subset of $X$ as $X^\co := \Lambda\setminus X$.

We consider a general class of the Hamiltonian beyond the two-body interaction~\eqref{Ham:general_power} as 
\begin{align}
H= \sum_{Z\subseteq \Lambda} h_Z, \label{def:Ham_main}
\end{align}
where each of the interaction terms $\{h_Z\}_{Z\subseteq \Lambda}$ acts on the sites in $Z \subseteq \Lambda$.
Notably, we \textit{do not assume} the few-body interaction here, i.e., $|Z|$ can be arbitrarily large up to $|Z|=n$.
Therefore, the Hamiltonian $H$ includes macroscopic interactions such as $\sigma_1^z \otimes \sigma_2^z \otimes \cdots \otimes \sigma_n^z$.
However, the assumption~\eqref{alternative_basic_assump_power_main} below restricts the amplitude of such interactions as ${\rm poly}(1/n)$.
Here, the only assumption is the following power-law decay of the interactions:
\begin{align}
&\sup_{i \in \Lambda}\sum_{Z:Z\ni i, \diam(Z) \ge r } \| h_Z\| \le  g r^{-\alpha+D}, \notag \\
&\sup_{i,j \in \Lambda}\sum_{Z: Z\ni \{i,j\}} \| h_Z\| \le  g_0 (\dist_{i,j}+1)^{-\alpha}  \label{alternative_basic_assump_power_main}
\end{align}
with
\begin{align}
\alpha>2D+1
\end{align}
for an arbitrary site pair of $\{i,j\} \subset\Lambda$. Here, $ \sum_{Z: Z\ni \{i,j\}}$ denotes the 
summation that encompasses all the interaction terms $\{h_Z\}_{Z\subseteq\Lambda}$, including the sites $i$ and $j$, and
$\|\cdots\|$ is the operator norm. By considering an appropriate energy unit, we set $g=1$.

To formulate our main theorem, we first define the coarse-grained subsets (see Fig.~\ref{fig:extended_coarse_grain}).
For a subset $X\subseteq \Lambda$, we first define $\bal{X}{r}$ as the extended subset:
\begin{align}
\bal{X}{r}:= \{i\in \Lambda| \dist_{X,i} \le r \} , \label{def:bal_X_r}
\end{align}
where $\bal{X}{0}=X$ and $r$ is an arbitrary positive number.
We also define the coarse-grained total set $\Lambda^{(\xi)}$ as the minimum subset such that $\bal{\Lambda^{(\xi)}}{\xi}= \Lambda$, namely
\begin{align}
\Lambda^{(\xi)} := \arg\min_{Z \subseteq \Lambda| \bal{Z}{\xi} = \Lambda} |Z|, \label{mthd:def:coarse_graining_Lambda^(r)}
\end{align}
where $\Lambda^{(0)}=\Lambda$.
Similarly, for an arbitrary subset $X\subseteq \Lambda$, we define $X^{(\xi)}\subseteq \Lambda^{(\xi)}$ as follows:
\begin{align}
X^{(\xi)} := \arg\min_{Z \subseteq \Lambda^{(\xi)}| \bal{Z}{\xi} \supset X} |Z|, \label{mthd:def:coarse_graining_X^(r)}
\end{align}
where $X^{(0)}=X$. 
From the definition, the cardinality of the subset $X^{(\xi)}$ is roughly $(1/\xi)^D$ times that of the original, namely $|X^{(\xi)}| \approx (1/\xi)^D |X|$.

\begin{figure*}
\centering
\subfigure[Definition of $\bal{X}{r}$
]
{\includegraphics[clip, scale=0.3]{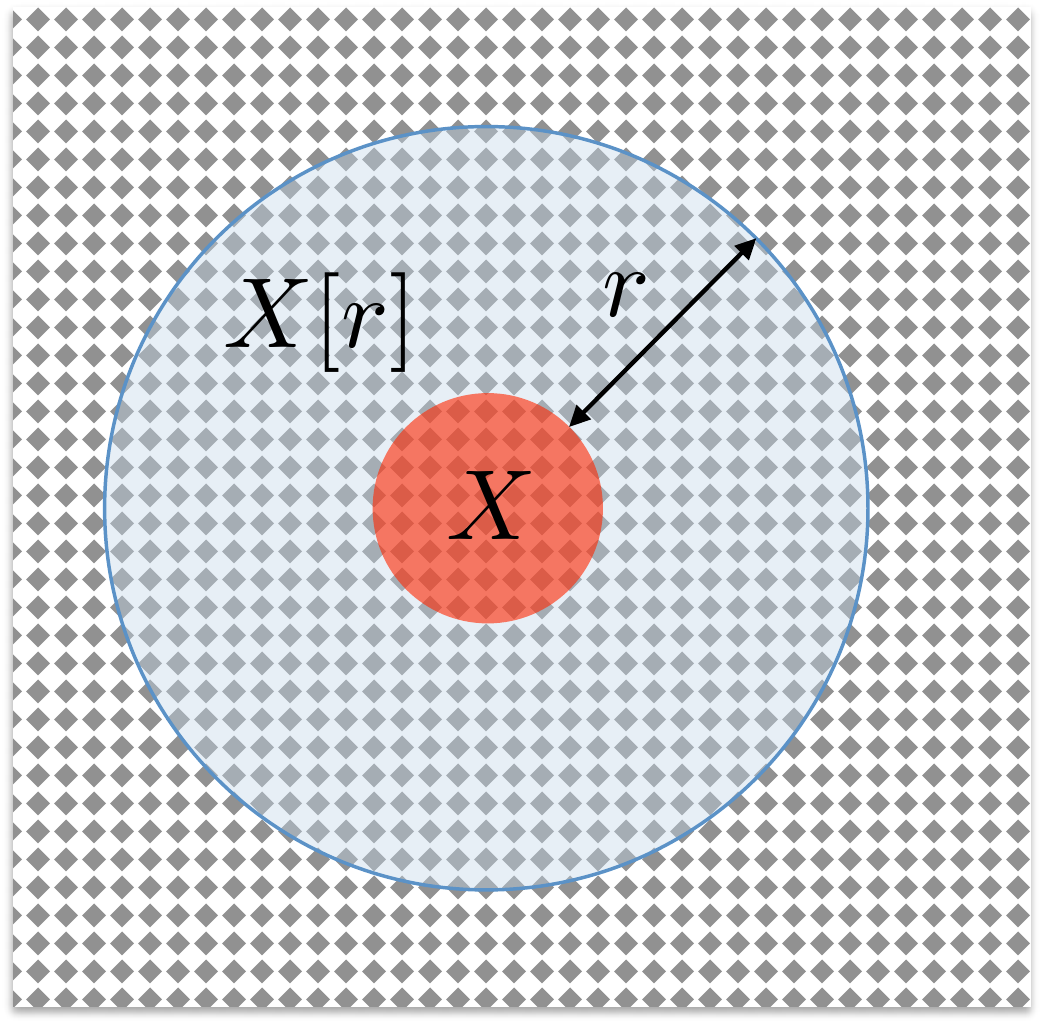}
}
\subfigure[Definition of $\Lambda^{(\xi)}$
]
{\includegraphics[clip, scale=0.3]{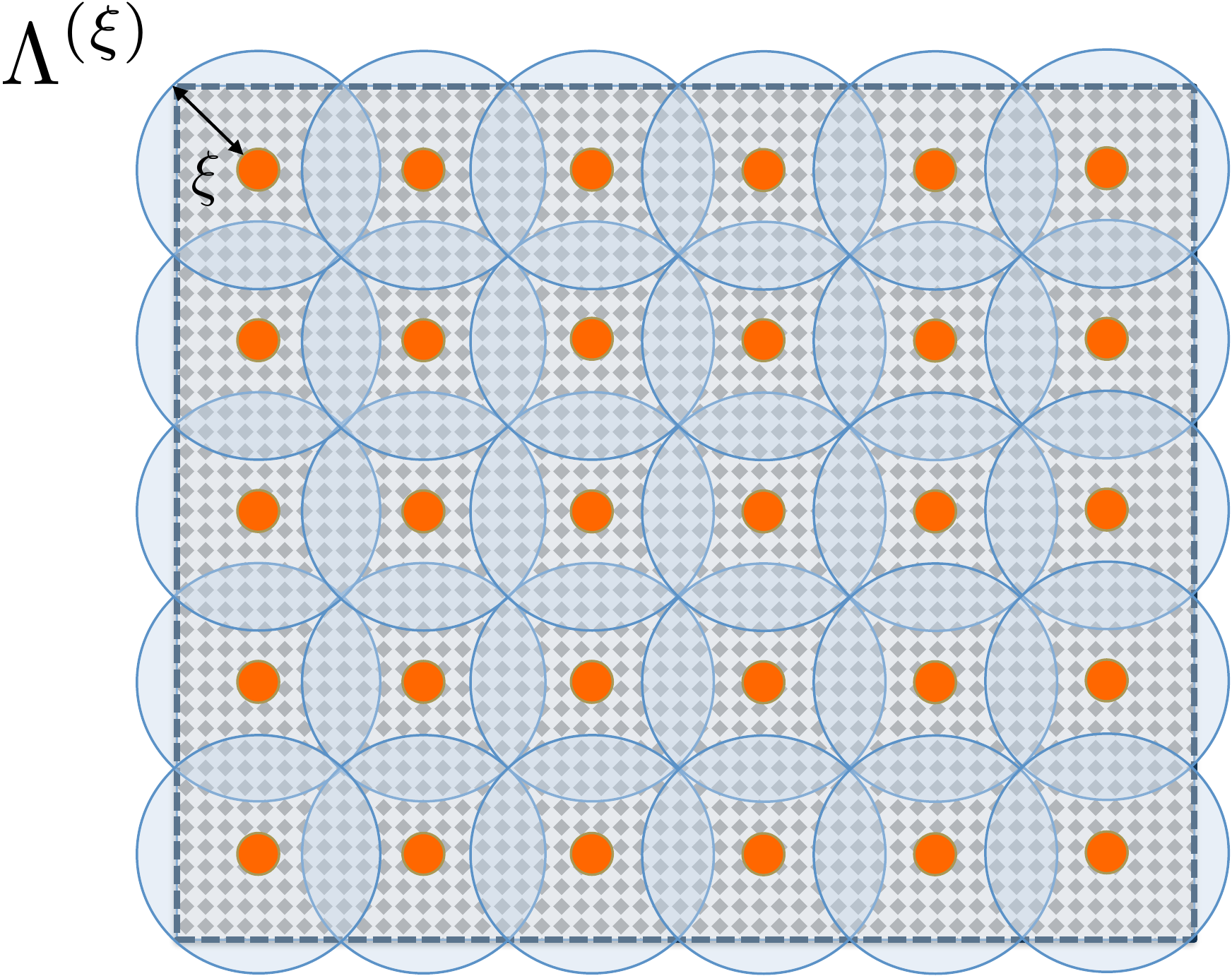}
}
\subfigure[Definitions of $X^{(\xi)}$
]
{\includegraphics[clip, scale=0.3]{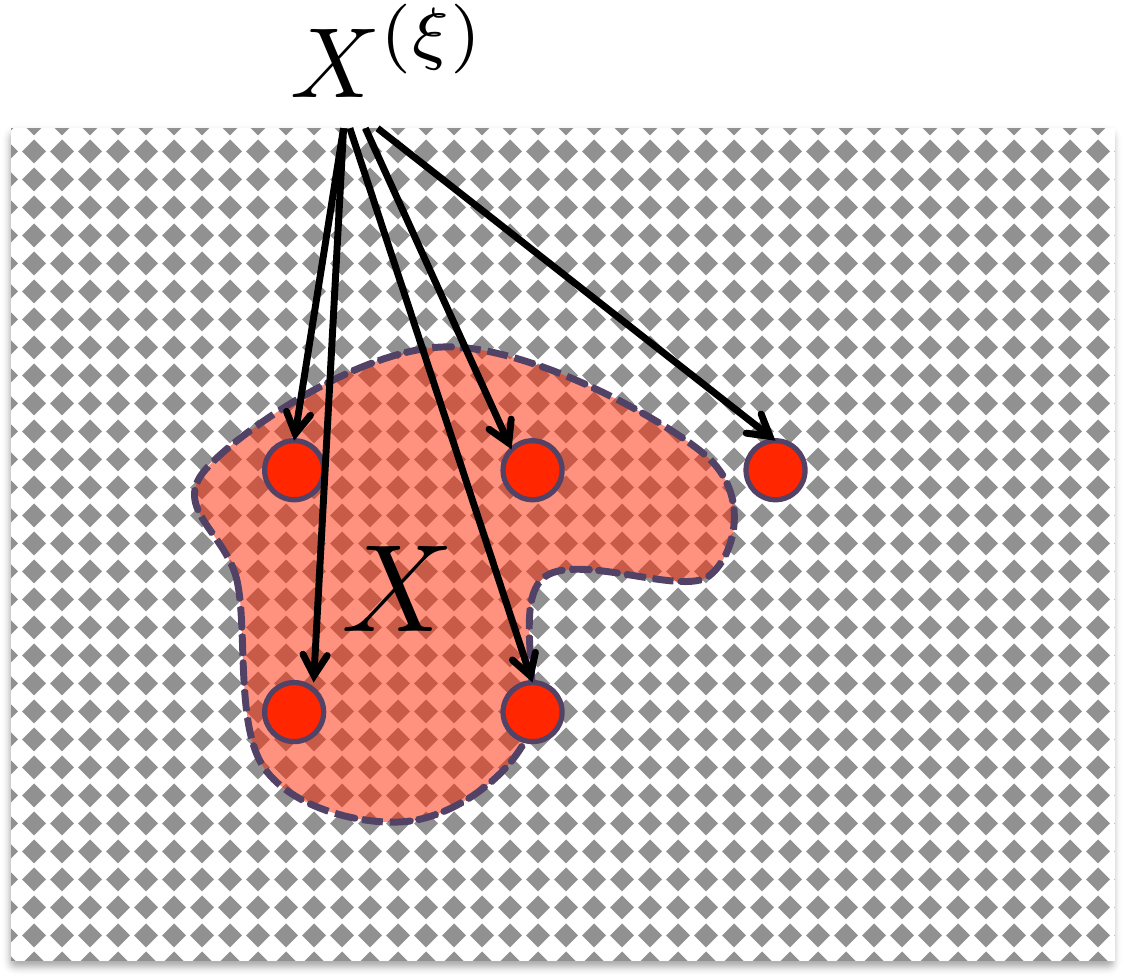}
}
\caption{(a) Extended subset. The subset $\bal{X}{r}$ is defined by extending the original subset $X$ by a distance $r$. 
(b) The subset $\Lambda^{(\xi)} \subseteq \Lambda$ is the coarse grained lattice (orange sites) which is defined as the minimum subset such that $\bal{\Lambda^{(\xi)}}{\xi}=\Lambda$.  
(c) For an arbitrary subset $X\subseteq \Lambda$ (pink region), $X^{(\xi)}\subseteq \Lambda^{(\xi)}$ is the coarse grained subset (red sites) which is defined as the minimum subset such that $\bal{X^{(\xi)}}{\xi}\supseteq X$. 
The cardinality of the subset $X^{(\xi)}$ is roughly $(1/\xi)^D$ times as the original one, namely $|X^{(\xi)}| \approx (1/\xi)^D |X|$.
}
\label{fig:extended_coarse_grain}
\end{figure*}

Using the notation $X^{(\xi)}$, we state our main theorem as follows:\\
{~}\\
{\bf Main theorem.} 
\textit{Let us consider the long-range interacting Hamiltonian $H$ of the form \eqref{def:Ham_main} with the assumption \eqref{alternative_basic_assump_power_main}.
For $|t|\ge 1$, this Hamiltonian $H$ satisfies the Lieb--Robinson bound for arbitrary operators $O_X$ and $O_Y$ that are supported on $X\subseteq \Lambda$ and $Y\subseteq \Lambda$, respectively:
\begin{align}
\frac{\|[O_X(t),O_Y]\|}{\|O_X\|\cdot \|O_Y\|} \le  \mathcal{C}_H \abb X^{(\vH  |t| )} \abb \cdot \abb Y^{(\vH  |t| )} \abb  \frac{  |t|^{2D+1}\log^{2D} (x +1)}{(x-\vH  |t| )^\alpha} 
\notag 
\end{align}  
and
\begin{align}
\frac{\|[O_X(t),O_Y]\|}{\|O_X\|\cdot \|O_Y\|} \le  \mathcal{C}'_H \abb X^{(\vH  |t| )} \abb^2   \frac{  |t|^{D+1}\log^{2D} (x +1)}{(x-\vH  |t| )^{\alpha-D}}, 
\notag 
\end{align}  
where $\mathcal{C}_H$, $\mathcal{C}'_H$, and $\vH$ are constants that depend only on parameters $\{D, g_0, \alpha\}$ and a geometric constant that is determined by the lattice structure.
Note that we use the notation from Eq.~\eqref{mthd:def:coarse_graining_X^(r)} with $\xi=\vH  |t|$.}
 \\
{~}\\
Here, the coefficient $\log^{2D} (x+1)$ exists because of a technical reason, which results from the macroscopic interactions in Eq.~\eqref{def:Ham_main} (e.g., $\orderof{n}$-body interactions). 
If we restrict ourselves to the few-body (or $k$-local) Hamiltonians with $k=\orderof{1}$, namely
\begin{align}
H= \sum_{Z\subseteq \Lambda,\ |Z|\le k} h_Z, \label{def:Ham_main_k-local}
\end{align}
the Lieb--Robinson bound is slightly improved as follows: \\
{~}\\
{\bf Main theorem ($\boldsymbol{k}$-local Hamiltonians).} 
\textit{Let us consider the long-range interacting Hamiltonian $H$ of the form \eqref{def:Ham_main_k-local} with the assumption \eqref{alternative_basic_assump_power_main}.
For $|t|\ge 1$, this Hamiltonian $H$ satisfies the Lieb--Robinson bound as follows:
\begin{align}
\frac{\|[O_X(t),O_Y]\|}{\|O_X\|\cdot \|O_Y\|} \le  \mathcal{C}^{(k)}_H \abb X^{(\vH^{(k)}  |t| )} \abb \cdot \abb Y^{(\vH^{(k)}  |t| )} \abb  \frac{  |t|^{2D+1}}{(x-\vH^{(k)}  |t| )^\alpha} 
\notag 
\end{align}  
and
\begin{align}
\frac{\|[O_X(t),O_Y]\|}{\|O_X\|\cdot \|O_Y\|} \le  \mathcal{C}^{(k)'}_H \abb X^{(\vH^{(k)}  |t| )} \abb^2   \frac{  |t|^{D+1}}{(x-\vH^{(k)}  |t| )^{\alpha-D}}, 
\notag 
\end{align}  
where $\mathcal{C}^{(k)}_H$, $\mathcal{C}^{(k)'}_H$ and $\vH^{(k)}$ are constants that depend only on the parameters $\{D, g_0, \alpha,k\}$ and a geometric constant that is determined by the lattice structure.}
 \\
{~}\\
The first and second inequalities in the above theorem reduce to the inequalities~\eqref{main_thm_ineq1_main} and \eqref{main_thm_ineq2_main} in the main part, respectively; in deriving the inequality~\eqref{main_thm_ineq2_main}, we use the discussion in Ref.~\cite{PhysRevLett.97.050401} to relate the commutator inequality to the local approximation.

%
%

\begin{figure*}[t]
\centering
{
\includegraphics[clip, scale=0.35]{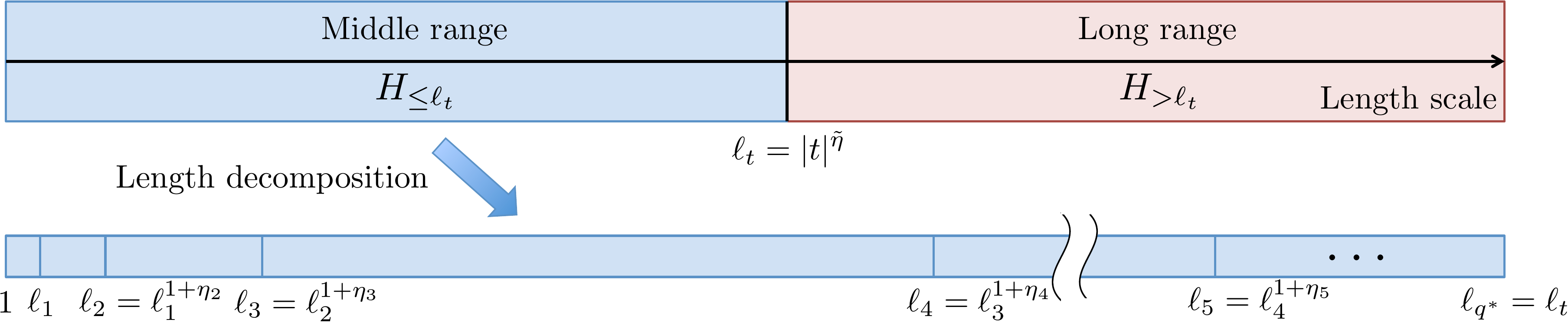}
}
\caption{Decomposition of the length scale. We first decompose the total Hamiltonian into two regimes: $H_{\le \ell_t}$ and $H_{>\ell_t}$.
The Hamiltonian $H_{\le \ell_t}$ includes all the interactions up to the length scale $\ell_t$ that is dependent on time, and $H_{>\ell_t}$ includes the other interactions.
To obtain the middle-range Lieb--Robinson bound, we further decompose the range $[1,\ell_t]$ into $q^\ast$ segments.
We start from the Hamiltonian that only includes the length scale $\ell_1$ and iteratively consider the increasing length scales.
}
\label{fig:Fig_scale}
\end{figure*}

\section{Sketch of the Proof}\label{appendixB}
We herein show the essential ideas, and further details are provided in the supplementary materials~\cite{Supplement_long_range_LR}. 
For the proof, we first decompose the length scale into $\le \ell_t$ and $>\ell_t$ for a fixed $t$ and consider the following decomposition for the total Hamiltonian:
$
H=H_{\le \ell_t} + H_{> \ell_t} ,
$
where we define $H_{\le \ell}$ for arbitrary $\ell \in \mathbb{N}$ as the operator that includes all the interaction terms whose length scales are less than $\ell$:
$
H_{\le \ell}= \sum_{\diam(Z) \le \ell} h_Z. 
$
In the case where the length $\ell$ is short range or $\ell=\orderof{1}$, the Hamiltonian $H_{\le \ell}$ provides the Lieb--Robinson bound with a finite velocity.
However, we consider the case of $\ell=\ell_t$ with $\ell_t$ depending on time $t$ here. 
In the following computations, we choose $\ell_t = |t|^{\tilde{\eta}}$ with $\tilde{\eta}:= 1-\frac{\alpha -2D -1}{2(\alpha-D)} <1$.
When the length scale $\ell$ is in the middle-range between $\ell=\orderof{t^0}$ and $\ell=\infty$, 
it is no longer considered trivial if the light-cone for the dynamics by $H_{\le \ell_t}$ is retained.

To obtain the Lieb--Robinson bound for $H_{\le \ell}$ with a generic $\ell$, we further decompose $H_{\le \ell}$ as follows (Fig.~\ref{fig:Fig_scale}):
\begin{align}
H_{\le \ell} = \sum_{q=1}^{q^\ast} H_{q},\quad H_q:=\sum_{\ell_{q-1}\le \diam(Z) < \ell_q} h_Z.  \notag
\end{align} 
We define a set of length scales $\{\lsc_q\}_{q=1}^{q^\ast}$ as 
$\lsc_q =\ell_{q-1}^{1+\eta_q}$ with  $\eta_q >0$,
where $\{\eta_q\}_{q=2}^{q^\ast}$ is appropriately chosen such that $\{\ell_q\}_{q=1}^{q^\ast}$ is an integer and there exists an integer $q^\ast \in \mathbb{N}$ satisfying 
$
\ell_{q^\ast}= \ell_1^{(1+\eta_2)(1+\eta_3)\cdots (1+\eta_{q^\ast}) } = \ell.
$
In this case, $\ell_q$ increases by the double exponential function with respect to $q$.

In deriving the Lieb--Robinson bound for $e^{-iH_{\le \ell}t}$, we iteratively take in the increasing length scales. 
We begin with the unitary operator $e^{-iH_1t}$;
as long as $\ell_1$ is independent of $t$, the Lieb--Robinson bound for $e^{-iH_1t}$ is in the shorter range with a velocity of $v_1=\orderof{1}$. 
Then, using this Lieb--Robinson bound, we derive a new Lieb--Robinson bound for $e^{-i (H_1+H_2)t}= e^{-iH_{1:2}t}$, where we define $H_{1:q}:=\sum_{s=1}^q H_s$ with $1\le q\le q^\ast$. 
This process is repeated to extend the length scales for $\ell_1\to \ell_2 \to \cdots \to \ell_{q^\ast}= \ell$.
In each of these steps, based on the Lieb--Robinson bound for $e^{-iH_{1:q-1}t}$, we update the Lieb--Robinson bound for $e^{-iH_{1:q}t}$.

In the first update, we start from the following decomposition of the unitary operator $e^{-iH_{1:2}t}$:
\begin{align}
&U_{2,t}:= e^{- i (H_1+H_2) t}=  e^{-i H_1t} \mathcal{T} e^{-\int_0^{t} H_{2}(H_1,\tau) d\tau } . \notag 
\end{align}  
The operator spreading by $e^{-i H_1t}$ is of the order of $\orderof{v_1t}$; hence, as long as $t\le \Delta t_2\approx \ell_2/v_1$, $H_{2}(H_1,\tau)$ ($\tau \le \Delta t_2$) has the same interaction length as the original, namely $\ell_2$.
Thus, we can obtain the Lieb--Robinson bound with a linear light cone for $\mathcal{T} e^{-\int_0^{\Delta t_2} H_{2}(H_1,\tau) d\tau }$.
For any unspecified time $t$, we consider 
$
U_{2,t}=  U_{2,\Delta t_2}^m U_{2,\Delta t'_2}  ,
$
with $t=m\Delta t_2 + \Delta t'_2$, where $\Delta t_2 \propto \ell_2/v_1$ and $\Delta t'_2< \Delta t_2$.
By appropriately connecting all the Lieb--Robinson bounds, this results in a Lieb--Robinson bound of the following form:
\begin{align}
\| [O_X(H_{\le \ell_2},t) , O_Y]\|\lesssim  (1+x/\ell_2)^{D-1}  e^{-2(x-v_2|t|)/\ell_2} \notag 
\end{align}  
with $x=\dist_{X,Y}$,
where the velocity $v_2>v_1$ is upper-bounded using the Lieb--Robinson velocity $v_1$ for $e^{-iH_1t}$.

For general $q-1$, we define $v_{q-1}$ as the Lieb--Robinson velocity for $e^{- i H_{1:q-1}t}$ and analyze the unitary operator $U_{q,t} :=e^{- i H_{1:q}t}$,
which we decompose as $U_{q,t} = U_{q,\Delta t_q}^m U_{q,\Delta t'_q} $ with 
$$
U_{q,\Delta t_q}= e^{-i H_{1:q-1}\Delta t_q}\mathcal{T} e^{-\int_0^{\Delta t_q} H_{q}(H_{1:q-1},\tau) d\tau} \notag 
$$
and $t=m\Delta t_q + \Delta t'_q$; here, $\Delta t_q \propto \ell_{q}/v_{q-1}$ and $\Delta t'_q < \Delta t_q $.
From the choice of $\Delta t_q$, we can ensure that $H_{q}(H_{1:q-1},\tau)$ has the same interaction length as that of $H_{q}$, namely $\ell_q$.
We then obtain the Lieb--Robinson bound for $H_{1:q}$ as follows:
\begin{align}
\| [O_X(H_{1:q},t) , O_Y]\|\lesssim  (1+x/\ell_q)^{D-1}  e^{-2(x-v_q|t|)/\ell_q} \notag 
\end{align}  
where $v_q$ depends on the $v_{q-1}$. 
Thus, we iteratively estimate the Lieb--Robinson velocity $v_q$ using $v_{q-1}$.
Further, we can derive the following recursion relation: 
\begin{align}
v_q = v_{q-1} \left(1+\frac{c \log (\ell_q)}{\ell_{q-1}^{\eta}}+ \frac{c'}{ \log(\ell_{q-1})} \right) ,  \notag 
\end{align}  
where $\eta = \sqrt{1+\frac{\alpha -(2D +1)}{D+2} } -1$ and $c,c'$ are constants. We provide the explicit form in \cite{Supplement_long_range_LR}.
The length scale $\ell_q$ is now lower-bounded by a double exponential function with respect to $q$.
Therefore, $\lim_{q\to \infty} v_q$ converges to a constant $v^\ast$, and we obtain the following ``middle-range Lieb--Robinson bound'' for $e^{-i H_{\le \ell} t}$:
\begin{align}
\| [O_X(H_{\le \ell},t) , O_Y]\|\lesssim  (1+x/\ell)^{D-1}  e^{-2(x-v^\ast |t|)/\ell} .
\label{middle-range Lieb-Robinson bound_main_part}
\end{align}  

In this manner, we can ensure that $H_{\le \ell}$ retains the linear light cone for $\ell \lesssim t$ from \eqref{middle-range Lieb-Robinson bound_main_part}, whereas no such confirmation is possible for $\ell \gtrsim t$. 
Thus, we consider the case of $\ell=\ell_t$ and decompose the total time evolution as 
\begin{align}
&e^{-iHt} = e^{-iH_{\le \ell_t} t}  U_{>\ell_t},   \notag 
\end{align}  
where $U_{>\ell_t}:=\mathcal{T} e^{-i\int_0^t H_{> \ell_t} (H_{\le \ell_t}, \tau) d\tau}$.
In order to estimate the quasi-locality of the interaction for $H_{> \ell_t} (H_{\le \ell_t}, \tau)$, we apply the middle-range Lieb--Robinson bound~\eqref{middle-range Lieb-Robinson bound_main_part}. Based on the quasi-locality, we utilize the standard recursion approach~\cite{ref:Hastings2006-ExpDec,ref:Nachtergaele2006-LR} to obtain the Lieb-Robinson bound for $U_{>\ell_t}$.
After intricate calculations, we obtain the Lieb--Robinson bound for $U_{>\ell_t}$ as 
\begin{align}
\| [U_{>\ell_t}^\dagger O_X U_{>\ell_t}  , O_Y]\| \lesssim \frac{  |t|^{2D+1}\log^{2D} (x+1)}{(x -\kappa_0 v^\ast |t|)^\alpha},
\label{long-range Lieb-Robinson bound_main_part}
\end{align}  
where $\kappa_0$ is a constant.
We then connect the two Lieb--Robinson bounds for 
$e^{-i H_{\le \ell_t} t}$~\eqref{middle-range Lieb-Robinson bound_main_part} and $U_{>\ell_t}$~\eqref{long-range Lieb-Robinson bound_main_part} to derive the total Lieb--Robinson bound in the main theorem.

\bibliography{Long_range_LR_arXiv}

\clearpage
\newpage

\setcounter{equation}{0}
\renewcommand{\theequation}{S.\arabic{equation}}

\begin{widetext}

\begin{center}
{\large \bf Supplementary Material for  \protect \\ 
  ``Strictly linear light cone long-range interacting systems of arbitrary dimensions'' }\\
\vspace*{0.3cm}
Tomotaka Kuwahara$^{1,2}$ and Keiji Saito$^{3}$ \\
\vspace*{0.1cm}
$^{1}${\small \it Mathematical Science Team, RIKEN Center for Advanced Intelligence Project (AIP),1-4-1 Nihonbashi, Chuo-ku, Tokyo 103-0027, Japan \protect \\
$^{2}$Interdisciplinary Theoretical \& Mathematical Sciences Program (iTHEMS) RIKEN 2-1, Hirosawa, Wako, Saitama 351-0198, Japan} \\
$^{3}${\small \it Department of Physics, Keio University, Yokohama 223-8522, Japan} 
\end{center}

\tableofcontents


%
%
%
%
%
%
%
%


\section{Setup}

\subsection{Definition of the lattice}
We here recall the setup in the main theorem.
We consider a quantum spin system with $n$ spins, where each of the spin sits on a vertex of the $D$-dimensional graph (or $D$-dimensional lattice) with $\Lambda$ the total spin set, namely $|\Lambda|=n$.
We assume that a finite dimensional Hilbert space is assigned to each of the spins. 
We note that our Lieb-Robinson bound does not depend on the spin dimensions.
For a partial set $X\subseteq \Lambda$, we denote the cardinality, that is, the number of vertices contained in $X$, by $|X|$ (e.g. $X=\{i_1,i_2,\ldots, i_{|X|}\}$).
We also denote the complementary subset of $X$ by $X^\co := \Lambda\setminus X$.

For arbitrary subsets $X, Y \subseteq \Lambda$, we define $\dist_{X,Y}$ as the shortest path length on the graph that connects $X$ and $Y$; that is, if $X\cap Y \neq \emptyset$, $\dist_{X,Y}=0$. 
When $X$ is composed of only one element (i.e., $X=\{i\}$), we denote $\dist_{\{i\},Y}$ by $\dist_{i,Y}$ for the simplicity.
We also define $\diam(X)$ as follows: 
\begin{align}
\diam(X):  =1+ \max_{i,j\in X} (\dist_{i,j}).
\end{align}


\subsection{Definition of the long-range interacting Hamiltonian}

\begin{table}[b]
  \caption{Dependence of the $k$-locality of our main theorems. Theorem~\ref{thm:Middle-range Lieb-Robinson bound} does not depend on the $k$-locality.}
  \label{tab:Ham} 
\begin{ruledtabular}
\begin{tabular}{lrrr}
\textrm{\textbf{Hamiltonian}}&\textrm{\textbf{Theorem~\ref{thm:long-range Lieb-Robinson bound}}}&\textrm{\textbf{Theorem~\ref{thm:Middle-range Lieb-Robinson bound}}}&\textrm{\textbf{Theorem~\ref{thm :long-range contribution Lieb-Robinson bound}}}
\\
\colrule
 Generic Hamiltonians  in Eq.~\eqref{def:Ham} & \eqref{main_thm_ineq1} and \eqref{main_thm_ineq2}&\eqref{thm_middle_range_Lieb-Robinson} &  \eqref{ineq:long-range_contribution}\\
 $k$-local Hamiltonians  in Eq.~\eqref{def:Ham_k-local}  & \eqref{main_thm_ineq1_k-local} and \eqref{main_thm_ineq2_k-local}&\eqref{thm_middle_range_Lieb-Robinson}  & \eqref{ineq:long-range_contribution_k-local} \\
\end{tabular}
\end{ruledtabular}
\end{table}

We consider a Hamiltonian as 
\begin{align}
H= \sum_{Z\subseteq \Lambda} h_Z, \label{def:Ham}
\end{align}
where each of the interaction terms $\{h_Z\}_{Z\subseteq \Lambda}$ acts on the sites of $Z \subseteq \Lambda$.
Note that we here \textit{do not} need to assume the few-body interaction (i.e., $|Z|=\orderof{1}$).
We do not explicitly consider the time-dependence of the Hamiltonians, but all the analyses can be generalized to the time-dependent Hamiltonians.

When we discuss the few-body interactions, we impose the following additional assumption:
\begin{align}
H= \sum_{Z\subseteq \Lambda, |Z| \le k} h_Z. \label{def:Ham_k-local}
\end{align}
This Hamiltonian includes at most $k$-body interactions, and we refer to such a Hamiltonian as ``$k$-local Hamiltonian.''
For example, in the case of $|Z|\le 2$ (i.e., including up to two-body interactions), the Hamiltonian is given in the form of
\begin{align}
H= \sum_{Z: |Z|=2} h_Z  + \sum_{Z: |Z|=1} h_Z = \sum_{i<j} h_{i,j} + \sum_{i=1}^n h_i  .  
\label{eq:ham_1D_k=2}
\end{align}
In the main text, we considered the above Hamiltonian for the simplicity of the notation.
If we consider the $k$-local Hamiltonian instead of the generic Hamiltonian~\eqref{def:Ham}, our main results are slightly modified as shown in Table~\ref{tab:Ham}.

Throughout the paper, for arbitrary operators $A$ and $O$, we define
\begin{align}
O(A,t):= e^{iAt } O e^{-iAt }.\label{time-evolution:def}
\end{align}
In particular, for $A=H$, we denote 
\begin{align}
O(t):= e^{iHt } O e^{-iHt }\label{time-evolution:def_real_ham}
\end{align}
for the simplicity of the notation.

In order to characterize the long-range interaction of the Hamiltonian, we impose the following assumption for the Hamiltonian:
\begin{assump}[Power-law decaying interactions]  \label{lem:power_law_k_local}
We assume the power-law decay of the interaction in the following senses:
\begin{align}
&\sup_{i \in \Lambda}\sum_{Z:Z\ni i, \diam(Z) \ge r } \| h_Z\| \le  g r^{-\alpha+D},\label{basic_assump_power}\\
&\sup_{i,j \in \Lambda}\sum_{Z: Z\ni \{i,j\}} \| h_Z\| \le  g_0 (\dist_{i,j}+1)^{-\alpha}  \label{alternative_basic_assump_power}
\end{align}
with
\begin{align}
\alpha>2D+1,
\end{align}
where $\|\cdots\|$ denotes the operator norm and the parameters $g, g_0$ are $\orderof{1}$ constants which do not depend on the system size $n$.
By taking the energy unit appropriately, we set 
\begin{align}
g=1.\label{assump_g=1}
\end{align} 
Here, $\sum_{Z:Z\ni i, \diam (Z)\ge r}$ means the summation which picks up all the subsets $Z\subseteq \Lambda$ such that  $Z\ni i$ and $\diam (Z)\ge r$.
In the similar way, $\sum_{Z:Z\ni \{i,j\}}$ means the summation which picks up all the subsets $Z\subseteq \Lambda$ which include $\{i,j\}$
We notice that we actually need only the condition~\eqref{alternative_basic_assump_power} since it implies the condition~\eqref{basic_assump_power}.
\end{assump}
\noindent
From the assumption~\eqref{alternative_basic_assump_power}, we can also derive 
\begin{align}
\sum_{Z: Z \cap X \neq \emptyset , Z\cap Y\neq \emptyset} \| h_Z\| \le
\sum_{i\in X}  \sum_{j\in Y}\sum_{Z: Z\ni \{i,j\}} \| h_Z\|  \le  g_0 \sum_{i\in X}  \sum_{j\in Y} (\dist_{i,j}+1)^{-\alpha} 
\le g_0  |X| \cdot |Y|  (\dist_{X,Y}+1)^{-\alpha},
\label{alternative_basic_assump_power_2}
\end{align}
where we use $\dist_{X,Y} \le \dist_{i,j}$ for $i\in X$ and $j\in Y$.

\subsection{Coarse grained set (Fig.~\ref{fig:extended_coarse_grain})}

For a subset $X\subseteq \Lambda$, we define $\bal{X}{r}$ as
\begin{align}
\bal{X}{r}:= \{i\in \Lambda| \dist_{X,i} \le r \} , \label{def:bal_X_r}
\end{align}
where $\bal{X}{0}=X$ and $r$ is an arbitrary positive number (i.e., $r\in \mathbb{R}^+$).
We also define the coarse grained total set $\Lambda^{(\xi)}$ as the minimum subset such that $\bal{\Lambda^{(\xi)}}{\xi}= \Lambda$, namely
\begin{align}
\Lambda^{(\xi)} := \arg\min_{Z \subseteq \Lambda| \bal{Z}{\xi} = \Lambda} |Z|, \label{def:coarse_graining_Lambda^(r)}
\end{align}
where $\Lambda^{(0)}=\Lambda$.
Similarly, for an arbitrary subset $X\subseteq \Lambda$, we define $X^{(\xi)}$ ($\subseteq \Lambda^{(\xi)}$) as 
\begin{align}
X^{(\xi)} := \arg\min_{Z \subseteq \Lambda^{(\xi)}| \bal{Z}{\xi} \supseteq X} |Z|, \label{def:coarse_graining_X^(r)}
\end{align}
where $X^{(0)}=X$. 
From the definition, because of $X^{(\xi_2)}[\xi_1+\xi_2]=(X^{(\xi_2)}[\xi_2])[\xi_1] \supseteq X[\xi_1]$, we notice that 
\begin{align}
X[\xi_1] \subseteq  X^{(\xi_2)}[\xi_1+\xi_2] \quad {\rm and} \quad  |X[\xi_1]| \le  | X^{(\xi_2)}[\xi_1+\xi_2]| ,\label{coarse_grained_subset_ineq}
\end{align}
where we use the fact that $|X| \le |Y| $ for $X\subseteq Y$.

\begin{figure}
\centering
\subfigure[Definition of $\bal{X}{r}$
]
{\includegraphics[clip, scale=0.4]{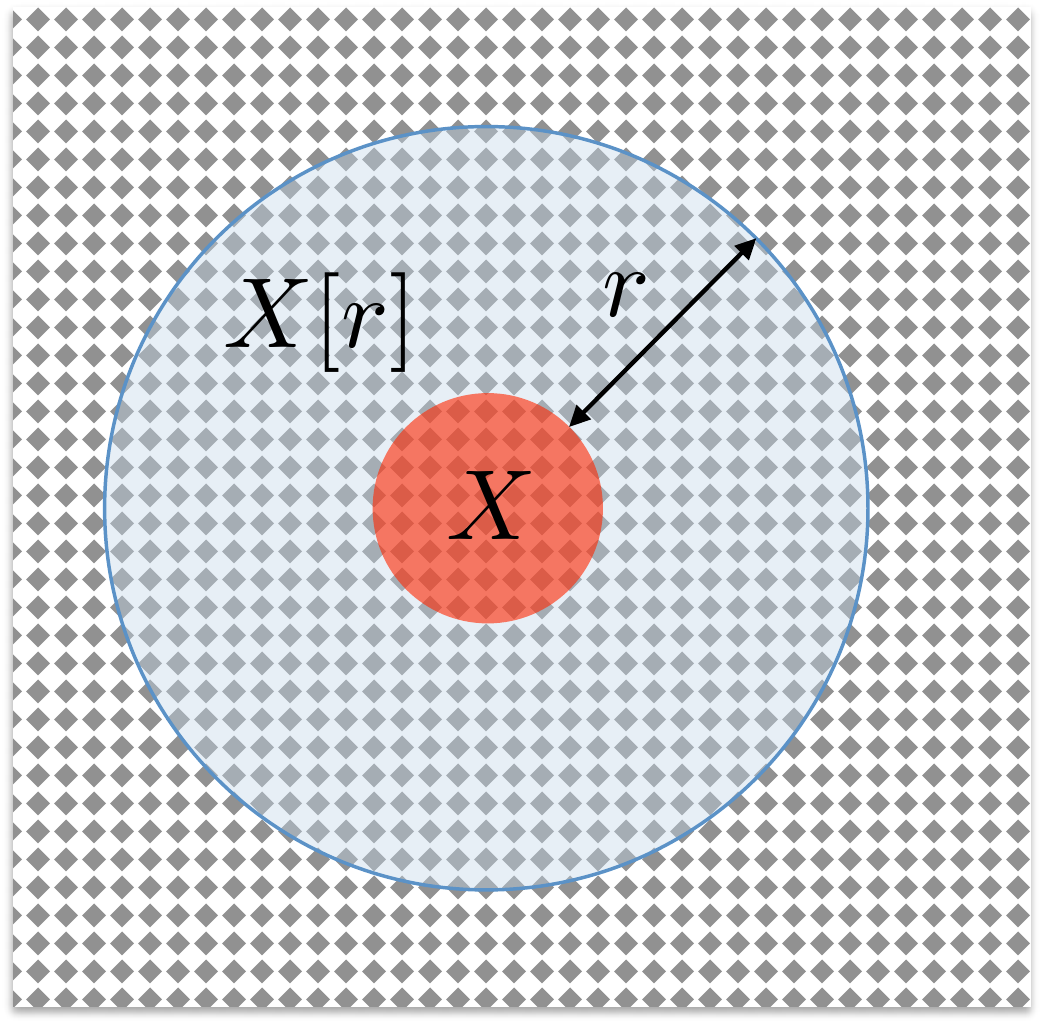}
}
\subfigure[Definition of $\Lambda^{(\xi)}$
]
{\includegraphics[clip, scale=0.4]{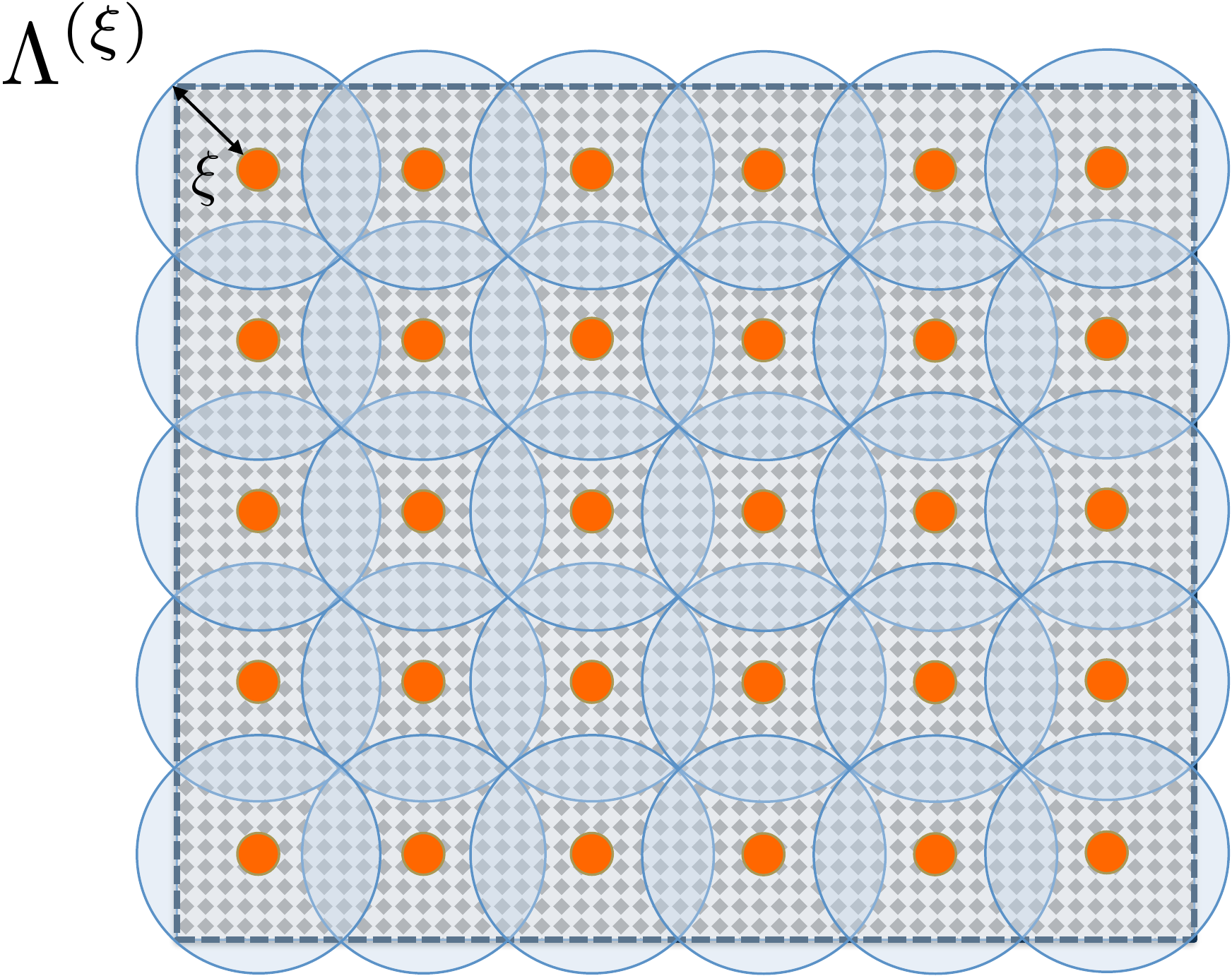}
}
\subfigure[Definitions of $X^{(\xi)}$
]
{\includegraphics[clip, scale=0.4]{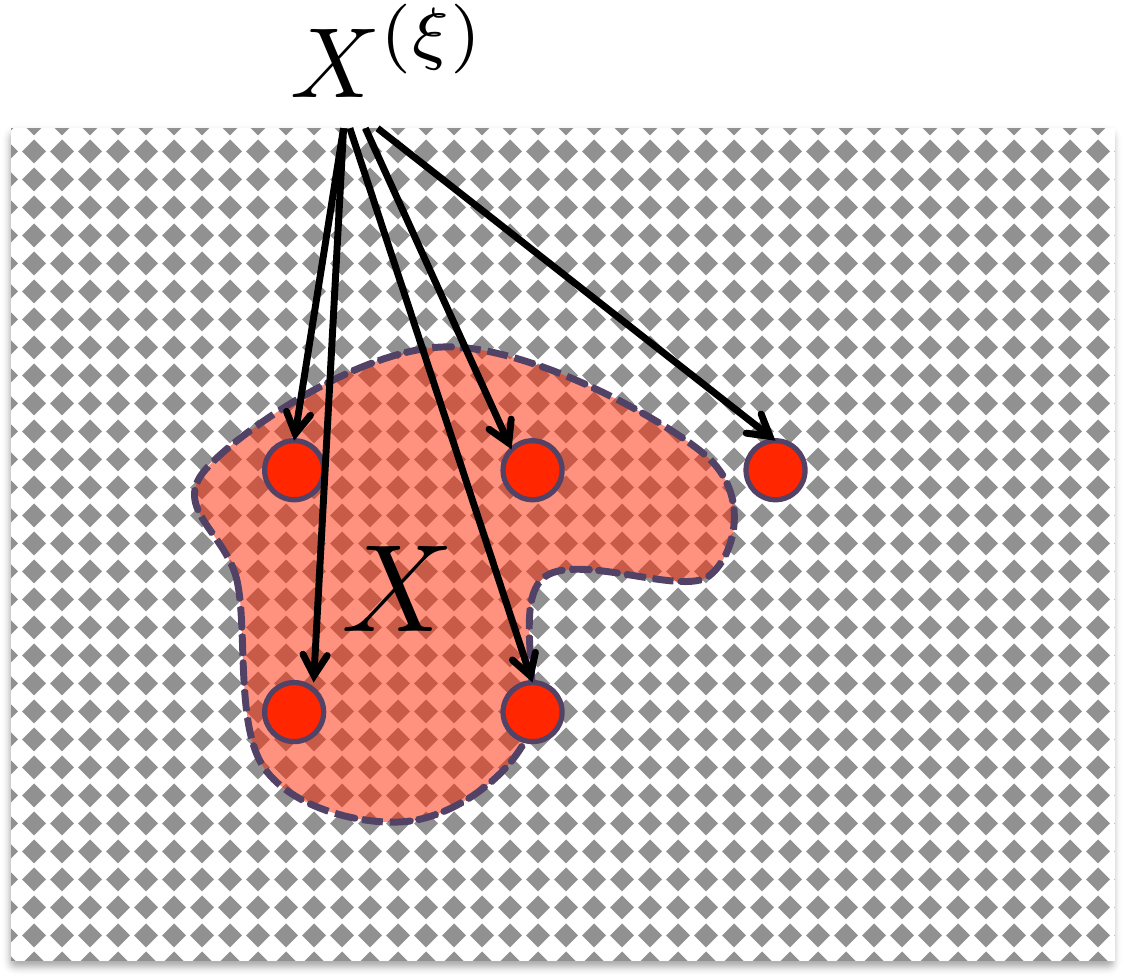}
}
\caption{(a) Extended subset. The subset $\bal{X}{r}$ is defined by extending the original subset $X$ by a distance $r$. 
(b) The subset $\Lambda^{(\xi)} \subseteq \Lambda$ is the coarse grained lattice (orange sites) which is defined as the minimum subset such that $\bal{\Lambda^{(\xi)}}{\xi}=\Lambda$.  
(c) For an arbitrary subset $X\subseteq \Lambda$ (pink region), $X^{(\xi)}\subseteq \Lambda^{(\xi)}$ is the coarse grained subset (red sites) which is defined as the minimum subset such that $\bal{X^{(\xi)}}{\xi}\supseteq X$. 
As shown in \eqref{geometric_parameter_gamma3}, the cardinality of the subset $X^{(\xi)}$ is roughly $(1/\xi)^D$ times as the original one, namely $|X^{(\xi)}| \approx (1/\xi)^D |X|$.
}
\label{fig:extended_coarse_grain}
\end{figure}

We introduce a geometric parameter $\gamma$ which is determined only by the lattice structure.
We define $\gamma \ge 1$ as a constant of $\orderof{1}$ which satisfies all the following inequalities for arbitrary $\xi \in \mathbb{R}^+$ and $X \subseteq \Lambda$:
\begin{align}
&|X|\le \gamma [\diam(X)]^D, \label{geometric_parameter_gamma1} \\ 
&|\bal{i}{r}|\le \gamma (2r)^D  \quad (r\ge 1), \label{geometric_parameter_gamma2} \\
&|X^{(\xi)}|\le \max\left(1, \gamma [\diam(X)/\xi]^D\right)  ,
\label{geometric_parameter_gamma3}\\
&\max_{i\in \Lambda}\left(\# \{j\in \Lambda^{(\xi)} | r \le \dist_{i,j} <  r+\xi\} \right)\le 2\gamma D (2r/\xi)^{D-1}
\for r\ge \xi \ge 1 ,\label{def:parameter_gamma}  
\end{align}
where we have defined $\diam(X) = \max_{i,j\in X}( \dist_{i,j})+1 \ge 1$.
Note that the inequality~\eqref{geometric_parameter_gamma2} implies
\begin{align}
&|\bal{X}{r}|= \left|\bigcup_{i\in X} \bal{i}{r} \right|\le \sum_{i\in X}|\bal{i}{r}| \le  \gamma (2r)^D|X|. \label{geometric_parameter_gamma4} 
\end{align}
Furthermore, we compare two subset $X^{(\xi)}$ and $X^{(c\xi)}$ with $c\ge1$. 
Because of $\bal{X^{(c\xi)}}{c\xi} \supseteq X$, we have $X^{(\xi)} \le (\bal{X^{(c\xi)}}{c\xi})^{(\xi)}$, and hence 
\begin{align}
\abb X^{(\xi)}\abb  \le \abb (\bal{X^{(c\xi)}}{c\xi})^{(\xi)} \abb \le \sum_{i\in X^{(c\xi)}}  \abb(\bal{i}{c\xi})^{(\xi)}\abb\le \sum_{i\in X^{(c\xi)}} \gamma (2c)^D  \le \gamma (2c)^D \abb X^{(c\xi)}\abb,
\label{ineq_coarse_grain_two}
\end{align} 
where in the third inequality we utilize~\eqref{geometric_parameter_gamma2}.

\section{Main results}

\subsection{Definitions}

We first define $\mathcal{G}(x,t,\fset{X},\fset{Y})$-Lieb-Robinson bound and $\mathcal{G}(x,t,\fset{X})$-Lieb-Robinson bound as follows:
\begin{definition}
Let $H_0$ be an arbitrary Hamiltonian.
Then, the Hamiltonian $H_0$ satisfies $\mathcal{G}(x,t,\fset{X},\fset{Y})$-Lieb-Robinson bound if it satisfies 
\begin{align}
\frac{\| [O_{\fset{X}}(H_0,t), O_{\fset{Y}} \|}{\|O_{\fset{X}}\|\cdot \|O_{\fset{Y}}\|} \le \mathcal{G}(x,t,\fset{X},\fset{Y}) ,\quad x= \dist_{\fset{X},\fset{Y}}
\label{Definition:LR_function}
\end{align} 
for arbitrary operators $O_{\fset{X}}$ and $O_{\fset{Y}}$ which are supported on $\fset{X}$ and $\fset{Y}$, respectively, 
where we define $O_{\fset{X}}(H_0,t)$ as in Eq.~\eqref{time-evolution:def}.
If the function $\mathcal{G}(x,t,\fset{X},\fset{Y})$ does not depend on the subset $\fset{Y}$, we simply denote it by $\mathcal{G}(x,t,\fset{X})$.
\end{definition}
\noindent
The definition clearly implies $\mathcal{G}(x,t,\fset{X},\fset{Y})  \le 2$ from $\| [O_{\fset{X}}(H_0,t), O_{\fset{Y}} \| \le 2 \|O_X\| \cdot\|O_Y\|$ and we assume that the cut off is automatically included in the definition.
For example, when $\mathcal{G}(x,t,\fset{X},\fset{Y}) $ is given in the form of 
\begin{align}
\mathcal{G}(x,t,\fset{X},\fset{Y}) = \min \left(2,|\fset{X}| \cdot|\fset{Y}| e^{-x + vt}\right),\label{actual_form_of_Lieb-Robinson}
\end{align}
we denote it by omitting the $ \min (2, \cdots )$ for the simplicity of the notation:
\begin{align}
\mathcal{G}(x,t,\fset{X},\fset{Y}) =|\fset{X}| \cdot|\fset{Y}| e^{-x + vt}.
\end{align}

The function $\mathcal{G}(x,t,\fset{X})$ is connected to an error in approximating time-evolved operator on a local region. 
We show the following lemma which has been given in Ref.~\cite{PhysRevLett.97.050401}:
\begin{lemma}[Bravyi, Hastings and Verstaete~\cite{PhysRevLett.97.050401}] \label{Bravyi, Hastings and Verstaete}
Let $H_0$ be an arbitrary Hamiltonian satisfying $\mathcal{G}(x,t,\fset{X})$-Lieb-Robinson bound.
For arbitrary subsets $X,\tilde{X}\subseteq \Lambda$, we define $O_X(H_0,t,\tilde{X})$ as the local approximation of $O_X(H_0,t)$ onto the subset $\tilde{X}$:
\begin{align}
O_X(H_0,t,\tilde{X})  := \frac{1}{\tr_{\tilde{X}^\co}(\hat{1})} \tr_{\tilde{X}^\co} \left[O_X(t)\right] \otimes \hat{1}_{\tilde{X}^\co},
\label{def:O_X_local_approx}
\end{align}
where $\tr_{\tilde{X}^\co}(\cdots)$ is the partial trace with respect to the subset $\tilde{X}^\co$.
The definition implies that the operator $O_X(H_0,t,\tilde{X})$ is supported on the subset $\tilde{X}$ and it also satisfies $\|O_X(H_0,t,\tilde{X})\|\le \|O_X\|$.
Then, by choosing $\tilde{X}=\bal{X}{r}$ ($r\in \mathbb{N}$), we obtain 
\begin{align}
\| O_X(H_0,t)- O_X(H_0,t, \bal{X}{r}) \| \le \mathcal{G}(r,t,X) , 
\label{lem:BHV_ineq}
\end{align} 
where $\|O_X\|=1$ and $\bal{X}{r}$ was defined in Eq.~\eqref{def:bal_X_r}.
Recall that we mean by $\mathcal{G}(x,t,\fset{X})$ the function $\mathcal{G}(x,t,\fset{X},\fset{Y})$ which does not depend on the subset $\fset{Y}$.
\end{lemma}

\subsection{Main theorem}

We here show our main theorems:

\begin{theorem}[Lieb-Robinson bound for long-range interacting systems] \label{thm:long-range Lieb-Robinson bound}
Let us consider the long-range interacting Hamiltonian $H$ satisfying the assumption~\ref{lem:power_law_k_local}.
For $|t|\ge 1$, the Hamiltonian $H$ satisfies the $\mathcal{G}_{H}(x,t,\fset{X},\fset{Y})$-Lieb-Robinson bound as 
\begin{align}
\mathcal{G}_{H}(x,t,\fset{X},\fset{Y}) = \mathcal{C}_H \abb \fset{X}^{(\vH  |t| )} \abb \cdot \abb \fset{Y}^{(\vH  |t| )} \abb  \frac{  |t|^{2D+1}\log^{2D} (x +1)}{(x-\vH  |t| )^\alpha} 
\quad (|t| \le x/\vH)
\label{main_thm_ineq1}
\end{align}  
and
\begin{align}
\mathcal{G}_{H}(x,t,\fset{X}) = \mathcal{C}'_H \abb \fset{X}^{(\vH  |t| )} \abb^2   \frac{  |t|^{D+1}\log^{2D} (x +1)}{(x-\vH  |t| )^{\alpha-D}}
\quad (|t| \le x/\vH) , 
\label{main_thm_ineq2}
\end{align}  
where $\mathcal{C}_H$, $\mathcal{C}'_H$ and $\vH$ are constants which depend only on the parameters $\{D, g_0, \alpha, \gamma\}$.
Note that we use the notation in Eq.~\eqref{def:coarse_graining_X^(r)} for $\fset{X}^{(\vH  |t| )}$ and $\fset{Y}^{(\vH  |t| )}$. 
\end{theorem}
{~}\\
\noindent
{\bf Theorem~\ref{thm:long-range Lieb-Robinson bound}' ($\boldsymbol{k}$-local Hamiltonian).}
\textit{Let us impose an additional assumption of the $k$-locality~\eqref{def:Ham_k-local} with $k=\orderof{1}$.
Then, the Hamiltonian $H$ satisfies the $\mathcal{G}^{(k)}_{H}(x,t,\fset{X},\fset{Y})$-Lieb-Robinson bound as 
\begin{align}
\mathcal{G}^{(k)}_{H}(x,t,\fset{X},\fset{Y}) = \mathcal{C}^{(k)}_H \abb \fset{X}^{(\vH^{(k)}  |t| )} \abb \cdot \abb \fset{Y}^{(\vH^{(k)}  |t| )} \abb  \frac{  |t|^{2D+1}}{(x-\vH^{(k)}  |t| )^\alpha} 
\quad (|t| \le x/\vH^{(k)})
\label{main_thm_ineq1_k-local}
\end{align}  
and
\begin{align}
\mathcal{G}^{(k)}_{H}(x,t,\fset{X}) = \mathcal{C}^{(k)'}_H \abb \fset{X}^{(\vH^{(k)}  |t| )} \abb^2   \frac{  |t|^{D+1}}{(x-\vH^{(k)}  |t| )^{\alpha-D}}
\quad (|t| \le x/\vH^{(k)}) , 
\label{main_thm_ineq2_k-local}
\end{align}  
where $\mathcal{C}^{(k)}_H$, $\mathcal{C}^{(k)'}_H$ and $\vH^{(k)}$ are constants which depend only on the parameters $\{D, g_0, \alpha, \gamma,k\}$.
}

{~}\\
{\bf Remark.}
From the theorem, outside the light-cone of $R \ge \vH |t|$, the commutator is as small as $\orderof{R^{-\alpha}}$.
The first inequality~\eqref{main_thm_ineq1} is stronger than the second one~\eqref{main_thm_ineq2} when subset $\fset{Y}$ is small. 
However, in the case where $\fset{Y}$ is infinitely large, the inequality~\eqref{main_thm_ineq1} is useless. 
For example, in order to obtain the local approximation~\eqref{lem:BHV_ineq} for time-evolved operators, we need the second inequality~\eqref{main_thm_ineq2}.

In the case of generic Hamiltonians, we need the logarithmic correction $\log^{2D} (x +1)$.
This condition arises from Theorem~\ref{thm :long-range contribution Lieb-Robinson bound} below, which characterizes a contribution from interactions with sufficiently long length scales.
This logarithmic correction can be deleted when we assume the $k$-locality of the Hamiltonian (see Theorem~\ref{thm :long-range contribution Lieb-Robinson bound}').

\clearpage

\begin{figure}[]
\centering
{
\includegraphics[clip, scale=0.6]{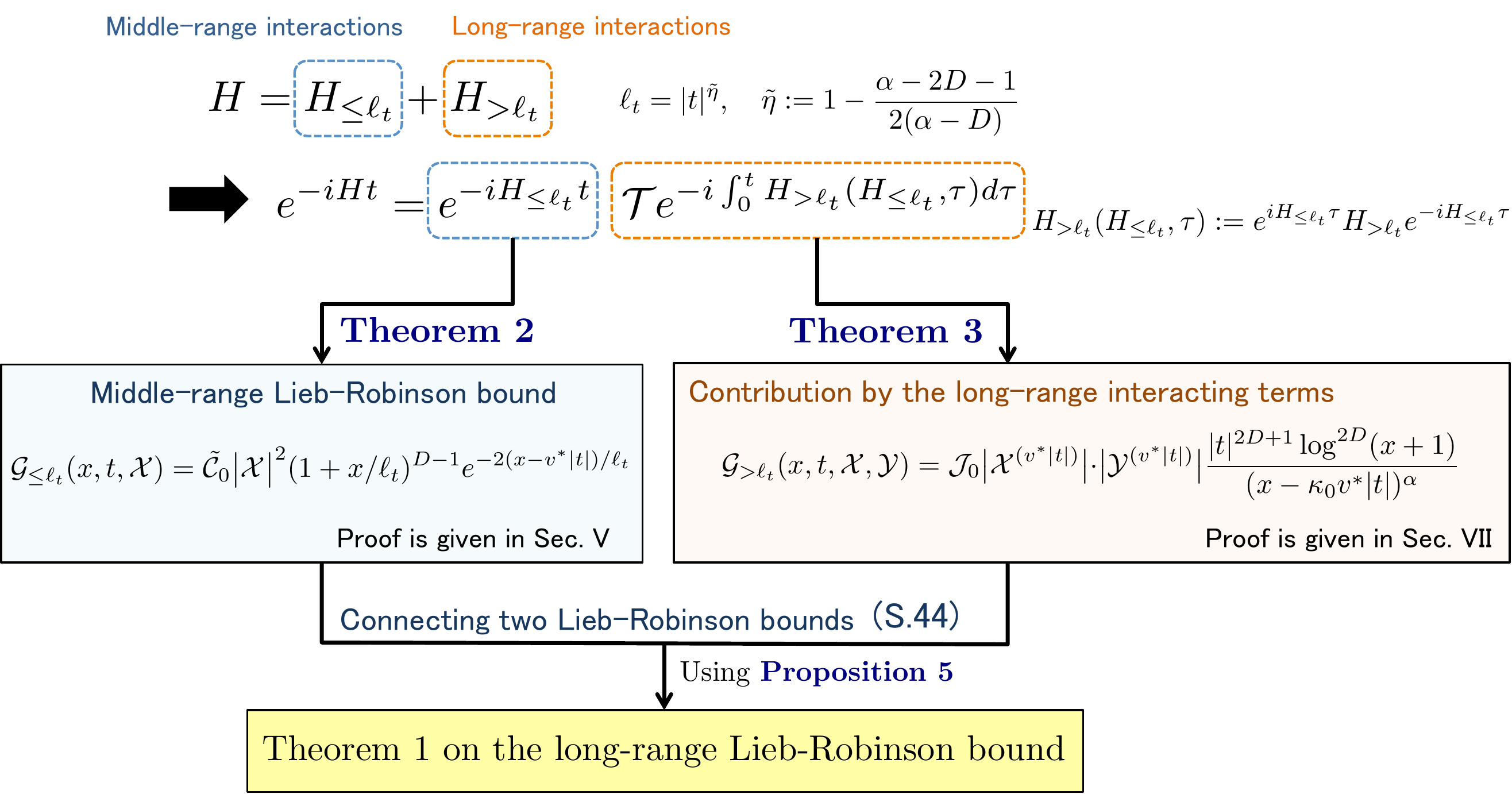}
}
\caption{Outline of the proof. In the proof, we decompose the Hamiltonian as in Eq.~\eqref{decompose_Hamiltonian_long_middle}. 
The unitary operator $e^{-iHt}$ also consists of $e^{-iH_{\le \ell_t}t}$ and $U_{>\ell_t}=\mathcal{T} e^{-i\int_0^t H_{> \ell_t} (H_{\le \ell_t}, \tau) d\tau}$ as in Eq.~\eqref{Eq:e_-iHt_decomp}.
The former one characterizes the middle-range interactions, while the latter one gives the contribution from the long-range interactions.
Then, the central tasks for the proof of Theorem~\ref{thm:long-range Lieb-Robinson bound} are 
the derivations of the Lieb-Robinson bounds for $e^{-iH_{\le \ell_t}t}$ and $U_{>\ell_t}$, which are given by Theorems~\ref{thm:Middle-range Lieb-Robinson bound} and \ref{thm :long-range contribution Lieb-Robinson bound} (or Theorem~\ref{thm :long-range contribution Lieb-Robinson bound}'), respectively.
The remaining task is to connect these two bounds, which are given by Eqs.~\eqref{estimate_the_commutator_O_X_t_O_Y}---\eqref{eqref_O_X_OY_commt}.
For a technical reason (see Sec.~\ref{sec:Relation between local Lieb-Robinson bound and global Lieb-Robinson bound}), we first restrict the subset $Y$ such that $\diam(Y) \le 2 v^\ast |t| +1$ as in Eq.~\eqref{restriction_condition_for_Y_diam}.  Afterward, we remove the restriction by using Proposition~\ref{lemma:Connection of local _ global Lieb-Robinson bounds_long_range}, which completes the proof of Theorem~\ref{thm:long-range Lieb-Robinson bound}.
}
\label{fig:Outline_main_theorem}
\end{figure}

\subsection{Proof of Theorem~\ref{thm:long-range Lieb-Robinson bound} (see also Fig.~\ref{fig:Outline_main_theorem} for the basic outline)} \label{sec:subsection_proof_middle_range}

Before discussing the long-range interaction, we consider the Lieb-Robinson bound which is obtained from the following Hamiltonian with an interaction truncation:
\begin{align}
H_{\le \ell}= \sum_{Z\subseteq \Lambda : \diam(Z) \le \ell} h_Z. \label{def:Ham_ell}
\end{align}  
As long as $\ell=\orderof{1}$, the Hamiltonian gives the Lieb-Robinson bound with a finite velocity (see Lemma~\ref{lemma:Lieb_Robinson_op_quasi_locality_short_range}).
However, when we choose the length $\ell$ depending on the time $t$, it is highly nontrivial whether the unitary operator $e^{-iH_{\le \ell}t}$ satisfies the Lieb-Robinson bound with a finite velocity or not. 
This kind of ``middle-range'' Lieb-Robinson bound is crucial in discussing the long-range Lieb-Robinson bound.
We here show the following theorem which we will prove in Sec.~\ref{sec: Proof of the Middle-range Lieb-Robinson bound}:

\begin{theorem}[Middle-range Lieb-Robinson bound] \label{thm:Middle-range Lieb-Robinson bound}
We consider a Hamiltonian $H$ which satisfies Assumption~\eqref{basic_assump_power} with $\alpha>2D +1$.
Then, for arbitrary $\ell$, the Hamiltonian $H_{\le \ell}$ satisfies the $\mathcal{G}_{\le \ell}(x,t,\fset{X})$-Lieb-Robinson bound with  
\begin{align}
&\mathcal{G}_{\le \ell}(x,t,\fset{X}) = \tilde{\mathcal{C}}_0 \abb\fset{X} \abb^2 (1+x/\ell)^{D-1}  e^{-2(x-v^\ast |t|)/\ell} ,\label{thm_middle_range_Lieb-Robinson}
\end{align}
where $v^\ast$ is a constant which depends only on the parameters $\{D, g_0, \alpha,\gamma\}$ and 
\begin{align} 
\tilde{\mathcal{C}}_0  := \frac{4}{3}e^{10/3} 15^D  D! \gamma   \zeta_2  ,\quad \zeta_2 :=16e^3 \gamma^2 \cdot 45^D. 
\label{def:tilde_mathcal_C_0}
\end{align} 
\end{theorem}

From Theorem~\ref{thm:Middle-range Lieb-Robinson bound}, we ensure that the middle-range Lieb-Robinson bound~\eqref{thm_middle_range_Lieb-Robinson} retains the linear light cone for $\ell \le \orderof{t}$.
Afterward, we prove the long-range Lieb-Robinson bound by using this middle-range Lieb-Robinson bound.
For the purpose, we consider the following decomposition of the total Hamiltonian into
\begin{align}
H=H_{\le \ell_t} + H_{> \ell_t} \label{decompose_Hamiltonian_long_middle}
\end{align}  
with 
\begin{align} 
\ell_t = |t|^{\tilde{\eta}},\quad  \tilde{\eta}:= 1-\frac{\alpha -2D -1}{2(\alpha-D)},
\label{choice_of_ell_t_long}
\end{align}
where we defined $H_{\le \ell_t}$ by Eq.~\eqref{def:Ham_ell}.
Note that under the condition of $\alpha >2D+1$ we have $\tilde{\eta} <1$ and the dynamics by $H_{\le \ell_t}$ retains the linear light cone.
We then decompose the unitary operator $e^{-iHt}$ into the form of 
\begin{align}
e^{-iHt} = e^{-iH_{\le \ell_t} t} \mathcal{T} e^{-i\int_0^t H_{> \ell_t} (H_{\le \ell_t}, \tau) d\tau} , 
\label{Eq:e_-iHt_decomp}
\end{align}  
where $\mathcal{T}$ is the time-ordering operator and we use the notation~\eqref{time-evolution:def}, namely $H_{> \ell_t} (H_{\le \ell_t}, \tau)=e^{iH_{\le \ell_t}\tau } H_{> \ell_t} e^{-iH_{\le \ell_t}\tau}$.
We have already obtained the Lieb-Robinson bound for $e^{-iH_{\le \ell_t} t}$ from Theorem~\ref{thm:Middle-range Lieb-Robinson bound}. 
Hence, we need to consider the Lieb-Robinson bound for the unitary operator of 
\begin{align}
U_{>\ell_t}:= \mathcal{T} e^{-i\int_0^t H_{> \ell_t} (H_{\le \ell_t}, \tau) d\tau}. \label{long_range_effective_unitary}
\end{align}  
We can prove the following theorem on the Lieb-Robinson bound for the unitary operator $U_{>\ell_t}$ (see Sec.~\ref{Contribution of the Long-range interactions to Lieb-Robinson bound} for the proof):

\begin{theorem}[Contribution by the long-range interacting terms] \label{thm :long-range contribution Lieb-Robinson bound}
Under the choice of $\ell_t$ by Eq.~\eqref{choice_of_ell_t_long}, the unitary operator~\eqref{long_range_effective_unitary} satisfies the $\mathcal{G}_{> \ell_t}(x,t,\fset{X},\fset{Y})$-Lieb-Robinson bound as 
\begin{align}
\mathcal{G}_{> \ell_t}(x,t,\fset{X},\fset{Y})= &
\mathcal{J}_0 \abb \fset{X}^{(v^\ast |t|)} \abb \cdot \abb \fset{Y}^{(v^\ast |t|)}\abb \frac{  |t|^{2D+1}\log^{2D} (x+1)}{(x -\kappa_0 v^\ast |t|)^\alpha} ,
\label{ineq:long-range_contribution}
\end{align}  
where $\mathcal{J}_0$ and $\kappa_0$ are constants which depend only on the parameters $\{D, g_0, \alpha,\gamma\}$.
\end{theorem}
\noindent 
The logarithmic term $\log^{2D} (x+1)$ appears from the contribution by the macroscopic interactions (e.g., $\orderof{n}$-body interactions).
It vanishes under the assumption of the $k$-locality (i.e., few-body interactions) as follows:

{~}\\
{\bf Theorem~\ref{thm :long-range contribution Lieb-Robinson bound}' ($\boldsymbol{k}$-local Hamiltonian).}
\textit{
Let us assume the $k$-locality of the Hamiltonian [see Eq.~\eqref{def:Ham_k-local} for the definition].
Then, under the same setup as that of  Theorem~\ref{thm :long-range contribution Lieb-Robinson bound}, we obtain the Lieb-Robinson bound  
$\mathcal{G}^{(k)}_{> \ell_t}(x,t,\fset{X},\fset{Y})$ 
\begin{align}
\mathcal{G}^{(k)}_{> \ell_t}(x,t,\fset{X},\fset{Y})= &
\mathcal{J}_0^{(k)} \abb \fset{X}^{(v^\ast |t|)} \abb \cdot \abb \fset{Y}^{(v^\ast |t|)}\abb \frac{  |t|^{2D+1}}{(x -\kappa_0 v^\ast |t|)^\alpha} ,
\label{ineq:long-range_contribution_k-local}
\end{align} 
where $\mathcal{J}_0^{(k)}$ and $\kappa_0$ are constants which depend only on the parameters $\{D, g_0, \alpha,\gamma,k\}$.
}

{~}

By using Theorems~\ref{thm:Middle-range Lieb-Robinson bound} and \ref{thm :long-range contribution Lieb-Robinson bound}, we prove the main theorem as follows.
We here estimate the norm of the commutator
\begin{align}
\| [O_X(t),O_Y]\|=\| [ U_{>\ell_t}^\dagger  O_XU_{>\ell_t}, O_Y(H_{\le \ell_t},-t)] \|, \quad R=\dist_{X,Y}
\label{estimate_the_commutator_O_X_t_O_Y}
\end{align}
for operators $O_X$ and $O_Y$ with $\|O_X\|=\|O_Y\|=1$, where we impose the restriction of 
\begin{align}
\diam(Y) \le 2 v^\ast |t| +1.\label{restriction_condition_for_Y_diam}
\end{align}
This restriction is removed afterward.
Note that we impose no restrictions on $X$.
By using the notation of Eq.~\eqref{def:O_X_local_approx}, we first decompose $O_Y(H_{\le \ell_t},-t)$ as 
\begin{align}
&O_Y(H_{\le \ell_t},-t) =O_Y(H_{\le \ell_t},-t, \bal{Y}{\xi_1})+O_{t,\bal{Y}{\xi_{j_\ast}}}+  \sum_{j=2}^{j_\ast} O_{t,\bal{Y}{\xi_j}},  \notag \\
&O_{t,\bal{Y}{\xi_{j}}} := O_Y(H_{\le \ell_t},-t, \bal{Y}{\xi_{j}})- O_Y(H_{\le \ell_t},-t, \bal{Y}{\xi_{j-1}}) , \notag \\
&O_{t,\bal{Y}{\xi_{j_\ast}}} := O_Y(H_{\le \ell_t},-t)- O_Y(H_{\le \ell_t},-t, \bal{Y}{\xi_{j_\ast}}), \quad \xi_{j_\ast}:= \lfloor R/2 \rfloor,  \label{decomp_O_i_H_t_long_range}
\end{align}
where $\xi_1< \xi_2< \cdots <\xi_{j_\ast} =\lfloor R/2 \rfloor $ and we will appropriately determine $\{\xi_j\}_{j=1}^{j_\ast-1}$ afterwards.
From the inequality~\eqref{lem:BHV_ineq} in Lemma~\ref{Bravyi, Hastings and Verstaete}, we have 
\begin{align}
\| O_Y(H_{\le \ell_t},-t, \bal{Y}{\xi_1}) \| &\le 1, \notag \\
\|O_{t,\bal{Y}{\xi_{j_\ast}}}\| &\le \mathcal{G}_{\le \ell_t}(\xi_{j_\ast},t,Y)= \mathcal{G}_{\le \ell_t}(\lfloor R/2 \rfloor,t,Y) , \notag \\
\|O_{t,\bal{Y}{\xi_{j}}}\|&=  \|O_Y(H_{\le \ell_t},-t, \bal{Y}{\xi_{j}})- O_Y(H_{\le \ell_t},-t, \bal{Y}{\xi_{j-1}}) \|  \notag \\
&\le  \|O_Y(H_{\le \ell_t},-t, \bal{Y}{\xi_{j}})-O_Y(H_{\le \ell_t},-t)  \|  + \|O_Y(H_{\le \ell_t},-t) - O_Y(H_{\le \ell_t},-t, \bal{Y}{\xi_{j-1}}) \|  \notag \\
&\le 2 \mathcal{G}_{\le \ell_t}(\xi_{j-1},t,Y),
\label{estiamtion_O_Y_approx}
\end{align}
where we use $\xi_{j_\ast}:= \lfloor R/2 \rfloor$ and $\mathcal{G}_{\le \ell_t}(\xi_j,t,Y) \le \mathcal{G}_{\le \ell_t}(\xi_{j-1},t,Y)$.
By using the above decomposition, we obtain the upper bound of the commutator~\eqref{estimate_the_commutator_O_X_t_O_Y} as follows:
\begin{align}
&\| [ U_{>\ell_t}^\dagger  O_XU_{>\ell_t}, O_Y(H_{\le \ell_t},-t)] \|  \notag \\
\le& \| [ U_{>\ell_t}^\dagger  O_XU_{>\ell_t}, O_Y(H_{\le \ell_t},-t, \bal{Y}{\xi_1})] \| +\| [ U_{>\ell_t}^\dagger  O_XU_{>\ell_t},O_{t,\bal{Y}{\xi_{j_\ast}}} ] \| +
\sum_{j=2}^{\infty}  \| [ U_{>\ell_t}^\dagger  O_XU_{>\ell_t}, O_{t,\bal{Y}{\xi_{j}}} ] \|   \notag \\
\le& \mathcal{G}_{> \ell_t}(R- \xi_1 ,t,X,\bal{Y}{\xi_1}) \cdot\| O_Y(H_{\le \ell_t},-t, \bal{Y}{\xi_1}) \|  +2 \|O_X\| \cdot \|O_{t,\bal{Y}{\xi_{j_\ast}}}\| + \sum_{j=2}^{j_\ast}  \mathcal{G}_{> \ell_t}(R- \xi_j ,t,X,\bal{Y}{\xi_{j}}) \cdot\|O_{t,\bal{Y}{\xi_{j}}}\|  \notag \\
\le &\mathcal{G}_{> \ell_t}(R- \xi_1 ,t,X,\bal{Y}{\xi_1})   +2\mathcal{G}_{\le \ell_t}(\lfloor R/2 \rfloor,t,Y) +2 \sum_{j=2}^{j_\ast}  \mathcal{G}_{> \ell_t}(R- \xi_j ,t,X,\bal{Y}{\xi_{j}}) \cdot \mathcal{G}_{\le \ell_t}(\xi_{j-1},t,Y) , 
\label{uppe_bound_O_X_O_Y_long_Fig_used}
\end{align} 
where we use Theorems~\ref{thm :long-range contribution Lieb-Robinson bound} from the second line to the third line, and use the inequalities in~\eqref{estiamtion_O_Y_approx} and $\|O_X\|=1$  from the third line to the fourth line.

We, in the following, determine $\{\xi_j\}_{j=1}^{j_\ast -1}$ such that  
\begin{align}
\mathcal{G}_{\le \ell_t}(\xi_{j},t,Y) = \tilde{\mathcal{C}}_0 \abb Y \abb^2 (1+\xi_{j}/\ell_t)^{D-1}  e^{-2(\xi_{j}-v^\ast |t|)/\ell_t} \le  e^{-j}
\label{G_le_ell_t_1}
\end{align} 
for each of $j \ge 1$, where we use the form of Eq.~\eqref{thm_middle_range_Lieb-Robinson} for $\mathcal{G}_{\le \ell_t}(\xi_{j},t,Y)$.
First, because of $\diam(Y) \le 2 v^\ast |t| +1$, we obtain 
\begin{align}
\abb Y \abb^2\le \gamma^2 (2v^\ast |t|+1)^{2D}, 
\end{align} 
where we use the inequality~\eqref{geometric_parameter_gamma1}. 
Thus, we need to choose $\xi_j$ as 
\begin{align}
\xi_j =c_1 v^\ast |t| +c_2 \ell_t j  =  c_1 v^\ast |t| +c_2 |t|^{\tilde{\eta}}j  ,
\label{choice of xi_j_thm}
\end{align} 
where $c_1$ and $c_2$ are constants which depend on $v^\ast$, $\tilde{\mathcal{C}}_0$ and $\tilde{\eta}$; that is, they depend on $\{D, g_0, \alpha,\gamma\}$.
Remember that we have chosen $\ell_t$ as in Eq.~\eqref{choice_of_ell_t_long}.
By applying the inequality~\eqref{G_le_ell_t_1} to \eqref{uppe_bound_O_X_O_Y_long_Fig_used}, we obtain
\begin{align}
&\| [ U_{>\ell_t}^\dagger  O_XU_{>\ell_t}, O_Y(H_{\le \ell_t},-t)] \| \le 2\mathcal{G}_{\le \ell_t}( \lfloor R/2 \rfloor ,t,Y)+2 \sum_{j=1}^{j_\ast}   e^{-j+1} \cdot  \mathcal{G}_{> \ell_t}(R- \xi_j ,t,X,\bal{Y}{\xi_{j}}),
\label{uppe_bound_O_X_O_Y_long_2}
\end{align} 
where the first term in \eqref{uppe_bound_O_X_O_Y_long_Fig_used} is now involved in the summation.

We then consider the values of $ \mathcal{G}_{> \ell_t}(R- \xi_j ,t,X,\bal{Y}{\xi_j}) $ for $j\le j_\ast$.
Because of $\xi_j \le \lfloor R/2 \rfloor \le R/2$, we obtain 
\begin{align}
 \mathcal{G}_{> \ell_t}(R- \xi_j ,t,X,\bal{Y}{\xi_j}) 
 &= \mathcal{J}_0 \abb X^{(v^\ast |t|)} \abb \cdot \abb \bal{Y}{\xi_j}^{(v^\ast |t|)}\abb \frac{  |t|^{2D+1}\log^{2D} (R -\xi_j+1)}{(R -\xi_j-\kappa_0 v^\ast |t|)^\alpha}   \notag \\
 &\le \mathcal{J}_0 \abb X^{(v^\ast |t|)} \abb \gamma (3+ 2c_1+ 2c_2j|t|^{-1+\tilde{\eta}}/v^\ast)^D \frac{  |t|^{2D+1}\log^{2D} (R +1)}{(R/2-\kappa_0 v^\ast |t|)^\alpha} ,
 \label{G_ge_ell_t_1}
\end{align} 
where we estimate $\abb \bal{Y}{\xi_j}^{(v^\ast |t|)}\abb $ by using the inequality~\eqref{geometric_parameter_gamma3} as follows:
\begin{align}
\abb \bal{Y}{\xi_j}^{(v^\ast |t|)}\abb \le  \gamma \left( \frac{\diam(Y) +2 \xi_j }{v^\ast |t|} \right)^D \le  \gamma (3+ 2c_1+ 2c_2j|t|^{-1+\tilde{\eta}}/v^\ast)^D,
\end{align} 
where in the second inequality we use $(v^\ast |t|)^{-1}\diam(Y) \le 2 +(v^\ast |t|)^{-1}\le 3$ and the form of Eq.~\eqref{choice of xi_j_thm}.
By applying the inequality~\eqref{G_ge_ell_t_1} to \eqref{uppe_bound_O_X_O_Y_long_2}, we obtain
\begin{align}
&\| [ U_{>\ell_t}^\dagger  O_XU_{>\ell_t}, O_Y(H_{\le \ell_t},-t)] \|  \notag \\
\le&2 \mathcal{G}_{\le \ell_t}( \lfloor R/2 \rfloor ,t,Y)+2 \mathcal{J}_0 \gamma \abb X^{(v^\ast |t|)} \abb  \frac{  |t|^{2D+1}\log^{2D} (R +1)}{(R/2-\kappa_0 v^\ast |t|)^\alpha}  
\sum_{j=1}^{j_\ast}   e^{-j+1} (3+ 2c_1+ 2c_2j|t|^{-1+\tilde{\eta}}/v^\ast)^D \notag \\
\le &2\mathcal{G}_{\le \ell_t}( \lfloor R/2 \rfloor ,t,Y)+2^{\alpha+1}c_3 \mathcal{J}_0 \gamma \abb X^{(v^\ast |t|)} \abb  \frac{  |t|^{2D+1}\log^{2D} (R +1)}{(R-2 \kappa_0 v^\ast |t|)^\alpha} .
\label{uppe_bound_O_X_O_Y_long_3}
\end{align} 
We here define $c_3$ as
\begin{align}
c_3:= \sum_{j=1}^{\infty}   e^{-j+1}  (3+2c_1 +2 c_2j/v^\ast)^D \ge \sum_{j=1}^{j_\ast}   e^{-j+1}  (3+2c_1 + 2c_2j|t|^{-1+\tilde{\eta}}/v^\ast )^D  ,
\end{align}
where the inequality $ c_2j/v^\ast  \ge c_2j|t|^{-1+\tilde{\eta}}/v^\ast $ is satisfied from the condition of $|t| \ge 1$ in the theorem.

We define the parameter $\tilde{\kappa}_0$ ($>\kappa_0$) so that that it satisfies the following inequality for $R> \tilde{\kappa}_0 v^\ast |t| $: 
\begin{align}
\mathcal{G}_{\le \ell_t}( \lfloor R/2 \rfloor ,t,Y) \le 2^{\alpha}c_3 \mathcal{J}_0 \gamma \abb X^{(v^\ast |t|)} \abb  \frac{  |t|^{2D+1}\log^{2D} (R +1)}{(R-2 \tilde{\kappa}_0 v^\ast |t|)^\alpha} .
\label{definition_of_tilde_rfrak_0}
\end{align} 
More specifically, from the inequality~\eqref{thm_middle_range_Lieb-Robinson}, $\tilde{\kappa}_0$ is defined by 
\begin{align}
 \tilde{\mathcal{C}}_0 \abb Y \abb^2 (1+\lfloor R/2 \rfloor/|t|^{\tilde{\eta}})^{D-1}  e^{-2|t|^{-\tilde{\eta}}(\lfloor R/2 \rfloor -v^\ast |t|)} 
\le 2^{\alpha}c_3 \mathcal{J}_0 \gamma \abb X^{(v^\ast |t|)} \abb  \frac{  |t|^{2D+1}\log^{2D} (R +1)}{(R-2 \tilde{\kappa}_0 v^\ast |t|)^\alpha}.
\end{align}  
Because of $\tilde{\eta}<1$ and $\diam(Y) \le 2v^\ast|t|+1$, we can always find such a parameter $\tilde{\kappa}_0$  which depends only on $\{D, g_0, \alpha,\gamma\}$.
The inequality~\eqref{definition_of_tilde_rfrak_0} reduces the inequality~\eqref{uppe_bound_O_X_O_Y_long_3} to 
\begin{align}
&\| [O_X(t),O_Y]\|=\| [ U_{>\ell_t}^\dagger  O_XU_{>\ell_t}, O_Y(H_{\le \ell_t},-t)] \|  \le
2^{\alpha+2}c_3 \mathcal{J}_0 \gamma \abb X^{(v^\ast |t|)} \abb  \frac{  |t|^{2D+1}\log^{2D} (R +1)}{(R-2\tilde{\kappa}_0 v^\ast  |t| )^\alpha} .
\label{eqref_O_X_OY_commt}
\end{align} 
We note that the subset $Y$ is now restricted by the condition~\eqref{restriction_condition_for_Y_diam}, namely $\diam(Y) \le 2 v^\ast |t| +1$.

In order to remove the restriction, we use Proposition~\ref{lemma:Connection of local _ global Lieb-Robinson bounds_long_range} in the subsequent section.
Here, the inequality~\eqref{eqref_O_X_OY_commt} is given in the form of \eqref{Lieb_Robinson_O_fsetX_assump_long} by choosing 
\begin{align}
&\mathcal{F}(t,\fset{X},\fset{Y}) =2^{\alpha+2}c_3 \mathcal{J}_0 \gamma \abb \fset{X}^{(v^\ast |t|)} \abb\cdot |t|^{2D+1} ,\quad p=2D,\quad
\xi=v^\ast |t|, \quad \alpha_0=\alpha,\quad \kappa=2\tilde{\kappa}_0.
\end{align} 
Then, from the inequalities~\eqref{Lieb_L_L'_from_local_one_lemma_long_0} and \eqref{Lieb_Robinson_O_i_tildeF_long}, we obtain
\begin{align}
\mathcal{G}(x,t,\fset{X},\fset{Y})= 2^{\alpha+3}c_3 \mathcal{J}_0 \gamma \abb \fset{X}^{(v^\ast |t|)} \abb \cdot \abb \fset{Y}^{(v^\ast |t|)} \abb  \frac{  |t|^{2D+1}\log^{2D} (x +1)}{(x-2\tilde{\kappa}_0 v^\ast  |t| )^\alpha} 
\label{ineq:1_last_inequality}
\end{align}
and 
\begin{align}
\mathcal{G}(x,t,\fset{X},\fset{Y})=2^{\alpha+2}  c_3 \mathcal{J}_0 \gamma  \mathcal{C}'_{2\tilde{\kappa}_0,2D,\alpha} (v^\ast)^{-D} \abb \fset{X}^{(v^\ast |t|)} \abb ^2 \frac{  |t|^{D+1}\log^{2D} (x +1)}{(x-2\tilde{\kappa}_0 v^\ast |t| -4 v^\ast |t|)^{\alpha-D}} 
\label{ineq:2_last_inequality}
\end{align}
for $x\ge (2\tilde{\kappa}_0+5)v^\ast |t|$, where $\mathcal{C}'_{2\tilde{\kappa}_0,2D,\alpha}$ is defined by Eq.~\eqref{def_mathcal_C'_0_theorem}.
Therefore, the two inequalities~\eqref{ineq:1_last_inequality} and \eqref{ineq:2_last_inequality} 
reduce to the inequalities~\eqref{main_thm_ineq1} and \eqref{main_thm_ineq2} respectively by choosing $\vH= (2\tilde{\kappa}_0+5)v^\ast$; in order to upper-bound $\fset{X}^{(v^\ast |t|)}$ by $\fset{X}^{(\vH |t|)}$, we utilize the inequality~\eqref{ineq_coarse_grain_two}.
We note that all the constants depend only on $\{D, g_0, \alpha,\gamma\}$. 

In deriving the Lieb-Robinson bound for $k$-local Hamiltonians, we follow the same analytical steps.
The only difference is that we utilize Theorem~\ref{thm :long-range contribution Lieb-Robinson bound}' instead of Theorem~\ref{thm :long-range contribution Lieb-Robinson bound}.
After straightforward calculations, we can derive the inequalities~\eqref{main_thm_ineq1_k-local} and \eqref{main_thm_ineq2_k-local}.
This completes the proof of the main theorem. $\square$

\section{Extending the Lieb-Robinson bound for local operators to that for generic operators} \label{sec:Relation between local Lieb-Robinson bound and global Lieb-Robinson bound}
 
We first consider the Lieb-Robinson function $\mathcal{G}(x,t,\fset{X},\fset{Y})$ under the constraint that the local subsets $\fset{X},\fset{Y}$ satisfy $\diam(\fset{X}), \diam(\fset{Y}) \le 2\xi+1$ for a given $\xi$.
In our analyses, it is crucial to relate the Lieb-Robinson bound for local operators to that for generic operators, including global operators that have no restrictions on $\diam(\fset{X})$ and $\diam(\fset{Y})$.

To make the motivation clearer, we consider a connection of the Lieb-Robinson bounds from different time-evolutions $e^{-iH_1t}$ and $e^{-iH_2t}$.
We are now interested in the commutator of $\| [e^{iH_2t}e^{iH_1t}O_X e^{-iH_1t}e^{-iH_2t} , O_Y]\|$.
The difficulty in the connection lies in the subset dependence (i.e., $|X|$ or $|Y|$) of the Lieb-Robinson bound. 
We often obtain the Lieb-Robinson bounds for $e^{-iH_1t}$ and $e^{-iH_2t}$ in the form of
\begin{align}
\| [O_X(H_j,t), O_Y]\| \le |X| \mathcal{F}_j(x,t) \quad (\|O_X\|=\|O_Y\|=1)\label{LR_bound_U_j}
\end{align}  
for $j=1,2$ and $x=\dist_{X,Y}$ (see~\cite{nachtergaele2010lieb,ref:Hastings2006-ExpDec} for example), where $\mathcal{F}_j(x,t)$ is determined by details of $H_j$.
In order to connect the Lieb-Robinson bounds for $e^{-iH_1t}$ and $e^{-iH_2t}$ to obtain that for $e^{-iH_1t}e^{-iH_2t}$, we consider a similar decomposition to Eq.~\eqref{decomp_O_i_H_t_long_range}:
\begin{align}
&  O_X (H_1,t) = \sum_{s=1}^\infty\tilde{O}_{\bal{X}{s\xi}}\label{upper_bound_U_1_U_2_main}
\end{align}
with 
\begin{align}
&\tilde{O}_{\bal{X}{\xi}}= O_X(H_1,t,\bal{X}{\xi}) ,\quad 
\tilde{O}_{\bal{X}{s\xi}}= O_X(H_1,t,\bal{X}{s\xi}) - O_X(H_1,t,\bal{X}{(s-1)\xi})\quad (s\ge 2),
\end{align}  
where $\xi$ is appropriately chosen and we use the definition~\eqref{def:O_X_local_approx}.
Note that $\lim_{s\to\infty}O_X(H_1,t,\bal{X}{s\xi}) =O_X(H_1,t)$  and the operator $\tilde{O}_{\bal{X}{s\xi}}$ is supported on the subset $\bal{X}{s\xi}\subseteq \Lambda$.
By using the decomposition, we obtain
\begin{align}
\| [e^{iH_2t}e^{iH_1t}O_X e^{-iH_1t}e^{-iH_2t} , O_Y] \| \le  \sum_{s=1}^\infty\| [\tilde{O}_{\bal{X}{s\xi}}(H_2,t) , O_Y]\| \le  \sum_{s=1}^\infty \abb \bal{X}{s\xi}\abb \mathcal{F}_2(x-s\xi,t) \cdot \| \tilde{O}_{\bal{X}{s\xi}}\|, \label{upper_bound_U_1_U_2_main}
\end{align}
where we use $\dist_{\bal{X}{s\xi}, Y} \ge \dist_{X, Y} - s\xi=x-s\xi$ and the Lieb-Robinson bound~\eqref{LR_bound_U_j} with $j=2$.

From the inequality~\eqref{lem:BHV_ineq}, we can obtain $\| \tilde{O}_{\bal{X}{s\xi}}\| \le  2 |X| \mathcal{F}_1((s-1)\xi,t)$, where we use  the Lieb-Robinson bound~\eqref{LR_bound_U_j} with $j=1$.
Hence, the upper bound of \eqref{upper_bound_U_1_U_2_main} reduces to
\begin{align}
\| [e^{iH_2t}e^{iH_1t}O_X e^{-iH_1t}e^{-iH_2t} , O_Y] \| \le 2 |X|  \sum_{s=1}^\infty \abb \bal{X}{s\xi}\abb \mathcal{F}_2(x-s\xi,t)  \mathcal{F}_1((s-1)\xi,t) .
\end{align} 
Then, after connecting two unitary operators $e^{-iH_1t}$ and $e^{-iH_2t}$, the new Lieb-Robinson bound depends on the support $X$ in the form of $|X|^2$.
If we connect Lieb-Robinson bounds for $m$ unitary operators, $|X|$-dependence of the coefficient grows as $|X|^m$ and leads to a meaningless Lieb-Robinson bound in the process of the connections. 
In order to prevent it, we consider only the operators $O_X$ and $O_Y$ which are supported on local subsystems with $\diam(X),\diam(Y)\le 2\xi+1$ with appropriate $\xi$.
After the connection process, we remove the constraints on the subset sizes to obtain the Lieb-Robinson bound for generic operators.
The following theorem relates the Lieb-Robinson bound for local operators to that for generic operators:

\begin{theorem} \label{lemma:Connection of local _ global Lieb-Robinson bounds_2}
We consider a Hamiltonian $H_0$ which satisfies the $\mathcal{G}(x,t,\fset{X},\fset{Y})$-Lieb-Robinson bound in the form of
\begin{align}
\mathcal{G}(x,t,\fset{X},\fset{Y})= \mathcal{F}(t,\fset{X},\fset{Y}) \fset{L}(x) \quad {\rm with} \quad  \fset{L}(x):=e^{-x/\xi_0} \label{Lieb_Robinson_O_fsetX_assump}
\end{align}
for $\forall \fset{X},\fset{Y} \subseteq \Lambda$ such that $\diam(\fset{X}),\diam(\fset{Y}) \le 2\xi+1$, where we assume $\mathcal{F}(t,\fset{X},\fset{Y}) = \mathcal{F}(-t,\fset{X},\fset{Y}) $.
Then, for arbitrary subsets $X\subset \Lambda$ with $\diam(X)\le 2\xi+1$ and $L\subseteq \Lambda$, we obtain the Lieb-Robinson function $\mathcal{G}(x,t,X,L)$ which is given by
\begin{align}
&\mathcal{G}(x,t,X,L) = 2 \bigl| L^{(\xi)}\bigr| \tilde{\mathcal{F}}(t)\fset{L}(x) , \quad x=\dist_{X,L} ,\label{Lieb_L_X_from_local_one_lemma} \\
&\tilde{\mathcal{F}}(t):=\sup_{\substack{\fset{X},\fset{Y}\subseteq \Lambda \\ \diam(X),\diam(Y)\le 2\xi+1}}[\mathcal{F}(t,\fset{X},\fset{Y})] . \label{Lieb_Robinson_O_i_tildeF}
\end{align}
Furthermore, for arbitrary two subsets $L,L'\subseteq \Lambda$, we obtain the Lieb-Robinson function $\mathcal{G}(x,t,L,L')$ which is given by
\begin{align}
\mathcal{G}(x,t,L,L')  =
 \mathcal{C}_{\xi,\xi_0}  \abb L^{(\xi)}\abb^2 \tilde{\mathcal{F}}(t)  (1+ x/\xi )^{D-1}  e^{-x/\xi_0} ,  \quad x=\dist_{L,L'},
  \label{Lieb_L_L'_from_local_one_lemma} 
\end{align}
where we define
\begin{align}
\mathcal{C}_{\xi,\xi_0} :=2^{D+2} D! \gamma   (1+\xi_0/\xi)^D  e^{5 \xi/\xi_0} .
\label{def_mathcal_C_0_theorem}
\end{align}
Note that we here do not impose any restrictions on $L$ and $L'$.
\end{theorem}

We obtain a similar statement for the case where $\mathcal{G}(x,t,\fset{X},\fset{Y})$ decays polynomially with respect to $x$.
We have utilized the following proposition in deriving the inequalities~\eqref{ineq:1_last_inequality} and \eqref{ineq:2_last_inequality} in the proof of Theorem~\ref{thm:long-range Lieb-Robinson bound}:
\begin{prop} \label{lemma:Connection of local _ global Lieb-Robinson bounds_long_range}
We consider a Hamiltonian $H_0$ which satisfies the $\mathcal{G}(x,t,\fset{X},\fset{Y})$-Lieb-Robinson bound in the form of
\begin{align}
\mathcal{G}(x,t,\fset{X},\fset{Y})= \mathcal{F}(t,\fset{X},\fset{Y}) \fset{L}(x) \quad {\rm with} \quad 
\fset{L}(x):=\frac{\log^p(x+1)}{(x-\kappa\xi)^{-\alpha_0}}  \quad (\alpha_0>D+1)
\label{Lieb_Robinson_O_fsetX_assump_long}
\end{align}
for $\forall \fset{X} \subseteq \Lambda$ without any restrictions and $\forall \fset{Y} \subseteq \Lambda$ with $\diam(\fset{Y}) \le 2\xi+1$, where we assume $\mathcal{F}(t,\fset{X},\fset{Y}) = \mathcal{F}(-t,\fset{X},\fset{Y}) $.
Then, for arbitrary two subsets $L,L'\subseteq \Lambda$, we obtain the Lieb-Robinson function $\mathcal{G}(x,t,L,L')$ which is given by
\begin{align}
&\mathcal{G}(x,t,L,L')  = 2\bigl| L'^{(\xi)}\bigr| \tilde{\mathcal{F}}(t,L)\fset{L}(x)  ,  \quad x=\dist_{L,L'},
  \label{Lieb_L_L'_from_local_one_lemma_long_0} 
\end{align}
or 
\begin{align}
&\mathcal{G}(x,t,L,L')  =
  \mathcal{C}'_{\kappa,p,\alpha_0} \abb L^{(\xi)}\abb \tilde{\mathcal{F}}(t,L) \xi^{-D}\frac{\log^p(x+1) }{[x-(4+\kappa)\xi]^{\alpha_0-D}} ,  \quad x=\dist_{L,L'}, \label{Lieb_L_L'_from_local_one_lemma_long} 
\end{align}
where we assume $x\ge (\kappa+5)\xi$, and define $\tilde{\mathcal{F}}(t,L)$ and $\mathcal{C}'_{\kappa,p,\alpha_0}$ as
\begin{align}
&\tilde{\mathcal{F}}(t,L):= \sup_{\substack{\fset{Y}\subseteq \Lambda \\ \diam(\fset{Y})\le 2\xi+1}}[\mathcal{F}(t,L,\fset{Y})] ,
 \label{Lieb_Robinson_O_i_tildeF_long} \\
& \mathcal{C}'_{\kappa,p,\alpha_0}:= \tilde{C}_{\kappa,p,\alpha_0} 2^{D+2} \gamma D (\kappa+3)^{\alpha_0-1}, \quad \tilde{C}_{\kappa,p,\alpha_0}:=\sup_{z\in \mathbb{R}| z\ge \kappa+3}
 \left(\frac{z^{D+1} [2\log(z) +1]^p }{(z-\kappa-2)^{\alpha_0}}   \right).
\label{def_mathcal_C'_0_theorem}
\end{align}
Note that we here do not impose any restrictions on $L$ and $L'$.
\end{prop}

\subsection{Proof of Theorem~\ref{lemma:Connection of local _ global Lieb-Robinson bounds_2}}
We first prove the inequality~\eqref{Lieb_L_X_from_local_one_lemma}. For the proof, we need to consider the norm of
 \begin{align}
\| [ O_X(H_0,t), O_{L}]\|  \quad (\diam(X)\le 2\xi+1) ,
\end{align}
where $O_X$ and $O_L$ are arbitrary operators with the unit norm (i.e., $\|O_X\|=\|O_L\|=1$) supported on $X$ and $L$, respectively.
Now, we do not assume any constraints on the subset $L$. 
Because $O_X(H_0,t,L^\co) $ is supported on $L^\co$ from the definition~\eqref{def:O_X_local_approx}, we have $ [O_X(H_0,t,L^\co), O_{L}]=0$, and hence 
 we have
 \begin{align}
\| [ O_X(H_0,t), O_{L}]\| &= \| [O_X(H_0,t)- O_L(H_0,t,L^\co), O_{L}] + [O_X(H_0,t,L^\co), O_{L}] \| \notag \\
&= \| [O_X(H_0,t)- O_X(H_0,t,L^\co), O_{L}]  \|  \notag \\
&\le 2 \|  O_X(H_0,t)- O_X(H_0,t,L^\co) \|, \label{Lieb_L_L'_from_local_one_1st}
\end{align}
where we use $\|O_L\|=1$ in the last inequality.
The partial trace with respect to an arbitrary subset $X_0\subseteq \Lambda$ is given by
 \begin{align}
\frac{1}{\tr_{X_0}(\hat{1}_{X_0} )} \tr_{X_0} (O)  \otimes \hat{1}_{X_0}=   \int d\mu(U_{X_0})  U_{X_0}^\dagger O U_{X_0}
\end{align}
for an arbitrary operator $O$, where $U_{X_0}$ is a unitary operator acting on ${X_0}$ and $\mu(U_{X_0})$ be the Haar measure for $U_{X_0}$. 

We here estimate an upper bound of the norm of $\| O_X(H_0,t)- O_X(H_0,t,L^\co) \|$, which immediately gives the upper bound of $\| [ O_X(H_0,t), O_{L}]\| $ by Ineq.~\eqref{Lieb_L_L'_from_local_one_1st}.
For the purpose, we consider the coarse grained subset of $L$ [see Eq.~\eqref{def:coarse_graining_X^(r)}].
We here denote $L^{(\xi)}$ by  $\{j_s\}_{s=1}^{n_L} \subseteq \Lambda^{(\xi)}$, where we define $n_L:=\abb L^{(\xi)}\abb$.
We then define the following subsets $\{L_s\}_{s=1}^{n_L}$ recursively: 
\begin{align}
&L_1:=  \bal{j_1}{\xi} \cap L \for s=1 ,\notag \\
&L_s :=(\bal{j_s}{\xi} \cap L)\setminus (L_1\cup L_2 \cup\cdots \cup L_{s-1}) \for  s\ge 2. \label{Defi:L_i_Bal}
\end{align}
Note that for arbitrary $s\neq s'$ we have $L_s \cap L_{s'}=\emptyset$.
From the definition~\eqref{def:coarse_graining_X^(r)},  we have $\bal{L^{(\xi)}}{\xi} \supseteq L$ and hence
 \begin{align}
\bigcup_{s=1}^{n_L} L_s =  \bal{L^{(\xi)}}{\xi} \cap L   = L.
\label{properties_of_L_i_X_0}
\end{align}
Also, we notice that
 \begin{align}
&\dist_{X,L_s}\le \dist_{X,L} ,\quad  \diam(L_s) \le \diam(\bal{j_s}{\xi} ) \le 2 \xi+1 .
\label{properties_of_L_i_X}
\end{align}

By using the notations of \eqref{Defi:L_i_Bal}, we obtain 
 \begin{align}
O_X(H_0,t,L^\co) =\frac{1}{\tr_{L}(\hat{1}_{L} )} \tr_{L} [O_X(H_0,t)]  \otimes \hat{1}_{L}
= \int d\mu(U_{L_1}) \int d\mu(U_{L_2}) \cdots \int d\mu(U_{L_{n_L}})  U_{L}^\dagger O_X(H_0,t) U_{L},
\label{partial_trace_first_O_x_h_0_L_co}
\end{align}
where we define $U_{L}:= \prod_{s=1}^{n_L} U_{L_s}$ with $U_{L_s}=\hat{1}$ for $L_s=\emptyset$.  
From the assumption of $\diam(X)\le 2\xi+1$ and the inequality~\eqref{properties_of_L_i_X}, we can apply the Lieb-Robinson bound~\eqref{Lieb_Robinson_O_fsetX_assump} to $\left \|[O_X(H_0,t),   U_{L_s}] \right\|$ as follows:
 \begin{align}
 \left \|[O_X(H_0,t),   U_{L_s}] \right\| ) \le \mathcal{F}(t,X, L_s) \fset{L}(\dist_{X,L_s}) , 
\end{align}
where $ U_{L_s}$ can be an arbitrary unitary operator.
Therefore, we calculate 
 \begin{align}
\| O_X(H_0,t)- O_X(H_0,t,L^\co)) \| &\le \int d\mu(U_{L_1}) \int d\mu(U_{L_2}) \cdots \int d\mu(U_{L_{n_L}}) \left \|[O_X(H_0,t),   U_L   ] \right\| \notag \\
&\le \sum_{s=1}^{n_L} \sup_{U_{L_s}} ( \left \|[O_X(H_0,t),   U_{L_s}] \right\| ) \notag \\
&\le \sum_{s=1}^{n_L} \mathcal{F}(t,X, L_s) \fset{L}(\dist_{X,L_s})  \notag \\
&\le n_L  \tilde{\mathcal{F}}(t)\fset{L}(\dist_{X,L}) = \abb L^{(\xi)}\abb \tilde{\mathcal{F}}(t)\fset{L}(\dist_{X,L})  ,
\end{align}
where we use the definition~\eqref{Lieb_Robinson_O_i_tildeF} of $\tilde{\mathcal{F}}(t)$ and 
$\fset{L}(\dist_{X,L_s}) \le\fset{L}(\dist_{X,L})$ for $\dist_{X,L_s}\ge\dist_{X,L}$ in the fourth inequality.
By applying the above inequality to \eqref{Lieb_L_L'_from_local_one_1st}, we prove the inequality~\eqref{Lieb_L_X_from_local_one_lemma}.

We then prove the second inequality~\eqref{Lieb_L_L'_from_local_one_lemma}. 
For the proof, we need to estimate the norm of 
 \begin{align}
\| [ O_L(H_0,t), O_{L'}]\|,
\end{align}
where $O_L$ and $O_{L'}$ are arbitrary operators supported on $L,L'\subseteq \Lambda$ with  $\|O_L\|=\|O_{L'}\|=1$.
In order to estimate the commutator norm, we calculate
 \begin{align}
&\| O_L(H_0,t)- O_L(H_0,t,\bal{L}{\dist_{L,L'}-1}) \| ,
\end{align}
which gives the upper bound of
 \begin{align}
\| [ O_L(H_0,t), O_{L'}]\| &= \| [O_L(H_0,t)- O_L(H_0,t,\bal{L}{\dist_{L,L'}-1}), O_{L'}] + [O_L(H_0,t,\bal{L}{\dist_{L,L'}-1}), O_{L'}] \| \notag \\
&= \| [O_L(H_0,t)- O_L(H_0,t,\bal{L}{\dist_{L,L'}-1}), O_{L'}] \| \notag \\
&\le 2 \| O_L(H_0,t)- O_L(H_0,t,\bal{L}{\dist_{L,L'}-1})\|. \label{Lieb_L_L'_from_local_one}
\end{align}

In the following, we estimate an upper bound of the norm of $\| O_L(H_0,t)- O_L(H_0,t,\bal{L}{R}) \|$ for an arbitrary $R$.
We will set $R=\dist_{L,L'}-1$ (or $R+1=\dist_{L,L'}$) afterward.  
Similar to the definition~\eqref{Defi:L_i_Bal},  we consider a set of spins $(\bal{L}{R}^\co)^{(\xi)}:= \{j_s\}_{s=1}^{\tilde{n}} \subseteq \Lambda^{(\xi)}$ with $\tilde{n}=|(\bal{L}{R}^\co)^{(\xi)}|$,  
and define a set of $\{X_s\}_{s=1}^{\tilde{n}}$ as follows: 
 \begin{align}
X_s:=(\bal{j_s}{\xi} \cap \bal{L}{R}^\co) \setminus (X_1\cup X_2 \cup\cdots \cup X_{s-1}) .
\label{decomposition_of_total_xi}
\end{align}
We obtain the similar relations to \eqref{properties_of_L_i_X_0} and \eqref{properties_of_L_i_X} as follows:
 \begin{align}
\bigcup_{s=1}^{\tilde{n}} X_s =   \bal{L}{R}^\co
,\quad \dist_{X_s,L}\ge R+1  ,\quad  {\rm and} \quad   \diam(X_s) \le \diam(\bal{j_s}{\xi} ) \le 2 \xi+1  ,
\end{align}
where we use $\dist_{X_s,L}\ge \dist_{\bal{L}{R}^\co,L}=R+1$ in the second inequality.
By following the derivation of Eq.~\eqref{partial_trace_first_O_x_h_0_L_co}, the partial trace with respect to $\bal{L}{R}^\co$ is given by 
 \begin{align}
O_L(H_0,t,\bal{L}{R}) =\int d\mu(U_{L_1}) \int d\mu(U_{L_2}) \cdots \int d\mu(U_{L_{\tilde{n}}})  U_{\bal{L}{R}^\co}^\dagger O_L(H_0,t) U_{\bal{L}{R}^\co},
\end{align}
where we define $U_{\bal{L}{R}^\co}:= \prod_{s=1}^{\tilde{n}} U_{X_s}$. 
Because of $\diam(X_s) \le 2 \xi+1$, we can apply the inequality~\eqref{Lieb_L_X_from_local_one_lemma} to $\|[O_L(H_0,t),   U_{X_s}] \|$, which yields
 \begin{align}
\left \|[O_L(H_0,t),   U_{X_s}] \right\|=\left \|[ U_{X_s}(H_0,-t),O_L] \right\|  \le 2 \bigl| L^{(\xi)}\bigr| \tilde{\mathcal{F}}(t)  \fset{L}(\dist_{X_s,L}) ,
\end{align}
and hence
 \begin{align}
\| O_L(H_0,t)- O_L(H_0,t,\bal{L}{R}) \| &\le\int d\mu(U_{L_1}) \int d\mu(U_{L_2}) \cdots \int d\mu(U_{L_{\tilde{n}}}) \left \|[O_L(H_0,t),   U_{\bal{L}{R}^\co}    ] \right\| \notag \\
&\le \sum_{s=1}^{\tilde{n}}  \sup_{U_{X_s}} ( \left \|[O_L(H_0,t),   U_{X_s}] \right\| ) \notag \\
&\le2  \tilde{\mathcal{F}}(t) \sum_{s=1}^{\tilde{n}}  \bigl| L^{(\xi)}\bigr| \fset{L}(\dist_{L,X_s})  . \label{L_bal_L_R_ineq_1}
\end{align}
Thus, the remaining task is to estimate the summation with respect to $s$.

By using the relation $L^{(\xi)}[\xi] \supseteq L$ and the definition of $\fset{L}(x):=e^{-x/\xi_0}$, we obtain 
 \begin{align}
\dist_{L,X_s}\ge \dist_{\bal{L^{(\xi)}}{\xi},X_s}  \quad {\rm  and} \quad  
\fset{L}\left(\dist_{\bal{L^{(\xi)}}{\xi},X_s} \right) \le \max_{i\in L^{(\xi)}}\left[ \fset{L}\left(\dist_{\bal{i}{\xi},X_s} \right) \right]\le \sum_{i\in L^{(\xi)}} \fset{L}\left(\dist_{\bal{i}{\xi},X_s} \right)   , \label{L_bal_L_R_ineq_002}
\end{align}
which yields 
 \begin{align}
\sum_{s=1}^{\tilde{n}}  \fset{L}\left(\dist_{L,X_s}\right)   \le \sum_{s=1}^{\tilde{n}} \sum_{i\in L^{(\xi)}}  
\fset{L}\left(\dist_{\bal{i}{\xi},X_s} \right)  . \label{L_bal_L_R_ineq_2}
\end{align}
Moreover, the definition~\eqref{decomposition_of_total_xi} implies  $X_s \subseteq \bal{j_s}{\xi}$ and we obtain for $i\in L^{(\xi)}$
  \begin{align}
&\dist_{\bal{i}{\xi},X_s} \ge \dist_{\bal{i}{\xi},\bal{j_s}{\xi}} \ge \dist_{i,j_s} -2 \xi .
\label{dist_ba_i_xi_X_s}
\end{align}
We therefore obtain 
 \begin{align}
 \sum_{s=1}^{\tilde{n}} \sum_{i\in L^{(\xi)}}\fset{L}\left(\dist_{\bal{i}{\xi},X_s}\right)
 \le   \sum_{s=1}^{\tilde{n}} \sum_{i\in L^{(\xi)}} \fset{L}\left(\dist_{i,j_s} -2 \xi \right)
  &=\sum_{i\in L^{(\xi)}} \sum_{j\in (\bal{L}{R}^{\co})^{(\xi)}}\fset{L}\left(\dist_{i,j} -2 \xi \right)   , \label{L_bal_L_R_ineq_2.5}
\end{align}
where in the equation we use the definition of $\{j_s\}_{s=1}^{\tilde{n}}:=(\bal{L}{R}^\co)^{(\xi)}$.
The conditions $i\in L^{(\xi)}$, $j\in (\bal{L}{R}^{\co})^{(\xi)}$\ and $\dist_{L,\bal{L}{R}^\co}=R+1$ imply 
  \begin{align}
\dist_{i,j} \ge R+1 -2 \xi ,
\label{dist_ba_i_js_2}
\end{align}
which yields
 \begin{align}
{\rm for}\quad  \forall i\in L^{(\xi)}, \quad \sum_{j\in (\bal{L}{R}^{\co})^{(\xi)}}\fset{L}\left(\dist_{i,j} -2 \xi \right)  \le \sum_{j\in \Lambda^{(\xi)}: \dist_{i,j} \ge R+1 -2 \xi} \fset{L}\left(\dist_{i,j} -2 \xi \right) .
\end{align}
By applying the above inequality to \eqref{L_bal_L_R_ineq_2.5}, we obtain 
 \begin{align}
 \sum_{s=1}^{\tilde{n}} \sum_{i\in L^{(\xi)}}\fset{L}\left(\dist_{\bal{i}{\xi},X_s}\right)
 \le    \sum_{i\in L^{(\xi)}} \sum_{j\in \Lambda^{(\xi)}: \dist_{i,j} \ge R+1 -2 \xi} \fset{L}\left(\dist_{i,j} -2 \xi \right)  . \label{L_bal_L_R_ineq_3}
\end{align}

By using the inequality~\eqref{def:parameter_gamma}, the summation with respect to $j$ is bounded from above by
 \begin{align}
\sum_{j\in \Lambda^{(\xi)}: \dist_{i,j} \ge R+1 -2 \xi}  \fset{L}\left(\dist_{i,j} -2 \xi \right) 
&\le \sum_{s=0}^\infty \sum_{\substack{j\in \Lambda^{(\xi)} \\  R+1 -2 \xi +s\xi\le \dist_{i,j} < R+1 -2 \xi  +(s+1)\xi}} 
  \fset{L}\left(\dist_{i,j} -2 \xi \right)  \notag \\ 
&\le 2\gamma D\sum_{s=0}^\infty \fset{L}\left(R+1 -4 \xi +s\xi\right)   [2(R+1 -2 \xi +s\xi)/\xi]^{D-1}.
\label{general_form_summation_fset_L_s0}
\end{align}
The form of $\fset{L}(x)$ is now given by
 \begin{align}
\fset{L}(x)= e^{-x/\xi_0} ,
\end{align}
and hence we obtain
 \begin{align}
 &2\gamma D \sum_{s=0}^\infty \fset{L}\left(R+1 -4 \xi +s\xi\right)   [2(R+1 -2 \xi +s\xi)/\xi]^{D-1}\notag \\
&\le 2^D\gamma D e^{-(R+1-4 \xi) /\xi_0}  \int_{0}^\infty  e^{- (x-1)\xi/\xi_0}  [(R+1-2 \xi )/\xi + x ]^{D-1} dx \notag \\
&\le 2^D\gamma D  e^{5 \xi /\xi_0}(D-1)! (\xi_0/\xi)  [(R+1-2 \xi +\xi_0)/\xi ]^{D-1}  e^{-(R+1) /\xi_0}  \notag \\
&\le 2^D D! \gamma  (1+\xi_0/\xi)^D  e^{5 \xi/\xi_0}  [1+(R+1)/\xi]^{D-1} e^{-(R+1) /\xi_0} =(\mathcal{C}_{\xi,\xi_0}/4)  [1+ (R+1)/\xi]^{D-1} e^{-(R+1) /\xi_0}, \label{L_bal_L_R_ineq_4}
\end{align}
where we use the inequalities 
 \begin{align}
\int_{0}^\infty (x+x_0)^{D-1} e^{-x/x_1}  dx \le (D-1)! x_1 (x_1+x_0)^{D-1}  \label{integral_convenient_1}
\end{align}
and
 \begin{align}
(R+1-2 \xi +\xi_0)/\xi =[(R+1+\xi)/\xi] \cdot [1+(-3 \xi +\xi_0)/(R+1+\xi)] \le[ 1+ (R+1)/\xi] \cdot (1+\xi_0/\xi).
\end{align}
Also, the definition of $\mathcal{C}_{\xi,\xi_0}$ has been given in Eq.~\eqref{def_mathcal_C_0_theorem}.
We thus arrive at the inequality of 
 \begin{align}
\sum_{j\in \Lambda^{(\xi)}: \dist_{i,j} \ge R+1 -2 \xi}  \fset{L}\left(\dist_{i,j} -2 \xi \right) 
&\le (\mathcal{C}_{\xi,\xi_0}/4)  [1+(R+1)/\xi]^{D-1} e^{-(R+1) /\xi_0} .
\label{general_form_summation_fset_L_s}
\end{align}

By combining the inequalities~\eqref{L_bal_L_R_ineq_2},~\eqref{L_bal_L_R_ineq_3} and \eqref{general_form_summation_fset_L_s}, we obtain
 \begin{align}
\sum_{s=1}^{\tilde{n}}  \fset{L}\left(\dist_{L,X_s}\right)   &\le \sum_{i\in L^{(\xi)}} (\mathcal{C}_{\xi,\xi_0}/4) [1+(R+1)/\xi]^{D-1}e^{-(R+1) /\xi_0} \notag \\
&= \abb L^{(\xi)}\abb (\mathcal{C}_{\xi,\xi_0}/4) [1+(R+1)/\xi]^{D-1}e^{-(R+1) /\xi_0} .
\end{align} 
By applying the above inequality to~\eqref{L_bal_L_R_ineq_1}, we obtain the upper bound of $\| O_L(H_0,t)- O_L(H_0,t,\bal{L}{R}) \| $.
We combine the upper bound of $\| O_L(H_0,t)- O_L(H_0,t,\bal{L}{R}) \| $ with the inequality \eqref{Lieb_L_L'_from_local_one} by choosing $R=\dist_{L,L'}-1$.
Then, we arrive at the inequality~\eqref{Lieb_L_L'_from_local_one_lemma}.
This completes the proof. $\square$

\subsection{Proof of Proposition~\ref{lemma:Connection of local _ global Lieb-Robinson bounds_long_range}}
First of all, 
because the proof of \eqref{Lieb_L_X_from_local_one_lemma} does not rely on $\diam(\fset{X})$ and the form of $\fset{L}(x)$, 
we obtain the inequality~\eqref{Lieb_L_L'_from_local_one_lemma_long_0} in exactly the same way.

Our task is to prove the inequality~\eqref{Lieb_L_L'_from_local_one_lemma_long}.
We start from the inequality~\eqref{Lieb_L_L'_from_local_one}:
 \begin{align}
\| [ O_L(H_0,t), O_{L'}]\| & \le 2 \| O_L(H_0,t)- O_L(H_0,t,\bal{L}{\dist_{L,L'}-1})\|. \label{Lieb_L_L'_from_local_one_long_range}
\end{align}
We obtain the same inequality as \eqref {L_bal_L_R_ineq_1} for an arbitrary integer $R$ as follows:
\begin{align}
\| O_L(H_0,t)- O_L(H_0,t,\bal{L}{R}) \| &\le  \int d\mu(U_{X_1})  \int d\mu(U_{X_2})\cdots   \int d\mu(U_{X_{\tilde{n}}}) \left \|[O_L(H_0,t),   U_{\bal{L}{R}^\co}    ] \right\| \notag \\
&\le \sum_{s=1}^{\tilde{n}}  \sup_{U_{X_s}} ( \left \|[O_L(H_0,t),   U_{X_s}] \right\| ) \notag \\
&\le \sum_{s=1}^{\tilde{n}}  \mathcal{F}(t,L,X_s)  \fset{L}(\dist_{L,X_s}) 
\le \tilde{\mathcal{F}}(t,L) \sum_{s=1}^{\tilde{n}}\fset{L}(\dist_{L,X_s})  , \label{L_bal_L_R_ineq_long_range}
\end{align}
where $\tilde{n}=|(\bal{L}{R}^\co)^{(\xi)}|$, the third inequality is derived from the Lieb-Robinson bound~\eqref{Lieb_Robinson_O_fsetX_assump_long}, and the last inequality is derived from the definition of $\tilde{\mathcal{F}}(t,L)$ in Eq.~\eqref{Lieb_Robinson_O_i_tildeF_long}.

For the summation of $\fset{L}(\dist_{L,X_s})$, we can utilize the same inequality as~\eqref{L_bal_L_R_ineq_2} and \eqref{L_bal_L_R_ineq_3}.
We then obtain 
 \begin{align}
\sum_{s=1}^{\tilde{n}}  \fset{L}\left(\dist_{L,X_s}\right)  & \le    \sum_{i\in L^{(\xi)}} \sum_{j\in \Lambda^{(\xi)}: \dist_{i,j} \ge R+1 -2 \xi} \fset{L}\left(\dist_{i,j} -2 \xi \right)  .
\label{L_bal_L_R_ineq_long_range_2nd}
\end{align}
In order to estimate the summation with respect to $j$, we utilize the following lemma (see Sec.~\ref{summation_discrete_f_Proofsec}):
\begin{lemma} \label{summation_discrete_f}
Let $f(z)$ be an arbitrary function and $C_{f,x_0}$ $(x_0\ge \xi)$ be a constant such that 
 \begin{align}
C_{f,x_0} = \sup_{z\in \mathbb{R}| z\ge x_0} \left(\frac{z^{D+1} f(z)}{x_0^{D+1} f(x_0)} \right). \label{definition_C_f_x_0}
\end{align}
Then, for an arbitrary $i\in \Lambda$, we obtain the following upper bound: 
 \begin{align}
\sum_{j\in \Lambda^{(\xi)}:\dist(i,j) \ge x_0} f(\dist_{i,j})  \le  2^{D+1} C_{f,x_0}  \gamma D \xi^{-D}x_0^{D}  f(x_0). \label{summation_discrete_f_main_ineq}
\end{align}
\end{lemma}
\noindent
In the following, we set 
 \begin{align}
&x_0= R+1 -2 \xi ,\quad f(z) =\fset{L}\left(z-2 \xi \right) = \frac{\log^p(z-2\xi+1) }{[z-(2+\kappa)\xi]^{\alpha_0}},  \notag \\
&C_{f,x_0} = \sup_{z\in \mathbb{R}| z\ge R+1 -2 \xi}
 \left(\frac{z^{D+1}}{(R+1 -2 \xi)^{D+1}}\cdot \frac{\fset{L}\left(z -2 \xi \right)}{\fset{L}\left(R+1 -4 \xi \right)}  \right)
\end{align}
in the above lemma.  
Note that $x_0\ge \xi$ is satisfied from the condition of $R\ge (\kappa+5)\xi$.
We then obtain from the inequality~\eqref{summation_discrete_f_main_ineq}
 \begin{align}
 \sum_{j\in \Lambda^{(\xi)}: \dist_{i,j} \ge R+1 -2 \xi} \fset{L}\left(\dist_{i,j} -2 \xi \right)   
 &\le  2^{D+1}\xi^{-D} C_{f,x_0}  \gamma D  (R+1 -2 \xi)^{D} \frac{\log^p(R-4\xi+2) }{[R+1-(4+\kappa)\xi]^{\alpha_0}}  \notag \\
 &\le 2^{D+1}\xi^{-D} C_{f,x_0}C'_{R,\kappa,D}  \gamma D   \frac{\log^p(R+2) }{[R+1-(4+\kappa)\xi]^{\alpha_0-D}} ,
  \label{summation_discrete_f_main_ineq_use}
\end{align}
where we define 
 \begin{align}
C'_{R,\kappa,D} := \frac{ (R+1 -2 \xi)^D }{[R+1 -(4+\kappa) \xi]^D }.
\end{align}

We then discuss the values of $C'_{R,\kappa,D}$ and $C_{f,x_0}$ in more details.
First, from the condition of $R\ge (\kappa+5)\xi$ in this proposition, we have
 \begin{align}
C'_{R,\kappa,D} \le \frac{ (\kappa\xi+3\xi +1)^D }{(\xi+1)^D }\le (\kappa+3)^D. \label{upp_tildeCRrD}
\end{align}
Note that for $x\le y$ we have $(y/x)^D \le [(y-\delta)/(x-\delta)]^D$ ($\delta >0$).
Second, we notice that $x_0=R+1 -2 \xi  \ge (\kappa+3)\xi $ due to $R\ge (\kappa+5)\xi$.
Also, because $C_{f,x_0}$ monotonically decreases with $x_0$, we have  
 \begin{align}
C_{f,x_0} \le C_{f,(\kappa+3)\xi} 
&= \sup_{z\in \mathbb{R}| z\ge (\kappa+3)\xi}
 \left(\frac{z^{D+1}}{[(\kappa+3)\xi]^{D+1}}\cdot \frac{\fset{L}\left(z -2 \xi \right)}{\fset{L}\left(\kappa\xi+\xi \right)}  \right) \notag \\
&=\sup_{z\in \mathbb{R}| z\ge (\kappa+3)\xi}
 \left(\frac{z^{D+1}}{[(\kappa+3)\xi]^{D+1}}\cdot \frac{\log^p(z-2\xi+1)}{[z-(\kappa+2)\xi]^{\alpha_0}} \cdot \frac{[(\kappa+1)\xi]^{\alpha_0}}{\log^p(\kappa\xi+\xi+1)}  \right)  \notag \\
 &\le (\kappa+3)^{\alpha_0-D-1}\sup_{z\in \mathbb{R}| z\ge (\kappa+3)\xi}
 \left(\frac{(z/\xi)^{D+1} \log^p(z-2\xi+1) }{[z/\xi-\kappa-2]^{\alpha_0}\log^p(\kappa\xi+\xi+1)}   \right)  .
\end{align}
Then, by upper-bounding
 \begin{align}
\frac{\log(z-2\xi+1) }{\log(\kappa\xi+\xi+1)} \le \frac{\log(z/\xi-2+1/\xi) +\log(\xi) }{\log(\xi+1)}\le \frac{\log(z/\xi)}{\log(2)}+1 \le 2\log(z/\xi) +1,
\label{logarithmic_upper_bound_r_z_2xi}
\end{align}
we have 
 \begin{align}
C_{f,x_0} \le C_{f,(\kappa+3)\xi} 
\le (\kappa+3)^{\alpha_0-D-1}\sup_{z\in \mathbb{R}| z\ge \kappa+3}
 \left(\frac{z^{D+1} [2\log(z) +1]^p }{(z-\kappa-2)^{\alpha_0}}   \right)=: (\kappa+3)^{\alpha_0-D-1} \tilde{C}_{\kappa,p,\alpha_0}. \label{upp_C_f_x_0}
\end{align}

By applying the inequalities~\eqref{upp_tildeCRrD} and \eqref{upp_C_f_x_0} to the inequality~\eqref{summation_discrete_f_main_ineq_use}, 
we finally obtain
 \begin{align}
 \sum_{j\in \Lambda^{(\xi)}: \dist_{i,j} \ge R+1 -2 \xi} \fset{L}\left(\dist_{i,j} -2 \xi \right)   
 &\le   \tilde{C}_{\kappa,p,\alpha_0} 2^{D+1}\xi^{-D} \gamma D (\kappa+3)^{\alpha_0-1} \frac{\log^p(R+2) }{[R+1-(4+\kappa)\xi]^{\alpha_0-D}}.
 \end{align}
By combining the above inequality with \eqref{L_bal_L_R_ineq_long_range_2nd}, we have
 \begin{align}
\sum_{s=1}^{\tilde{n}}  \fset{L}\left(\dist_{L,X_s}\right)  & \le   \abb L^{(\xi)}\abb  \tilde{C}_{\kappa,p,\alpha_0} 2^{D+1}\xi^{-D} \gamma D (\kappa+3)^{\alpha_0-1} \frac{\log^p(R+2) }{[R+1-(4+\kappa)\xi]^{\alpha_0-D}} .
\end{align}
The above inequality with \eqref{L_bal_L_R_ineq_long_range} reduces the inequality \eqref{Lieb_L_L'_from_local_one_long_range} to the main inequality~\eqref{Lieb_L_L'_from_local_one_lemma_long} by choosing $R=\dist_{L,L'}-1$.
This completes the proof. $\square$

\subsubsection{Proof of Lemma~\ref{summation_discrete_f}}  \label{summation_discrete_f_Proofsec}

From the definition of $C_{f,x_0}$, the following summation with respect to $j$ is bounded from above by
 \begin{align}
\sum_{j\in \Lambda^{(\xi)}: \dist_{i,j} \ge x_0 }  f(\dist_{i,j}) 
=f(x_0) x_0^{D+1} \sum_{j\in \Lambda^{(\xi)}: \dist_{i,j} \ge x_0 }  \frac{f(\dist_{i,j}) \dist_{i,j}^{D+1} }{f(x_0) x_0^{D+1}}  \dist_{i,j}^{-D-1}
&\le C_{f,x_0} f(x_0) x_0^{D+1} \sum_{j\in \Lambda^{(\xi)}: \dist_{i,j} \ge x_0 }  \dist_{i,j}^{-D-1} ,
\label{summation_discrete_f_pf_ineq1}
\end{align}
where $i$ is an arbitrary site ($i\in \Lambda$).
Then, by using the inequality~\eqref{def:parameter_gamma}, we obtain 
 \begin{align}
\sum_{j\in \Lambda^{(\xi)}: \dist_{i,j} \ge x_0 }  \dist_{i,j}^{-D-1} 
&\le \sum_{s=0}^\infty \sum_{\substack{j\in \Lambda^{(\xi)} \\ x_0+s\xi\le \dist_{i,j} < x_0+(s+1)\xi}} 
\dist_{i,j}^{-D-1}  \notag \\ 
&\le 2\gamma D\sum_{s=0}^\infty \left( x_0 +s\xi\right)^{-D-1}   [2(x_0+s\xi)/\xi]^{D-1} \notag \\
&\le 2^D \gamma D \xi^{-D+1}\sum_{s=0}^\infty \left( x_0 +s\xi\right)^{-2}\le  2^{D+1} \gamma D \xi^{-D}x_0^{-1},
\label{summation_discrete_f_pf_ineq2}
\end{align}
where the last inequality is derived from
 \begin{align}
\sum_{s=0}^\infty \left( x_0 +s\xi\right)^{-2} &= x_0^{-2} + \sum_{s=1}^\infty \left( x_0 +s\xi\right)^{-2}\le x_0^{-2}  + \int_0^\infty \left( x_0 +x\xi\right)^{-2} dx  = x_0^{-2} + (\xi x_0)^{-1} \le 2 (\xi x_0)^{-1}.
\end{align} 
Note that $x_0\ge \xi$ from the assumption.
By combining the inequalities~\eqref{summation_discrete_f_pf_ineq1} and \eqref{summation_discrete_f_pf_ineq2}, we obtain the inequality~\eqref{summation_discrete_f_main_ineq}.
This completes the proof. $\square$

\section{Proof of Theorem~\ref{thm:Middle-range Lieb-Robinson bound}: Middle-range Lieb-Robinson bound} \label{sec: Proof of the Middle-range Lieb-Robinson bound}

\subsection{statement}
{~}\\
{\bf Theorem~\ref{thm:Middle-range Lieb-Robinson bound}} (Middle-range Lieb-Robinson bound)
\textit{
We consider the Hamiltonian $H$ which satisfies Assumption~\eqref{basic_assump_power} with $\alpha>2D +1$.
Then, for arbitrary $\ell$, the Hamiltonian $H_{\le \ell}$ satisfies the $\mathcal{G}_{\le \ell}(x,t,\fset{X})$-Lieb-Robinson bound with  
\begin{align}
&\mathcal{G}_{\le \ell}(x,t,\fset{X}) = \tilde{\mathcal{C}}_0 \abb\fset{X} \abb^2 (1+x/\ell)^{D-1}  e^{-2(x-v^\ast |t|)/\ell} ,\label{thm_middle_range_Lieb-Robinson_proof}
\end{align}
where $v^\ast$ is a constant which depends only on the parameters $\{D, g_0, \alpha,\gamma\}$ and 
\begin{align} 
\tilde{\mathcal{C}}_0  := \frac{4}{3}e^{10/3} 15^D  D! \gamma   \zeta_2  ,\quad \zeta_2 :=16e^3 \gamma^2 \cdot 45^D. 
\label{def:tilde_mathcal_C_0_proof}
\end{align} 
}

\begin{figure}[bb]
\centering
{
\includegraphics[clip, scale=0.45]{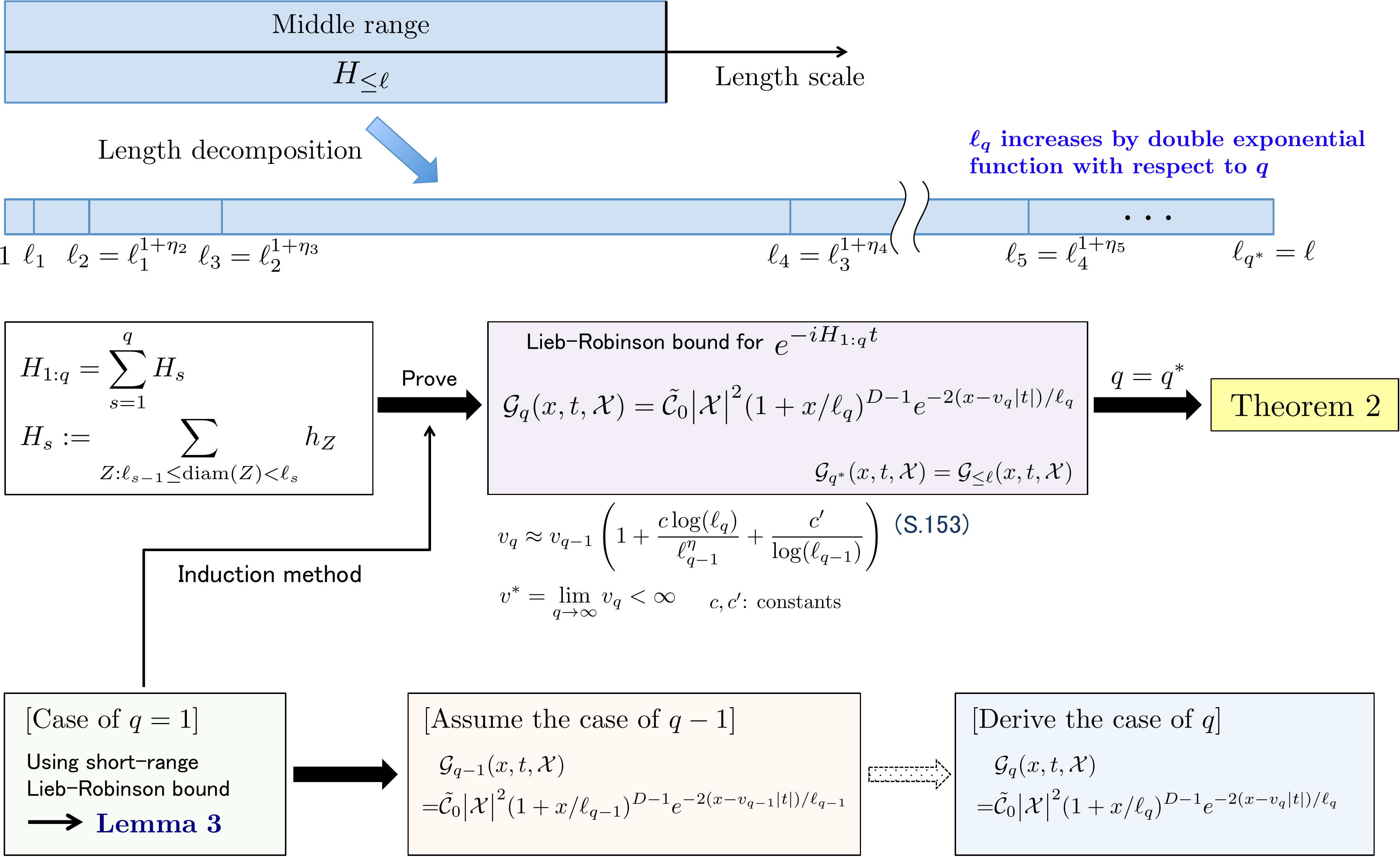}
}
\caption{Outline of the proof on the middle-range Lieb-Robinson bound~1. For the proof, we first decompose the length $\ell$ into $q^\ast$ pieces as $\{\lsc_q\}_{q=1}^{q^\ast}$ with $\ell_{q^\ast}=\ell$.
We choose them so that the length scale $\lsc_q $ increases double exponentially with respect to $q$ [see Ineq.~\eqref{doubly_exponential_ell_q_bound}]. 
Our purpose is to estimate the Lieb-Robinson bound for $e^{-iH_{1:q^\ast}t}$, which is equal to $e^{-iH_{\le \ell}t}$ from the definition~\eqref{def_H_1:_q_middle}.
The main task is to derive the Lieb-Robinson bound for $e^{-iH_{1:q}t}$, which we denote by $\mathcal{G}_{q}(x,t,\fset{X})$, in the form of Eq.~\eqref{LR_func_G_Q_aim_to_prove}
 with \eqref{definitions_v_1_v_q_delta_Fig_used}.
If we can derive the bound, Theorem~\ref{thm:Middle-range Lieb-Robinson bound} is immediately derived by choosing $q=q^\ast$.
In order to prove Eq.~\eqref{LR_func_G_Q_aim_to_prove}, we use the induction method. 
The case of $q=1$ is derived by using the standard technique to obtain the short-range Lieb-Robinson bound (see Lemma~\ref{lemma:Lieb_Robinson_op_quasi_locality_short_range}
with its proof in Sec.~\ref{Sec:Lieb-Robinson bound for Hamiltonian with a finite length scale}).
The derivation of the case $q$ from the assumption of the case $q-1$ is a bit intricate and we give the outline in Fig.~\ref{fig:Outline_middle_range_theorem2}. 
}
\label{fig:Outline_middle_range_theorem1}
\end{figure}

\clearpage

\begin{figure}[tt]
\centering
{
\includegraphics[clip, scale=0.4]{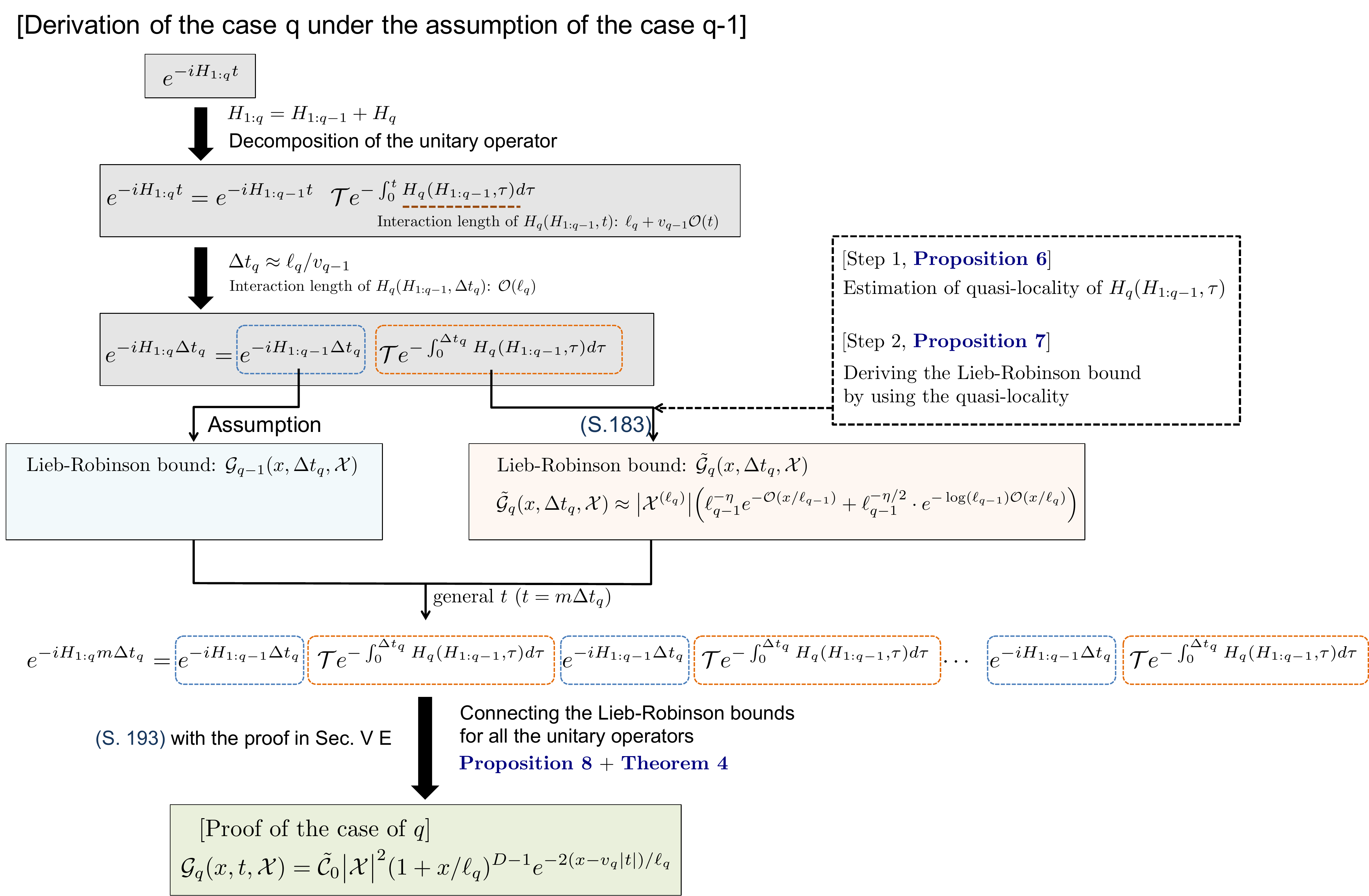}
}
\caption{Outline of the proof on the middle-range Lieb-Robinson bound~2. In order to derive the case $q$ from the assumption~\eqref{LR_func_G_Q-1_start} for the case of $q-1$, we decompose the unitary operator $e^{-iH_{1:q}t}$ as in Eq.~\eqref{time_evolution_decomp_unitary}: $e^{-i H_{1:q-1} t}$ and $\mathcal{T} e^{-\int_0^t H_q (H_{1:q-1}, \tau) d\tau}$.
From the assumption, the unitary operator $e^{-i H_{1:q-1} t}$ satisfies the $\mathcal{G}_{q-1}(x,t,\fset{X})$-Lieb-Robinson bound in the form of Eq.~\eqref{LR_func_G_Q_aim_to_prove}.
Then, the time-evolved Hamiltonian $H_q (H_{1:q-1}, t)$ has the interaction length of $\ell_q + v_{q-1} \orderof{t}$, which is of order of $\orderof{\ell_q}$ for $t\lesssim \ell_q/v_{q-1}$.
We thus consider the time range $t\le \Delta t_q \approx \ell_q/v_{q-1}$ [see Eq.~\eqref{def:delta_t_q} in detail] for the first, and extend the time scale afterward [see Eq.~\eqref{genenral_time_decompostion_unitary}].
In order to obtain the Lieb-Robinson bound for $\mathcal{T} e^{-\int_0^t H_q (H_{1:q-1}, \tau) d\tau}$ for $t\le \Delta t_q$, we take two steps: 
i) in the first step, we estimate the quasi-locality of $H_q (H_{1:q-1}, \tau)$ by using Proposition~\ref{prop:LR_op_quasi_locality} (the proof is given in Sec.~\ref{Sec:Quasi-locality of operator after time-evolution}), ii) in the second step, based on the quasi-locality, we derive the Lieb-Robinson bound~\eqref{Lieb_Robinson_tilde_H_t_practical_2_2_Fig_used} for $\mathcal{T} e^{-\int_0^t H_q (H_{1:q-1}, \tau) d\tau}$ 
by using Proposition~\ref{lemma:Lieb_Robinson_op_quasi_locality} (the proof is given in Sec.~\ref{proof_sec_Lieb-Robinson bound by time-evolved Hamiltonian}).
After obtaining the Lieb-Robinson bounds for $e^{-i H_{1:q-1} t}$ and $\mathcal{T} e^{-\int_0^t H_q (H_{1:q-1}, \tau) d\tau}$ with $t\le \Delta t_q$, 
we extend the time scale to generic $t$. 
In order to connect all the Lieb-Robinson bounds appropriately, we first restrict the subsets $X$ and $Y$ such that  $\diam(\fset{X})  \le \ell_q$ and $\diam(\fset{Y})\le \ell_q$, 
and remove the restrictions afterward by using Theorem~\ref{lemma:Connection of local _ global Lieb-Robinson bounds_2}. 
The connections rely on Proposition~\ref{lemma:Connection of two Lieb-Robinson bounds_1}, which yields the inequality~\eqref{local_Lieb_Robinson_q_Fig_used} (see Sec.~\ref{connect_repeat_essential_ineq} for the proof).
By combining all the ingredients, we derive the Lieb-Robinson bound for the case of $q$. 
}
\label{fig:Outline_middle_range_theorem2}
\end{figure}

\clearpage

\subsection{Decomposition of the length scale}

We define a set of the length scales $\{\lsc_q\}_{q=1}^{q^\ast}$.
First, $\ell_1$ is defined as a constant which depends only on the parameters $\{D, g_0, \alpha,\gamma\}$. 
We show the conditions for $\ell_1$ in Sec.~\ref{Conditions for ell_1}.
Other length scales $\{\lsc_q\}_{q=2}^{q^\ast}$ are defined as
\begin{align}
&\lsc_q :=(\ell_{q-1})^{1+\eta_q}   \quad ( \eta \le \eta_q \le \bar{\eta}) , \label{doubly_exponential_ell_q}   \\
&\bar{\eta}:= \frac{\alpha -(2D +1)}{D+2}, \quad \eta := \sqrt{1+\bar{\eta}} -1 , \label{Def:eta_setup}
\end{align} 
where we choose $\{\eta_q\}_{q=2}^{q^\ast}$ appropriately such that $\{\ell_q\}_{q=2}^{q^\ast}$ become integer and for $q^\ast \ge 2$ the length $\ell_{q^\ast}$ satisfies
\begin{align}
\ell_{q^\ast}= \ell_1^{(1+\eta_2)(1+\eta_3)\cdots (1+\eta_{q^\ast}) } = \ell:=\ell_1^{p_\ell}.
\end{align}
From the condition $\eta \le \eta_q \le \bar{\eta}$ ($1\le q\le q^\ast $), we have 
\begin{align}
\frac{\log(p_\ell)}{\log(1+\bar{\eta})} \le  q^{\ast}-1 \le \frac{\log(p_\ell)}{\log(1+\eta)} = \frac{2\log(p_\ell)}{\log(1+\bar{\eta})}.
\end{align}
Thus, as long as 
\begin{align}
p_\ell \ge 1+ \bar{\eta}\Or \ell \ge \ell_1^{1+\bar{\eta}},
\end{align}
there exists parameters $\{\eta_q\}_{q=2}^{q^\ast}$ such that $\ell_{q^\ast}= \ell$.
From the definition, we note that 
\begin{align}
\ell_{q-1}^{\eta} \le \frac{\ell_q}{\ell_{q-1}} = \ell_{q-1}^{\eta_q} \le \ell_{q-1}^{\bar{\eta}} 
\end{align}
and 
 \begin{align}
\lsc_q  \ge  \ell_1^{(1+\eta)^{q-1}}.\label{doubly_exponential_ell_q_bound}
\end{align} 
Therefore, the length scale $\lsc_q $ increases double exponentially with respect to $q$.

We then define Hamiltonian $H_q$ which picks up all the interaction terms with interaction length from $\lsc_{q-1}$ to $\lsc_{q}$:
\begin{align}
&H_q := \sum_{Z: \lsc_{q-1} \le  \diam(Z) < \lsc_{q}} h_Z \quad  (q=1,2,\ldots, q^\ast -1) , \notag \\
&H_{q^\ast} := \sum_{Z: \lsc_{q^\ast-1} \le  \diam(Z) \le \lsc_{q^\ast }} h_Z ,
\end{align}
where we set $\lsc_0=1$.
From the assumption~\eqref{basic_assump_power}, we obtain 
\begin{align}
\sup_{i \in \Lambda}\sum_{\substack{Z:Z\ni i \\ \lsc_{q-1} \le \diam(Z) < \lsc_{q}} } \| h_Z\| \le 
\sup_{i \in \Lambda}\sum_{\substack{Z:Z\ni i \\   \diam(Z) \ge  \lsc_{q-1}} } \| h_Z\| \le  g_q\label{basic_assump_power_q}
\end{align}
with 
\begin{align}
g_q:= g \lsc_{q-1}^{-\alpha+D}.
\end{align}
From the definition~\eqref{Def:eta_setup}, we have 
\begin{align}
\lsc_q^{D+1} g_q \le  g \lsc_{q-1}^{(D+1)(1+\eta_q)-\alpha+D } = g \lsc_{q-1}^{(D+2)\eta_q - \eta_q -\alpha+2D+1 }
 \le  g \lsc_{q-1}^{-\eta_q}\le  g \lsc_{q-1}^{-\eta} , \label{extensiveness_time_evolve}
\end{align}
where in the second inequality we use $\eta_q \le \bar{\eta}=\frac{\alpha -(2D +1)}{D+2}$.

\subsection{Conditions for $\ell_1$} \label{Conditions for ell_1}

In the proof, we adopt various kinds of conditions for $\ell_1$. 
We choose $\ell_1$ so that the following conditions for $\ell_q$ ($q\ge 1$) are satisfied.  
Some of the conditions include $\ell_q$, but they reduce to lower bounds for $\ell_1$ because of Eq.~\eqref{doubly_exponential_ell_q}.
We note that all the conditions depend only on the parameters $\{D, g_0, \alpha,\gamma\}$.
We summarize all the conditions in the following: 
\begin{enumerate}

\item{} Used in  the inequality~\eqref{bound_on_v_q-1_1}:
\begin{align} 
&\frac{1}{1-\delta_{q-1}}  \le 2,\quad 
{\rm with} \quad \delta_{q-1} = \frac{6\log [ \tilde{\mathcal{C}}_0 \gamma^2 (2\ell_q)^{2D}]+6(D-1) \log (1+\ell_{q-1}^{\eta_q}/12) }{\ell_{q-1}^{\eta_q}} ,
\label{condition1_for_v_q-1_1}
\end{align} 
where $\tilde{\mathcal{C}}_0$ is defined in Eq.~\eqref{def:tilde_mathcal_C_0_proof}.

\item{} Used in  the inequalities~\eqref{ell_q-1_eta_4_gamma_2_result} and \eqref{ell_q-1_eta_4_gamma_2_result_2}:
\begin{align}
\ell_{q-1}^{\eta_q} \ge \ell_{q-1}^{\eta} \ge \ell_1^{\eta} \ge  \max \left( 24  \log(2\gamma^2),48\right)  . \label{ell_q-1_eta_4_gamma_2}
\end{align}

\item{}Used in  the inequality~\eqref{result:eta_condition_ell_1}:
\begin{align} 
 \frac{e^2 g \gamma^2 (20D)^D}{3\ell_1 v_0} \le 1 , \label{eta_condition_ell_1}
\end{align} 
where $v_0= \frac{2 e^3 g \gamma (\alpha-2D)}{\alpha-2D -1}$ as defined in Eq.~\eqref{Lieb-Robinson_velog}.

\item{} Used in the inequality~\eqref{upper_bound_xi_q_mathal_R_q}:
\begin{align} 
\frac{\ell_{q-1} \log ( \zeta_1)}{\xi_q}    \le 1 \quad {\rm or} \quad  \frac{\ell_{q-1}^{\eta}}{\log(\ell_{q-1})} \ge \frac{\eta \log ( \zeta_1) }{16},
\label{zeta_condition_ell_4_connect}
\end{align} 
where $\zeta_1=2^{D+1}e^2 \gamma^2 D!$ and $\xi_q:=\frac{16\ell_q }{\eta \log(\ell_{q-1})}$ as defined in Eq.~\eqref{def:zeta_1} and Eq.~\eqref{def_xi_q_1}, respectively.

\item{} Used in the inequalities~\eqref{eta_condition_ell_2_LR_result1} and \eqref{Lieb_Robinson_tilde_H_t_practical_3}.
\begin{align} 
\frac{1}{18\ell_{q-1}} \ge \frac{\eta \log (\ell_{q-1})}{8\ell_q} = \frac{2}{\xi_q} \quad {\rm or} \quad \frac{\ell_{q-1}^{\eta}}{\log(\ell_{q-1})} \ge \frac{9\eta}{4},
\label{eta_condition_ell_2_LR}
\end{align} 
where $\xi_q:=\frac{16\ell_q }{\eta \log(\ell_{q-1})}$ as defined in Eq.~\eqref{def_xi_q_1}.

\item{} Used in the inequality~\eqref{condition_nu_q_ge2_result}:
\begin{align} 
&\frac{\ell_q}{\xi_q}= \frac{\eta \log(\ell_{q-1})}{16} \ge3  \quad {\rm or} \quad  \log(\ell_{q-1}) \ge \log(\ell_1) \ge \frac{48}{\eta} ,  \label{cond_for_nu_q_ge_2}
\end{align} 
where $\xi_q:=\frac{16\ell_q }{\eta \log(\ell_{q-1})}$ as defined in Eq.~\eqref{def_xi_q_1}.

\item{} Used in the inequality~\eqref{r_q_1_ast_ineq_2}:
\begin{align} 
\frac{2 \ell_{q-1} \log ( \zeta_1)}{\ell_q} =\frac{2\log ( \zeta_1)}{\ell_{q-1}^{\eta}} \le \frac{2\log ( \zeta_1)}{\ell_1^\eta} \le 1,  
\label{zeta_condition_ell_2}
\end{align} 
where $\zeta_1=2^{D+1}e^2 \gamma^2 D!$ as defined in Eq.~\eqref{def:zeta_1}.

\item{} Used in the inequality~\eqref{ineq_ell_1_ge_3}:
\begin{align} 
\ell_1 \ge 3.
\label{ell_1_ge_3}
\end{align}

\item{} Used in the inequality~\eqref{cond_r_q_2_LR_connect}:
\begin{align} 
\frac{2}{e} \ell_{q-1}^{-\eta/2} \le \frac{1}{\zeta_1}  \quad {\rm or} \quad  \ell_{q-1} \ge \ell_1 \ge  \left(\frac{4\zeta_1}{e} \right)^{2/\eta},
\label{zeta_condition_ell_3_connect}
\end{align} 
where $\zeta_1=2^{D+1}e^2 \gamma^2 D!$ as defined in Eq.~\eqref{def:zeta_1}.

\item{} Used in the inequality~\eqref{upper_bound_for_delta_t}:
\begin{align} 
\frac{\log (\tilde{\mathcal{C}}_0) +(D-1) \log (v_0 \ell_1+1)}{2v_0 \ell_1} \le \frac{1}{2},
\label{upper_bound_for_delta_t_cond}
\end{align} 
where $v_0= \frac{2 e^3 g \gamma (\alpha-2D)}{\alpha-2D -1}$ as defined in Eq.~\eqref{Lieb-Robinson_velog}.

\end{enumerate}

\subsection{Proof of Theorem~\ref{thm:Middle-range Lieb-Robinson bound} (see also Figs.~\ref{fig:Outline_middle_range_theorem1} and \ref{fig:Outline_middle_range_theorem2})}

We first consider the case of $\ell \le \ell_1^{1+\bar{\eta}}$, where $\bar{\eta}$ was defined in Eq.~\eqref{Def:eta_setup}. 
In this case, we cannot define $q^\ast$ $(\ge2)$ such that $\lsc_{q^\ast} =\ell$ in Eq.~\eqref{Def:eta_setup}, but the Hamiltonian $H_{\le \ell}$ has a finite interaction length which is independent of $t$.
As a general setup, we consider the Lieb-Robinson bound by a Hamiltonian which has a length scale at most $\xi$.
As long as $\xi=\orderof{1}$, we trivially obtain the Lieb-Robinson velocity of order $\orderof{1}$, whereas the $\xi$-dependence of the velocity may be non-linear with respect to $\xi$. 
Our purpose here is to prove that the Lieb-Robinson velocity is at most linear to $\xi$ as long as $\alpha>2D+1$.
We prove the following lemma (the proof is given in Sec.~\ref{Sec:Lieb-Robinson bound for Hamiltonian with a finite length scale}):

\begin{lemma} \label{lemma:Lieb_Robinson_op_quasi_locality_short_range}
Let $\tilde{H}$ be a Hamiltonian such that 
\begin{align} 
\tilde{H} = \sum_{Z\subset \Lambda:\diam(Z)\le \xi } h_Z
\end{align} 
with the same condition as~\eqref{basic_assump_power}, namely
\begin{align}
\sup_{i \in \Lambda}\sum_{\substack{Z:Z\ni i \\  r \le \diam(Z) \le \xi}} \| h_Z\| \le  g r^{-\alpha+D} \for  r\le\xi .\label{g-extensive_with_decay}
\end{align}
Then, the Hamiltonian $\tilde{H}$ satisfies $\mathcal{G}_{\tilde{H}}(x,t,\fset{X})$-Lieb-Robinson bound with 
\begin{align} 
\mathcal{G}_{\tilde{H}}(x,t,\fset{X})= \min \left(2 |\fset{X}|  \left( \frac{v_0 |t| }{e^2 \lceil x/ \xi \rceil }\right)^{\lceil x/ \xi \rceil},2 \right)
\le 2 |\fset{X}|  e^{- 2(x- \xi v_0 |t| )/ \xi }
 \label{Lieb_Robinson_tilde_H_t_short}
\end{align} 
with 
\begin{align} 
v_0:= \frac{2 e^3 g \gamma (\alpha-2D)}{\alpha-2D -1} . \label{Lieb-Robinson_velog}
\end{align} 
We note that the same inequality holds for time-dependent Hamiltonians.
\end{lemma}
\noindent
From Lemma~\ref{lemma:Lieb_Robinson_op_quasi_locality_short_range} or the inequality~\eqref{Lieb_Robinson_tilde_H_t_short}, we obtain for $\ell \le \ell_1^{1+\bar{\eta}}$
\begin{align}
\mathcal{G}_{\le \ell} (x,t,\fset{X})=2 |\fset{X}|  e^{- (x- \ell v_0 |t| )/ \ell} \le 2 |\fset{X}|  e^{-2 (x- \ell_1^{1+\bar{\eta}} v_0 |t| )/ \ell}.
\label{case_ell_small}
\end{align}
Because $\ell_1^{1+\bar{\eta}} v_0 $ depends only on the parameters $\{D, g_0, \alpha,\gamma\}$, we obtain the inequality~\eqref{thm_middle_range_Lieb-Robinson_proof} by choosing $v^\ast=\ell_1^{1+\bar{\eta}} v_0$ for the case of $\ell \le \ell_1^{1+\bar{\eta}}$.

We next focus on the case of $\ell > \ell_1^{1+\bar{\eta}}$, where we can define $q^\ast \ge 2$ such that $\lsc_{q^\ast} =\ell$.
We here define the Hamiltonian $H_{1:q}$ as follows: 
\begin{align}
H_{1:q}:= \sum_{s=1}^q H_s. \label{def_H_1:_q_middle}
\end{align}
For the proof, we need to estimate the Lieb-Robinson bound for the Hamiltonian $H_{1:q^\ast}$, which is equal to $H_{\le \ell}$.
We define that $\{H_{1:q}\}_{q=1}^{q^\ast}$ satisfy the $\mathcal{G}_q(x,t,\fset{X})$-Lieb-Robinson bound for $q=1,2,\ldots, q^\ast$, namely, 
\begin{align}
\frac{\| [e^{iH_{1:q}t} O_L e^{-iH_{1:q}t}, O_{L'}]\|}{\|O_L\|\cdot\|O_{L'}\|} \le \mathcal{G}_q(x,t,L), \quad x=\dist_{L,L'}
\end{align}
for arbitrary operators $O_L$ and $O_{L'}$.
We aim to prove the Lieb-Robinson bound for $H_{1:q}$ in the form of
\begin{align} 
&\mathcal{G}_{q}(x,t,\fset{X}) = \tilde{\mathcal{C}}_0 \abb\fset{X} \abb^2 (1+x/\ell_{q})^{D-1}  e^{-2(x-v_{q}|t|)/\ell_{q}},   \label{LR_func_G_Q_aim_to_prove}
\end{align}
where $\tilde{\mathcal{C}}_0$ is defined in Eq.~\eqref{def:tilde_mathcal_C_0_proof}, and $v_q$ is given by the recurrence relation as follows:
\begin{align} 
&v_1=\ell_1 v_0 ,\quad v_{q} = v_{q-1}\left (\frac{1}{1-\delta_{q-1}} + \frac{192 \log(\zeta_2)}{\eta \log(\ell_{q-1})} \right) ,  \notag \\
&\delta_{q-1} = \frac{6\log [ \tilde{\mathcal{C}}_0 \gamma^2 (2\ell_q)^{2D}]+6(D-1) \log (1+\ell_{q-1}^{\eta_q}/12) }{\ell_{q-1}^{\eta_q}}  ,\quad \zeta_2 =16e^3 \gamma^2 \cdot 45^D
\label{definitions_v_1_v_q_delta_Fig_used}
\end{align} 
with $v_0= \frac{2 e^3 g \gamma (\alpha-2D)}{\alpha-2D -1}$ as defined in Eq.~\eqref{Lieb-Robinson_velog}.
By remembering that the $\ell_q$ is lower-bounded by a double exponential function with respect to $q$ as in \eqref{doubly_exponential_ell_q_bound}, we can ensure  
\begin{align} 
v_\infty= \lim_{q\to \infty} v_{q}  < \infty,
\label{case_ell_large}
\end{align} 
and $v_\infty$ depends only on the parameters $\{D, g_0, \alpha,\gamma\}$.
From the inequality~\eqref{case_ell_small} and Eq.~\eqref{case_ell_large}, we can choose $v^\ast = \max(\ell_1^{1+\bar{\eta}} v_0, v_\infty)$.
We thus prove the theorem.

In the following, we prove the inequality~\eqref{LR_func_G_Q_aim_to_prove} by induction method.
We first consider the case of $q=1$.
From the inequality~\eqref{Lieb_Robinson_tilde_H_t_short}, we obtain
\begin{align}
\mathcal{G}_1(x,t,\fset{X})=2 |\fset{X}|  e^{- 2(x- \ell_1 v_0 |t| )/ \ell_1 } ,
\end{align}
which clearly reduces to the form of~\eqref{LR_func_G_Q_aim_to_prove}.
We then adopt the assumption that the Hamiltonian $H_{1:q-1}$ ($q\ge 2$) satisfies the Lieb-Robinson bound of
\begin{align} 
&\mathcal{G}_{q-1}(x,t,\fset{X}) = \tilde{\mathcal{C}}_0 \abb\fset{X} \abb^2 (1+x/\ell_{q-1})^{D-1}  e^{-2(x-v_{q-1}|t|)/\ell_{q-1}}. \label{LR_func_G_Q-1_start} 
\end{align}
By using the assumption \eqref{LR_func_G_Q-1_start}, we aim to derive the Lieb-Robinson bound for $H_{1:q}$ in the form of \eqref{LR_func_G_Q_aim_to_prove}.
We first restrict ourselves to 
\begin{align} 
t \le \Delta t_{q} := \frac{\ell_{q}}{12v_{q-1}}, \label{def:delta_t_q}
\end{align}
and consider the generic $t$ afterward [see Eq.~\eqref{genenral_time_decompostion_unitary}].
Before going to the proof, we derive the following lemma:
\begin{lemma} \label{Lieb_Robinson_reformationG_qminsu1}
For arbitrary $q\ge 2$, the Lieb-Robinson function $\mathcal{G}_{q-1}(x,t,\fset{X})$ in~\eqref{LR_func_G_Q-1_start} is reduced to the form of
\begin{align} 
\mathcal{G}_{q-1}(x,t,\fset{X}) =  \abb\fset{X}^{(\ell_{q})} \abb^2 e^{-(x-\bar{v}_{q-1} \Delta t_{q})/\ell_{q-1}} \quad (t\le \Delta t_q),
\label{LR_func_G_Q-1_start_2}
\end{align}
where we define $\bar{v}_{q-1}$ as 
\begin{align} 
\bar{v}_{q-1}:= \frac{v_{q-1}}{1-\delta_{q-1}} \le 2v_{q-1} 
\label{bound_on_v_q-1_1}
\end{align} 
with
\begin{align} 
\delta_{q-1} = \frac{6\log [ \tilde{\mathcal{C}}_0 \gamma^2 (2\ell_q)^{2D}]+6(D-1) \log (1+\ell_{q-1}^{\eta_q}/12) }{\ell_{q-1}^{\eta_q}} .
\end{align} 
Note that $\bar{v}_{q-1}\le 2v_{q-1}$ is a consequence from the condition~\eqref{condition1_for_v_q-1_1}. 
\end{lemma}

{~}

\noindent
\textit{Proof of Lemma~\ref{Lieb_Robinson_reformationG_qminsu1}.}
From Eq.~\eqref{actual_form_of_Lieb-Robinson} and the definition of $\delta_{q-1}$, the Lieb-Robinson bound~\eqref{LR_func_G_Q-1_start_2} gives a trivial bound for $x \le v_{q-1} \Delta t_{q}$.
Hence, we only have to consider the range of $x \ge v_{q-1} \Delta t_{q}$.

We start from the form of $\mathcal{G}_{q-1}(x,t,\fset{X})$ as in Eq.~\eqref{LR_func_G_Q-1_start}:
\begin{align} 
\mathcal{G}_{q-1}(x,t,\fset{X}) = \tilde{\mathcal{C}}_0 \abb\fset{X} \abb^2 (1+x/\ell_{q-1})^{D-1}  e^{-2(x-v_{q-1}|t|)/\ell_{q-1}}. 
\end{align}
First, from the definition~\eqref{def:coarse_graining_X^(r)}, we have $X \subseteq (X^{(\ell_q)})[\ell_q]$ for an arbitrary subset $X\subseteq \Lambda$, and hence $|X|$ is upper-bounded as follows:
\begin{align} 
|X| \le \abb(X^{(\ell_q)})[\ell_q]\abb \le \sum_{i\in X^{(\ell_q)}} \abb\bal{i}{\ell_q} \abb\le  \gamma (2\ell_q)^D \abb X^{(\ell_q)} \abb,
\end{align} 
where we use the inequality~\eqref{geometric_parameter_gamma2}.
Then, we have
\begin{align} 
\mathcal{G}_{q-1}(x,t,\fset{X})  &\le  \mathcal{G}_{q-1}(x,\Delta t_q,\fset{X})   \le    \tilde{\mathcal{C}}_0 \gamma^2 (2\ell_q)^{2D} \abb X^{(\ell_q)} \abb^2  (1+x/\ell_{q-1})^{D-1}e^{-2(x-v_{q-1}\Delta t_q)/\ell_{q-1}}   ,
 \label{lemma_LR_simple_ineq_1_1}
\end{align} 
where we use the condition $t \le \Delta t_q$ in the first inequality.

Second, by using $x \ge v_{q-1} \Delta t_{q}$, we obtain the upper bound of
\begin{align} 
&\tilde{\mathcal{C}}_0 \gamma^2 (2\ell_q)^{2D} \abb X^{(\ell_q)} \abb^2  (1+x/\ell_{q-1})^{D-1}e^{-2(x-v_{q-1}\Delta t_q)/\ell_{q-1}}  \notag \\
=& \exp\left\{\frac{-2x}{\ell_{q-1}} \left[1- \frac{\ell_{q-1} \log [\tilde{\mathcal{C}}_0 \gamma^2 (2\ell_q)^{2D}  (1+x/\ell_{q-1})^{D-1}] }{2x} \right]\right\} e^{2v_{q-1}\Delta t_q/\ell_{q-1}} \notag \\
\le& \exp\left\{\frac{-2x}{\ell_{q-1}} \left[1- \frac{\ell_{q-1} \log [\tilde{\mathcal{C}}_0 \gamma^2 (2\ell_q)^{2D}  (1+v_{q-1}\Delta t_q /\ell_{q-1})^{D-1}] }{2v_{q-1}\Delta t_q} \right]\right\} e^{2v_{q-1}\Delta t_q/\ell_{q-1}} \notag \\
=& \exp\left\{\frac{-2 (1-\delta_{q-1})}{\ell_{q-1}} \Bigl[ x -(1-\delta_{q-1})^{-1} v_{q-1} \Bigr]   \right\},
 \label{lemma_LR_simple_ineq_1_2}
\end{align} 
where in the last equation we use the definitions~$\Delta t_q=\ell_q/(12v_{q-1})$ and $\ell_q=\ell_{q-1}^{1+\eta_q}$ to derive
\begin{align} 
\frac{\ell_{q-1} \log [\tilde{\mathcal{C}}_0 \gamma^2 (2\ell_q)^{2D}  (1+v_{q-1}\Delta t_q /\ell_{q-1})^{D-1}] }{2v_{q-1}\Delta t_q}  
&=\frac{6\log [\tilde{\mathcal{C}}_0 \gamma^2 (2\ell_q)^{2D} ]+6(D-1) \log[ 1+\ell_q /(12\ell_{q-1})] }{\ell_q/\ell_{q-1}} \notag \\
&=\frac{6\log [\tilde{\mathcal{C}}_0 \gamma^2 (2\ell_q)^{2D} ]+6(D-1) \log( 1+\ell_{q-1}^{\eta_q} /12) }{\ell_{q-1}^{\eta_q}} = \delta_{q-1}.
\end{align} 

From the condition~\eqref{condition1_for_v_q-1_1}, we have $1/(1-\delta_{q-1}) \le 2$, which yields $2(1-\delta_{q-1})\ge 1$.
Therefore, the inequality~\eqref{lemma_LR_simple_ineq_1_2} reduces to 
\eqref{LR_func_G_Q-1_start_2}.  This completes the proof. $\square$

{~}\\
%



We now consider the unitary operator $e^{-i H_{1:q} t}$ for $t\le \Delta t_q$ and start from the decomposition of 
\begin{align} 
e^{-i H_{1:q} t} = e^{-i H_{1:q-1} t} \mathcal{T} e^{-\int_0^t e^{i H_{1:q-1} \tau} H_q e^{i H_{1:q-1} -\tau} d\tau}  = e^{-i H_{1:q-1} t} \mathcal{T} e^{-\int_0^t H_q (H_{1:q-1}, \tau) d\tau} .
\label{time_evolution_decomp_unitary}
\end{align} 
We aim to derive the Lieb-Robinson bound for $\mathcal{T} e^{-\int_0^t H_q (H_{1:q-1}, \tau) d\tau}$, which we characterize by the function $\tilde{\mathcal{G}}_q(x,t,\fset{X})$.
Note that the Lieb-Robinson bound for $e^{-i H_{1:q-1} t}$ has been already given by the function $\mathcal{G}_{q-1}(x,t,\fset{X})$ as in Eq.~\eqref{LR_func_G_Q-1_start_2}.
In order to estimate $\tilde{\mathcal{G}}_q(x,t,\fset{X})$, we first estimate the quasi-locality\footnote{An interaction is said to be quasi-local if the interaction decays rapidly (exponentially or polynomially) with the interaction length.} of $H_q (H_{1:q-1}, \tau)$.
We prove the following proposition (see Sec.~\ref{Sec:Quasi-locality of operator after time-evolution} for the proof):

\begin{prop} \label{prop:LR_op_quasi_locality}
Let $\tilde{\ell}_q$ be a length scale such that 
\begin{align}
2\gamma^2 e^{\bar{v}_{q-1} \tau /\ell_{q-1}} e^{-\tilde{\ell}_q/(2\ell_{q-1})}  \le 1 
\label{condition_for_tilde_ell_q}
\end{align}
for $\tau \le \Delta t_q$.
Then, under the assumption of \eqref{LR_func_G_Q-1_start_2}, we can give a decomposition of the time-evolved Hamiltonian $H_q (H_{1:q-1}, \tau)$  in the form of 
\begin{align}
&H_q (H_{1:q-1}, \tau)=\sum_{Z: \diam(Z) \le \ell_q} \sum_{s=1}^\infty h_{\tau,\bal{Z}{s\tilde{\ell}_q}} 
\label{prop_LR_op_quasi_locality_1_seq1_simple0}
\end{align}
with
\begin{align}
\sum_{\substack{Z: \diam(Z) \le \ell_q \\ \bal{Z}{s\tilde{\ell}_q}\cap X \neq \emptyset}} \| h_{\tau,\bal{Z}{s\tilde{\ell}_q}} \| \le  
(2\tilde{\ell}_q)^Dg_q \gamma  \bigl | X^{(s_0\tilde{\ell}_q)} \bigl | (s+s_0) ^D  e^{-( s -1)\tilde{\ell}_q/(2\ell_{q-1})}
\label{prop_LR_op_quasi_locality_1_seq1_simple}
\end{align}
for an arbitrary subset $X \subseteq \Lambda$, 
where $s_0$ can be arbitrarily chosen and $h_{\tau,\bal{Z}{s\tilde{\ell}_q}}$ is an operator which acts on the subset $\bal{Z}{s\tilde{\ell}_q}$.
\end{prop}
\noindent
In order that the condition~\eqref{condition_for_tilde_ell_q} is satisfied for $\tau \le \Delta t_q=\ell_{q}/(12v_{q-1})$, we choose 
\begin{align}
\tilde{\ell}_q=5 \ell_q/12.
\end{align}
This choice implies the condition~\eqref{condition_for_tilde_ell_q} as follows:
\begin{align}
2\gamma^2 e^{\bar{v}_{q-1} \tau /\ell_{q-1}} e^{-\tilde{\ell}_q/(2\ell_{q-1})}  \le  
2\gamma^2 e^{2v_{q-1} [\ell_{q}/(12v_{q-1})]/\ell_{q-1}} e^{-(5\ell_q/12)/(2\ell_{q-1})} = e^{-\ell_{q-1}^{\eta_q}/24 + \log(2\gamma^2)}<1,
\label{ell_q-1_eta_4_gamma_2_result}
\end{align}
where  in the first inequality we use $\bar{v}_{q-1} \le 2v_{q-1}$ which is given in \eqref{bound_on_v_q-1_1}, and  the second inequality $\ell_{q-1}^{\eta_q}/24 \ge \ell_{q-1}^{\eta}/24 \ge  \log(2\gamma^2) $ is derived from the condition~\eqref{ell_q-1_eta_4_gamma_2}.
From the choice of $\tilde{\ell}_q=5\ell_q/12$, we have
 \begin{align}
\diam(\bal{Z}{s\tilde{\ell}_q}) = \diam(Z) + 2s\tilde{\ell}_q \le 11 s\ell_q/6  \for \diam(Z) \le \ell_q  .\label{diam_Z_s_tilde_ell_q}
\end{align}
Also, by replacing $s_0\to22s_0/5$, we have
\begin{align}
(2\tilde{\ell}_q)^Dg_q \gamma  \bigl | X^{(s_0\tilde{\ell}_q)} \bigl | (s+s_0) ^D &\to 
(2\tilde{\ell}_q)^Dg_q \gamma  \bigl | X^{(22s_0\tilde{\ell}_q/5)} \bigl | (s+22s_0/5) ^D \le (10\tilde{\ell}_q)^Dg_q \gamma  \bigl | X^{(11s_0\ell_q/6)} \bigl | (s+s_0) ^D .
\label{diam_Z_s_tilde_ell_q_11s0/6}
\end{align}
Note that $s_0$ can be arbitrarily chosen. 

Moreover, we can prove the following proposition on the Lieb-Robinson bound for quasi-local operators in the form of Eq.~\eqref{prop_LR_op_quasi_locality_1_seq1_simple0} with \eqref{prop_LR_op_quasi_locality_1_seq1_simple} (see Sec.~\ref{proof_sec_Lieb-Robinson bound by time-evolved Hamiltonian} for the proof):
\begin{prop} \label{lemma:Lieb_Robinson_op_quasi_locality}
Let $\tilde{H}$ be a Hamiltonian such that 
\begin{align} 
\tilde{H} = \sum_{Z:Z\subseteq \Lambda} h_Z = \sum_{s=1}^\infty \sum_{\diam(Z_s)\le s\xi } h_{Z_s}.
\end{align} 
For an arbitrary subset $X$ and an arbitrary positive $s_0$, we assume 
\begin{align} 
&\sum_{ \substack{\diam(Z_s)\le s\xi \\ Z_s \cap X \neq \emptyset } } \| h_{Z_s}\| \le 
\tilde{g} \bigl | X^{(s_0\xi)} \bigl | (s+s_0) ^D e^{-\mu (s-1)} \quad (\mu>1).
\label{extensiveness_estimation_tilde_H0}
\end{align} 
Then, the Hamiltonian $\tilde{H}$ satisfies $\mathcal{G}_{\tilde{H}}(x,t,\fset{X})$-Lieb-Robinson bound with 
\begin{align} 
\mathcal{G}_{\tilde{H}}(x,t,\fset{X})=e \abb \fset{X}^{(\xi)}\abb  \left[ \tilde{v}  |t|\exp\left( - \frac{\mu-1}{2}\frac{x}{\xi} +\tilde{v}  |t|\right) +\left( \frac{e^2 \tilde{v}|t|}{m_x^\ast} \right)^{m_x^\ast} \right]
 \label{Lieb_Robinson_tilde_H_t0}
\end{align} 
with 
\begin{align} 
\tilde{v}:=4\tilde{g}\gamma (4D)^D ,\quad  m_x^\ast :=\left\lfloor \frac{\mu-1}{2\mu}\frac{x}{\xi} \right \rfloor +1.
\label{def:tilde_v_m_ast}
\end{align} 
We note that the same inequality holds for time-dependent Hamiltonians.
\end{prop}

\noindent
From Proposition~\ref{prop:LR_op_quasi_locality}, in order to apply Proposition~\ref{lemma:Lieb_Robinson_op_quasi_locality} to $H_q(H_{1:q-1},\tau)$ ($\tau\le \Delta t_q$), 
we choose $\{\xi,\mu,\tilde{g}\}$ as 
\begin{align} 
\xi= 11\ell_q/6 \le 2\ell_q, \quad \mu= \frac{\tilde{\ell}_q}{2\ell_{q-1}} \ge \frac{\ell_q}{4\ell_{q-1}} \quad  {\rm and}\quad \tilde{g}=  (10\tilde{\ell}_q)^Dg_q \gamma 
\le (5\ell_q)^Dg_q \gamma  ,
\end{align} 
where we use the inequalities~\eqref{diam_Z_s_tilde_ell_q} and \eqref{diam_Z_s_tilde_ell_q_11s0/6} in determining the parameters $\xi$ and $\tilde{g}$ as in \eqref{extensiveness_estimation_tilde_H0}.
Note that $\mu \ge \ell_q/(4\ell_{q-1})=\ell_{q-1}^{\eta_q}/4 >1$ is ensured from the condition~\eqref{ell_q-1_eta_4_gamma_2}.
From Eq.~\eqref{Lieb_Robinson_tilde_H_t0}, we obtain the Lieb-Robinson function $\tilde{\mathcal{G}}_{q}(x,t,\fset{X})$ for dynamics by the time-evolved Hamiltonian $H_q (H_{1:q-1}, \tau)$ ($0\le \tau\le t $ with $t\le \Delta t_q$) as follows:
\begin{align} 
\tilde{\mathcal{G}}_{q}(x,t,\fset{X}) & \le  e \abb \fset{X}^{(11\ell_q/6)}\abb  \left[ \tilde{v}_q  |t|\exp\left( - \frac{1-4\ell_{q-1}/\ell_q}{16\ell_{q-1}}x +\tilde{v}_q  |t|\right) +\left( \frac{e^2 \tilde{v}_q|t|}{m_x^\ast} \right)^{m_x^\ast} \right]
 \label{Lieb_Robinson_tilde_H_t_practical}
\end{align} 
with 
\begin{align} 
\tilde{v}_q:=4\gamma^2 (20D\ell_q)^D g_q  ,\quad  m_x^\ast :=\left\lfloor \frac{(1-4\ell_{q-1}/\ell_q)x}{11\ell_q/3} \right \rfloor +1 \ge
\frac{x}{4\ell_q}  ,
\label{ell_q-1_eta_4_gamma_2_result_2}
\end{align} 
where we use the condition~\eqref{ell_q-1_eta_4_gamma_2} to obtain $\ell_q/\ell_{q-1}=\ell_{q-1}^{\eta_q} \ge 48$, which yields $1-4\ell_{q-1}/\ell_q \ge 11/12$.

We further simplify the bound~\eqref{Lieb_Robinson_tilde_H_t_practical} as 
\begin{align} 
\tilde{\mathcal{G}}_{q}(x,t,\fset{X})&\le e \abb \fset{X}^{(\ell_q)}\abb  \left[\tilde{v}_q  |t|\exp\left( - \frac{11}{192\ell_{q-1}}x +\tilde{v}_q  |t|\right) +
(e^2 \tilde{v}_q|t|)^{m_x^\ast/2} \cdot (e^2 \tilde{v}_q|t|)^{m_x^\ast/2}  \right] \notag \\
&\le e \abb \fset{X}^{(\ell_q)}\abb  \left[\tilde{v}_q  |t|\exp\left( - \frac{1}{18\ell_{q-1}}x +\tilde{v}_q  |t|\right) +
(e^2 \tilde{v}_q|t|)^{1/2} \cdot \exp\left( \frac{\log (e^2 \tilde{v}_q|t|)}{8\ell_q}x   \right) \right] 
 \label{Lieb_Robinson_tilde_H_t_practical_2}
\end{align} 
for $t\le \Delta t_q$, where we use $\abb \fset{X}^{(11\ell_q/6)}\abb \le \abb \fset{X}^{(\ell_q)}\abb $.
Note that from $\Delta t_q= \ell_{q}/(12v_{q-1})$ and Def.~\eqref{ell_q-1_eta_4_gamma_2_result_2} for $\tilde{v}_q$, we have 
\begin{align} 
e^2 \tilde{v}_q\Delta t_q =4e^2 \gamma^2 (20D\ell_q)^D g_q \cdot\ell_{q}/(12v_{q-1}) = \frac{e^2 \gamma^2 (20D)^D}{3v_{q-1}} \ell_q^{D+1} g_q 
\le \frac{e^2 g \gamma^2 (20D)^D}{3\ell_1 v_0}\ell_{q-1}^{-\eta} \le \ell_{q-1}^{-\eta} ,
\label{result:eta_condition_ell_1}
\end{align}
where we use $v_{q-1}\ge v_1 := \ell_1 v_0$ [see Eq.~\eqref{definitions_v_1_v_q_delta_Fig_used} for the definition of $v_1$] and the inequality~\eqref{extensiveness_time_evolve} for $\ell_q^{D+1} g_q $ in the first inequality and the condition~\eqref{eta_condition_ell_1} in the second inequality.
This reduces the inequality~\eqref{Lieb_Robinson_tilde_H_t_practical_2} to 
\begin{align} 
\tilde{\mathcal{G}}_{q}(x,t,\fset{X})
&\le e \abb \fset{X}^{(\ell_q)}\abb  \left[ \ell_{q-1}^{-\eta} \exp\left( - \frac{1}{18\ell_{q-1}}x \right) +
\ell_{q-1}^{-\eta/2}\cdot \exp\left( \frac{-\eta \log(\ell_{q-1})}{8\ell_q}x   \right) \right] \for t\le \Delta t_q, \label{Lieb_Robinson_tilde_H_t_practical_2_2_Fig_used}
\end{align} 
where we use the following bound in deriving the first term:
\begin{align} 
\tilde{v}_q  |t| \exp(\tilde{v}_q  |t|) \le \tilde{v}_q \Delta t_q \exp(\tilde{v}_q  \Delta t_q ) \le \ell_{q-1}^{-\eta}\frac{\exp(\ell_{q-1}^{-\eta}/e^2)}{e^2} \le\ell_{q-1}^{-\eta}\frac{\exp(1/e^2)}{e^2} \le \ell_{q-1}^{-\eta}.
\end{align} 

In the following,  by combining $\mathcal{G}_{q-1}(x,t,\fset{X})$ and $\tilde{\mathcal{G}}_{q}(x,t,\fset{X})$, we obtain the Lieb-Robinson function $\mathcal{G}_q(x,t ,\fset{X},\fset{Y})$ for generic time $t$.
For the purpose, we first estimate $\mathcal{G}_q(x,t ,\fset{X},\fset{Y})$ under the constraints of 
\begin{align} 
\diam(\fset{X})  \le \ell_q,\quad \diam(\fset{Y})\le \ell_q.
\label{restriction_X_and_Y_ell_q}
\end{align} 
We then remove the constraints by using Theorem~\ref{lemma:Connection of local _ global Lieb-Robinson bounds_2}.

We first decompose the unitary operator $e^{-iH_{1:q} t}$ into a sequence of the small time unitary operators.
For the purpose, for a fixed $t$, we define the sequence of times $\{t_0, t_1,t_2,\ldots, t_{m_0}\}$ as
\begin{align}  
&t_0:=0,\quad t_m := (m-1) \Delta t_q + \Delta t'_q \quad (m\ge 1), \notag \\
&m_0=\lceil t/\Delta t_q \rceil,\quad \Delta t'_q=t-(m_0-1) \Delta t_q ,
\end{align}
where the definition implies $\Delta t'_q\le \Delta t_q$.
We aim to derive the Lieb-Robinson bound for the unitary time evolution $e^{-iH_{1:q} t_m}$ which is decomposed as 
\begin{align} 
e^{-i H_{1:q}t_m} &=e^{-i H_{1:q}\Delta t'_q} [e^{-i H_{1:q}\Delta t_q}]^m   \notag \\
&=e^{-i H_{1:q-1} \Delta t'_q} \mathcal{T} e^{-\int_0^{\Delta t'_q} H_q (H_{1:q-1}, \tau) d\tau}  \left[e^{-i H_{1:q-1} \Delta t_q} \mathcal{T} e^{-\int_0^{\Delta t_q} H_q (H_{1:q-1}, \tau) d\tau}\right]^{m-1} .
\label{genenral_time_decompostion_unitary}
\end{align} 
We note that $e^{-i H_{1:q-1} \Delta t_q}$ and $e^{-i H_{1:q-1} \Delta t'_q}$ satisfies the $\mathcal{G}_{q-1}(x,t,\fset{X})$-Lieb-Robinson bound, while $\mathcal{T} e^{-\int_0^{\Delta t_q} H_q (H_{1:q-1}, \tau) d\tau}$ and $\mathcal{T} e^{-\int_0^{\Delta t'_q} H_q (H_{1:q-1}, \tau) d\tau}$ satisfies the $\tilde{\mathcal{G}}_{q}(x,t,\fset{X})$-Lieb-Robinson bound~\eqref{Lieb_Robinson_tilde_H_t_practical_2_2_Fig_used}.
Our task is to connect all the Lieb-Robinson bounds appropriately. 
We here utilize the following proposition (see Sec.~\ref{connection of two Lieb-Robinson bounds from different unitary operators} for the proof):

\begin{prop} \label{lemma:Connection of two Lieb-Robinson bounds_1}
We consider a Hamiltonian $H_0$ which satisfies the Lieb-Robinson bound for a fixed $t$ as 
\begin{align}
\mathcal{G}(x,t ,\fset{X},\fset{Y}) \le \mathcal{C} e^{-(x- v|t|)/\xi}   \label{Lieb_Robinson_O_i_assump}
\end{align}
for $\diam(\fset{X}) , \diam(\fset{Y}) \le \nu \xi$ ($\nu \ge 3$).
Also, let $H'_0$ be a Hamiltonian which satisfies the Lieb-Robinson bound for a fixed $t'$ as 
\begin{align}
&\mathcal{G}' (x,t',\fset{X}) = \abb \fset{X}^{(\nu \xi)}\abb^2 \mathcal{F}' (x,t'),  \label{Lieb_Robinson_O_i_assump2}
\end{align}
where the function $\mathcal{F}' (x,t')$ is assumed to satisfy
\begin{align}
\mathcal{F}' (x,t') \le   \mathcal{F}' (x_0,t') e^{-2(x-x_0 )/\xi} \for x\ge 0, \quad 0\le x_0 \le x.
\label{Lieb_Robinson_O_i_assump2_2}
\end{align}
Note that we impose no assumptions on $\diam(\fset{X})$ for $\mathcal{G}' (x,t',\fset{X})$. We then obtain the Lieb-Robinson function $\mathcal{G}''(x,t,t' ,\fset{X},\fset{Y})$  for $e^{-iH_0t}e^{-iH'_0 t'}$ as 
\begin{align}
\mathcal{G}''(x,t,t' ,\fset{X},\fset{Y})
= 4\mathcal{C}\gamma (3/\nu)^D  (2\kappa^\ast +\nu)^D  e^{-(x-\kappa^\ast\xi - v|t|)/\xi} \label{lem:connection_Lieb_Robinson}
\end{align}
for $\diam(\fset{X}) , \diam(\fset{Y}) \le \nu\xi$,
where $\kappa^\ast$ is defined as an integer ($\kappa^\ast \in \mathbb{N}$) which satisfies 
\begin{align}
\mathcal{F}' (\kappa^\ast\xi,t') \le \frac{1}{2^{D+1}e^2 \gamma^2 D!} = : \zeta_1^{-1}.
\label{def:zeta_1}
\end{align}
Note that $t'$-dependence of $\mathcal{G}''(x,t,t' ,\fset{X},\fset{Y})$ is included in $\kappa^\ast$.
\end{prop}

\noindent
By using Proposition~\ref{lemma:Connection of two Lieb-Robinson bounds_1}, we prove the form of 
\begin{align} 
\mathcal{G}_q(x, t_m ,\fset{X},\fset{Y}) =\mathcal{R}_q^{m}  e^{-(x- \bar{v}_{q-1} t_m)/\xi_q}  \quad (m=0,1,2,\ldots,m_0)
\label{local_Lieb_Robinson_q_Fig_used}
\end{align} 
under the constraint of $\diam(\fset{X}) \le \ell_q$ and $\diam(\fset{Y}) \le \ell_q$ [see Sec.~\ref{connect_repeat_essential_ineq} for the proof of Ineq.~\eqref{local_Lieb_Robinson_q_Fig_used}], where
\begin{align} 
\xi_q:=  \frac{16\ell_q}{\eta \log(\ell_{q-1})}  \label{def_xi_q_1}
\end{align} 
and 
\begin{align} 
\mathcal{R}_q:=  4e \gamma 9^{D}  e^{\ell_{q-1} \log ( \zeta_1)/\xi_q}  \cdot  4 e\gamma 5^D = 16e^2 \gamma^2 \cdot 45^D \cdot  e^{\ell_{q-1} \log ( \zeta_1)/\xi_q} .
\label{definition:mathcal_R_q}
\end{align} 
For the parameters $\xi_q$ and $\mathcal{R}_q$, we can derive
\begin{align} 
\xi_q \le \frac{\ell_q}{3}, \quad \mathcal{R}_q\le 16e^3 \gamma^2 \cdot 45^D:=\zeta_2,
\label{upper_bound_xi_q_mathal_R_q}
\end{align} 
where in the first inequality  we use the condition~\eqref{cond_for_nu_q_ge_2}, and in the second inequality we use $e^{\ell_{q-1} \log ( \zeta_1)/\xi_q} \le e$ which is a direct consequence from the condition~\eqref{zeta_condition_ell_4_connect}.

From the inequality~\eqref{local_Lieb_Robinson_q_Fig_used}, we can reduce $\mathcal{G}_q(x, t ,\fset{X},\fset{Y})$ ($=\mathcal{G}_q(x, t_{m_0} ,\fset{X},\fset{Y})$) to
\begin{align} 
\mathcal{G}_q(x, t ,\fset{X},\fset{Y}) &=\mathcal{R}_q \exp\left[- \left( \frac{x- \bar{v}_{q-1} t_{m_0} - \xi_q (m_0-1)\Delta t_q \Delta t_q^{-1} \log(\mathcal{R}_q)}{\xi_q} \right)\right] \notag \\
&\le \zeta_2 \exp\left[- \left( \frac{x- t [\bar{v}_{q-1} + \xi_q \Delta t_q^{-1} \log(\zeta_2) ]}{\ell_q/2} \right)\right] =\zeta_2 e^{-2(x-v_{q}t)/\ell_q} ,
\label{final_form_but_constrained_x_y}
\end{align} 
where we use $(m_0-1)\Delta t_q \le t_{m_0} =t$ and $\xi_q \le \ell_q/3\le \ell_q/2$ from \eqref{upper_bound_xi_q_mathal_R_q} for the inequality and define $v_{q}$ as 
\begin{align} 
v_{q} = \bar{v}_{q-1} + \frac{\xi_q \log (  \zeta_2)}{\Delta t_q} 
 =  \bar{v}_{q-1} +v_{q-1} \frac{192 \log(\zeta_2)}{\eta \log(\ell_{q-1})}  ,
\end{align} 
where we have defined $\Delta t_q$ and $\xi_q$ in Eqs.~\eqref{def:delta_t_q} and \eqref{def_xi_q_1}, respectively.
By remembering that $\bar{v}_{q-1}=(1-\delta_{q-1})^{-1}v_{q-1}$ from Eq.~\eqref{bound_on_v_q-1_1}, we obtain the recurrence equation~\eqref{definitions_v_1_v_q_delta_Fig_used} for $v_q$. 
In this way, we obtain the $\mathcal{G}_q(x, t ,\fset{X},\fset{Y}) $-Lieb-Robinson bound under the constraints of $\diam(\fset{X})$ and $\diam(\fset{Y})\le \ell_q$.

Finally,  in order to remove the constraints of $\diam(\fset{X}),\diam(\fset{Y})\le \ell_q$, we apply Theorem~\ref{lemma:Connection of local _ global Lieb-Robinson bounds_2} to $\mathcal{G}_q(x, t ,\fset{X},\fset{Y}) $ in \eqref{final_form_but_constrained_x_y} by setting 
\begin{align}
\xi_0=\ell_q/2,\quad \xi=(\ell_q-1)/2 \ge \ell_q/3, \label{ineq_ell_1_ge_3}
\end{align}
where we use $\ell_q \ge \ell_1 \ge 3$ from the condition~\eqref{ell_1_ge_3}.
From the inequality~\eqref{Lieb_L_L'_from_local_one_lemma}, we obtain the Lieb-Robinson bound for generic operators without the constraints as follows:
\begin{align}
\mathcal{G}_q(x,t,\fset{X})  \le
 \mathcal{C}_{\ell_q/3 ,\ell_q/2}   \abb \fset{X}^{(\ell_q/3)}\abb^2 (1+3x/\ell_q)^{D-1} \zeta_2 e^{-2(x-v_{q}t)/\ell_q}
 \le  \tilde{\mathcal{C}}_0  \abb \fset{X}\abb^2 (1+x/\ell_q)^{D-1}  e^{-2(x-v_{q}t)/\ell_q}, \label{Lieb_L_L'_from_local_one_lemma_q} 
\end{align}
where $\mathcal{C}_{\ell_q/3 ,\ell_q/2}$ is defined by Eq.~\eqref{def_mathcal_C_0_theorem}, and $ \tilde{\mathcal{C}}_0$ was defined as
\begin{align}
 \tilde{\mathcal{C}}_0  :=\frac{4}{3}e^{10/3} 15^D  D! \gamma   \zeta_2.
\end{align}
We thus prove the inequality~\eqref{LR_func_G_Q_aim_to_prove}. 
This completes the proof of Theorem~\ref{thm:Middle-range Lieb-Robinson bound}. $\square$

%

\subsection{Proof of the inequality~\eqref{local_Lieb_Robinson_q_Fig_used}}\label{connect_repeat_essential_ineq}
We prove the inequality~\eqref{local_Lieb_Robinson_q_Fig_used} by induction.
For $m=0$, we have  $\mathcal{G}_q(x, t_0 ,\fset{X},\fset{Y}) =0$ because of $t_0=0$ and the inequality is trivially obtained.

Then, we assume the form of \eqref{local_Lieb_Robinson_q_Fig_used} for a fixed $m$ and prove the case of $m+1$. 
We here consider the Lieb-Robinson bound for 
\begin{align} 
e^{-i H_{1:q}t_{m+1}}= e^{-i H_{1:q}t_{m+1}} e^{-i H_{1:q}\Delta t_q} = e^{-i H_{1:q}t_m} e^{-i H_{1:q-1} \Delta t_q} \mathcal{T} e^{-\int_0^{\Delta t_q} H_q (H_{1:q-1}, \tau) d\tau},
\end{align} 
where we use Eq.~\eqref{time_evolution_decomp_unitary} for the decomposition of $e^{-i H_{1:q}\Delta t_q}$.
In order to obtain the Lieb-Robinson bound for $e^{-i H_{1:q}t_{m+1}}$, we first connect the two Lieb-Robinson bounds of $e^{-i H_{1:q}t_{m}}$ and $e^{-i H_{1:q-1}\Delta t_q} $. 
We then connect the two Lieb-Robinson bounds of $e^{-i H_{1:q}t_m} e^{-i H_{1:q-1} \Delta t_q} $ and $\mathcal{T} e^{-\int_0^{\Delta t_q} H_q (H_{1:q-1}, \tau) d\tau}$.
In each of the connections, we utilize Proposition~\ref{lemma:Connection of two Lieb-Robinson bounds_1}.

For the first, we consider the Lieb-Robinson bound for $e^{-i H_{1:q}t_m} e^{-i H_{1:q-1} \Delta t_q}$, which we characterize by the function of
\begin{align} 
\mathcal{G}_q'(x, t_m ,\Delta t_q, \fset{X},\fset{Y}).
\end{align} 
The Lieb-Robinson bound for $e^{-i H_{1:q}t_m}$ is given by the assumption~\eqref{local_Lieb_Robinson_q_Fig_used}, namely
\begin{align} 
\mathcal{G}_q(x, t_m ,\fset{X},\fset{Y}) =\mathcal{R}_q^{m}  e^{-(x- \bar{v}_{q-1} t_m)/\xi_q} ,
\end{align} 
where we have defined $\xi_q$ as in Eq.~\eqref{def_xi_q_1}.
Also, the Lieb-Robinson bound for $e^{-i H_{1:q-1} t}$ is given by $\mathcal{G}_{q-1}(x,t,\fset{X})$ in Eq.~\eqref{LR_func_G_Q-1_start_2},
which has the form of
 \begin{align} 
&\mathcal{G}_{q-1}(x,t,\fset{X}) =  \abb\fset{X}^{(\ell_{q})}\abb^2 \mathcal{F}_{q-1}(x,t) ,\quad \mathcal{F}_{q-1}(x,t) :=  e^{-(x-\bar{v}_{q-1} t)/\ell_{q-1}} .
\end{align}
We notice that $ \mathcal{F}_{q-1}(x,t)$ satisfies the condition~\eqref{Lieb_Robinson_O_i_assump2_2} in  Proposition~\ref{lemma:Connection of two Lieb-Robinson bounds_1} as follows:
 \begin{align} 
\mathcal{F}_{q-1}(x,t) =  e^{-(x-\bar{v}_{q-1}t)/\ell_{q-1}} 
&\le  e^{-(x_0-\bar{v}_{q-1} t)/\ell_{q-1}}  e^{-(x-x_0)/\ell_{q-1}} \notag\\ 
&\le   \mathcal{F}_{q-1}(x_0,t)e^{-2(x-x_0)/\xi_q} 
\label{eta_condition_ell_2_LR_result1}
\end{align}
for $0\le x_0 \le x$, where $1/\ell_{q-1} \ge 2/\xi_q$ is a consequence from the condition~\eqref{eta_condition_ell_2_LR}, namely $1/(18\ell_{q-1}) \ge 2/\xi_q$.

In Proposition~\ref{lemma:Connection of two Lieb-Robinson bounds_1}, we set 
\begin{align}
\mathcal{G}(x,t ,\fset{X},\fset{Y}) = \mathcal{G}_q(x, t_m ,\fset{X},\fset{Y}) , \quad \mathcal{G}' (x,t',\fset{X})= \mathcal{G}_{q-1}(x,\Delta t_q,\fset{X}), 
\quad \mathcal{F}' (x,t')= \mathcal{F}_{q-1}(x,\Delta t_q), 
\end{align}
and the parameters therein are given by
\begin{align}
t=t_m,\quad t'=\Delta t_q,\quad \mathcal{C}=\mathcal{R}_q^{m},\quad v=\bar{v}_{q-1},\quad \xi=\xi_q,\quad \nu=\nu_q := \frac{\ell_q}{\xi_q} \ge3, \label{condition_nu_q_ge2_result}
\end{align}
where the inequality $\nu_q\ge 3$ is given by the condition~\eqref{cond_for_nu_q_ge_2}.
Then, from Eq.~\eqref{lem:connection_Lieb_Robinson}, the Lieb-Robinson function $\mathcal{G}_q'(x, t_m ,\Delta t_q, \fset{X},\fset{Y})$ is given by
\begin{align}
\mathcal{G}_q'(x, t_m ,\Delta t_q, \fset{X},\fset{Y})
=4\mathcal{R}_q^{m} \gamma (3/\nu_q)^D (2\kappa_{q,1}^\ast +\nu_q)^D  e^{-(x-\kappa_{q,1}^\ast \xi_q - \bar{v}_{q-1} |t_m|)/\xi_q}, \label{lem:connection_Lieb_Robinson_proof_1}
\end{align}
where $\kappa_{q,1}^\ast$ is chosen so that the following inequality holds
\begin{align}
\mathcal{F}_{q-1}(\kappa_{q,1}^\ast \xi_q ,\Delta t_q) = e^{-(\kappa_{q,1}^\ast \xi_q-\bar{v}_{q-1} \Delta t_{q})/\ell_{q-1}}\le \zeta_1^{-1}, \label{condition_r_q_1ast}
\end{align}
where we have defined $\zeta_1$ in \eqref{def:zeta_1}.
The parameter $\kappa_{q,1}^\ast$ is given by
\begin{align}
\kappa_{q,1}^\ast =\left\lceil \frac{ \bar{v}_{q-1} \Delta t_{q} + \ell_{q-1} \log ( \zeta_1)}{ \xi_q}\right \rceil \le \frac{ \bar{v}_{q-1} \Delta t_{q} + \ell_{q-1} \log ( \zeta_1)}{ \xi_q} +1.
\label{r_q_1_ast_ineq}
\end{align}
We also obtain
\begin{align}
(3/\nu_q)(2\kappa_{q,1}^\ast + \nu_q)&\le  3\left(1 +\frac{2}{\nu_q} + \frac{2\bar{v}_{q-1} \ell_q/(12v_{q-1}) +2 \ell_{q-1} \log ( \zeta_1)}{\ell_q} \right) \le 9
\label{r_q_1_ast_ineq_2}
\end{align}
with $\Delta t_q = \ell_q /(12v_{q-1})$,
where we use $\xi_q \nu_q= \ell_q$, the condition~\eqref{zeta_condition_ell_2}, $\bar{v}_{q-1} \le 2v_{q-1}$ [see Eq.~\eqref{bound_on_v_q-1_1}] and $\nu_q\ge3$.
The inequalities~\eqref{r_q_1_ast_ineq} and \eqref{r_q_1_ast_ineq_2} reduce the inequality~\eqref{lem:connection_Lieb_Robinson_proof_1} to 
\begin{align}
\mathcal{G}_q'(x, t_m ,\Delta t_q, \fset{X},\fset{Y})
&\le \mathcal{R}_q^{m} \cdot 4\gamma  e^{-1/\xi_q } 9^{D}  e^{-(x-\bar{v}_{q-1} \Delta t_{q} - \ell_{q-1} \log ( \zeta_1) - \bar{v}_{q-1} |t_m|)/\xi_q}  \notag \\
&=:\mathcal{R}_q^{m} \mathcal{R}_{q,1}  e^{-(x- \bar{v}_{q-1} |t_{m+1}|)/\xi_q}  , \label{lem:first_connection_Lieb_Robinson}
\end{align}
where we define
\begin{align}
\mathcal{R}_{q,1} := 4\gamma  e^{-1/\xi_q } 9^{D}  e^{\ell_{q-1} \log ( \zeta_1)/\xi_q} \le 4e \gamma 9^{D}  e^{\ell_{q-1} \log ( \zeta_1)/\xi_q}.
\end{align}

In the next step, we connect the Lieb-Robinson bounds for the unitary operators
\begin{align} 
e^{-i H_{1:q}t_m} e^{-i H_{1:q-1} \Delta t_q} \quad {\rm and} \quad  \mathcal{T} e^{-\int_0^{\Delta t_q} H_q (H_{1:q-1}, \tau) d\tau},
\label{two_unitary_second_try}
\end{align} 
which gives the Lieb-Robinson bound for $e^{-i H_{1:q}t_{m+1}}$, namely 
$\mathcal{G}_q(x, t_{m+1}, \fset{X},\fset{Y})$.
In order to apply Proposition~\ref{lemma:Connection of two Lieb-Robinson bounds_1} to this case, we first simplify the Lieb-Robinson bound~\eqref{Lieb_Robinson_tilde_H_t_practical_2_2_Fig_used} for $\mathcal{T} e^{-\int_0^{\Delta t_q} H_q (H_{1:q-1}, \tau) d\tau}$ as follows:
\begin{align} 
\tilde{\mathcal{G}}_{q}(x,\Delta t_q,\fset{X})
&\le e \abb \fset{X}^{(\ell_q)}\abb  \left[ \ell_{q-1}^{-\eta} \exp\left( - \frac{1}{18\ell_{q-1}}x \right) +
\ell_{q-1}^{-\eta/2}\cdot \exp\left( \frac{-\eta \log(\ell_{q-1})}{8\ell_q}x   \right) \right]  \notag \\
&\le 2e  \abb \fset{X}^{(\ell_q)}\abb^2 \ell_{q-1}^{-\eta/2}\cdot e^{-2x/\xi_q}=\abb \fset{X}^{(\ell_q)}\abb^2 \tilde{\mathcal{F}}_{q}(x,\Delta t_q) 
\label{Lieb_Robinson_tilde_H_t_practical_3}
\end{align} 
with
\begin{align} 
\tilde{\mathcal{F}}_{q}(x,\Delta t_q)= 2e \ell_{q-1}^{-\eta/2}\cdot e^{-2x/\xi_q}, \label{eta_condition_ell_2_LR_result}
\end{align} 
where we use the condition~\eqref{eta_condition_ell_2_LR} and the trivial inequality of $\abb \fset{X}^{(\ell_q)} \abb \le \abb \fset{X}^{(\ell_q)}\abb^2$.
We notice that $\tilde{\mathcal{F}}_{q}(x,\Delta t_q)$ satisfies the condition~\eqref{Lieb_Robinson_O_i_assump2_2} in  Proposition~\ref{lemma:Connection of two Lieb-Robinson bounds_1}.

In order to connect two unitary operators~\eqref{two_unitary_second_try}, we set in Proposition~\ref{lemma:Connection of two Lieb-Robinson bounds_1} 
\begin{align}
\mathcal{G}(x,t ,\fset{X},\fset{Y}) = \mathcal{G}_q'(x, t_m ,\Delta t_q, \fset{X},\fset{Y}) 
, \quad \mathcal{G}' (x,t',\fset{X})= \tilde{\mathcal{G}}_{q}(x,\Delta t_q,\fset{X}) , 
\quad \mathcal{F}' (x,t')= \tilde{\mathcal{F}}_{q}(x,\Delta t_q), 
\end{align}
and the parameters therein are given by
\begin{align}
t=t_m+\Delta t_q =t_{m+1},\quad t'=\Delta t_q,\quad \mathcal{C}=\mathcal{R}_q^{m} \mathcal{R}_{q,1} ,\quad v=\bar{v}_{q-1},\quad \xi=\xi_q,\quad \nu=\nu_q := \frac{\ell_q}{\xi_q} \ge3.
\end{align}
We then obtain from Eq.~\eqref{lem:connection_Lieb_Robinson}
\begin{align}
\mathcal{G}_q(x, t_{m+1}, \fset{X},\fset{Y})
=4\mathcal{R}_q^{m} \mathcal{R}_{q,1}\gamma(3/\nu_q)^D  ( 2\kappa_{q,2}^\ast +\nu_q)^D  e^{-(x-\kappa_{q,2}^\ast \xi_q - \bar{v}_{q-1} |t_{m+1}|)/\xi_q}, \label{lem:connection_Lieb_Robinson_proof_2}
\end{align}
where $\kappa_{q,2}^\ast$ is chosen such that it satisfies
\begin{align}
\tilde{\mathcal{F}}_{q}(\kappa_{q,2}^\ast\xi_q,\Delta t_q)= 2e \ell_{q-1}^{-\eta/2}\cdot e^{-2\kappa_{q,2}^\ast} \le \zeta_1^{-1}.  \label{cond_r_q_2_LR_connect}
\end{align}
From the condition~\eqref{zeta_condition_ell_3_connect}, the above condition is satisfied by choosing $\kappa^\ast_{q,2}=1$.
Hence, we have 
\begin{align}
\mathcal{G}_q(x, t_{m+1}, \fset{X},\fset{Y})
=4\mathcal{R}_q^{m} \mathcal{R}_{q,1}\gamma  (6/\nu_q+3)^D  e^{-(x-\xi_q - \bar{v}_{q-1} |t_{m+1}|)/\xi_q} = 
\mathcal{R}_q^{m} \mathcal{R}_{q,1} \mathcal{R}_{q,2} e^{-(x - \bar{v}_{q-1} |t_{m+1}|)/\xi_q}
 \label{lem:second_connection_Lieb_Robinson}
\end{align}
with 
\begin{align}
\mathcal{R}_{q,2} =4 e\gamma  (6/\nu_q+3)^D \le 4 e\gamma 5^D,
\end{align}
where we use $\nu_q\ge 3$.
Because of $\mathcal{R}_{q,1} \mathcal{R}_{q,2}\le \mathcal{R}_q$ from the definition~\eqref{definition:mathcal_R_q}, we prove the inequality~\eqref{local_Lieb_Robinson_q_Fig_used} for the case of $m+1$.
This completes the proof.
$\square$


\section{Detailed proof of the complementary statements for the middle-range Lieb-Robinson bound}

\subsection{Proof of Proposition~\ref{prop:LR_op_quasi_locality}: Quasi-locality of operator after time-evolution} \label{Sec:Quasi-locality of operator after time-evolution}

\subsubsection{Statement}
{~}\\
{\bf Proposition~\ref{prop:LR_op_quasi_locality}.}
\textit{
Let $\tilde{\ell}_q$ be a length scale such that 
\begin{align}
2\gamma^2 e^{\bar{v}_{q-1} \tau /\ell_{q-1}} e^{-\tilde{\ell}_q/(2\ell_{q-1})}  \le 1 
\label{gamma_condition_tau_ell_tilde_q}
\end{align}
for $\tau \le \Delta t_q$.
Then, under the assumption of \eqref{LR_func_G_Q-1_start_2}, we obtain a decomposition the time-evolved Hamiltonian $H_q (H_{1:q-1}, \tau)$  in the form of 
\begin{align}
&H_q (H_{1:q-1}, \tau)=\sum_{Z: \diam(Z) \le \ell_q} \sum_{s=1}^\infty h_{\tau,\bal{Z}{s\tilde{\ell}_q}} 
\end{align}
with
\begin{align}
\sum_{\substack{Z: \diam(Z) \le \ell_q \\ \bal{Z}{s\tilde{\ell}_q}\cap X \neq \emptyset}} \| h_{\tau,\bal{Z}{s\tilde{\ell}_q}} \| \le  (2\tilde{\ell}_q)^Dg_q \gamma  \bigl | X^{(s_0\tilde{\ell}_q)} \bigl | (s+s_0) ^D  e^{-( s -1)\tilde{\ell}_q/(2\ell_{q-1})} 
\label{lemma_LR_op_quasi_locality_1_seq1_simple_proof}
\end{align}
for an arbitrary subset $X \subseteq \Lambda$, 
where $s_0$ can be arbitrarily chosen and $h_{\tau,\bal{Z}{s\tilde{\ell}_q}}$ is an operator which acts on the subset $\bal{Z}{s\tilde{\ell}_q}$.
}

\subsubsection{Proof of Proposition~\ref{prop:LR_op_quasi_locality}}

For the proof, we consider a more general setup. 
We define two operators $A_1$ and $A_2$. 
We assume that the operator $A_1$ satisfies $\mathcal{G}_{A_1}(x,t,\fset{X})$-Lieb-Robinson bound with
\begin{align}
\mathcal{G}_{A_1}(x,t,\fset{X}) = \mathcal{C} \abb \fset{X}^{(\xi_2)}\abb^2 e^{-(x-v|t|)/\xi_1 }. \label{assumption_A_1_lieb_robinson}
\end{align}
Also, the operator $A_2$ consists of terms with interaction length at most of $\xi_2$, namely
\begin{align}
A_2= \sum_{Z: \diam(Z) \le \xi_2} a_Z ,\quad \sum_{Z:Z\ni i} \|a_Z\| \le g_2.  \label{extensive_condition_power_law}
\end{align}


We here consider 
\begin{align}
A_2(A_1,t)= e^{iA_1t} A_2 e^{-iA_1t} = \sum_{Z: \diam(Z) \le \xi_2} a_Z(A_1,t),
\end{align}
and denote $A_2(A_1,t)$ by
\begin{align}
&A_2(A_1,t)=\sum_{Z: \diam(Z) \le \xi_2} \sum_{s=1}^\infty a_{t,\bal{Z}{s\tilde{\xi}_2}} , 
\end{align}
where $\tilde{\xi}_2$ is an arbitrary positive integer such that $\tilde{\xi}_2\ge \xi_2$. We here define 
\begin{align}
&a_{t,\bal{Z}{\tilde{\xi}_2}} :=a_Z(A_1,t,\bal{Z}{\tilde{\xi}_2}),\notag \\
&a_{t,\bal{Z}{s\tilde{\xi}_2 }} := a_Z(A_1,t,\bal{Z}{s\tilde{\xi}_2}) - a_Z(A_1,t,\bal{Z}{\tilde{\xi}_2(s-1)}) \for s\ge 2,   
\label{def:time_ham_decomp}
\end{align}
where $a_Z(A_1,t,\bal{Z}{s\tilde{\xi}_2})$ is defined by using the notation of \eqref{def:O_X_local_approx}.
Notice that $\lim_{s\to \infty} a_Z(A_1,t,\bal{Z}{s\tilde{\xi}_2})=a_Z(A_1,t) $.

We then derive the following lemma for the time-evolved operator $A_2(A_1,t)$ which satisfies the quasi-locality:
\begin{lemma} \label{lemma:LR_op_quasi}
Let $s_0$ be an arbitrary positive number. Then, we have
\begin{align}
\sum_{\substack{Z: \diam(Z) \le \xi_2\\\bal{Z}{s\tilde{\xi}_2} \cap X \neq \emptyset}}   \| a_{t,\bal{Z}{s\tilde{\xi}_2}} \| 
\le 
2(2\tilde{\xi}_2)^{D} \mathcal{C}\gamma^3 g_2  e^{v|t|/\xi_1}\bigl |  X^{(s_0\tilde{\xi}_2)}\bigl |  (s+s_0)^D  e^{-( s -1)\tilde{\xi}_2/\xi_1} 
\label{lemma_LR_op_quasi_locality_1}
\end{align}
for $s\ge 2$, where $\diam(\bal{Z}{s\tilde{\xi}_2})= \diam(Z) + 2s\tilde{\xi}_2\le  \xi_2 + 2s\tilde{\xi}_2 $ from $\diam(Z) \le \xi_2$.
In particular, for $s=1$,  we have,
\begin{align}
\sum_{\substack{Z: \diam(Z) \le \xi_2\\\bal{Z}{\tilde{\xi}_2} \cap X \neq \emptyset}}    \|a_{t,\bal{Z}{\tilde{\xi}_2}} \|  
\le (2\tilde{\xi}_2)^D g_2 \gamma \bigl |  X^{(s_0\tilde{\xi}_2)}\bigl |  ( s_0  + 1)^D .
\label{lemma_LR_op_quasi_locality_2}
\end{align}
\end{lemma}

\noindent
In this lemma, we further simplify the upper bound \eqref{lemma_LR_op_quasi_locality_1} as follows. 
By choosing $\tilde{\xi}_2$ as 
\begin{align}
2\gamma^2 \mathcal{C}e^{v|t|/\xi_1} e^{-\tilde{\xi}_2/(2\xi_1)}  \le 1,
\label{gamma_condition_tilde_xi}
\end{align}
we obtain for $s\ge 2$
\begin{align}
2(2\tilde{\xi}_2)^{D} \mathcal{C} \gamma^3 g_2  e^{v|t|/\xi_1}  (s+s_0)^D  e^{-( s -1)\tilde{\xi}_2/\xi_1} 
=&2\gamma^2  \mathcal{C} e^{v|t|/\xi_1} e^{-(s-1)\tilde{\xi}_2/(2\xi_1)}\cdot (2\tilde{\xi}_2)^{D} g_2 \gamma (s+s_0) ^D  e^{-( s -1)\tilde{\xi}_2/(2\xi_1)}    \notag \\
\le &2\gamma^2  \mathcal{C} e^{v|t|/\xi_1} e^{-\tilde{\xi}_2/(2\xi_1)}\cdot (2\tilde{\xi}_2)^{D} g_2 \gamma (s+s_0) ^D  e^{-( s -1)\tilde{\xi}_2/(2\xi_1)}    \notag \\
\le & (2\tilde{\xi}_2)^{D} g_2 \gamma (s+s_0) ^D  e^{-( s -1)\tilde{\xi}_2/(2\xi_1)},
\label{upp_bound_s_ge_2_norm_sum}
\end{align}
which reduces the inequality~\eqref{lemma_LR_op_quasi_locality_1} to 
\begin{align}
\sum_{\substack{Z: \diam(Z) \le \xi_2\\\bal{Z}{s\tilde{\xi}_2} \cap X \neq \emptyset}}   \| a_{t,\bal{Z}{s\tilde{\xi}_2}} \| \le  (2\tilde{\xi}_2)^Dg_2 \gamma  \bigl | X^{(s_0\tilde{\xi}_2)} \bigl | (s+s_0) ^D  e^{-( s -1)\tilde{\xi}_2/(2\xi_1)}. 
\label{lemma_LR_op_quasi_locality_1_seq1_simple}
\end{align}
We note that the inequality~\eqref{lemma_LR_op_quasi_locality_2} also reduces to \eqref{lemma_LR_op_quasi_locality_1_seq1_simple} for $s=1$.

Proposition~\ref{prop:LR_op_quasi_locality} is immediately given by choosing $|t|=\tau$, $A_1= H_{1:q-1}$, $A_2=H_q$.
Note the Lieb-Robinson bound $\mathcal{G}_{A_1}(x,t,\fset{X})$ in \eqref{assumption_A_1_lieb_robinson} is replaced by $\mathcal{G}_{q-1}(x,t,\fset{X})$ which has been given in \eqref{LR_func_G_Q-1_start_2}, namely $\mathcal{G}_{q-1}(x,t,\fset{X}) =  \abb\fset{X}^{(\ell_{q})} \abb^2 e^{-(x-\bar{v}_{q-1} \Delta t_{q})/\ell_{q-1}} $.
Now, the parameters $\{g_2,\xi_1,\xi_2,\tilde{\xi}_2, \mathcal{C}, v\}$ in Lemma~\ref{lemma:LR_op_quasi} are given by
\begin{align}
g_2= g_q,\quad  \xi_1=\ell_{q-1} ,\quad \xi_2=\ell_{q}, \quad \tilde{\xi}_2=\tilde{\ell}_q, \quad \mathcal{C}=1, \quad v= \bar{v}_{q-1} .
\end{align}
Then, the condition~\eqref{gamma_condition_tilde_xi} and the inequality~\eqref{lemma_LR_op_quasi_locality_1_seq1_simple} reduces 
to the condition~\eqref{gamma_condition_tau_ell_tilde_q} and the inequality~\eqref{lemma_LR_op_quasi_locality_1_seq1_simple_proof}, respectively.
This completes the proof. $\square$

\subsubsection{Proof of Lemma~\ref{lemma:LR_op_quasi}}
In the proof, we aim to obtain the upper bound of 
\begin{align}
\sum_{\substack{Z: \diam(Z) \le \xi_2\\\bal{Z}{s\tilde{\xi}_2} \cap X \neq \emptyset}}  \|a_{t,\bal{Z}{s\tilde{\xi}_2}} \|  . \label{sumZ_LR_bal1_start}
\end{align}
First, from the Lieb-Robinson bound in the form of of Eq.~\eqref{assumption_A_1_lieb_robinson} and Lemma~\ref{Bravyi, Hastings and Verstaete}, we obtain
\begin{align}
\left \| a_Z(A_1,t) - a_Z(A_1,t,\bal{Z}{s\tilde{\xi}_2})\right \| \le \mathcal{G}_{A_1}(s\xi_2,t,Z) = \mathcal{C} \abb \fset{X}^{(\xi_2)}\abb^2 e^{-(s\xi_2-v|t|)/\xi_1 },
\end{align}
which upper-bounds the norm of $a_{t,\bal{Z}{s\tilde{\xi}_2}}$ in Eq.~\eqref{def:time_ham_decomp} as follows:
\begin{align}
\left \| a_{t,\bal{Z}{s\tilde{\xi}_2}}\right \| &\le 
\left \| a_Z(A_1,t) - a_Z(A_1,t,\bal{Z}{s\tilde{\xi}_2})\right \| + \left \| a_Z(A_1,t) - a_Z(A_1,t,\bal{Z}{s\tilde{\xi}_2 -\tilde{\xi}_2})\right \|   \notag \\
&\le \|a_Z\| \cdot \mathcal{C} e^{v|t|/\xi_1}  \abb Z^{(\xi_2)}\abb^2 e^{-s \tilde{\xi}_2/\xi_1}  + \|a_Z\| \cdot \mathcal{C} e^{v|t|/\xi_1}  \abb Z^{(\xi_2)}\abb^2 e^{-( s -1)\tilde{\xi}_2/\xi_1}   \notag \\
&\le\|a_Z\| \cdot 2\gamma^2 \mathcal{C} e^{v|t|/\xi_1}  e^{-( s -1)\tilde{\xi}_2/\xi_1}  ,
\label{upper_bound_each_of_time_term}
\end{align}
where we use the inequality~\eqref{geometric_parameter_gamma3} with the condition $\diam(Z) \le \xi_2$ to obtain $\abb Z^{(\xi_2)} \abb \le \gamma [\diam(Z)/\xi_2]^D =\gamma$. 
For $s=1$, from the definitions~\eqref{def:O_X_local_approx} and \eqref{def:time_ham_decomp}, we have
\begin{align}
\left \| a_{t,\bal{Z}{\tilde{\xi}_2}}\right \| & \le \|a_Z\| . \label{s_eq_1_h_t_Z}
\end{align}

By applying the inequality~\eqref{upper_bound_each_of_time_term} to the summation~\eqref{sumZ_LR_bal1_start}, we have
\begin{align}
\sum_{\substack{Z: \diam(Z) \le \xi_2\\\bal{Z}{s\tilde{\xi}_2} \cap X \neq \emptyset}}      \|a_{t,\bal{Z}{s\tilde{\xi}_2}} \| 
\le 2\gamma^2\mathcal{C} e^{v|t|/\xi_1}  \sum_{\substack{Z: \diam(Z) \le \xi_2\\\bal{Z}{s\tilde{\xi}_2} \cap X \neq \emptyset}}  
\|a_Z\| e^{-( s -1)\tilde{\xi}_2/\xi_1} \for s\ge2. \label{sumZ_LR_bal1}
\end{align}
On the summation with respect to $Z$, we obtain 
\begin{align}
\sum_{\substack{Z: \diam(Z) \le \xi_2\\\bal{Z}{s\tilde{\xi}_2} \cap X \neq \emptyset}}   \|a_Z\| 
=\sum_{\substack{Z: \diam(Z) \le \xi_2\\Z\cap \bal{X}{s\tilde{\xi}_2}  \neq \emptyset}}     \|a_Z\| 
 &\le \sum_{j\in X[s\tilde{\xi}_2]} \sum_{Z:Z \ni j} \|a_Z\|  
 \le  \sum_{j\in X[s\tilde{\xi}_2]}  g_2 \le g_2 \left|\bal{X}{ s\tilde{\xi}_2}\right|   \notag \\
 &\le g_2 \bigl | \bal{X^{(s_0\tilde{\xi}_2)}}{ (s+s_0)\tilde{\xi}_2}\bigl | \le 
g_2 \gamma \bigl |  X^{(s_0\tilde{\xi}_2)}\bigl | \cdot   \left[ 2 (s+s_0)\tilde{\xi}_2 \right]^D
  \label{sumZ_LR_bal2}
\end{align}
for arbitrary $s_0\ge 0$, where we use the inequalities~\eqref{extensive_condition_power_law}, \eqref{coarse_grained_subset_ineq} and \eqref{geometric_parameter_gamma4} in the second, fourth and fifth inequalities, respectively.
Note that the condition $\bal{Z}{s\tilde{\xi}_2} \cap X \neq \emptyset$ for $Z\subseteq \Lambda$ is equivalent to $Z \cap \bal{X}{s\tilde{\xi}_2} \neq \emptyset$.

By combining the inequalities~\eqref{sumZ_LR_bal1} and \eqref{sumZ_LR_bal2}, we have
\begin{align}
\sum_{Z: \bal{Z}{s\tilde{\xi}_2} \cap X \neq \emptyset}  \|a_{t,\bal{Z}{s\tilde{\xi}_2}} \| 
\le 2(2\tilde{\xi}_2)^{D} \mathcal{C}\gamma^3 g_2  e^{v|t|/\xi_1}\bigl |  X^{(s_0\tilde{\xi}_2)}\bigl |   (s+s_0)^D  e^{-( s -1)\tilde{\xi}_2/\xi_1} . \label{sumZ_LR_bal3}
\end{align} 
We thus prove the inequality~\eqref{lemma_LR_op_quasi_locality_1}. 
In the same way, we can prove the case of $s=1$. By using the inequality~\eqref{s_eq_1_h_t_Z}, we have
\begin{align}
\sum_{\substack{Z: \diam(Z) \le \xi_2\\\bal{Z}{\tilde{\xi}_2} \cap X \neq \emptyset}}    \|a_{t,\bal{Z}{\tilde{\xi}_2}} \| 
\le\sum_{\substack{Z: \diam(Z) \le \xi_2\\\bal{Z}{\tilde{\xi}_2} \cap X \neq \emptyset}}   \|a_Z\| \le (2\tilde{\xi}_2)^D g_2 \gamma \bigl |  X^{(s_0\tilde{\xi}_2)}\bigl |  ( s_0  + 1)^D , 
\end{align}
where in the last inequality we use \eqref{sumZ_LR_bal2} with $s=1$.
This completes the proof of Lemma~\ref{prop:LR_op_quasi_locality}. $\square$

\subsection{Proof of Proposition~\ref{lemma:Lieb_Robinson_op_quasi_locality}: Lieb-Robinson bound by time-evolved Hamiltonian} \label{proof_sec_Lieb-Robinson bound by time-evolved Hamiltonian}

\subsubsection{Statement}
{~}\\
{\bf Proposition~\ref{lemma:Lieb_Robinson_op_quasi_locality}.}
\textit{
Let $\tilde{H}$ be a Hamiltonian such that 
\begin{align} 
\tilde{H} = \sum_{Z:Z\subseteq \Lambda} h_Z = \sum_{s=1}^\infty \sum_{\diam(Z_s)\le s\xi } h_{Z_s}.
\label{Def_tilde_H_proof}
\end{align} 
For an arbitrary subset $X$ and an arbitrary positive $s_0$, we assume 
\begin{align} 
&\sum_{ \substack{\diam(Z_s)\le s\xi \\ Z_s \cap X \neq \emptyset } } \| h_{Z_s}\| \le 
\tilde{g} \bigl | X^{(s_0\xi)} \bigl | (s+s_0) ^D e^{-\mu (s-1)} \quad (\mu>1).
\label{extensiveness_estimation_tilde_H}
\end{align} 
Then, the Hamiltonian $\tilde{H}$ satisfies $\mathcal{G}_{\tilde{H}}(x,t,\fset{X})$-Lieb-Robinson bound with 
\begin{align} 
\mathcal{G}_{\tilde{H}}(x,t,\fset{X})=e \abb \fset{X}^{(\xi)}\abb  \left[ \tilde{v}  |t|\exp\left( - \frac{\mu-1}{2}\frac{x}{\xi} +\tilde{v}  |t|\right) +\left( \frac{e^2 \tilde{v}|t|}{m_x^\ast} \right)^{m_x^\ast} \right]
 \label{Lieb_Robinson_tilde_H_t}
\end{align} 
with 
\begin{align} 
\tilde{v}:=4\tilde{g}\gamma (4D)^D ,\quad  m_x^\ast :=\left\lfloor \frac{\mu-1}{2\mu}\frac{x}{\xi} \right \rfloor +1.
\label{def:tilde_v_m_ast_proof}
\end{align} 
We note that the same inequality holds for time-dependent Hamiltonians.
}

\subsubsection{Proof of Proposition~\ref{lemma:Lieb_Robinson_op_quasi_locality}}

We estimate the norm of $\| [ O_L (\tilde{H},t) , O_{L'}] \|$,
where $O_L$ and $O_{L'}$ are arbitrary operators with the unit norm (i.e., $\|O_L\|=\|O_{L'}\|=1$) which are supported on $L$ and $L'$, respectively.
Following Ref.~\cite{ref:Hastings2006-ExpDec}, we start from the inequality as 
\begin{align} 
\frac{d}{dt} \| [ O_L (\tilde{H},t) , O_{L'}] \|  = \frac{d}{dt} \| [e^{i\tilde{H}t}O_Le^{-i\tilde{H}t}, O_{L'}] \|  \le
 2\|O_L\|\cdot \|[\tilde{H}_L(\tilde{H},\tau) , O_{L'} ]\| = 2\|[\tilde{H}_L(\tilde{H},\tau) , O_{L'} ]\| ,
\label{LR_start_inequality_commutator}
\end{align} 
where we define $\tilde{H}_L$ as follows:
\begin{align} 
\tilde{H}_L := \sum_{s=1}^\infty \sum_{ \substack{\diam(Z_s)\le s\xi \\ Z_s \cap L \neq \emptyset } } h_{Z_s}.
\end{align}
The operator $\tilde{H}_L$ picks up all the interaction terms $\{h_Z\}_{Z\subseteq \Lambda}$ in \eqref{Def_tilde_H_proof} that act on the subset $L$.

By using the inequality~\eqref{LR_start_inequality_commutator}, we first obtain
\begin{align} 
\| [ O_L (\tilde{H},t)  , O_{L'}] \| &\le \| [ O_L,O_{L'}] \| + 2 \int_0^t \|[\tilde{H}_L(\tilde{H},\tau) , O_{L'}]\| d\tau  \notag \\
&\le 2  \sum_{s_1=1}^\infty \sum_{ \substack{\diam(Z_{s_1})\le s_1\xi \\ Z_{s_1} \cap L \neq \emptyset } }  \int_0^t \|[h_{Z_{s_1}} (\tilde{H},\tau_1) , O_{L'} ]\| d\tau_1   ,
\label{commutator_first_expansion_1}
\end{align} 
where we use $d_{L,L'}\ge 1$ which implies $[ O_L,O_{L'}] =0$.
By applying the inequality~\eqref{LR_start_inequality_commutator} to $\|[h_{Z_{s_1}} (\tilde{H},\tau_1) , O_{L'} ]\| $, we have
\begin{align} 
\|[h_{Z_{s_1}} (\tilde{H},\tau_1) , O_{L'} ]\| \le \| [ h_{Z_{s_1}} , O_{L'}] \| + 2\|h_{Z_{s_1}} \| \sum_{s_2=1}^\infty 
 \sum_{ \substack{\diam(Z_{s_2})\le s_2\xi \\ Z_{s_2} \cap Z_{s_1} \neq \emptyset } } \int_0^{\tau_1} \|[h_{Z_{s_2}} (\tilde{H},\tau_2) , O_{L'} ]\| d\tau _2 .
\label{commutator_first_expansion_2_pre}
\end{align} 
In the following, for the simplicity of the notation, we simply describe as 
\begin{align} 
\sum_{ \substack{\diam(Z_{s_2})\le s_2\xi \\ Z_{s_2} \cap Z_{s_1} \neq \emptyset } } = \sum_{ Z_{s_2} \cap Z_{s_1} \neq \emptyset} 
\end{align} 
by omitting the constraint of $\diam(Z_{s_2})\le s_2\xi $.
By combining the inequalities~\eqref{commutator_first_expansion_1} and \eqref{commutator_first_expansion_2_pre}, we have
\begin{align} 
&\| [ O_L (\tilde{H},t)  , O_{L'}] \| 
\le  2|t| \sum_{s_1=1}^\infty  \sum_{\substack{Z_{s_1} \cap L \neq \emptyset \\ Z_{s_1} \cap L' \neq \emptyset}}   \| [ h_{Z_{s_1}}  , O_{L'}] \| \notag \\
& +2^2 \sum_{s_1=1}^\infty \sum_{Z_{s_1} \cap L \neq \emptyset} \|h_{Z_{s_1}}\| \sum_{s_2=1}^\infty \sum_{Z_{s_2} \cap Z_{s_1} \neq \emptyset}  \int_0^t \int_0^{\tau_1} \|[h_{Z_{s_2}} (\tilde{H},\tau_2) , O_{L'}]\| d\tau _2 d\tau_1   ,
\end{align} 
where in the summation of the first term we add the restriction of $Z_{s_1} \cap L'\neq \emptyset$ because of $ \| [ h_{Z_{s_1}}  , O_{L'}] \| = 0$ for $Z_{s_1} \cap L' = \emptyset$.

By repeating this process $(m^\ast-1)$ times, we obtain
\begin{align} 
&\| [ O_L (\tilde{H},t) , O_{L'}] \|  \notag \\
\le&2|t| \sum_{s_1=1}^\infty  \sum_{\substack{Z_{s_1} \cap L \neq \emptyset \\ Z_{s_1} \cap L' \neq \emptyset}} \| [ h_{Z_{s_1}}  , O_{L'}] \| 
 +\frac{(2|t|)^2}{2!}\sum_{s_1=1}^\infty \sum_{Z_{s_1} \cap L \neq \emptyset} \|h_{Z_{s_1}}\|  \sum_{s_2=1}^\infty  \sum_{\substack{Z_{s_2} \cap Z_{s_1} \neq \emptyset \\ Z_{s_2} \cap L' \neq \emptyset}}\| [ h_{Z_{s_2}}  , O_{L'}] \| \notag \\
 +& \frac{(2|t|)^3}{3!}\sum_{s_1=1}^\infty \sum_{Z_{s_1} \cap L \neq \emptyset} \|h_{Z_{s_1}}\|  \sum_{s_2=1}^\infty \sum_{Z_{s_2} \cap Z_{s_1} \neq \emptyset}  \|h_{Z_{s_2}} \| 
\sum_{s_3=1}^\infty  \sum_{\substack{Z_{s_3} \cap Z_{s_2} \neq \emptyset \\ Z_{s_3} \cap L' \neq \emptyset}}  \| [ h_{Z_{s_3}}  , O_{L'}] \|   \notag \\
 +&\cdots + 2^{m^\ast}  \sum_{s_1=1}^\infty \sum_{Z_{s_1} \cap L \neq \emptyset} \|h_{Z_{s_1}}\|  
\sum_{s_2=1}^\infty \sum_{Z_{s_2} \cap Z_{s_1} \neq \emptyset}  \|h_{Z_{s_2}} \|   \cdots \sum_{s_m^\ast=1}^\infty \sum_{Z_{s_{m^\ast}} \cap Z_{s_{m^\ast-1}} \neq \emptyset}  \notag \\
&\quad\quad \quad \quad\quad \quad \int_0^t\int_0^{\tau_1} \cdots \int_0^{\tau_{m^\ast-1}} \|[ h_{Z_{s_{m^\ast}}}(\tilde{H},\tau_{m^\ast}) ,O_{L'} ]\|  d\tau_1  d\tau_2\cdots  d\tau_{m^\ast}\notag \\
 \le &\sum_{m=1}^{m^\ast}\mathcal{L}_m.
 \label{Lieb-Robinson_expansion_0}
\end{align} 
In the above inequality, we define $\{\mathcal{L}_m\}_{m=1}^{m^\ast}$ as 
 \begin{align} 
\mathcal{L}_m:=&\frac{2(2|t|)^m}{m!} \sum_{s_1=1}^\infty \sum_{Z_{s_1} \cap L \neq \emptyset} \|h_{Z_{s_1}}\|  
\sum_{s_2=1}^\infty \sum_{Z_{s_2} \cap Z_{s_1} \neq \emptyset}  \|h_{Z_{s_2}} \|\cdots 
 \sum_{s_m=1}^\infty  \sum_{\substack{Z_{s_m} \cap Z_{s_{m-1}} \neq \emptyset \\ Z_{s_m} \cap L' \neq \emptyset}} \| h_{Z_{s_m}} \| 
\label{mathdal_L_m_def}
\end{align} 
for $m=1,2,\ldots,m^\ast-1$, where we use $ \| [ h_{Z_{s_m}}  , O_{L'}] \| \le 2 \|h_{Z_{s_m}}\|$ from $\|O_{L'}\|=1$.
For $m=m^\ast$, we define $\mathcal{L}_{m^\ast}$ as
 \begin{align} 
\mathcal{L}_{m^\ast}:= \frac{2(2|t|)^{m^\ast}}{m^\ast!}  \sum_{s_1=1}^\infty \sum_{Z_{s_1} \cap L \neq \emptyset} \|h_{Z_{s_1}}\|  
&\sum_{s_2=1}^\infty \sum_{Z_{s_2} \cap Z_{s_1} \neq \emptyset}  \|h_{Z_{s_2}} \|  \cdots \sum_{s_m^\ast=1}^\infty \sum_{Z_{s_{m^\ast}} \cap Z_{s_{m^\ast-1}} \neq \emptyset}  \|h_{Z_{s_{m^\ast}}} \| .
\label{mathdal_L_m_ast__def}
\end{align} 
We notice that for $\mathcal{L}_{m^\ast}$ the summation with respect to $Z_{s_{m^\ast}}$ is not restricted to $\{Z_{s_{m^\ast}} |Z_{s_{m^\ast}}\cap L'\neq \emptyset\}$.

We, in the following, aim to upper-bound the $m$th term $\mathcal{L}_m$ in Eq.~\eqref{mathdal_L_m_def}.
By using the condition~\eqref{extensiveness_estimation_tilde_H} with $s_0=s_{j-1}$, we obtain the upper bound of
 \begin{align} 
\sum_{Z_{s_j} \cap Z_{s_{j-1}} \neq \emptyset} \|h_{Z_{s_j}}\| \le  \tilde{g} \abb Z_{s_{j-1}}^{(s_{j-1} \xi)}\abb (s_j+s_{j-1})^D e^{-\mu (s_j-1)} \le \tilde{g} \gamma  (s_j+s_{j-1})^D e^{-\mu (s_j-1)}
\label{sum_Z_s_j_j_ge2}
\end{align} 
for $j\ge 2$, where we use the inequality~\eqref{geometric_parameter_gamma3} with the condition $\diam(Z_{s_{j-1}}) \le s_{j-1}\xi$ to obtain 
 \begin{align} 
\abb Z_{s_{j-1}}^{(s_{j-1} \xi)}\abb \le  \gamma [\diam(Z_{s_{j-1}})/(s_{j-1}\xi)]^D = \gamma.
\end{align} 
Also, from the condition~\eqref{extensiveness_estimation_tilde_H} with $s_0=1$, we obtain
 \begin{align} 
\sum_{Z_{s_1} \cap L \neq \emptyset} \|h_{Z_{s_j}}\| \le  \tilde{g} \abb L^{(\xi)}\abb (1+s_1)^D e^{-\mu (s_1-1)} \le \tilde{g} \gamma  \abb L^{(\xi)}\abb (1+s_1)^D e^{-\mu (s_1-1)},
\label{sum_Z_s_j_j_eq1}
\end{align} 
where we use $\gamma\ge 1$. 
By combining the inequalities~\eqref{sum_Z_s_j_j_ge2} and \eqref{sum_Z_s_j_j_eq1}, we have 
 \begin{align} 
\sum_{Z_{s_1} \cap L \neq \emptyset} \|h_{Z_{s_1}}\|\sum_{Z_{s_2} \cap Z_{s_1} \neq \emptyset}  \|h_{Z_{s_2}} \|  
\cdots \sum_{Z_{s_m} \cap Z_{s_{m-1}} \neq \emptyset}  \|h_{Z_{s_m}} \| 
\le \prod_{j=1}^m g_{s_{j-1},s_j},
\label{product_g_s_j-1_g_j}
\end{align} 
where $g_{s_{j-1},s_j}:=\tilde{g} \gamma (s_j+s_{j-1})^D e^{-\mu (s_j-1)}$ for $j\ge 2$ and $g_{s_{0},s_1} =\tilde{g} \gamma \abb L^{(\xi)}\abb (1+s_1)^D e^{-\mu (s_1-1)}$. 
For the product of $\{g_{s_{j-1},s_j}\}_{j=1}^m$, the following inequality holds as
 \begin{align}  
\prod_{j=1}^m g_{s_{j-1},s_j}&= (\tilde{g}\gamma )^m e^{-\mu(S_m-m)} \abb L^{(\xi)}\abb (1+s_1)^D    \prod_{j=2}^m (s_j+s_{j-1})^D   \le \abb L^{(\xi)}\abb (\tilde{g}\gamma )^m \left(\frac{2S_m}{m} \right)^{mD} e^{-\mu(S_m-m)},
\label{product_g_s_j-1_g_j_ineq}
\end{align} 
where $S_m :=\sum_{j=1}^m s_j \ge m$ and we use 
 \begin{align}  
 (1+s_1)^D \prod_{j=2}^m (s_j+s_{j-1})^D   \le \left(\frac{2S_m}{m} \right)^{mD} .
\end{align} 
Under the constraints of $\{Z_{s_1} \cap L \neq \emptyset, Z_{s_2} \cap Z_{s_1} \neq \emptyset, \ldots, Z_{s_m} \cap Z_{s_{m-1}} \neq \emptyset\}$, a necessary condition for $Z_{s_m} \cap L' \neq \emptyset$ is given by
 \begin{align}  
S_m \xi \ge \dist_{L,L'},
\end{align} 
and hence by combining \eqref{mathdal_L_m_def}, \eqref{product_g_s_j-1_g_j} and \eqref{product_g_s_j-1_g_j_ineq}, the $m$th term $\mathcal{L}_m$ is bounded from above by
 \begin{align} 
\mathcal{L}_m\le  
\frac{2(2\tilde{g} \gamma |t|)^m}{m!} \abb L^{(\xi)}\abb  \sum_{\substack{s_1\ge 1, s_2 \ge 1 ,\cdots , s_m\ge 1 \\ S_m \xi \ge \dist_{L,L'}} }  \left(\frac{2S_m}{m} \right)^{mD} e^{-\mu(S_m-m)} .
\label{upper_bound_mathcal_L_m_formal}
\end{align} 
We upper-bound the summation as follows:
 \begin{align} 
\sum_{\substack{s_1\ge 1, s_2 \ge 1 ,\cdots , s_m\ge 1 \\ S_m \xi \ge \dist_{L,L'}} }  \left(\frac{2S_m}{m} \right)^{mD} e^{-\mu(S_m-m)}
=&\sum_{S \ge \max(m, \dist_{L,L'}/ \xi)} \sum_{\substack{s_1\ge 1, s_2 \ge 1 ,\cdots , s_m\ge 1 \\ s_1+ s_2 +\cdots s_m=S}}   \left(\frac{2S}{m} \right)^{mD} e^{-\mu(S-m)} \notag \\
=&\sum_{S \ge \max(m, \dist_{L,L'}/ \xi)} \multiset{m}{S-m}  \left(\frac{2S}{m} \right)^{mD} e^{-\mu(S-m)} \notag \\
=&\sum_{S \ge \max(m, \dist_{L,L'}/ \xi)} \binom{S-1}{m-1}  \left(\frac{2S}{m} \right)^{mD} e^{-\mu(S-m)} \notag \\
\le &\frac{2^{mD} e^{\mu m}}{m^{mD} (m-1)!}\sum_{S \ge \dist_{L,L'}/ \xi} S^{mD +m -1} e^{-\mu S}, \label{sum_S_m_s_1___s_m}
\end{align} 
where the summation with respect to $\{s_1, s_2, \cdots , s_m\}$ such that $s_1+s_2+\cdots +s_{m}=S$ is equal to the $(S-m)$-multicombination from a set of $m$ elements, and we use $\binom{S-1}{m-1} \le S^{m-1}/(m-1)!$ in the inequality.
For an arbitrary constant $S_0 \ge 0$ and $z \in \mathbb{N}$, we obtain
 \begin{align} 
\sum_{S \ge S_0}  S^z e^{-\mu S} =\sum_{S \ge S_0}  S^z e^{-(\mu -1) S}  e^{-S} &\le e^{-(\mu-1) S_0} \sum_{S \ge S_0}  S^z e^{-S}  \notag \\
&\le e^{-(\mu-1) S_0} \int_{S_0}^\infty x^z e^{-x+1} dx \le e^{-(\mu-1) S_0+1} z!,
\end{align} 
where  in the first inequality we use the condition $\mu>1$.
By using the above inequality with $S_0=\dist_{L,L'}/ \xi$ and $z=mD+m-1$, the inequality~\eqref{sum_S_m_s_1___s_m} reduces to 
 \begin{align} 
\sum_{\substack{s_1\ge 1, s_2 \ge 1 ,\cdots , s_m\ge 1 \\ S_m \xi \ge \dist_{L,L'}} }  \left(\frac{2S_m}{m} \right)^{mD} e^{-\mu(S_m-m)} 
\le &\frac{2^{mD} e^{1+\mu m-(\mu-1)\dist_{L,L'}/ \xi }}{m^{mD} (m-1)!}(mD +m -1)!  \notag \\
=&\frac{2^{mD} e^{1+\mu m-(\mu-1)\dist_{L,L'}/ \xi }}{m^{mD}} (mD)!\binom{mD+m-1}{m-1}  \notag \\
\le &e(2D)^{mD} e^{\mu m-(\mu-1)\dist_{L,L'}/ \xi }\binom{mD+m-1}{m-1} \notag \\
\le &e(4D)^{mD}  2^{m-1} e^{\mu m-(\mu-1)\dist_{L,L'}/ \xi }, \label{sum_S_m_s_1___s_m_final_form}
\end{align} 
where in the second inequality we use $(mD)! \le (mD)^{mD}$ and in the last inequality we use $\binom{mD+m-1}{m-1} \le 2^{mD+m-1}$.

Therefore, by combining the inequalities~\eqref{upper_bound_mathcal_L_m_formal} and \eqref{sum_S_m_s_1___s_m_final_form},
we finally arrive at the inequality for $\mathcal{L}_m$ as 
 \begin{align} 
\mathcal{L}_m\le  
\frac{e[4\tilde{g} \gamma (4D)^D |t|]^m }{m!}  \abb L^{(\xi)}\abb e^{\mu m-(\mu-1)\dist_{L,L'}/ \xi } 
=\frac{e(\tilde{v} |t|)^m }{m!}  \abb L^{(\xi)}\abb e^{\mu m-(\mu-1)\dist_{L,L'}/ \xi } ,
\end{align} 
where we use the definition of $\tilde{v}$ in Eq.~\eqref{def:tilde_v_m_ast_proof}.
In the same way, we can derive the similar upper bound for $\mathcal{L}_{m^\ast}$.
For $\mathcal{L}_{m^\ast}$, only the difference is that the constraint of $S_m \xi \ge \dist_{L,L'}$ in \eqref{upper_bound_mathcal_L_m_formal} is removed.
Hence, the same calculations as \eqref{sum_S_m_s_1___s_m} and \eqref{sum_S_m_s_1___s_m_final_form} give
 \begin{align} 
\sum_{s_1\ge 1, s_2 \ge 1 ,\cdots , s_m\ge 1}  \left(\frac{2S_m}{m} \right)^{mD} e^{-\mu(S_m-m)} 
=&\sum_{S \ge m} \binom{S-1}{m-1}  \left(\frac{2S}{m} \right)^{mD} e^{-\mu(S-m)} \notag \\
\le &\frac{2^{mD} e^{\mu m}}{m^{mD} (m-1)!}\sum_{S \ge m} S^{mD +m -1} e^{-\mu S} \notag \\
\le& \frac{2^{mD} e^{\mu m}}{m^{mD} (m-1)!} e^{-(\mu-1) m+1} (mD +m -1)!   \le e(4 D)^{mD} 2^{m-1} e^m .
\end{align} 
We thus obtain from Eq.~\eqref{mathdal_L_m_ast__def}
 \begin{align} 
\mathcal{L}_{m^\ast} \le  
\frac{e[4e\tilde{g} \gamma (4D)^D |t|]^{m^\ast}}{m^\ast!} \abb L^{(\xi)}\abb
=\frac{e(e\tilde{v} |t|)^{m^\ast}}{m^\ast!} \abb L^{(\xi)}\abb
 \le e\left( \frac{e^2 \tilde{v}|t|}{m^\ast} \right)^{m^\ast}  \abb L^{(\xi)}\abb ,
\end{align} 
where we use $m^\ast! \ge (m^\ast/e)^{m^\ast}$.

We here choose $m^\ast$ such that 
\begin{align} 
&\mu (m^\ast-1)-(\mu-1)\frac{\dist_{L,L'}}{\xi}\le - \frac{\mu-1}{2}\frac{\dist_{L,L'}}{\xi}\notag \\
&\Or m^\ast -1 =\left\lfloor \frac{\mu-1}{2\mu}\frac{\dist_{L,L'}}{\xi} \right \rfloor.
\end{align} 
Then,  by using the upper-bounds for $\{\mathcal{L}_m\}_{m=1}^{m^\ast}$, we arrive at the following inequality:
\begin{align} 
\| [ O_L (t) , O_{L'}] \|  \le\mathcal{L}_{m^\ast}+ \sum_{m=1}^{m^\ast-1}\mathcal{L}_m
&\le e  \abb L^{(\xi)}\abb  \left[ \tilde{v}  |t|\exp\left(\mu (m^\ast-1)-(\mu-1)\frac{\dist_{L,L'}}{\xi} +\tilde{v}  |t|\right) +\left( \frac{e^2 \tilde{v}|t|}{m^\ast} \right)^{m^\ast} \right] \notag \\
&\le e  \abb L^{(\xi)}\abb  \left[ \tilde{v}  |t|\exp\left( - \frac{\mu-1}{2}\frac{\dist_{L,L'}}{\xi} +\tilde{v}  |t|\right) +\left( \frac{e^2 \tilde{v}|t|}{m^\ast} \right)^{m^\ast} \right],
\end{align} 
where we use $\sum_{m=1}^\infty x^m/m! \le x\sum_{m=0}^\infty x^m/(m+1)! \le x\sum_{m=0}^\infty x^m/m! \le xe^x$. 
We thus prove the inequality~\eqref{Lieb_Robinson_tilde_H_t}. This completes the proof.
$\square$

\subsection{Proof of Lemma~\ref{lemma:Lieb_Robinson_op_quasi_locality_short_range}: Lieb-Robinson bound for Hamiltonian with a finite interaction length} \label{Sec:Lieb-Robinson bound for Hamiltonian with a finite length scale}

\subsubsection{Statement}
{~}\\
{\bf Lemma~\ref{lemma:Lieb_Robinson_op_quasi_locality_short_range}.}
\textit{
Let $\tilde{H}$ be a Hamiltonian such that 
\begin{align} 
\tilde{H} = \sum_{Z\subset \Lambda:\diam(Z)\le \xi } h_Z \label{tilde_H_xi_finite}
\end{align} 
with the same condition as~\eqref{basic_assump_power}, namely
\begin{align}
\sup_{i \in \Lambda}\sum_{\substack{Z:Z\ni i \\  r \le \diam(Z) \le \xi}} \| h_Z\| \le  g r^{-\alpha+D} \for  r\le\xi .\label{g-extensive_with_decay_proof}
\end{align}
Then, the Hamiltonian $\tilde{H}$ satisfies $\mathcal{G}_{\tilde{H}}(x,t,\fset{X})$-Lieb-Robinson bound with 
\begin{align} 
\mathcal{G}_{\tilde{H}}(x,t,\fset{X})= \min \left(2 |\fset{X}|  \left( \frac{v_0 |t| }{e^2 \lceil x/ \xi \rceil }\right)^{\lceil x/ \xi \rceil},2 \right)
\le 2 |\fset{X}|  e^{- 2(x- \xi v_0 |t| )/ \xi }
 \label{Lieb_Robinson_tilde_H_t_short_proof}
\end{align} 
with 
\begin{align} 
v_0:= \frac{2 e^3 g \gamma (\alpha-2D)}{\alpha-2D -1} . \label{Lieb-Robinson_velog_proof}
\end{align} 
We note that the same inequality holds for time-dependent Hamiltonians.
}

{~}\\
{\bf Remark.}
When we consider a Hamiltonian with $k$-body interactions ($k=\orderof{1}$) (i.e., $k$-local Hamiltonian), the inequality~\eqref{Lieb_Robinson_tilde_H_t_short_proof} is trivially obtained for $\alpha>D$; for example, please see Sec. 2.3 in Ref.~\cite{Phd_kuwahara} or Appendix C in Ref.~\cite{PhysRevA.80.052104}.
However, the $k$-dependence of the Lieb-Robinson bound is roughly given by
\begin{align} 
\mathcal{G}_{\tilde{H}}(x,t,\fset{X}) \sim  \left( \frac{\orderof{gk} |t|}{x/\xi}\right)^{x/\xi}.
\end{align} 
This estimation gives the Lieb-Robinson velocity of order $\orderof{1/(k\xi)}$.
The Hamiltonian~\eqref{tilde_H_xi_finite} includes up to $\orderof{\xi^D}$-body interactions, namely $k=\orderof{\xi^D}$.
Hence, the Lieb-Robinson velocity is not given by $\orderof{\xi}$, but given by $\orderof{\xi^{D+1}}$.
Therefore, in order to obtain the Lieb-Robinson velocity of $\orderof{\xi}$, we need to utilize the power-law decay as in \eqref{g-extensive_with_decay_proof} with $\alpha>2D+1$. 

\subsubsection{Proof of Lemma~\ref{lemma:Lieb_Robinson_op_quasi_locality_short_range}}

The proof is almost the same as Proposition~\ref{lemma:Lieb_Robinson_op_quasi_locality}. 
We note that $\min(\cdot,2)$ is given by the trivial upper bound of $\mathcal{G}_{\tilde{H}}(x,t,\fset{X})\le 2$.
We start from the same inequality as~\eqref{Lieb-Robinson_expansion_0}:
\begin{align} 
&\| [ O_L (\tilde{H}, t) , O_{L'}] \| \le \sum_{m=1}^{m^\ast}\mathcal{L}_m
 \label{Lieb-Robinson_expansion_0_finite}
\end{align} 
with $\|O_L \| = \| O_{L'}\|=1$, where $\mathcal{L}_m$ ($m<m^\ast$) and $\mathcal{L}_{m^\ast}$ are defined as
 \begin{align} 
\mathcal{L}_m= \frac{2(2|t|)^m}{m!} \sum_{Z_1 \cap L \neq \emptyset} \|h_{Z_1}\|  
\sum_{Z_{2} \cap Z_{1} \neq \emptyset}  \|h_{Z_{2}} \|\cdots    \sum_{\substack{Z_{m} \cap Z_{m-1} \neq \emptyset\\ Z_{m}\cap L'\neq \emptyset}}  \| h_{Z_m} \| 
\end{align} 
and 
\begin{align} 
\mathcal{L}_{m^\ast}= \frac{2(2|t|)^{m^\ast}}{m^\ast!} \sum_{Z_1 \cap L \neq \emptyset} \|h_{Z_1}\|  
\sum_{Z_{2} \cap Z_{1} \neq \emptyset}  \|h_{Z_{2}} \|\cdots    \sum_{Z_{m^\ast} \cap Z_{m^\ast-1} \neq \emptyset}  \| h_{Z_{m^\ast}} \| .
\label{mathdal_L_m_def_finite_________1}
\end{align} 
Now, due to the finite interaction length of $\xi$, we have $\mathcal{L}_m=0$ as long as $\xi m < \dist_{L,L'}$. Hence, if we choose 
 \begin{align} 
m^\ast = \lceil \dist_{L,L'}/ \xi \rceil \label{def_of_m_ast_d},
\end{align} 
we obtain $\mathcal{L}_{m}=0$ for $m\le m^\ast-1 < \dist_{L,L'}/ \xi $.  
This choice of $m^\ast$ gives the inequality of
\begin{align} 
&\| [ O_L (\tilde{H}, t) , O_{L'}] \| \le \mathcal{L}_{m^\ast}.
\end{align} 
Our task is to estimate the upper bound of $\mathcal{L}_{m^\ast}$.

In order to estimate $\mathcal{L}_{m^\ast}$, we first decompose the Hamiltonian as 
\begin{align} 
\tilde{H} = \sum_{\diam(Z)\le \xi } h_Z = \sum_{r=1}^{\xi} \sum_{Z_r: \diam(Z_r)=r } h_{Z_r}.
\end{align} 
We notice that the condition~\eqref{g-extensive_with_decay_proof} gives 
 \begin{align} 
\sum_{Z_r: \diam(Z_r)=r  , Z_r \ni i} \| h_{Z_r}\| \le\sum_{Z: \diam(Z) \ge r  , Z \ni i} \| h_{Z}\| \le g r^{-\alpha+D} \for \forall i\in \Lambda .
\end{align} 
Therefore, we have for arbitrary subset $L\subseteq\Lambda$ 
\begin{align} 
\sum_{Z_r: \diam(Z_r)=r , Z_r \cap L\neq \emptyset} \| h_{Z_r} \| \le 
\sum_{i\in L}\sum_{Z_r: \diam(Z_r)=r , Z_r \ni i}\| h_{Z_r} \|   \le g|L| r^{-\alpha+D} , 
\end{align} 
which gives the following upper bound:
\begin{align} 
&\sum_{Z_{m^\ast} \cap Z_{m^\ast-1} \neq \emptyset}  \| h_{Z_{m^\ast}} \|
\le  g |Z_{m^\ast-1}| \sum_{r_{m^\ast}=1}^\xi r_{m^\ast}^{-\alpha+D}.
\end{align} 
From the inequality~\eqref{geometric_parameter_gamma1}, we have $|Z| \le \gamma [\diam(Z)]^D$, which yields the inequality of 
\begin{align} 
 \sum_{Z_{m^\ast-1} \cap Z_{m^\ast-2} \neq \emptyset}  \| h_{Z_{m^\ast-1}} \| \cdot |Z_{m^\ast-1}|
&\le g |Z_{m^\ast-2}| \sum_{r_{m^\ast-1}=1}^\xi  (\gamma r_{m^\ast-1}^D)  r_{m^\ast-1}^{-\alpha+D}.
\end{align} 
By repeating the same process, we obtain 
\begin{align} 
\sum_{Z_1 \cap L \neq \emptyset} \|h_{Z_1}\|  
\sum_{Z_{2} \cap Z_{1} \neq \emptyset}  \|h_{Z_{2}} \|\cdots    \sum_{Z_{m^\ast} \cap Z_{m^\ast-1} \neq \emptyset}  \| h_{Z_{m^\ast}} \| 
&\le |L| \sum_{r_1,r_2,\ldots,r_{m^\ast}=1}^\xi  g r_{m^\ast}^{-\alpha+D} \prod_{s=1}^{m^\ast-1} g (\gamma r_{s}^D ) r_s^{-\alpha+D} \notag \\
& \le |L| \left(\sum_{r=1}^\xi  g \gamma r^{-\alpha+2D} \right)^{m^\ast},
\label{mathdal_L_m_def_finite_ele_1}
\end{align} 
where we use $r_{m^\ast}^{-\alpha+D} \le \gamma r_{m^\ast}^{-\alpha+2D}$ for $r_{m^\ast}\ge 1$ from $\gamma\ge 1$. 
For $\alpha>2D+1$, we obtain
\begin{align} 
\sum_{r=1}^\xi  r^{-\alpha+2D} \le 1+ \sum_{r=2}^\infty  r^{-\alpha+2D} \le 1+ \int_1^\infty x^{-\alpha+2D} dx = \frac{\alpha-2D}{\alpha-2D -1}.
\label{mathdal_L_m_def_finite_ele_2}
\end{align} 
By applying the inequalities~\eqref{mathdal_L_m_def_finite_ele_1} and~\eqref{mathdal_L_m_def_finite_ele_2} to Eq.~\eqref{mathdal_L_m_def_finite_________1}, we obtain the upper bound of $\mathcal{L}_{m^\ast}$ by
\begin{align} 
\mathcal{L}_{m^\ast}\le  \frac{2(2|t|)^{m^\ast}}{m^\ast!} |L|  \left(\frac{g \gamma (\alpha-2D)}{\alpha-2D -1} \right)^{m^\ast} 
\le 2 |L|  \left(\frac{2 e^3 g \gamma (\alpha-2D)}{\alpha-2D -1}  \frac{ |t| }{e^2 m^\ast }\right)^{m^\ast} = 
2 |L|  \left( \frac{v_0 |t| }{e^2 m^\ast }\right)^{m^\ast},
\end{align} 
where we use $m^\ast! \ge (m^\ast/e)^{m^\ast}$ and $v_0$ was defined in Eq.~\eqref{Lieb-Robinson_velog_proof}.
We thus obtain the first inequality in~\eqref{Lieb_Robinson_tilde_H_t_short_proof} from the choice of $m^\ast$ as in Eq.~\eqref{def_of_m_ast_d}.

Finally, we need to prove
\begin{align} 
\min \left(2 |L|  \left( \frac{v_0 |t| }{e^2 \lceil x/ \xi \rceil }\right)^{\lceil x/ \xi \rceil},2 \right)
\le 2 |L|  e^{- 2(x- \xi v_0 |t| )/ \xi }.
 \label{Lieb_Robinson_tilde_H_t_short_exp_form_prof}
\end{align} 
For $x<\xi v_0 |t| $, the inequality trivially holds, and hence we only need to consider $x\ge \xi v_0 |t| $.
For $x\ge \xi v_0 |t| $, we use the following upper bounds: 
\begin{align} 
\left( \frac{v_0 |t| }{e^2 \lceil x/ \xi \rceil }\right)^{\lceil x/ \xi \rceil} \le \left( \frac{v_0 |t| }{e^2 (x/ \xi) }\right)^{x/ \xi} \le e^{-2x/ \xi } \le e^{- 2(x- \xi v_0 |t| )/ \xi },
\end{align} 
where the last inequality is a trivial inequality. 
We thus obtain the inequality~\eqref{Lieb_Robinson_tilde_H_t_short_exp_form_prof}.  This completes the proof. $\square$

\subsection{Proof of Proposition~\ref{lemma:Connection of two Lieb-Robinson bounds_1}: connection of two Lieb-Robinson bounds from different unitary operators}
\label{connection of two Lieb-Robinson bounds from different unitary operators}


\subsubsection{Statement}
{~}\\
{\bf Proposition~\ref{lemma:Connection of two Lieb-Robinson bounds_1}.}
\textit{
We consider a Hamiltonian $H_0$ which satisfies the Lieb-Robinson bound for a fixed $t$ as 
\begin{align}
\mathcal{G}(x,t ,\fset{X},\fset{Y}) \le \mathcal{C} e^{-(x- v|t|)/\xi}   \label{Lieb_Robinson_O_i_assump_sup}
\end{align}
for $\diam(\fset{X}) , \diam(\fset{Y}) \le \nu \xi$ ($\nu \ge 3$).
Also, let $H'_0$ be a Hamiltonian which satisfies the Lieb-Robinson bound for a fixed $t'$ as 
\begin{align}
&\mathcal{G}' (x,t',\fset{X}) = \abb \fset{X}^{(\nu \xi)}\abb^2 \mathcal{F}' (x,t'),  \label{Lieb_Robinson_O_i_assump2_1_sup}
\end{align}
where the function $\mathcal{F}' (x,t')$ is assumed to satisfy
\begin{align}
\mathcal{F}' (x,t') \le   \mathcal{F}' (x_0,t') e^{-2(x-x_0 )/\xi} \for x\ge 0, \quad 0\le x_0 \le x.
\label{Lieb_Robinson_O_i_assump2_2_sup}
\end{align}
Note that we impose no assumptions on $\diam(\fset{X})$ for $\mathcal{G}' (x,t',\fset{X})$. We then obtain the Lieb-Robinson function $\mathcal{G}''(x,t,t' ,\fset{X},\fset{Y})$  for $e^{-iH_0t}e^{-iH'_0 t'}$ as 
\begin{align}
\mathcal{G}''(x,t,t' ,\fset{X},\fset{Y})
= 4\mathcal{C}\gamma  (3/\nu)^D (2\kappa^\ast +\nu)^D  e^{-(x-\kappa^\ast\xi - v|t|)/\xi} \label{lem:connection_Lieb_Robinson_sup}
\end{align}
for $\diam(\fset{X}) , \diam(\fset{Y}) \le \nu\xi$,
where $\kappa^\ast$ is defined as an integer ($\kappa^\ast \in \mathbb{N}$) which satisfies 
\begin{align}
\mathcal{F}' (\kappa^\ast\xi,t') \le \frac{1}{2^{D+1}e^2 \gamma^2 D!} = : \zeta_1^{-1}.
\label{def:zeta_1_sup}
\end{align}
Note that $t'$-dependence of $\mathcal{G}''(x,t,t' ,\fset{X},\fset{Y})$ is included in $\kappa^\ast$.
}

\subsubsection{Proof of Proposition~\ref{lemma:Connection of two Lieb-Robinson bounds_1}}

We need to consider the norm of the commutator
\begin{align}
[e^{iH'_0 t'} e^{iH_0t}O_Xe^{-iH_0t} e^{-iH'_0 t'}, O_Y] 
\end{align}
for arbitrary operators $O_X$ and $O_Y$ with $\diam(X),\diam(Y) \le \nu \xi$ ($\nu \ge 1$).
In the following, we aim to estimate the upper bound of 
\begin{align}
\| [e^{iH'_0 t'} e^{iH_0t}O_Xe^{-iH_0t} e^{-iH'_0 t'} , O_Y]\| = \| [O_X(H_0, t) , O_Y(H'_0,-t')]\| .
\end{align}

We first decompose $O_Y(H'_0,-t')$ as 
\begin{align}
&O_Y(H'_0,-t') =O_Y(H'_0,-t', \bal{Y}{\kappa^\ast \xi})+ \sum_{s=\kappa^\ast+1}^{\infty} O_{\bal{Y}{s\xi}} , \notag \\
&O_{\bal{Y}{s\xi}} : = O_Y(H'_0,-t', \bal{Y}{s\xi})-O_Y(H'_0,-t', \bal{Y}{(s-1)\xi}) ,  \label{decomp_O_i_H_t}
\end{align}
where we determine $\kappa^\ast$ afterward.
From the Lieb-Robinson bound~\eqref{Lieb_Robinson_O_i_assump2_1_sup} and Lemma~\ref{Bravyi, Hastings and Verstaete}, 
the norm of $O_{\bal{Y}{s\xi}}$ is bounded from above by 
\begin{align}
\left \|O_{\bal{Y}{s\xi}}\right \| &\le 
\left \| O_Y(H'_0,-t')-  O_Y(H'_0,-t', \bal{Y}{s\xi}) \right \| + \left \|  O_Y(H'_0,-t')- O_Y(H'_0,-t', \bal{Y}{(s-1)\xi})\right \|   \notag \\
&\le \mathcal{G}' (s\xi,t',Y)+ \mathcal{G}' ((s-1)\xi,t',Y)   \notag \\
&\le 2\abb Y^{(\nu\xi)}\abb^2 \mathcal{F}' ((s-1)\xi,t') \le 2\gamma^2 \mathcal{F}' ((s-1)\xi,t'), \label{Lieb_Robinson_spread_op_O_i}
\end{align}
where we use $\abb Y^{(\nu\xi)}\abb \le  \gamma[\diam (Y)/(\nu\xi)]^D \le \gamma$ which is derived from the inequality~\eqref{geometric_parameter_gamma3}.
On the other hand, we have the trivial upper bound of
\begin{align}
\|O_Y(H'_0,-t', \bal{Y}{\kappa^\ast \xi}) \|\le \|O_Y\|=1. \label{Lieb_Robinson_spread_op_O_i_2}
\end{align}

By using the decomposition~\eqref{decomp_O_i_H_t},  we obtain
\begin{align}
\| [O_X(H_0, t) , O_Y(H'_0,-t')]\| \le \| [O_X(H_0, t) ,O_Y(H'_0,-t', \bal{Y}{\kappa^\ast \xi})]\| + \sum_{s=\kappa^\ast+1}^{\infty}  \| [O_X(H_0, t) ,  O_{\bal{Y}{s\xi}} ]\| .\label{Decomposition_commutator_X_Y_1}
\end{align}
In the estimation of the commutators $[O_X(H_0, t) , O_{\bal{Y}{s\xi}}]$, 
 we cannot directly use the Lieb-Robinson bound~\eqref{Lieb_Robinson_O_i_assump} because the subset $\bal{Y}{s\xi}$ may no longer satisfy 
the condition $\diam\bigl(\bal{Y}{s\xi}\bigr) \le \nu\xi $. 
We thus utilize Theorem~\ref{lemma:Connection of local _ global Lieb-Robinson bounds_2}.
From the definition of $\mathcal{G}(x,t ,\fset{X},\fset{Y})$ in Eq.~\eqref{Lieb_Robinson_O_i_assump_sup}, we set 
\begin{align}
\mathcal{F}(t,\fset{X},\fset{Y})= \mathcal{C}e^{v|t|/\xi } ,\quad   \fset{L}(x) =e^{-x/\xi} ,
\end{align}
in Eq.~\eqref{Lieb_Robinson_O_fsetX_assump}. 
Because the assumption $\diam(Y) \le \nu\xi$ ($\nu\ge3$) implies $\diam(Y) \le 2 (\nu \xi/3) +1$, the inequality~\eqref{Lieb_L_X_from_local_one_lemma} in the theorem gives
\begin{align}
&\| [O_X(H_0, t) ,O_Y(H'_0,-t', \bal{Y}{\kappa^\ast \xi})]\| \le  2
\mathcal{C}  \bigl| \bal{Y}{\kappa^\ast\xi}^{(\nu \xi/3)}\bigr| e^{-(\dist_{X,Y}-\kappa^\ast\xi - v|t|)/\xi} , \notag \\
&\| [O_X(H_0, t) , O_{\bal{Y}{s\xi}}]\| \le \| O_{\bal{Y}{s\xi}}\| \cdot 2\mathcal{C}  \bigl| \bal{Y}{s\xi}^{(\nu \xi/3)}\bigr| e^{-(\dist_{X,Y}-s\xi - v|t|)/\xi} ,
\end{align}
where we use \eqref{Lieb_Robinson_spread_op_O_i_2} in the first inequality.
Note that $\dist_{X,\bal{Y}{s\xi}}=\dist_{X,Y}-s\xi $.
Because of $\diam\bigl(\bal{Y}{s\xi}\bigr) \le  2s \xi +\diam(Y) \le 2s \xi +\nu \xi$, we have
\begin{align}
 \bigl| \bal{Y}{s\xi}^{(\nu \xi/3)}\bigr| \le  \gamma 3^D ( 1+2s/\nu)^D
\end{align}
from the inequality~\eqref{geometric_parameter_gamma3}.
We thus reduce the inequality~\eqref{Decomposition_commutator_X_Y_1} to 
\begin{align}
&\| [O_X(H_0, t) , O_Y(H'_0,-t')]\|  \notag \\
\le&  2\mathcal{C}\gamma 3^D (1+2\kappa^\ast/\nu)^D  e^{-(\dist_{X,Y}-\kappa^\ast\xi - v|t|)/\xi} 
+2\mathcal{C} \gamma 3^D \sum_{s=\kappa^\ast+1}^{\infty} \| O_{\bal{Y}{s\xi}}\| \cdot (1+ 2s/\nu)^D e^{-(\dist_{X,Y}-s\xi - v|t|)/\xi} \notag \\
\le&2\mathcal{C}\gamma 3^D e^{-(\dist_{X,Y}-\kappa^\ast\xi - v|t|)/\xi}\left [(1+2\kappa^\ast /\nu)^D + 
  2\gamma^2 \sum_{s=\kappa^\ast+1}^{\infty} (1+2s/\nu)^D e^{s - \kappa^\ast}  \mathcal{F}' ((s-1)\xi,t') \right] ,\label{Decomposition_commutator_X_Y_2}
\end{align}
where we use the upper bound~\eqref{Lieb_Robinson_spread_op_O_i} for $\| O_{\bal{Y}{s\xi}}\| $.

We then calculate an upper bound of 
\begin{align}
\sum_{s=\kappa^\ast+1}^{\infty} (1+ 2s/\nu)^D e^{s - \kappa^\ast}  \mathcal{F}' ((s-1)\xi,t')  
=\sum_{s=0}^{\infty}  [1+2(\kappa^\ast +1+s)/\nu ]^D  e^{s+1}  \mathcal{F}' ((\kappa^\ast +s)\xi,t') .\label{upp_s_r_ast_F'_1}
\end{align}
By using the inequality~\eqref{Lieb_Robinson_O_i_assump2_2_sup} with $x_0=\kappa^\ast \xi$, we obtain
\begin{align}
\mathcal{F}' ((\kappa^\ast +s)\xi,t') \le \mathcal{F}' (\kappa^\ast\xi,t') e^{-2s} ,
\end{align}
which upper-bounds \eqref{upp_s_r_ast_F'_1} as 
\begin{align}
\sum_{s=\kappa^\ast+1}^{\infty} ( 1+2s/\nu)^D e^{s - \kappa^\ast}  \mathcal{F}' ((s-1)\xi,t')  
&\le \mathcal{F}' (\kappa^\ast\xi,t')  \sum_{s=0}^{\infty}  [1+2(\kappa^\ast +1+s)/\nu ]^D   e^{-s+1}  \notag \\
&\le\mathcal{F}' (\kappa^\ast\xi,t') e^2 (2/\nu)^{D}\int_0^\infty (x+\kappa^\ast + 1+\nu/2)^D e^{-x}dx   \notag \\
&\le \mathcal{F}' (\kappa^\ast\xi,t')  \cdot  e^2 (2/\nu)^{D} D! (\kappa^\ast + 2+\nu/2)^D ,
\label{upp_s_r_ast_F'_2}
\end{align}
where in the last inequality we use \eqref{integral_convenient_1}. 
By applying the inequality~\eqref{upp_s_r_ast_F'_2} to \eqref{Decomposition_commutator_X_Y_2}, we finally obtain
\begin{align}
&\| [O_X(H_0, t) , O_Y(H'_0,-t')]\|  \notag \\
\le &2\mathcal{C}\gamma (3/\nu)^D e^{-(\dist_{X,Y}-\kappa^\ast\xi - v|t|)/\xi}\left [(2\kappa^\ast + \nu)^D + 
  2e^2 \gamma^2  D! (2\kappa^\ast + 4+ \nu )^D \mathcal{F}' (\kappa^\ast\xi,t')    \right] .
\end{align}
By choosing $\kappa^\ast$ in the above inequality such that 
\begin{align}
\mathcal{F}' (\kappa^\ast\xi,t') \le \frac{(2\kappa^\ast + \nu )^D }{2e^2 \gamma^2 D! (2\kappa^\ast + 4+ \nu )^D }, \label{cond_F_r_ast_t'}
\end{align}
we obtain the inequality~\eqref{lem:connection_Lieb_Robinson_sup}.
Because of $\kappa^\ast \ge 1$ and $\nu\ge 3$, we have
\begin{align}
\frac{(2\kappa^\ast + \nu )^D}{(2\kappa^\ast + 4+ \nu )^D } = \left(1+ \frac{4}{2\kappa^\ast + \nu}\right)^{-D} \ge (9/5)^{-D} > 2^{-D},
\end{align}
and hence the condition~\eqref{cond_F_r_ast_t'} is simplified as 
\begin{align}
\mathcal{F}' (\kappa^\ast\xi,t') \le \frac{1}{2^{D+1}e^2 \gamma^2 D!}.
\end{align}
 This completes the proof. $\square$

\section{Contribution of the Long-range interactions to Lieb-Robinson bound: Proof of Theorem~\ref{thm :long-range contribution Lieb-Robinson bound}}
\label{Contribution of the Long-range interactions to Lieb-Robinson bound}

\subsection{Statement}
{~}\\
{\bf Theorem~\ref{thm :long-range contribution Lieb-Robinson bound}} (Contribution by the long-range interacting terms).
\textit{
Under the choice of $\ell_t$ by Eq.~\eqref{choice_of_ell_t_long}, the unitary operator~\eqref{long_range_effective_unitary} satisfies the $\mathcal{G}_{> \ell_t}(x,t,\fset{X},\fset{Y})$-Lieb-Robinson bound as 
\begin{align}
\mathcal{G}_{> \ell_t}(x,t,\fset{X},\fset{Y})= &
\mathcal{J}_0 \abb \fset{X}^{(v^\ast |t|)} \abb \cdot \abb \fset{Y}^{(v^\ast |t|)}\abb \frac{  |t|^{2D+1}\log^{2D} (x+1)}{(x -\kappa_0 v^\ast |t|)^\alpha} ,
\label{ineq:long-range_contribution_proof}
\end{align} 
where $\mathcal{J}_0$ and $\kappa_0$ are constants which depend only on the parameters $\{D, g_0, \alpha,\gamma\}$.
}

{~}\\
{\bf Theorem~\ref{thm :long-range contribution Lieb-Robinson bound}' ($\boldsymbol{k}$-local Hamiltonian).}
\textit{
Let us assume the $k$-locality of the Hamiltonian [see Eq.~\eqref{def:Ham_k-local} for the definition].
Then, under the same setup as that of  Theorem~\ref{thm :long-range contribution Lieb-Robinson bound}, we obtain the Lieb-Robinson bound  
$\mathcal{G}^{(k)}_{> \ell_t}(x,t,\fset{X},\fset{Y})$ 
\begin{align}
\mathcal{G}^{(k)}_{> \ell_t}(x,t,\fset{X},\fset{Y})= &
\mathcal{J}_0^{(k)} \abb \fset{X}^{(v^\ast |t|)} \abb \cdot \abb \fset{Y}^{(v^\ast |t|)}\abb \frac{  |t|^{2D+1}}{(x -\kappa_0 v^\ast |t|)^\alpha} ,
\label{ineq:long-range_contribution_proof_k-local}
\end{align} 
where $\mathcal{J}_0^{(k)}$ and $\kappa_0$ are constants which depend only on the parameters $\{D, g_0, \alpha,\gamma,k\}$.
}

\clearpage 

\begin{figure}[]
\centering
{
\includegraphics[clip, scale=0.5]{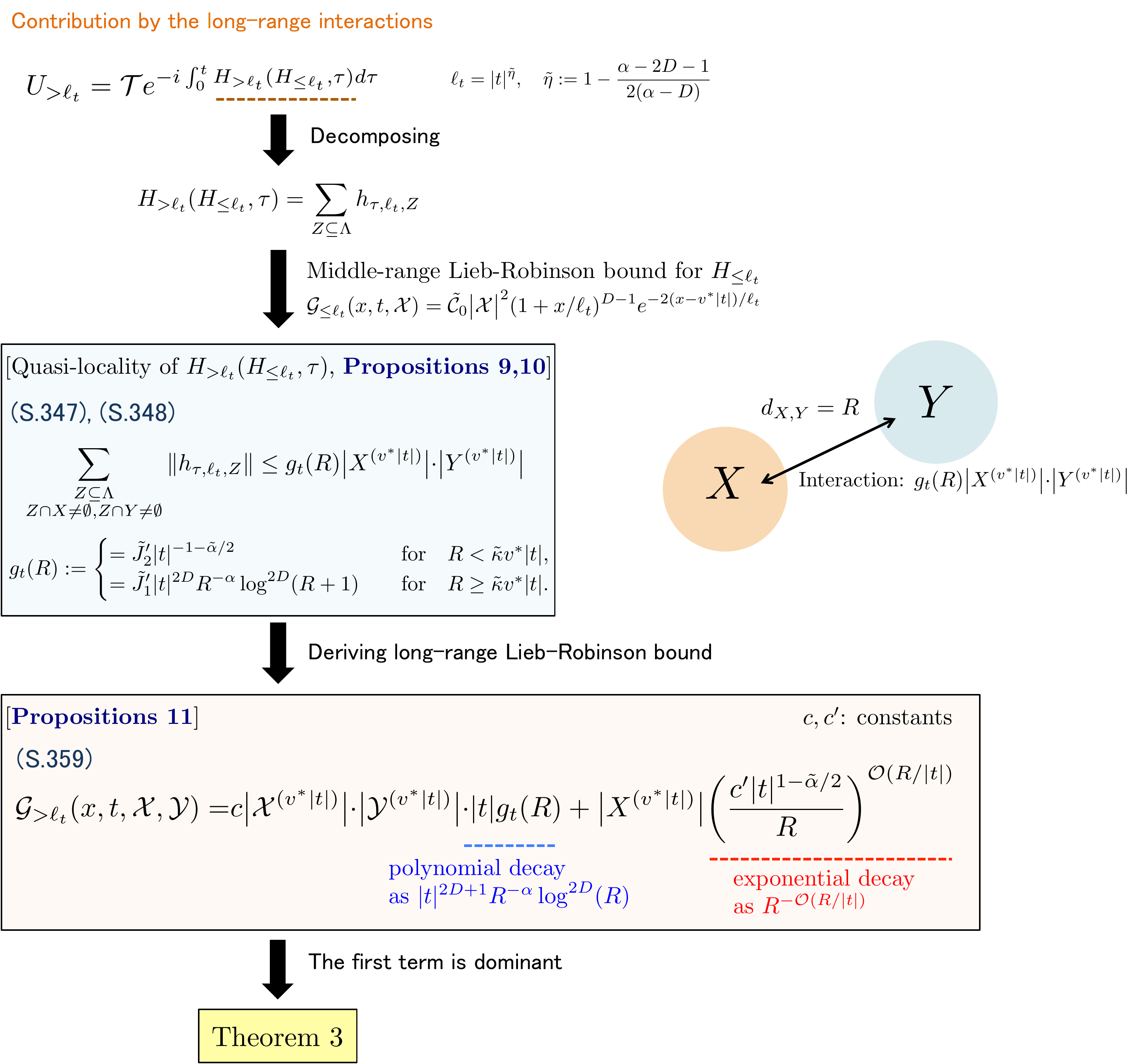}
}
\caption{Outline of the proof. In the proof of Theorem~\ref{thm :long-range contribution Lieb-Robinson bound}, we need to derive the Lieb-Robinson bound for $U_{>\ell_t}$ in Eq.~\eqref{long_range_effective_unitary2}. 
The proof consists of the following two steps: i) estimation of the quasi-locality of $H_{> \ell_t} (H_{\le \ell_t}, \tau)$, ii) derivation of the Lieb-Robinson bound under the estimated quasi-locality.
In the first step, we can utilize Theorem~\ref{thm:Middle-range Lieb-Robinson bound} for the time evolution by $H_{\le \ell_t}$, and aim to prove Proposition~\ref{prop:quasi_local_long_range} (the proof is given in Sec.~\ref{Sec:prop:quasi_local_long_range}).
By using the proposition, we can derive the quasi-locality of  $H_{> \ell_t} (H_{\le \ell_t}, \tau)$ in the form of Eq.~\eqref{math:ineq_quasi_local_using_paractical_form_for_prop} with \eqref{math:ineq_quasi_local_using_paractical_form_for_prop_g_t_Fig_used}. 
Based on this estimation of the quasi-locality, we derive Proposition~\ref{prop:quasi_local_long_range_Lieb-Robinson} (the proof is given in Sec.~\ref{Sec:prop:quasi_local_long_range_Lieb-Robinson}),
which yields the Lieb-Robinson bound for $U_{>\ell_t}$ as in Ineq.~\eqref{mathcal_G_ell_t_x_t_upp_two_term_Fig_used}.
By appropriately transform the inequality~\eqref{mathcal_G_ell_t_x_t_upp_two_term_Fig_used}, we arrive at the main inequality~\eqref{ineq:long-range_contribution_proof}.
In the case where the $k$-locality is assumed, the only difference arises from the point that Proposition~\ref{prop:quasi_local_long_range} is replaced by Proposition~\ref {prop:quasi_local_long_range_k-local}.}
\label{fig:Outline_long_range_theorem}
\end{figure}

\subsection{Proof of the theorem (see Fig.~\ref{fig:Outline_long_range_theorem} for the outline)}

In the following, we define $\tilde{\alpha}$ as  
\begin{align}
\tilde{\alpha}: = \alpha -2D -1 , \label{def:alpha_tilde}
\end{align}
which gives $\tilde{\eta}$ in Eq.~\eqref{choice_of_ell_t_long} as 
\begin{align}
\tilde{\eta} =1 - \frac{\tilde{\alpha}}{2( \alpha-D)}  .
\end{align}
From the assumption of $\alpha>2D+1$, we have $\tilde{\alpha} >0$.
Also, the assumption~\eqref{basic_assump_power} implies
\begin{align}
\sum_{\substack{Z\subseteq \Lambda , Z\ni i \\ \diam(Z)> \ell_t} } \| h_Z\| \le 
g \ell_t^{-\alpha+D} =g|t|^{-\alpha+D+\tilde{\alpha}/2},
\end{align}  
where we use the definition of $\ell_t:=|t|^{\tilde{\eta}}$ [see Eq.~\eqref{choice_of_ell_t_long}].

Here, the primary task is to derive the Lieb-Robinson bound for the unitary operator of 
\begin{align}
U_{>\ell_t}:= \mathcal{T} e^{-i\int_0^t H_{> \ell_t} (H_{\le \ell_t}, \tau) d\tau}, \label{long_range_effective_unitary2}
\end{align}  
which appears from the decomposition of the unitary operator $e^{-iHt}$:
\begin{align}
e^{-iHt} = e^{-iH_{\le \ell_t} t} \mathcal{T} e^{-i\int_0^t H_{> \ell_t} (H_{\le \ell_t}, \tau) d\tau} ,
\end{align}  
Notice that we have already obtained the Lieb-Robinson bound for $e^{-iH_{\le \ell_t} t}$ from Theorem~\ref{thm:Middle-range Lieb-Robinson bound}. 

%

From Theorem~\ref{thm:Middle-range Lieb-Robinson bound}, $H_{\le \ell_t}$ satisfies the $\mathcal{G}_{\le \ell_t}(x,t,\fset{X})$-Lieb-Robinson bound as 
\begin{align}
\mathcal{G}_{\le \ell_t}(x,t,\fset{X})= \tilde{\mathcal{C}}_0 \abb\fset{X} \abb^2 (1+x/\ell_t)^{D-1}  e^{-2  (x- v^\ast  |t|)/\ell_t} .
\end{align}
We first rewrite the function $\mathcal{G}_{\le \ell_t}(x,t,\fset{X})$ in the form of 
\begin{align} 
&\mathcal{G}_{\le \ell_t}(x,t,\fset{X}) =   \abb\fset{X}\abb^2 e^{-2 (1- \delta_t) [x- (1- \delta_t)^{-1}v^\ast  |t|]/\ell_t} .
\label{mathcal_G_function_remon_le_ell}
\end{align} 
By following the same calculations in the proof of Lemma~\ref{Lieb_Robinson_reformationG_qminsu1}, we obtain $\delta_{t}$ as follows:
\begin{align} 
\delta_{t} = \frac{\log (\tilde{\mathcal{C}}_0)+(D-1) \log (v^\ast |t|/\ell_t+1)}{2(v^\ast |t|/\ell_t)} 
\le \frac{\log (\tilde{\mathcal{C}}_0) +(D-1) \log (v_0 \ell_1+1)}{2v_0 \ell_1} \le \frac{1}{2},
\label{upper_bound_for_delta_t}
\end{align} 
where in the first inequality we use $v^\ast |t|/\ell_t =v^\ast |t|^{\tilde{\eta}}  \ge v^\ast \ge v_0 \ell_1$ [see Eq.~\eqref{definitions_v_1_v_q_delta_Fig_used}], and in the second inequality we use the condition~\eqref{upper_bound_for_delta_t_cond} for $\ell_1$.
We thus obtain
\begin{align} 
1-\delta_{t} \ge 1/2,
\end{align} 
which reduces \eqref{mathcal_G_function_remon_le_ell} to 
\begin{align} 
&\mathcal{G}_{\le \ell_t}(x,t,\fset{X}) \le  \abb\fset{X}\abb^2 e^{-(x- 2v^\ast  |t|)/\ell_t}. 
\end{align}

For the derivation of \eqref{ineq:long-range_contribution_proof}, we first show the quasi-locality of $H_{> \ell_t} (H_{\le \ell_t}, \tau)$ for $\tau<t$.
We prove the following propositions (see Sec.~\ref{Sec:prop:quasi_local_long_range} for the proof):
\begin{prop}\label{prop:quasi_local_long_range}
Let us consider an operator $A$ as follows:
\begin{align}
A= \sum_{Z \subseteq \Lambda} a_{Z},
\end{align}  
where $A$ satisfies the assumptions~\eqref{basic_assump_power} and \eqref{alternative_basic_assump_power} with $g=g_{a}$ and $g_{0}=g_{a,0}$.
We also consider an operator $A_0$ which satisfies $\mathcal{G}_0(x,t,\fset{X})$-Lieb-Robinson bound as follows:
 \begin{align}
&\mathcal{G}_0(x,t,\fset{X}) = \mathcal{C}  |\fset{X}|^2 e^{-(x- v|t|)/\xi} \quad  ( \mathcal{C}\ge 1)  .
\end{align}  
For arbitrary $\tau \le |t|$ such that $v|t|/\xi \ge 1$, we obtain the decomposition of 
\begin{align}
&A(A_0, \tau)= \sum_{Z\subseteq \Lambda } a_{\tau, Z}
\end{align}  
such that
\begin{align}
\sum_{\substack{Z\subseteq \Lambda\\ Z \cap X \neq \emptyset, Z \cap Y \neq \emptyset}} \| a_{\tau, Z} \| 
\le &\min \left[ J_1  g_{a,0}  \abb X^{(v|t|)} \abb\cdot \abb Y^{(v|t|)} \abb (\tilde{f}_{R}\xi)^{2D} \left(R^{-\alpha} + e^{-(R-4\tilde{f}_{R}\xi)/(8\xi)}\right), 
J_2g_a \abb X^{(v|t|)} \abb (\tilde{f}_{R}\xi)^D\right]  \notag \\
&+ g_a J_3 \abb X^{(v|t|)} \abb (v|t|)^D R^{-\alpha}  \quad   {\rm with}\quad  \tilde{f}_R:= \log (2^{2D+1}e\gamma^2 \mathcal{C} R^{2\alpha}) + v|t|/\xi ,
\label{math:ineq_quasi_local_proof}
\end{align}  
where $R = \dist_{X,Y}$ and we define $J_1:= 2^{6D+\alpha} e (2D)!   \gamma^2$, $J_2:=e 6^{D}  \gamma  D! $ and $J_3 =e \gamma D!  [8 \log (2e\gamma^2 \mathcal{C})]^D c_{\alpha,D}$ with $c_{\alpha,D}:=D^{-1} \sup_{r\ge 1} \left([1+2D \log(r)]^D/r^{\alpha-2D-1} \right)$.
\end{prop}

\noindent 
Because of $\tilde{f}_R\propto \log (R)$, this proposition implies that the interaction strength algebraically decays as $R^{-\alpha} \log^{2D} (R)$.
In particular, if we assume the $k$-locality for the Hamiltonian, we can obtain a stronger statement as follows:
\begin{prop}[{\bf $\boldsymbol{k}$-local operator}]\label{prop:quasi_local_long_range_k-local}
Let us consider an operator $A$ which includes at most $k$-body interactions (i.e., $k$-local operator), namely
\begin{align}
A= \sum_{Z \subset \Lambda, |Z|\le k} a_{Z}.
\end{align}  
Then, under the same setup as that of Proposition~\ref{prop:quasi_local_long_range}, we obtain the decomposition of 
\begin{align}
&A(A_0, \tau)= \sum_{Z\subseteq \Lambda } a_{\tau, Z}
\end{align}  
such that
\begin{align}
\sum_{\substack{Z\subseteq \Lambda\\ Z \cap X \neq \emptyset, Z \cap Y \neq \emptyset}} \| a_{\tau, Z} \| 
\le \min \left[J_1^{(k)}g_{a,0} \abb X^{(v|t|)} \abb\cdot \abb Y^{(v|t|)} \abb  (v|t| )^{2D} \left(R^{-\alpha} + e^{-(R-4v|t|)/(8\xi)}\right),J_2^{(k)}   g_a  \abb X^{(v|t|)} \abb (v |t|)^D\right],
\label{math:ineq_quasi_local_k-local}
\end{align} 
where $R = \dist_{X,Y}$ and we define 
$J_1^{(k)}:= (2e\mathcal{C}k^2) 2^{6D+\alpha} e (2D)!  \gamma^2$ and  $J_2^{(k)}:=(2e\mathcal{C}k^2) e 6^{D}   \gamma D! $.
\end{prop}

{~}

\noindent
In Proposition~\ref{prop:quasi_local_long_range}, we choose $A= H_{> \ell_t}$ and $A_0= H_{\le \ell_t}$, which gives the parameters  
\begin{align} 
\xi=\ell_t, \quad  \mathcal{C}= 1, \quad g_{a,0}=g_0,\quad g_a =g  |t|^{-\alpha+D+\tilde{\alpha}/2} ,  \quad v= 2 v^\ast .
\end{align}
We note that these choices ensure the condition $v|t| \ge \xi$ because of $v^\ast>1$ and $\ell_t = |t|^{\tilde{\eta}} \le |t|$ ($|t|\ge1$).
Then, for $\tau \le |t|$, we have
\begin{align} 
H_{> \ell_t} (H_{\le \ell_t},\tau) = \sum_{Z\subseteq \Lambda } h_{\tau,\ell_t, Z} 
\end{align}
with 
\begin{align}
&\sum_{\substack{Z\subseteq \Lambda\\ Z \cap X \neq \emptyset, Z \cap Y \neq \emptyset}} \| h_{\tau,\ell_t, Z}  \| 
\le g  |t|^{-\alpha+D+\tilde{\alpha}/2} \tilde{J}_3 \abb X^{(2v^\ast |t|)} \abb (2v^\ast |t|)^D R^{-\alpha}  \notag \\
&+\min \left[ \tilde{J}_1  g_{0} \abb X^{(2v^\ast |t|)} \abb\cdot \abb Y^{(2v^\ast |t|)} \abb  (\tilde{f}_{t,R}\ell_t)^{2D} \left(R^{-\alpha} + e^{-(R-4\tilde{f}_{t,R}\ell_t)/(8\ell_t)}\right), 
\tilde{J}_2g  |t|^{-\alpha+D+\tilde{\alpha}/2}\abb X^{(2v^\ast |t|)} \abb (\tilde{f}_{t,R}\ell_t )^D\right]  
 \label{math:ineq_quasi_local_using}
\end{align}
and 
\begin{align}
\tilde{f}_{t,R}:= \log (2^{2D+1}e\gamma^2  R^{2\alpha}) +  2 v^\ast |t|/\ell_t,
\end{align}
where $\{\tilde{J}_1,\tilde{J}_2,\tilde{J}_3\}$ are constants which are defined by $\{J_1, J_2,J_3\}$ in Proposition~\ref{prop:quasi_local_long_range}.
Note that these parameters depend only on $\{D, g_0, \alpha, \gamma\}$. 
Furthermore, from $\tilde{f}_{t,R}\ell_t \ge 2 v^\ast |t|$, we can reduce the inequality~\eqref{math:ineq_quasi_local_using} to 
\begin{align}
&\sum_{\substack{Z\subseteq \Lambda\\ Z \cap X \neq \emptyset, Z \cap Y \neq \emptyset}} \| h_{\tau,\ell_t, Z}  \| \notag \\
&\le \tilde{c}_3 \min \left[ \tilde{J}_1  g_{0} \abb X^{(v^\ast |t|)} \abb\cdot \abb Y^{(v^\ast |t|)} \abb  (\tilde{f}_{t,R}\ell_t)^{2D} \left(R^{-\alpha} + e^{-(R-4\tilde{f}_{t,R}\ell_t)/(8\ell_t)}\right), 
\tilde{J}_2g  |t|^{-\alpha+D+\tilde{\alpha}/2}\abb X^{(v^\ast |t|)} \abb (\tilde{f}_{t,R}\ell_t )^D\right]  ,
 \label{math:ineq_quasi_local_using2}
\end{align}
where the first term in \eqref{math:ineq_quasi_local_using} is absorbed by taking an appropriate constant $\tilde{c}_3$ which depends on $g$, $g_0$ and $\{\tilde{J}_1,\tilde{J}_2,\tilde{J}_3\}$.
Note that we have $\abb X^{(2v^\ast |t|)} \abb \le \abb X^{(v^\ast |t|)} \abb $.

From $\ell_t=|t|^{\tilde{\eta}}$, we have $\tilde{f}_{t,R} \ell_t$ in the form of
\begin{align}
\tilde{f}_{t,R} \ell_t=|t|^{\tilde{\eta}}\log (2^{2D+1}e\gamma^2 R^{2\alpha} )  + 2 v^\ast |t|  .
\end{align}  
From the above expression, there always exists a constant $\tilde{\kappa}$ ($\ge 2$) such that it satisfies the following inequalities
\begin{align} 
R -4\tilde{f}_{t,R} \ell_t \ge \frac{R}{2} \quad {\rm and} \quad  e^{-R/(16\ell_t)}  \le R^{-\alpha}  \for R \ge \tilde{\kappa}  v^\ast |t| ,
 \label{def:tilde_r_ineq}
\end{align}  
\begin{align} 
\frac{d}{dR} [R^{-\alpha} \log^{2D} (R+1)] = -\alpha R^{-\alpha+1} \log^{2D}(R+1) + \frac{2DR^{-\alpha} \log^{2D-1}(R+1)}{R+1}  \le 0    \for R \ge \tilde{\kappa}  v^\ast |t| ,
 \label{def:tilde_r__monotonic_decrease}
\end{align}  
and 
\begin{align}
\tilde{f}_{t,R} \ell_t=2 v^\ast |t|  \left (\frac{\log (2^{2D+1}e\gamma^2 R^{2\alpha}) }{2 v^\ast |t|^{1-\tilde{\eta}} }+ 1\right)  
\le  
\begin{cases}
 c^\ast v^\ast |t|                     &\for    R<  \tilde{\kappa}  v^\ast |t| ,  \\
 c^\ast v^\ast |t|  \log(R+1) &\for    R\ge\tilde{\kappa}  v^\ast |t| , 
\end{cases}
\label{math:ineq_quasi_local_using_f_t_R_high}
\end{align}  
where $c_\ast$ depends on the parameters $\{D, g_0, \alpha, \gamma\}$. 
Note that the inequality~\eqref{def:tilde_r_ineq} implies $R\ge 8\tilde{f}_{t,R} \ell_t \ge 16 v^\ast |t|$ and hence $\tilde{\kappa}$ is at least larger than $16$, namely $\tilde{\kappa}\ge16$.

%
%
%
Under the above definition of $\tilde{\kappa}$, for $R \ge \tilde{\kappa}  v^\ast |t| $, the inequalities~\eqref{def:tilde_r_ineq} and \eqref{math:ineq_quasi_local_using_f_t_R_high} give
\begin{align}
  (\tilde{f}_{t,R}\ell_t)^{2D} \left(R^{-\alpha} + e^{-(R-4\tilde{f}_{t,R}\ell_t)/(8\ell_t)}\right) \le 2  ( c^\ast v^\ast |t|)^{2D} R^{-\alpha} \log^{2D} (R+1).
  \label{math:ineq_quasi_local_using_f_t_R_high_use}
 \end{align}
Also, for $R< \tilde{\kappa}  v^\ast |t| $, the inequality~\eqref{math:ineq_quasi_local_using_f_t_R_high} gives 
\begin{align}
 |t|^{-\alpha+D+\tilde{\alpha}/2} (\tilde{f}_{t,R}\ell_t )^D  
 \le ( c^\ast v^\ast)^D |t|^{-\alpha+2D+\tilde{\alpha}/2} 
 \le ( c^\ast v^\ast)^D |t|^{-1+\tilde{\alpha}/2}   ,
 \label{math:ineq_quasi_local_using_f_t_R_low_use}
\end{align}
where the last inequality is given by the condition $\alpha>2D+1$.
By using the inequalities~\eqref{math:ineq_quasi_local_using_f_t_R_high_use} and \eqref{math:ineq_quasi_local_using_f_t_R_low_use}, the inequality~\eqref{math:ineq_quasi_local_using2} reduces to 
\begin{align}
\sum_{\substack{Z\subseteq \Lambda\\ Z \cap X \neq \emptyset, Z \cap Y \neq \emptyset}} \| h_{\tau,\ell_t, Z}  \| &
\le   g_t(R)\abb X^{(v^\ast |t|)} \abb \cdot \abb Y^{(v^\ast |t|)}\abb 
\label{math:ineq_quasi_local_using_paractical_form_for_prop}
\end{align}  
with
\begin{align}
g_t(R):=\begin{cases}
\displaystyle
\tilde{c}_3g\tilde{J}_2( c^\ast v^\ast)^D |t|^{-1-\tilde{\alpha}/2}  =:\tilde{J}'_2 |t|^{-1-\tilde{\alpha}/2}  &\for R < \tilde{\kappa} v^\ast |t|,  \\
\displaystyle
 2 \tilde{c}_3 g_0 \tilde{J}_1  ( c^\ast v^\ast)^{2D}   
\frac{|t|^{2D} \log^{2D} (R+1) }{R^\alpha} =: \tilde{J}'_1 |t|^{2D} R^{-\alpha}\log^{2D} (R+1) & \for R \ge \tilde{\kappa}  v^\ast |t|, 
\end{cases}
\label{math:ineq_quasi_local_using_paractical_form_for_prop_g_t_Fig_used}
\end{align}  
where $ \tilde{J}'_1$ and $ \tilde{J}'_2$ depend on the parameters $\{D, g_0, \alpha, \gamma\}$. 
Note that the function $g_t(R)$ monotonically decreases with $R$ because of the condition~\eqref{def:tilde_r__monotonic_decrease}.

In the case where the Hamiltonian is $k$-local, we can apply the same analyses. Only the difference is that $\tilde{f}_{t,R}\ell_t$ is replaced with $2v^\ast |t|$ from Proposition~\ref{prop:quasi_local_long_range_k-local}.
Hence, the logarithmic dependence $\log^{2D} (R+1)$, which results from the inequality \eqref{math:ineq_quasi_local_using_f_t_R_high}, does not appear in this case. 
Then, the inequality~\eqref{math:ineq_quasi_local_using_paractical_form_for_prop} is replaced by 
\begin{align}
\sum_{\substack{Z\subseteq \Lambda \\ Z \cap X \neq \emptyset, Z \cap Y \neq \emptyset}} \| h_{\tau,\ell_t, Z}  \| &
\le  g^{(k)}_t(R)\abb X^{(v^\ast |t|)} \abb \cdot \abb Y^{(v^\ast |t|)}\abb 
\label{math:ineq_quasi_local_using_paractical_form_for_prop_k-local}
\end{align}  
with
\begin{align}
g_t^{(k)}(R):=\begin{cases}
\displaystyle
\tilde{J}_2^{(k)} |t|^{-1-\tilde{\alpha}/2}  &\for R < \tilde{\kappa} v^\ast |t|,  \\
\displaystyle
\tilde{J}^{(k)}_1  |t|^{2D}  R^{-\alpha} & \for R \ge \tilde{\kappa}  v^\ast |t|,
\end{cases}
\label{math:ineq_quasi_local_using_paractical_form_for_prop_g_t_k-local}
\end{align}  
where $\tilde{J}^{(k)}_1$ and $\tilde{J}^{(k)}_2$ are constants which depend on the parameters $\{D, g_0, \alpha, \gamma,k\}$.

%
%

We here show the Lieb-Robinson bound for time-evolved long-range Hamiltonians $H_{>\ell_t}(H_{\le \ell_t},\tau)$.
We can prove the following proposition (see Sec.~\ref{Sec:prop:quasi_local_long_range_Lieb-Robinson} for the proof):
\begin{prop}\label{prop:quasi_local_long_range_Lieb-Robinson}
Let us consider a Hamiltonian $\tilde{H}$ as follows:
\begin{align}
\tilde{H}= \sum_{Z \subseteq \Lambda} \tilde{h}_{Z}
\end{align}  
with 
\begin{align}
&\sum_{\substack{Z\subseteq \Lambda\\ Z \cap X \neq \emptyset, Z \cap Y \neq \emptyset}} \| \tilde{h}_{Z} \| \le \abb X^{(\xi_0)} \abb \cdot \abb Y^{(\xi_0)}\abb g(R)
,\quad R=\dist_{X,Y}, \notag \\
&g(R) = \begin{cases}
\tilde{g}  &\for R< \kappa\xi_0, \\
\tilde{g}_R R^{-\alpha} &\for R\ge \kappa\xi_0 ,
\end{cases} \quad  \tilde{g}_R:=\tilde{g}_0 \log^p (R+1) \    (p\in \mathbb{N}),
\label{def:f_R_LR_Long_statement}
\end{align}  
where $\kappa\ge 2$, $\xi_0\ge1$, and $g(R)$ monotonically decreases with $R$.
Then, the Hamiltonian $\tilde{H}$ satisfies $\mathcal{G}(x,t,\fset{X},\fset{Y})$-Lieb-Robinson bound with
\begin{align}
\mathcal{G}(x,t,\fset{X},\fset{Y})= \abb \fset{X}^{(\xi_0)} \abb \cdot \abb \fset{Y}^{(\xi_0)}\abb \frac{2^{\alpha+1} e\Gamma(\alpha+2) }{\tilde{\lambda}} \left( e^{2e \tilde{\lambda}\gamma^2 9^D |t|}-1 \right)  g(R)
+2\abb X^{(\xi_0)} \abb \left( \frac{4e\kappa \xi_0 \tilde{\lambda} \gamma^2 9^D |t|}{R} \right)^{R/(2\kappa\xi_0)},
\label{main_ineq_long_range_LR}
\end{align}  
where $\Gamma(x)$ is the gamma function and we define $\tilde{\lambda}$ as 
\begin{align}
 &\tilde{\lambda}:=\tilde{g}\gamma [2(\kappa+3)]^D + \frac{c_{p,\kappa}  \tilde{g}_0 \gamma D 2^{D+1} }{\kappa+2}  \frac{\log^p(\kappa\xi_0+1)}{\xi_0^{\alpha} }  ,\notag \\
 &c_{p,\kappa}:= \sup_{z\in \mathbb{R}| z\ge \kappa+2} \left[\frac{ (\log (z)+1)^p}{(z-2)^{\alpha-D-1}}\right]\label{def_para_constant_f} .
\end{align}  
\end{prop}


In the following, we apply Proposition~\ref{prop:quasi_local_long_range_Lieb-Robinson} to the unitary operator of
\begin{align}
\mathcal{T} e^{-i\int_0^t H_{> \ell_t} (H_{\le \ell_t}, \tau) d\tau} .\label{unitary_long_range_term}
\end{align} 
From the inequality~\eqref{math:ineq_quasi_local_using_paractical_form_for_prop} with Eq.~\eqref{math:ineq_quasi_local_using_paractical_form_for_prop_g_t_Fig_used}, the parameters in Proposition~\ref{prop:quasi_local_long_range_Lieb-Robinson} is now given by 
\begin{align}
\xi_0 = v^\ast |t|, \quad  \kappa=\tilde{\kappa} ,\quad p=2D,\quad   \tilde{g}= \tilde{J}'_2 |t|^{-1-\tilde{\alpha}/2}   ,\quad 
\tilde{g}_0=\tilde{J}'_1 |t|^{2D} .
\label{choice_of_the_parameters}
\end{align} 
Then, the definition~\eqref{def_para_constant_f} gives $\tilde{\lambda} |t|$ as follows:
\begin{align}
\tilde{\lambda} |t|&=\tilde{J}'_2\gamma [2(\tilde{\kappa}+3)]^D  |t|^{-\tilde{\alpha}/2}
+ \frac{c_{2D,\tilde{\kappa}} \tilde{J}'_1\gamma D 2^{D+1} }{\tilde{\kappa}+2} |t|^{2D+1} \frac{\log^{2D}(\tilde{\kappa} v^\ast |t|+1)}{(v^\ast |t|)^{\alpha} }     \notag \\
&=\tilde{J}'_2\gamma [2(\tilde{\kappa}+3)]^D  |t|^{-\tilde{\alpha}/2}
+ \frac{c_{2D,\tilde{\kappa}} \tilde{J}'_1\gamma D 2^{D+1} }{[v^\ast]^\alpha(\tilde{\kappa}+2)} |t|^{-\tilde{\alpha}} \log^{2D}(\tilde{\kappa} v^\ast |t|+1)   ,
\end{align} 
where $\tilde{\alpha}:=\alpha-2D-1$ as defined in Eq.~\eqref{def:alpha_tilde}. 
Because of $\tilde{\alpha}>0$ and $|t|\ge 1$, there exists a constant $\lambda_{0}$ which depends  only on the parameters $\{D, g_0, \alpha, \gamma\}$ such that
\begin{align}
\tilde{\lambda} |t|&\le\lambda_0   |t|^{-\tilde{\alpha}/2}  \le \lambda_0.  \label{def_para_constant_f_practical_use}
\end{align} 
Then, from the inequality~\eqref{main_ineq_long_range_LR} in Proposition~\ref{prop:quasi_local_long_range_Lieb-Robinson}, 
the unitary operator~\eqref{unitary_long_range_term} satisfies $\mathcal{G}_{> \ell_t}(x,t,\fset{X},\fset{Y})$-Lieb-Robinson bound with
\begin{align}
\mathcal{G}_{> \ell_t}(x,t,\fset{X},\fset{Y})= &
\abb \fset{X}^{(v^\ast |t|)} \abb \cdot \abb \fset{Y}^{(v^\ast |t|)}\abb\frac{2^{\alpha+1} e\Gamma (\alpha+2)\left( e^{2e \lambda_0\gamma^2 9^D}-1\right)}{\lambda_{0}} |t|  g_t(R) \notag \\
&+2\abb X^{(v^\ast |t|)} \abb  \left( \frac{4e\tilde{\kappa} v^\ast \lambda_0 \gamma^2 9^D |t|^{1-\tilde{\alpha}/2}}{R} \right)^{R/(2\tilde{\kappa} v^\ast |t|)}.
\label{mathcal_G_ell_t_x_t_upp_two_term_Fig_used}
\end{align}  

From the definition of $g_t(R)$ in Eq.~\eqref{math:ineq_quasi_local_using_paractical_form_for_prop_g_t_Fig_used}, the first term in \eqref{mathcal_G_ell_t_x_t_upp_two_term_Fig_used} decays as 
$|t|^{2D+1}R^{-\alpha} \log^{2D} (R+1)$ for $R\ge \tilde{\kappa} v^\ast |t|$. 
On the other hand, the second term exponentially decays as $(R^{-1}|t|^{1-\tilde{\alpha}/2})^{R/|t|}$.
Hence, there exists a constant $\kappa_0 \ge \tilde{\kappa}$ such that 
\begin{align}
\textrm{First term in \eqref{mathcal_G_ell_t_x_t_upp_two_term_Fig_used}} \ge  \textrm{Second term in \eqref{mathcal_G_ell_t_x_t_upp_two_term_Fig_used}} \for R\ge \kappa_0 v^\ast |t|.
\end{align}  
We then obtain 
\begin{align}
\mathcal{G}_{> \ell_t}(x,t,\fset{X},\fset{Y})= &
2 \abb \fset{X}^{(v^\ast |t|)} \abb \cdot \abb \fset{Y}^{(v^\ast |t|)}\abb\frac{2^{\alpha+1} e\Gamma (\alpha+2)\left( e^{2e \lambda_0\gamma^2 9^D}-1\right)}{\lambda_{0}} \tilde{J}'_1 |t|^{2D+1}R^{-\alpha}  \log^{2D} (R+1)
\end{align}  
for $R\ge \kappa_0 v^\ast |t|$.
Therefore, we finally obtain 
\begin{align}
\mathcal{G}_{> \ell_t}(x,t,\fset{X},\fset{Y})= &
\mathcal{J}_0 \abb \fset{X}^{(v^\ast |t|)} \abb \cdot \abb \fset{Y}^{(v^\ast |t|)}\abb \frac{  |t|^{2D+1}\log^{2D} (R+1)}{R^\alpha} \for R\ge \kappa_0 v^\ast |t| 
\label{mathcal_G_ell_t_x_t_upp_two_term_final_ippomae}
\end{align}  
by choosing 
\begin{align}
\mathcal{J}_0 =\frac{2^{\alpha+2} e\Gamma (\alpha+2)\left( e^{2e \lambda_0\gamma^2 9^D}-1\right)}{\lambda_{0}} \tilde{J}'_1.
\end{align}  
The inequality~\eqref{mathcal_G_ell_t_x_t_upp_two_term_final_ippomae} reduces to the main inequality~\eqref{ineq:long-range_contribution_proof}.

In the case where the Hamiltonian is $k$-local, from the inequality~\eqref{math:ineq_quasi_local_using_paractical_form_for_prop_k-local} with Eq.~\eqref{math:ineq_quasi_local_using_paractical_form_for_prop_g_t_k-local}, we set the parameters in Proposition~\ref{prop:quasi_local_long_range_Lieb-Robinson} as follows:
\begin{align}
\xi_0 = v^\ast |t|, \quad  \kappa=\tilde{\kappa} ,\quad p=0,\quad   \tilde{g}= \tilde{J}^{(k)}_2 |t|^{-1-\tilde{\alpha}/2}   ,\quad 
\tilde{g}_0=\tilde{J}^{(k)}_1 |t|^{2D} .
\label{choice_of_the_parameters_k-local}
\end{align} 
By following the same steps for the derivation of \eqref{mathcal_G_ell_t_x_t_upp_two_term_final_ippomae}, we obtain for the $k$-local Hamiltonians 
\begin{align}
\mathcal{G}_{> \ell_t}(x,t,\fset{X},\fset{Y})= &
\mathcal{J}_0^{(k)} \abb \fset{X}^{(v^\ast |t|)} \abb \cdot \abb \fset{Y}^{(v^\ast |t|)}\abb \frac{   |t|^{2D+1} }{R^\alpha} \for R\ge \kappa_0 v^\ast |t| ,
\label{mathcal_G_ell_t_x_t_upp_two_term_final_ippomae_k-local}
\end{align}  
where $\mathcal{J}_0^{(k)}$ is a constant which depends on $\{D, g_0, \alpha, \gamma,k\}$. 
This completes the proof of the theorem. $\square$

\subsection{Proofs of Propositions~\ref{prop:quasi_local_long_range} and \ref{prop:quasi_local_long_range_k-local}} \label{Sec:prop:quasi_local_long_range}

\subsubsection{Statement}
\noindent
{\bf Proposition~\ref{prop:quasi_local_long_range}.}
\textit{
Let us consider an operator $A$ as follows:
\begin{align}
A= \sum_{Z \subseteq \Lambda} a_{Z},
\end{align}  
where $A$ satisfies the assumptions~\eqref{basic_assump_power} and \eqref{alternative_basic_assump_power} with $g=g_{a}$ and $g_{0}=g_{a,0}$.
We also consider an operator $A_0$ which satisfies $\mathcal{G}_0(x,t,\fset{X})$-Lieb-Robinson bound as follows:
 \begin{align}
&\mathcal{G}_0(x,t,\fset{X}) = \mathcal{C}  |\fset{X}|^2 e^{-(x- v|t|)/\xi} \quad  ( \mathcal{C}\ge 1)  .
\end{align}  
For arbitrary $\tau\le |t|$ such that $v|t|/\xi \ge 1$, we obtain the decomposition of 
\begin{align}
&A(A_0, \tau)= \sum_{Z\subseteq \Lambda } a_{\tau, Z}
\end{align}  
such that
\begin{align}
\sum_{\substack{Z\subseteq \Lambda\\ Z \cap X \neq \emptyset, Z \cap Y \neq \emptyset}} \| a_{\tau, Z} \| 
\le &\min \left[ J_1  g_{a,0}  \abb X^{(v|t|)} \abb\cdot \abb Y^{(v|t|)} \abb (\tilde{f}_{R}\xi)^{2D} \left(R^{-\alpha} + e^{-(R-4\tilde{f}_{R}\xi)/(8\xi)}\right), 
J_2g_a \abb X^{(v|t|)} \abb (\tilde{f}_{R}\xi)^D\right]  \notag \\
&+ g_a J_3 \abb X^{(v|t|)} \abb (v|t|)^D R^{-\alpha}  \quad   {\rm with}\quad  \tilde{f}_R:= \log (2^{2D+1}e\gamma^2 \mathcal{C} R^{2\alpha}) + v|t|/\xi ,
\label{math:ineq_quasi_local_proof}
\end{align}  
where $R = \dist_{X,Y}$ and we define $J_1:= 2^{6D+\alpha} e (2D)!   \gamma^2$, $J_2:=e 6^{D}  \gamma  D! $ and $J_3 =e \gamma D!  [8 \log (2e\gamma^2 \mathcal{C})]^D c_{\alpha,D}$ with $c_{\alpha,D}:=D^{-1} \sup_{r\ge 1} \left([1+2D \log(r)]^D/r^{\alpha-2D-1} \right)$.
}

{~}

\noindent
{\bf Proposition~\ref{prop:quasi_local_long_range_k-local} ({\bf $\boldsymbol{k}$-local operator}).}
\textit{
Let us consider an operator $A$ which includes at most $k$-body interactions (i.e., $k$-local operator), namely
\begin{align}
A= \sum_{Z \subset \Lambda, |Z|\le k} a_{Z}.
\end{align}  
Then, under the same setup as that of Proposition~\ref{prop:quasi_local_long_range}, we obtain the decomposition of 
\begin{align}
&A(A_0, \tau)= \sum_{Z\subseteq \Lambda } a_{\tau, Z}
\end{align}  
such that
\begin{align}
\sum_{\substack{Z\subseteq \Lambda\\ Z \cap X \neq \emptyset, Z \cap Y \neq \emptyset}} \| a_{\tau, Z} \| 
\le \min \left[J_1^{(k)}g_{a,0} \abb X^{(v|t|)} \abb\cdot \abb Y^{(v|t|)} \abb  (v|t| )^{2D} \left(R^{-\alpha} + e^{-(R-4v|t|)/(8\xi)}\right),J_2^{(k)}   g_a  \abb X^{(v|t|)} \abb (v |t|)^D\right],
\label{math:ineq_quasi_local_proof_k-local}
\end{align} 
where $R = \dist_{X,Y}$ and we define 
$J_1^{(k)}:= (2e\mathcal{C}k^2) 2^{6D+\alpha} e (2D)!  \gamma^2$ and  $J_2^{(k)}:=(2e\mathcal{C}k^2) e 6^{D}   \gamma D! $.
 }

\subsubsection{Proofs of Propositions~\ref{prop:quasi_local_long_range} and \ref{prop:quasi_local_long_range_k-local}.}

\noindent
{\bf [Few-body ($k$-local) operator]}

We first consider the case where the operator $A$ includes up to $k$-body interacting terms:
\begin{align}
A= \sum_{Z \subset \Lambda, |Z| \le k} a_{Z}.
\end{align}  
In this case, the proof is much simpler than that for generic $A$. 
We here denote
\begin{align}
d_{X,Y} = R
\end{align}  
for the simplicity.

For the proof, we decompose $A$ as 
\begin{align}
A= \sum_{r=1}^\infty  \sum_{\substack{Z_r\subset \Lambda \\|Z_r| \le k,  \diam(Z_r) =r}} a_{Z_r}. \label{A_decomp_prop_quasi_long_range}
\end{align}  
We then obtain $A(A_0, \tau)$ in the following form:
\begin{align}
A(A_0, \tau)= \sum_{r=1}^\infty  \sum_{\substack{Z_r\subset \Lambda \\|Z_r| \le k,  \diam(Z_r) =r}} a_{Z_r}(A_0, \tau).
\label{A_decomp_prop_quasi_long_range_time}
\end{align}  
We further decompose each of $\{a_{Z_r}(A_0, \tau)\}_{Z_r \subset \Lambda}$ as follows:
\begin{align}
a_{Z_r}(A_0, \tau)=\sum_{s=0}^\infty a_{\tau, \bal{Z_r}{s\xi + \xi_r}} \with \diam(Z_r) =r, \label{h_Z_m_tau_A_0}
\end{align}  
where  we define 
\begin{align}
a_{\tau, \bal{Z_r}{s\xi+ \xi_r}}:= \begin{cases} 
a_{Z_r}(A_0, \tau, \bal{Z_r}{\xi_r})  & \for  s=0,\\
a_{Z_r}(A_0, \tau, \bal{Z_r}{s\xi+\xi_r}) - a_{Z_r}(A_0, \tau, \bal{Z_r}{(s-1)\xi+\xi_r}) & \for s\ge 1.
\end{cases}
\label{Def_a_ta_bal_bal_z}
\end{align}  
The parameter $\xi_r$ is appropriately chosen afterward.
By combining Eqs.~\eqref{A_decomp_prop_quasi_long_range_time} and \eqref{h_Z_m_tau_A_0}, we have
\begin{align}
A(A_0, \tau)= \sum_{r=1}^\infty  \sum_{\substack{Z_r\subset \Lambda \\|Z_r| \le k,  \diam(Z_r) =r}} \sum_{s=0}^\infty a_{\tau, \bal{Z_r}{s\xi + \xi_r}}  . \label{A_decomp_prop_quasi_long_range}
\end{align}  

In order to obtain the inequality~\eqref{math:ineq_quasi_local_proof}, we need to estimate the following quantity:
\begin{align}
\sum_{r=1}^\infty  \sum_{s=0}^\infty  \sum_{\substack{Z_r\subset \Lambda , |Z_r|\le k, \diam(Z_r) =r\\ \bal{Z_r}{s\xi+\xi_r} \cap X\neq \emptyset,\bal{Z_r}{s\xi+\xi_r} \cap Y\neq \emptyset }} \| a_{\tau, \bal{Z_r}{s\xi+\xi_r}}\|.
\label{h_Z_m_tau_A_0_decomposition_Lieb_R_upp_0}
\end{align} 
We notice  that the condition $\bal{Z_r}{s\xi+\xi_r} \cap X\neq \emptyset$ for $Z_r\subset \Lambda$ is equivalent to $Z_r \cap \bal{X}{s\xi+\xi_r}\neq \emptyset$, and hence 
\begin{align}
 \sum_{r=1}^\infty  \sum_{s=0}^\infty\sum_{\substack{Z_r\subset \Lambda, |Z_r| \le k , \diam(Z_r) =r\\ \bal{Z_r}{s\xi+\xi_r} \cap X\neq \emptyset,\bal{Z_r}{s\xi+\xi_r} \cap Y\neq \emptyset }} \| a_{\tau, \bal{Z_r}{s\xi+\xi_r}} \|
&=\sum_{r=1}^\infty  \sum_{s=0}^\infty\sum_{\substack{Z_r\subset \Lambda, |Z_r| \le k , \diam(Z_r) =r\\ Z_r \cap \bal{X}{s\xi+\xi_r}\neq \emptyset,Z_r \cap \bal{Y}{s\xi+\xi_r}\neq \emptyset }} \| a_{\tau, \bal{Z_r}{s\xi+\xi_r}} \|.
\label{ineq_sum_Z_r_X_Y}
\end{align} 
By using the $\mathcal{G}_0(x,t,\fset{X})$-Lieb-Robinson bound of $A_0$ and Lemma~\ref{Bravyi, Hastings and Verstaete}, we obtain
\begin{align}
\|a_{\tau, \bal{Z_r}{s\xi+\xi_r}}\|&= \| a_{Z_r}(A_0, \tau, \bal{Z_r}{s\xi+\xi_r}) - a_{Z_r}(A_0, \tau, \bal{Z_r}{(s-1)\xi+\xi_r })  \|  \notag \\
&\le  \| a_{Z_r} \|  \mathcal{G}_0(s\xi+\xi_r, \tau, Z_r) + \| a_{Z_r} \|  \mathcal{G}_0((s-1)\xi+\xi_r, \tau, Z_r)  \notag \\
&\le  2\| a_{Z_r} \|  \mathcal{G}_0((s-1)\xi+\xi_r, |t|, Z_r) ,
\label{Lieb_R_assump_upp_srge20}
\end{align}  
where we use $\mathcal{G}_0(x,\tau,\fset{X}) \le \mathcal{G}_0(x,|t|,\fset{X})$ for $\tau \le |t|$. 
By using the $k$-locality of the operator $A$, namely $|Z_r|\le k$ for arbitrary $r$ and choosing $\xi_r$  as 
\begin{align}
\xi_r=v|t| \for \forall r \in \mathbb{N},
\end{align} 
we reduce the inequality~\eqref{Lieb_R_assump_upp_srge20} to
\begin{align}
\|a_{\tau, \bal{Z_r}{s\xi+\xi_r}}\| &\le  2\mathcal{C} |Z_r|^2  \| a_{Z_r} \| e^{-(s-1) - (\xi_r - v|t|)/\xi}  \notag \\
&\le 2e\mathcal{C}k^2 e^{- (\xi_r - v|t|)/\xi}  \| a_{Z_r} \|  e^{-s} = 2e\mathcal{C}k^2\| a_{Z_r} \|  e^{-s} .
\label{Lieb_R_assump_upp_srge2}
\end{align}  
For $\|a_{\tau, \bal{Z_r}{\xi_r}}\|$, we obtain the following trivial inequality:
\begin{align}
\|a_{\tau, \bal{Z_r}{\xi_r}}\|&= \| a_{Z_r}(A_0, \tau, \bal{Z_r}{\xi_r}) \| \le  \| a_{Z_r}\| \le 2e\mathcal{C}k^2\| a_{Z_r} \|.
\label{Lieb_R_assump_upp_sr}
\end{align}  

Thus, we obtain the following upper bound for the summation~\eqref{ineq_sum_Z_r_X_Y}:
\begin{align}
\sum_{r=1}^\infty  \sum_{s=0}^\infty \sum_{\substack{Z_r\subset \Lambda, |Z_r| \le k , \diam(Z_r) =r\\ Z_r \cap \bal{X}{s\xi+\xi_r}\neq \emptyset,Z_r \cap \bal{Y}{s\xi+\xi_r}\neq \emptyset }} \| a_{\tau, \bal{Z_r}{s\xi+\xi_r}} \| 
\le&  2e\mathcal{C}k^2 \sum_{s=0}^\infty  e^{-s} \sum_{r=1}^\infty \sum_{\substack{Z_r\subset \Lambda, |Z_r| \le k , \diam(Z_r) =r\\ Z_r \cap \bal{X}{s\xi+\xi_r}\neq \emptyset,Z_r \cap \bal{Y}{s\xi+\xi_r}\neq \emptyset }}   \| a_{Z_r} \|   \notag \\
=&2e\mathcal{C}k^2 \sum_{s=0}^\infty  e^{-s} \sum_{\substack{Z\subset \Lambda, |Z|\le k \\ Z \cap \bal{X}{s\xi+v|t| }\neq \emptyset,Z \cap \bal{Y}{s\xi+v|t| }\neq \emptyset }}   \| a_{Z} \|   
\notag \\
\le &2e\mathcal{C}k^2 \sum_{s=0}^\infty  e^{-s}  \frac{g_{a,0} \abb\bal{X}{s\xi+v|t| } \abb \cdot \abb\bal{Y}{s\xi+v|t| } \abb }{\left[\max(R- 2s\xi -2v|t|, 0) +1 \right]^{\alpha}}
\label{ineq_sum_Z_r_X_Y_ressumation}
\end{align} 
with $R=\dist_{X,Y}$,
where the last inequality is derived from the assumption~\eqref{alternative_basic_assump_power_2} with $g_0=g_{a,0}$ for the operator $A$. 
Note that we have $\dist_{\bal{X}{s\xi+v|t| },\bal{Y}{s\xi+v|t| }} \ge \dist_{X,Y} - 2(s\xi+v|t| ) = R-  2(s\xi+v|t| )$.

We define $s_R$ as $s_R:=(R-4v|t|)/(4\xi)$, which satisfies $R-2s_R \xi -2v|t| = R/2$. 
We first consider the case of $s_R> 0$ (i.e., $R>4v|t|$). 
From the inequalities~\eqref{coarse_grained_subset_ineq} and \eqref{geometric_parameter_gamma4},  we obtain 
\begin{align}
\abb \bal{X}{s\xi+v|t|}  \abb \le  \bigl |  \bal{X^{(v|t|)}}{s\xi+2v|t|} \bigl |  \le \gamma   \abb X^{(v|t|)} \abb \cdot   \left[2(s\xi +2v|t|)\right]^D .
\label{abb_X_s_xi_tilde_f_R}
\end{align} 
Then, for $s< s_R$, we obtain the following upper bound for the summation of
\begin{align}
\sum_{s<s_R}  e^{-s}  \frac{g_{a,0} \abb\bal{X}{s\xi+v|t| } \abb \cdot \abb\bal{Y}{s\xi+v|t| } \abb }{\left[\max(R- 2s\xi -2v|t|, 0) +1 \right]^{\alpha}}
&\le\frac{g_{a,0} \gamma^2\abb X^{(v|t|)} \abb \cdot\abb Y^{(v|t|)} \abb }{(R/2)^\alpha} \sum_{s< s_R} [2( s\xi+2v|t|)]^{2D} e^{-s}   \notag \\
&\le \frac{e 2^{2D+\alpha} (2D)!g_{a,0}\gamma^2\abb X^{(v|t|)} \abb \cdot\abb Y^{(v|t|)} \abb}{R^\alpha} (\xi+2v|t| )^{2D},
\end{align} 
where we use $R-2s \xi -2v|t| \ge R/2$ for $s< s_R$ in the first inequality.
Also, the second inequality is derived from
\begin{align}
\sum_{s< s_R} [2( s\xi+2v|t|)]^{2D} e^{-s}&\le (2\xi)^{2D} \sum_{s= 0}^\infty (s+2v|t|/\xi)^{2D} e^{-s}  \notag \\
&\le (2\xi)^{2D} \int_{0}^\infty (x+2v|t|/\xi)^{2D} e^{-x+1} dx 
\le e(2\xi)^{2D} (2D)!(1+2v|t|/\xi)^{2D} , 
\label{upper/_bound_s_s_R_d}
\end{align} 
where we utilize the inequality~\eqref{integral_convenient_1}.
For $s\ge s_R$, we obtain 
\begin{align}
\sum_{s\ge s_R}  e^{-s}  \frac{g_{a,0} \abb\bal{X}{s\xi+v|t| } \abb \cdot \abb\bal{Y}{s\xi+v|t| } \abb }{\left[\max(R- 2s\xi -2v|t|, 0) +1 \right]^{\alpha}}
&\le g_{a,0} \gamma^2\abb X^{(v|t|)} \abb \cdot\abb Y^{(v|t|)} \abb  \sum_{s\ge s_R} [2( s\xi+2v|t|)]^{2D} e^{-s}   \notag \\
&\le g_{a,0} \gamma^2\abb X^{(v|t|)} \abb \cdot\abb Y^{(v|t|)} \abb e^{-s_R/2} \sum_{s\ge s_R} [2( s\xi+2v|t|)]^{2D} e^{-s/2}   \notag \\
&\le  e 2^{2D+1} (2D)!g_{a,0} \gamma^2 \abb X^{(v|t|)} \abb \cdot\abb Y^{(v|t|)} \abb (2\xi+2v|t|)^{2D} e^{-s_R/2}   , \label{s_ge_s_R_sestimation}
\end{align} 
where we use the same inequality as \eqref{upper/_bound_s_s_R_d} in estimating the summation in the third inequality.
By applying the above two upper bounds to \eqref{ineq_sum_Z_r_X_Y_ressumation}, we arrive at 
\begin{align}
&\sum_{r=1}^\infty  \sum_{s=0}^\infty \sum_{\substack{Z_r\subset \Lambda, |Z_r| \le k , \diam(Z_r) =r\\ Z_r \cap \bal{X}{s\xi+\xi_r}\neq \emptyset,Z_r \cap \bal{Y}{s\xi+\xi_r}\neq \emptyset }} \| a_{\tau, \bal{Z_r}{s\xi+\xi_r}} \| \notag \\
\le& 2e\mathcal{C}k^2  \left(\frac{e 2^{2D+\alpha} (2D)!g_{a,0} \gamma^2 \abb X^{(v|t|)} \abb \cdot\abb Y^{(v|t|)} \abb }{R^\alpha} (\xi+2v|t|)^{2D} 
+ e 2^{2D+1} (2D)!g_{a,0} \gamma^2\abb X^{(v|t|)} \abb \cdot\abb Y^{(v|t|)} \abb (2\xi+ 2v|t|)^{2D} e^{-s_R/2}  \right) \notag \\
\le &(2e\mathcal{C}k^2) 2^{2D+\alpha}e (2D)!   g_{a,0}\gamma^2 \abb X^{(v|t|)} \abb \cdot\abb Y^{(v|t|)} \abb (2\xi+2v|t| )^{2D} \left(R^{-\alpha} + e^{-(R-4v|t|)/(8\xi)}\right)  \notag \\
\le& (2e\mathcal{C}k^2) 2^{6D+\alpha} e (2D)!   g_{a,0}\gamma^2 \abb X^{(v|t|)} \abb \cdot\abb Y^{(v|t|)} \abb (v|t|)^{2D} \left(R^{-\alpha} + e^{-(R-4v|t|)/(8\xi)}\right) ,
\label{final_upp_k-local_proposition}
\end{align} 
where in the last inequality we use $2v|t| +2\xi \le 4 v|t|$ from the assumption of $v|t|\ge\xi $.

We second consider the case of $s_R\le 0$ (i.e., $R\le 4v|t|$). We take a different approach to estimate the summation~\eqref{ineq_sum_Z_r_X_Y_ressumation} as follows:
\begin{align}
\sum_{r=1}^\infty  \sum_{s=0}^\infty \sum_{\substack{Z_r\subset \Lambda, |Z_r| \le k , \diam(Z_r) =r\\ Z_r \cap \bal{X}{s\xi+\xi_r}\neq \emptyset,Z_r \cap \bal{Y}{s\xi+\xi_r}\neq \emptyset }} \| a_{\tau, \bal{Z_r}{s\xi+\xi_r}} \| 
\le&  2e\mathcal{C}k^2 \sum_{s=0}^\infty  e^{-s}   \sum_{\substack{Z\subset \Lambda, |Z|\le k \\ Z \cap \bal{X}{s\xi+v|t|}\neq \emptyset}}   \| a_{Z} \|  .
\label{upper_bound_s_R_0_1_rerfef}
\end{align} 
By applying the assumption~\eqref{basic_assump_power} with $g=g_a$ and the inequality~\eqref{abb_X_s_xi_tilde_f_R}, we have
\begin{align}
\sum_{\substack{Z\subset \Lambda, |Z|\le k \\ Z \cap \bal{X}{s\xi+v|t|}\neq \emptyset}}   \| a_{Z} \|  \le g_a\abb \bal{X}{s\xi+v|t|} \abb \le g_a \gamma  \abb X^{(v|t|)} \abb \cdot   \left[2(s\xi +2v|t|)\right]^D ,
\end{align} 
which gives the upper bound of
\begin{align}
\sum_{s=0}^\infty  e^{-s}   \sum_{\substack{Z\subset \Lambda, |Z|\le k \\ Z \cap \bal{X}{s\xi+v|t|}\neq \emptyset}}   \| a_{Z} \|  
 \le g_a \gamma\abb X^{(v|t|)} \abb \sum_{s=0}^\infty  e^{-s}[2 (s\xi+2v|t|)]^D \le e 2^D  g_a \gamma \abb X^{(v|t|)} \abb D! (\xi+2v|t|)^D ,
 \label{upper_bound_s_R_0_1_rerfef2}
\end{align} 
where we use the same upper bound as \eqref{upper/_bound_s_s_R_d}.
By combining the inequalities~\eqref{upper_bound_s_R_0_1_rerfef} and \eqref{upper_bound_s_R_0_1_rerfef2}, we obtain 
\begin{align}
\sum_{r=1}^\infty  \sum_{s=0}^\infty \sum_{\substack{Z_r\subset \Lambda, |Z_r| \le k , \diam(Z_r) =r\\ Z_r \cap \bal{X}{s\xi+\xi_r}\neq \emptyset,Z_r \cap \bal{Y}{s\xi+\xi_r}\neq \emptyset }} \| a_{\tau, \bal{Z_r}{s\xi+\xi_r}} \| 
&\le (2e\mathcal{C}k^2) e 2^D    g_a \gamma\abb X^{(v|t|)} \abb D! (\xi+2v|t|)^D \notag \\
&\le (2e\mathcal{C}k^2) e 6^{D}    g_a \gamma \abb X^{(v|t|)} \abb D! (v|t|)^D  ,
\label{final_upp_k-local_proposition2}
\end{align} 
where in the last inequality we use $v|t| +\xi \le 3 v|t|$ from the assumption of $v|t|\ge\xi $.

From the upper bounds of \eqref{final_upp_k-local_proposition} and \eqref{final_upp_k-local_proposition2},  we finally obtain
\begin{align}
&\sum_{r=1}^\infty  \sum_{s=0}^\infty \sum_{\substack{Z_r\subset \Lambda, |Z_r| \le k , \diam(Z_r) =r\\ Z_r \cap \bal{X}{s\xi+\xi_r}\neq \emptyset,Z_r \cap \bal{Y}{s\xi+\xi_r}\neq \emptyset }} \| a_{\tau, \bal{Z_r}{s\xi+\xi_r}} \| \notag \\
\le &\min \left [(2e\mathcal{C}k^2) 2^{6D+\alpha} e (2D)!   g_{a,0}\gamma^2 \abb X^{(v|t|)} \abb \cdot\abb Y^{(v|t|)} \abb (v|t|)^{2D} \left(R^{-\alpha} + e^{-(R-4v|t|)/(8\xi)}\right) , 
(2e\mathcal{C}k^2) e 6^{D}    g_a \gamma \abb X^{(v|t|)} \abb D! (v|t|)^D  \right] \notag \\
=&\min \left[J_1^{(k)}g_{a,0} \abb X^{(v|t|)} \abb \cdot\abb Y^{(v|t|)} \abb (v|t| )^{2D} \left(R^{-\alpha} + e^{-(R-4v|t|)/(8\xi)}\right),J_2^{(k)}   g_a  \abb X^{(v|t|)} \abb (v|t|)^D\right]
\label{final_upp_k-local_proposition_final_form}
\end{align} 
with
\begin{align}
J_1^{(k)}:= (2e\mathcal{C}k^2) 2^{6D+\alpha} e (2D)!   \gamma^2 ,\quad J_2^{(k)}:=(2e\mathcal{C}k^2)  e 6^{D}  \gamma  D!   .
\end{align} 
This completes the proof of Proposition~\ref{prop:quasi_local_long_range_k-local}. $\square$

{~}

\noindent
{\bf [generic cases]}

We here consider the general case without the assumption of the  $k$-locality:
\begin{align}
A= \sum_{Z \subseteq \Lambda} a_{Z}.
\end{align}  
In this case, we also start from the equation~\eqref{ineq_sum_Z_r_X_Y}. 
Here, in estimating the norm of $\|a_{\tau, \bal{Z_r}{s\xi+\xi_r}}\|$, we cannot utilize the upper bound of $ |Z_r|^2 \le k^2$ as in \eqref{Lieb_R_assump_upp_srge2}.
Instead, we utilize the following upper bound which is derived from the inequality~\eqref{geometric_parameter_gamma1}: 
\begin{align}
 |Z_r| \le \gamma[\diam(Z_r)]^D = \gamma r^D.
\end{align}  
This replaces the inequality~\eqref{Lieb_R_assump_upp_srge2} by
\begin{align}
\|a_{\tau, \bal{Z_r}{s\xi+\xi_r}}\|\le 2e\gamma^2\mathcal{C}  \| a_{Z_r} \| r^{2D}  e^{- (\xi_r - v|t|)/\xi}  e^{-s}  \quad (s\ge 1).
\label{Lieb_R_assump_upp_srge2_general}
\end{align}  
We here choose $\xi_r$ as 
\begin{align}
&\xi_r=f_r \xi    ,\quad  f_r:= \log (2e\gamma^2 \mathcal{C} r^{2D})+v|t|/\xi .
\label{choice_of_s_ast}
\end{align}
By applying the above choice of $\xi_r$ to \eqref{Lieb_R_assump_upp_srge2_general}, we obtain
\begin{align}
\|a_{\tau, \bal{Z_r}{s\xi+\xi_r}}\|\le  \| a_{Z_r} \|  e^{-s}   .
\end{align}  
We note that this inequality also holds for $s=0$ since $\|a_{\tau, \bal{Z_r}{\xi_r}}\|=\| a_{Z_r} \| $ from the definition~\eqref{Def_a_ta_bal_bal_z}.
We then estimate the same quantity as \eqref{ineq_sum_Z_r_X_Y_ressumation}: 
\begin{align}
&\sum_{r=1}^\infty  \sum_{s=0}^\infty \sum_{\substack{Z_r\subseteq \Lambda , \diam(Z_r) =r\\ Z_r \cap \bal{X}{s\xi+\xi_r}\neq \emptyset,Z_r \cap \bal{Y}{s\xi+\xi_r}\neq \emptyset }} \| a_{\tau, \bal{Z_r}{s\xi+\xi_r}} \| 
\le  \sum_{s=0}^\infty  e^{-s} \sum_{r=1}^\infty \sum_{\substack{Z_r\subseteq \Lambda , \diam(Z_r) =r\\ Z_r \cap \bal{X}{(s+f_r) \xi }\neq \emptyset,Z_r \cap \bal{Y}{(s+f_r) \xi }\neq \emptyset }}   \| a_{Z_r} \| .
\label{ineq_sum_Z_r_X_Y_ressumation_generic_case}
\end{align}


In order to estimate the summation with respect to $r$, we analyze the cases of $r\le R_0$ and $r>R_0$ in different ways, where we will choose $R_0$ afterward.
For $r\le R_0$, we use the assumption~\eqref{alternative_basic_assump_power_2} with $g_0=g_{a,0}$, which implies
\begin{align}
\sum_{r=1}^{R_0} \sum_{\substack{Z_r\subseteq \Lambda , \diam(Z_r) =r\\ Z_r \cap \bal{X}{(s+f_r) \xi }\neq \emptyset,Z_r \cap \bal{Y}{(s+f_r) \xi }\neq \emptyset }}   \| a_{Z_r} \| 
&\le\sum_{\substack{Z\subseteq \Lambda \\ Z \cap \bal{X}{(s+f_{R_0}) \xi }\neq \emptyset,Z \cap \bal{Y}{(s+f_{R_0}) \xi }\neq \emptyset }}   \| a_{Z} \| \notag \\
 &\le \frac{g_{a,0} \abb\bal{X}{(s+f_{R_0}) \xi } \abb \cdot \abb\bal{Y}{(s+f_{R_0}) \xi  } \abb }{\left[\max(R- 2s\xi -2f_{R_0}\xi, 0) +1 \right]^{\alpha}}  ,
\end{align} 
where we use $f_r \le f_{R_0}$ for $r\le R_0$ from the definition~\eqref{choice_of_s_ast}.
From this inequality, we obtain 
\begin{align}
&\sum_{s=0}^\infty  e^{-s}\sum_{r=1}^{R_0} \sum_{\substack{Z_r\subseteq \Lambda , \diam(Z_r) =r\\ Z_r \cap \bal{X}{(s+f_r) \xi }\neq \emptyset,Z_r \cap \bal{Y}{(s+f_r) \xi }\neq \emptyset }}   \| a_{Z_r} \|  \le 
g_{a,0}  \sum_{s=0}^\infty  e^{-s} \frac{\abb\bal{X}{(s+f_{R_0}) \xi } \abb \cdot \abb\bal{Y}{(s+f_{R_0}) \xi  } \abb }{\left[\max(R- 2s\xi -2f_{R_0}\xi, 0) +1 \right]^{\alpha}}   ,
\label{We thus arrive at the inequality of_0}
\end{align} 
and the remaining task is to estimate the summation with respect to $s$.
This summation has been already estimated in \eqref{ineq_sum_Z_r_X_Y_ressumation}, where we only need to replace $v|t|$ with $f_{R_0}\xi$.
Hence, we can obtain the similar upper bound to \eqref{final_upp_k-local_proposition_final_form}:
\begin{align}
&g_{a,0} \sum_{s=0}^\infty  e^{-s} \frac{\abb\bal{X}{(s+f_{R_0}) \xi } \abb \cdot \abb\bal{Y}{(s+f_{R_0}) \xi  } \abb }{\left[\max(R- 2s\xi -2f_{R_0}\xi, 0) +1 \right]^{\alpha}}  \notag \\
\le& \min \Bigl[ 2^{2D+\alpha} e (2D)!   g_{a,0}\gamma^2 \abb X^{(v|t|)} \abb \cdot\abb Y^{(v|t|)} \abb   (f_{R_0}\xi +2\xi+v|t|)^{2D} \left(R^{-\alpha} 
+ e^{-(R-4f_{R_0}\xi)/(8\xi)}\right),  \notag \\
&\quad \quad \ e 2^{D}    g_a \gamma \abb X^{(v|t|)} \abb D!   (f_{R_0}\xi+\xi+v|t|)^D\Bigr]  \notag \\
\le&\min \left[ 2^{6D+\alpha} e (2D)!   g_{a,0}\gamma^2  \abb X^{(v|t|)} \abb \cdot\abb Y^{(v|t|)} \abb  (f_{R_0}\xi)^{2D} \left(R^{-\alpha} + e^{-(R-4f_{R_0}\xi)/(8\xi)}\right), 
e6^{D}    g_a \gamma \abb X^{(v|t|)} \abb  D! (f_{R_0}\xi)^D\right],
\label{We thus arrive at the inequality of_1}
\end{align} 
where we use $f_{R_0} \xi \ge v|t| \ge \xi$ from the definition~\eqref{choice_of_s_ast} and the assumption $v|t| \ge \xi$.
By applying the inequality~\eqref{We thus arrive at the inequality of_1} to \eqref{We thus arrive at the inequality of_0}, we arrive at the inequality of
\begin{align}
&\sum_{s=0}^\infty  e^{-s}\sum_{r=1}^{R_0} \sum_{\substack{Z_r\subseteq \Lambda , \diam(Z_r) =r\\ Z_r \cap \bal{X}{(s+f_r) \xi }\neq \emptyset,Z_r \cap \bal{Y}{(s+f_r) \xi }\neq \emptyset }}   \| a_{Z_r} \|   \notag \\
\le & \min \left[ J_1  g_{a,0} \abb X^{(v|t|)} \abb \cdot\abb Y^{(v|t|)} \abb(f_{R_0}\xi)^{2D} \left(R^{-\alpha} + e^{-(R-4f_{R_0}\xi)/(8\xi)}\right), 
J_2g_a \abb X^{(v|t|)} \abb (f_{R_0}\xi)^D\right]
\label{R_0_smaller_r_inequality}
\end{align} 
with
\begin{align}
J_1:=  2^{6D+\alpha} e (2D)!   \gamma^2 ,\quad J_2:=e 6^{D}  \gamma  D!   .
\end{align}

We next consider the summation for $r> R_0$. We  here use the assumption~\eqref{basic_assump_power}, which yields 
\begin{align}
\sum_{r=R_0+1}^{\infty}  \sum_{\substack{Z_r\subseteq \Lambda , \diam(Z_r) =r\\ Z_r \cap \bal{X}{(s+f_r) \xi }\neq \emptyset,Z_r \cap \bal{Y}{(s+f_r) \xi }\neq \emptyset }}   \| a_{Z_r} \| 
 &\le \sum_{r=R_0+1}^{\infty}  \sum_{\substack{Z\subseteq \Lambda , \diam(Z) \ge r\\ Z \cap \bal{X}{(s+f_r) \xi }\neq \emptyset }}   \| a_{Z_r} \| \notag \\
&\le  \sum_{r=R_0+1}^{\infty}  \frac{g_a \abb\bal{X}{(s+f_r) \xi}\abb }{r^{\alpha-D}} \notag \\
&\le \sum_{r=R_0+1}^{\infty}  \frac{g_a \gamma \abb X^{(v|t|)} \abb [2(s\xi +f_r \xi + v|t|)]^{D}}{r^{\alpha-D}}  \notag \\
&\le g_a \gamma \abb X^{(v|t|)} \abb(2\xi)^{D}\sum_{r=R_0+1}^{\infty}   \frac{(s+f_r+v|t|/\xi)^D}{r^{\alpha-D}} ,
\end{align} 
where we use \eqref{abb_X_s_xi_tilde_f_R} in the third inequality.
We thus obtain 
\begin{align}
\sum_{s=0}^\infty  e^{-s}\sum_{r=R_0+1}^{\infty} \sum_{\substack{Z_r\subseteq \Lambda , \diam(Z_r) =r\\ Z_r \cap \bal{X}{(s+f_r) \xi }\neq \emptyset,Z_r \cap \bal{Y}{(s+f_r) \xi }\neq \emptyset }}   \| a_{Z_r} \| & \le 
g_a \gamma \abb X^{(v|t|)} \abb (2\xi)^{D} \sum_{r=R_0+1}^{\infty} \sum_{s=0}^\infty  e^{-s}  \frac{(s+f_r+v|t|/\xi)^D}{r^{\alpha-D}}  \notag \\
&\le g_a \gamma \abb X^{(v|t|)} \abb (2\xi)^{D}  eD! \sum_{r=R_0+1}^{\infty}  \frac{(1+f_r+v|t|/\xi)^D}{r^{\alpha-D}}   ,
\label{s_0_infty_R_0_1_larger_ineq}
 \end{align} 
where we use the inequality~\eqref{integral_convenient_1} to obtain the following upper bound:
\begin{align}
\sum_{s=0}^\infty  e^{-s} (s+f_r+v|t|/\xi)^D \le \int_0^\infty e^{-x+1} (x+f_r+v|t|/\xi)^D dx \le eD! (1+f_r+v|t|/\xi)^D .
 \end{align} 
From the definition of $f_r$ in Eq.~\eqref{choice_of_s_ast}, we have $f_r=f_1 +2D \log(r)$, and hence 
\begin{align}
\sum_{r=R_0+1}^{\infty}  \frac{(1+f_r+v|t|/\xi)^D}{r^{\alpha-D}} 
&=\sum_{r=R_0+1}^{\infty}  \frac{[1+f_1+v|t|/\xi +2D \log(r)]^D}{r^{\alpha-D}}  \notag \\
&\le (1+f_1+v|t|/\xi)^D \sum_{r=R_0+1}^{\infty}  \frac{[1+2D \log(r)]^D}{r^{\alpha-D}}  \notag \\
&\le (1+f_1+v|t|/\xi)^D \sup_{r\ge 1} \left(\frac{[1+2D \log(r)]^D}{r^{\alpha-2D-1}} \right) \sum_{r=R_0+1}^{\infty} \frac{1}{r^{D+1}} \notag \\
&\le \frac{(1+f_1+v|t|/\xi)^D/D}{R_0^D} \sup_{r\ge 1} \left(\frac{[1+2D \log(r)]^D}{r^{\alpha-2D-1}} \right) =: c_{\alpha,D} \frac{(1+f_1+v|t|/\xi)^D}{R_0^D},
 \end{align} 
where we define $c_{\alpha,D}:=D^{-1} \sup_{r\ge 1} \left([1+2D \log(r)]^D/r^{\alpha-2D-1} \right) $. Note that from $\alpha>2D+1$ the constant $c_{\alpha,D}$ has a finite value.
By applying the above inequality to \eqref{s_0_infty_R_0_1_larger_ineq}, we arrive at the following upper bound:
\begin{align}
&\sum_{s=0}^\infty  e^{-s}\sum_{r=R_0+1}^{\infty}  \sum_{\substack{Z_r\subseteq \Lambda , \diam(Z_r) =r\\ Z_r \cap \bal{X}{(s+f_r) \xi }\neq \emptyset,Z_r \cap \bal{Y}{(s+f_r) \xi }\neq \emptyset }}   \| a_{Z_r} \| 
\le g_a \gamma \abb X^{(v|t|)} \abb(2\xi)^{D}  eD!  c_{\alpha,D} \frac{(1+f_1+v|t|/\xi)^D}{R_0^D}  \notag \\
&\le  g_a \abb X^{(v|t|)} \abb  e \gamma D!  [8 \log (2e\gamma^2 \mathcal{C})]^D c_{\alpha,D} \frac{(v|t|)^D}{R_0^D} 
 = g_a J_3 \abb X^{(v|t|)} \abb  (v|t|)^D R_0^{-D}  ,
\label{R_0_larger_r_inequality}
\end{align} 
where we use $2\xi(1+f_1+v|t|/\xi)=2\xi+ 2\xi \log (2e\gamma^2 \mathcal{C}) +4 v|t| \le 8v|t| \log (2e\gamma^2 \mathcal{C})$ from the assumption of $v|t| \ge \xi$.
Note that $J_3$ is defined as 
\begin{align}
 J_3 =e \gamma D!  [8 \log (2e\gamma^2 \mathcal{C})]^D c_{\alpha,D} .
\end{align}

Finally, we choose $R_0$ as $R_0=\lceil R^{\alpha/D} \rceil \le 2R^{\alpha/D}$. 
The upper bounds in~\eqref{R_0_smaller_r_inequality} and \eqref{R_0_larger_r_inequality} reduce to 
\begin{align}
&\sum_{s=0}^\infty  e^{-s}\sum_{r=1}^{R_0} \sum_{\substack{Z_r\subseteq \Lambda , \diam(Z_r) =r\\ Z_r \cap \bal{X}{(s+f_r) \xi }\neq \emptyset,Z_r \cap \bal{Y}{(s+f_r) \xi }\neq \emptyset }}   \| a_{Z_r} \|   \notag \\
\le & \min \left[ J_1  g_{a,0}  \abb X^{(v|t|)} \abb \cdot\abb Y^{(v|t|)} \abb (\tilde{f}_{R}\xi)^{2D} \left(R^{-\alpha} + e^{-(R-4\tilde{f}_{R}\xi)/(8\xi)}\right), 
J_2g_a \abb X^{(v|t|)}  \abb(\tilde{f}_{R}\xi)^D\right]
\end{align} 
and 
\begin{align}
&\sum_{s=0}^\infty  e^{-s}\sum_{r=R_0+1}^{\infty}  \sum_{\substack{Z_r\subseteq \Lambda , \diam(Z_r) =r\\ Z_r \cap \bal{X}{(s+f_r) \xi }\neq \emptyset,Z_r \cap \bal{Y}{(s+f_r) \xi }\neq \emptyset }}   \| a_{Z_r} \| 
\le  g_a J_3 \abb X^{(v|t|)} \abb  (v|t|)^D R^{-\alpha}   ,
\end{align} 
respectively, where in the first inequality we use $f_{R_0} \le f_{2 R^{\alpha/D}} =: \tilde{f}_R$.
By combining the above two bounds with \eqref{ineq_sum_Z_r_X_Y_ressumation_generic_case}, we obtain the main inequality~\eqref{math:ineq_quasi_local_proof}.
This completes the proof of Proposition~\ref{prop:quasi_local_long_range}. $\square$

\subsection{Proof of Proposition~\ref{prop:quasi_local_long_range_Lieb-Robinson}} \label{Sec:prop:quasi_local_long_range_Lieb-Robinson}

\subsubsection{Statement}
{~}\\
{\bf Proposition~\ref{prop:quasi_local_long_range_Lieb-Robinson}.}
\textit{
Let us consider a Hamiltonian $\tilde{H}$ as follows:
\begin{align}
\tilde{H}= \sum_{Z \subseteq \Lambda} \tilde{h}_{Z}
\end{align}  
with 
\begin{align}
&\sum_{\substack{Z\subseteq \Lambda\\ Z \cap X \neq \emptyset, Z \cap Y \neq \emptyset}} \| \tilde{h}_{Z} \| \le \abb X^{(\xi_0)} \abb \cdot \abb Y^{(\xi_0)}\abb g(R)
,\quad R=\dist_{X,Y}, \notag \\
&g(R) = \begin{cases}
\tilde{g}  &\for R< \kappa\xi_0, \\
\tilde{g}_R R^{-\alpha} &\for R\ge \kappa\xi_0 ,
\end{cases} \quad  \tilde{g}_R:=\tilde{g}_0 \log^p (R+1) \    (p\in \mathbb{N}),
\label{def:f_R_LR_Long}
\end{align}  
where $\kappa\ge 2$, $\xi_0\ge1$, and $g(R)$ monotonically decreases with $R$.
Then, the Hamiltonian $\tilde{H}$ satisfies $\mathcal{G}(x,t,\fset{X},\fset{Y})$-Lieb-Robinson bound with
\begin{align}
\mathcal{G}(x,t,\fset{X},\fset{Y})= \abb \fset{X}^{(\xi_0)} \abb \cdot \abb \fset{Y}^{(\xi_0)}\abb \frac{2^{\alpha+1} e\Gamma(\alpha+2) }{\tilde{\lambda}} \left( e^{2e \tilde{\lambda}\gamma^2 9^D |t|}-1 \right)  g(R)
+2\abb X^{(\xi_0)} \abb \left( \frac{4e\kappa \xi_0 \tilde{\lambda} \gamma^2 9^D |t|}{R} \right)^{R/(2\kappa\xi_0)},
\label{main_ineq_long_range_LR_proof}
\end{align}  
where $\Gamma(x)$ is the gamma function and we define $\tilde{\lambda}$ as 
\begin{align}
 &\tilde{\lambda}:=\tilde{g}\gamma [2(\kappa+3)]^D + \frac{c_{p,\kappa}  \tilde{g}_0 \gamma D 2^{D+1} }{\kappa+2}  \frac{\log^p(\kappa\xi_0+1)}{\xi_0^{\alpha} } , \notag \\
 &c_{p,\kappa}:= \sup_{z\in \mathbb{R}| z\ge \kappa+2} \left[\frac{ (\log (z)+1)^p}{(z-2)^{\alpha-D-1}}\right]
  \label{def_para_constant_f_proof} .
\end{align}   }

{~}\\
{\bf Remark.}
The theorem is similar to the ones in~\cite{nachtergaele2010lieb,ref:Hastings2006-ExpDec}.
Indeed, we will utilize their proof technique, but the theorems therein cannot be directly applied.
The main reason is the following. 
The theorems in Ref.~\cite{nachtergaele2010lieb,ref:Hastings2006-ExpDec} roughly give the following bound: 
\begin{align}
\mathcal{G}(x,t,\fset{X},\fset{Y})= \abb \fset{X} \abb \cdot \abb \fset{Y}\abb \left( e^{{\rm const}. \lambda |t|}-1 \right)  g(R),
\end{align}  
where the parameter $\lambda$ is defined by the inequality of
\begin{align}
\max_{x,y \in \Lambda}\sum_{z\in \Lambda} g(\dist_{x,z}) g(\dist_{z,y})  \le \lambda g(\dist_{x,y}). \label{simple_inequality_eqref}
\end{align} 
In our case, from the definition~\eqref{def:f_R_LR_Long}, the parameter $\lambda$ is roughly upper-bounded by $\tilde{g} \xi_0^D$, and hence the exponential term is given by $e^{\orderof{\tilde{g} \xi_0^D|t|}}$.
On the other hand, the inequality \eqref{def_para_constant_f_proof} gives it by $$e^{\orderof{\tilde{g}|t|}+ \orderof{\tilde{g}_0\xi_0^{-\alpha} |t|}},$$ which is much better than $e^{\orderof{\tilde{g} \xi_0^D|t|}}$ when $\tilde{g}$, $\tilde{g}_0$ and $\xi_0$ depend on the time as in Eq.~\eqref{choice_of_the_parameters}.  
This improvement is crucial in proving the linear light cone for $\alpha>2D+1$.

\subsubsection{Proof of Proposition~\ref{prop:quasi_local_long_range_Lieb-Robinson}.}
We denote $\Lambda^{(\xi_0)}=\{i_s\}_{s=1}^{\tilde{n}}$ with $\tilde{n}=\abb\Lambda^{(\xi_0)}\abb $.
We also define $\{L_s\}_{s=1}^{\tilde{n}}$ as 
 \begin{align}
&L_s:=\bal{i_s}{\xi_0}   \for i_s \in \Lambda^{(\xi_0)}.
\label{decomposition_of_total_xi_LR}
\end{align}
Note that $L_1 \cup L_2 \cdots L_{\tilde{n}}=\Lambda$ and 
 \begin{align}
\diam(L_s) = \diam(\bal{i_s}{\xi_0} ) \le 2\xi_0+1. \label{diam_L_s_upp}
\end{align}
The essence of the proof is similar to the ones in~\cite{nachtergaele2010lieb,ref:Hastings2006-ExpDec}.
The key idea to improve the original results is to treat the coarse grained set $\Lambda^{(\xi_0)}$ instead of the original total set $\Lambda$.
For example, this point is reflected in the inequality~\eqref{summation_second_term_1.55}.
In order to simplify the inequality to a convenient form, we will utilize the similar inequality~\eqref{correpondence_lambda_tilde} to \eqref{simple_inequality_eqref}. 
The parameter $\tilde{\lambda}$ defined in Eq.~\eqref{def_para_constant_f_proof} will appear in order to upper-bound the summation~\eqref{upp_bound_lambda_ineq}.

We start from the same inequality as~\eqref{Lieb-Robinson_expansion_0}:
\begin{align} 
&\| [ O_X (\tilde{H},t) , O_{Y}] \| \le \sum_{m=1}^{m^\ast}\mathcal{L}_m 
 \label{Lieb-Robinson_expansion_0_finite}
\end{align} 
with $\|O_X \| = \| O_{Y}\|=1$ and $R=\dist_{X,Y}$, where $\{\mathcal{L}_m\}_{m=1}^{m^\ast-1}$ and $\mathcal{L}_{m^\ast}$ are defined as
 \begin{align} 
\mathcal{L}_m= \frac{2(2|t|)^m}{m!} \sum_{Z_1 \cap X \neq \emptyset} \|\tilde{h}_{Z_1}\|  
\sum_{Z_{2} \cap Z_{1} \neq \emptyset}  \|\tilde{h}_{Z_{2}} \|\cdots    \sum_{\substack{Z_{m} \cap Z_{m-1} \neq \emptyset\\ Z_{m}\cap Y\neq \emptyset}}  \| \tilde{h}_{Z_m} \| 
\label{mathdal_L_m_def_finite_m}
\end{align} 
and 
\begin{align} 
\mathcal{L}_{m^\ast}= \frac{2(2|t|)^{m^\ast}}{m^\ast!} \sum_{Z_1 \cap X \neq \emptyset} \|\tilde{h}_{Z_1}\|  
\sum_{Z_{2} \cap Z_{1} \neq \emptyset}  \|\tilde{h}_{Z_{2}} \|\cdots    \sum_{Z_{m^\ast} \cap Z_{m^\ast-1} \neq \emptyset}  \| \tilde{h}_{Z_{m^\ast}} \| .
\label{mathdal_L_m_def_finite_m_ast}
\end{align} 
In the proof, we choose $m^\ast$ such that for $m\le m^\ast-1$ the following inequalities are satisfied:
\begin{align} 
\frac{R -2(m+1)\xi_0}{m} \ge \frac{R}{2m}\quad  {\rm and} \quad \frac{R}{2m} \ge \kappa\xi_0.
\label{condition_for_m_ast_1}
\end{align}
We note that the above inequalities hold for $m\le R/(4\xi_0) -1$ and $m\le R/(2\kappa\xi_0)$, respectively. 
Because of $\kappa\ge 2$, we need to choose so that the parameter $m$ may satisfy $m\le R/(2\kappa\xi_0)$, and hence we adopt
\begin{align} 
m^\ast =   \left \lfloor  \frac{R}{2\kappa\xi_0}\right\rfloor +1 \ge \frac{R}{2\kappa\xi_0}.
\label{choice_of_mast_def}
\end{align}

We first consider $\mathcal{L}_1$:
 \begin{align} 
\mathcal{L}_1= 2(2|t|) \sum_{Z_1 \cap X \neq \emptyset, Z_1 \cap Y \neq \emptyset} \|\tilde{h}_{Z_1}\|  \le 2(2|t|) \abb X^{(\xi_0)} \abb \cdot \abb Y^{(\xi_0)}\abb g(R) ,
\label{mathdal_L_m_def_finite_first}
\end{align} 
where the inequality is given by the definition of the interaction~\eqref{def:f_R_LR_Long}.
We second consider $\mathcal{L}_2$:
 \begin{align} 
\mathcal{L}_2= \frac{2(2|t|)^2}{2!} \sum_{Z_1 \cap X \neq \emptyset} \|\tilde{h}_{Z_1}\|  \sum_{Z_{2} \cap Z_{1} \neq \emptyset, Z_{2}\cap Y\neq \emptyset} \|\tilde{h}_{Z_2}\|  .
\label{mathdal_L_m_def_finite_second}
\end{align} 
From $X^{(\xi_0)}[\xi_0] \supseteq X$, we obtain 
 \begin{align} 
\sum_{Z_1 \cap X \neq \emptyset} \|\tilde{h}_{Z_1}\|   \le 
\sum_{Z_1 \cap X^{(\xi_0)}[\xi_0] \neq \emptyset} \|\tilde{h}_{Z_1}\|   \le 
\sum_{i_{s} \in X^{(\xi_0)}} \sum_{Z_1 \cap \bal{i_{s}}{\xi_0}\neq \emptyset} \|\tilde{h}_{Z_1}\|
=\sum_{i_{s} \in X^{(\xi_0)}} \sum_{Z_1 \cap L_s\neq \emptyset} \|\tilde{h}_{Z_1}\|,
\end{align} 
where we use the notation of $L_s$ in \eqref{decomposition_of_total_xi_LR}. 
By using the above inequality, we have 
 \begin{align} 
\sum_{Z_1 \cap X \neq \emptyset} \|\tilde{h}_{Z_1}\|  \sum_{Z_{2} \cap Z_{1} \neq \emptyset, Z_{2}\cap Y\neq \emptyset} \|\tilde{h}_{Z_2}\|  
&\le 
\sum_{i_{s} \in X^{(\xi_0)}} \sum_{i_{s'} \in Y^{(\xi_0)}}  \sum_{Z_1 \cap L_{s} \neq \emptyset} \|\tilde{h}_{Z_1}\|  \sum_{Z_{2} \cap Z_{1} \neq \emptyset, Z_{2}\cap L_{s'}\neq \emptyset} \|\tilde{h}_{Z_2}\|  .
\label{summation_second_term_01}
\end{align} 
Furthermore, from $L_{1} \cup L_2 \cup \cdots \cup L_{\tilde{n}}=\Lambda$, we obtain
 \begin{align} 
\sum_{Z_1 \cap L_{s} \neq \emptyset} \|\tilde{h}_{Z_1}\|  \sum_{Z_{2} \cap Z_{1} \neq \emptyset, Z_{2}\cap L_{s'}\neq \emptyset} \|\tilde{h}_{Z_2}\|  
&\le
\sum_{s_1=1}^{\tilde{n}} \sum_{Z_1 \cap L_s \neq \emptyset, Z_1 \cap L_{s_1} \neq \emptyset} \|\tilde{h}_{Z_1}\|  \sum_{Z_{2} \cap L_{s_1} \neq \emptyset, Z_{2}\cap L_{s'}\neq \emptyset} \|\tilde{h}_{Z_2}\| .
\label{summation_second_term_001}
\end{align} 
The definition~\eqref{def:f_R_LR_Long} gives the upper bound of 
 \begin{align} 
\sum_{Z_1 \cap L_s \neq \emptyset, Z_1 \cap L_{s_1} \neq \emptyset} \|\tilde{h}_{Z_1}\|  \le \abb L_{s}^{(\xi_0)} \abb \cdot \abb L_{s_1}^{(\xi_0)}\abb g(\dist_{L_s,L_{s_1}}) 
\le  \gamma^29^D g(\dist_{L_s,L_{s_1}}) ,
\end{align} 
where we use the inequality~\eqref{geometric_parameter_gamma3} to obtain $\abb L_{s}^{(\xi_0)} \abb \le \gamma [\diam(L_s)/\xi_0]^D \le \gamma [(2\xi_0+1)/\xi_0]^D \le \gamma 3^D$ ($\xi_0\ge1$).
This upper bound reduces the inequality~\eqref{summation_second_term_001} to  
 \begin{align} 
\sum_{Z_1 \cap L_{s} \neq \emptyset} \|\tilde{h}_{Z_1}\|  \sum_{Z_{2} \cap Z_{1} \neq \emptyset, Z_{2}\cap L_{s'}\neq \emptyset} \|\tilde{h}_{Z_2}\|  
\le  (\gamma^2 9^D )^2 \sum_{s_1=1}^{\tilde{n}} g(\dist_{L_s,L_{s_1}}) g(\dist_{L_{s_1},L_{s'}}).
\label{summation_second_term_1}
\end{align} 
By combining \eqref{summation_second_term_01} and \eqref{summation_second_term_1}, we have
 \begin{align} 
\sum_{Z_1 \cap X \neq \emptyset} \|\tilde{h}_{Z_1}\|  \sum_{Z_{2} \cap Z_{1} \neq \emptyset, Z_{2}\cap Y\neq \emptyset} \|\tilde{h}_{Z_2}\|  
&\le (\gamma^29^D )^2
\sum_{i_{s} \in X^{(\xi_0)}} \sum_{i_{s'} \in Y^{(\xi_0)}} \sum_{s_1=1}^{\tilde{n}}  g(\dist_{L_s,L_{s_1}}) g(\dist_{L_{s_1},L_{s'}}) .
\label{summation_second_term_1.55}
\end{align}

Because of $i_{s} \in X^{(\xi_0)}$, $i_{s'} \in Y^{(\xi_0)}$, we have $L_s \cap X \neq \emptyset$ and $L_{s'} \cap Y \neq \emptyset$. Then, from $R=\dist_{X,Y}$, 
the distances $\dist_{L_s,L_{s_1}}$ and $\dist_{L_{s_1},L_{s'}}$ should satisfy 
 \begin{align} 
R \le \dist_{L_s,L_{s_1}}+\dist_{L_{s_1},L_{s'}} + \diam(L_s)-1+ \diam(L_{s'})-1+ \diam(L_{s_1}) -1\le \dist_{L_s,L_{s_1}}+\dist_{L_{s_1},L_{s'}}+6\xi_0,
\label{R_le_distance_meq2_6xi0}
\end{align}
where we use the inequality~\eqref{diam_L_s_upp} for $\diam(L_s)$.
The above inequality yields  
 \begin{align} 
\max( \dist_{L_s,L_{s_1}}, \dist_{L_{s_1},L_{s'}}) \ge \frac{R- 6\xi_0}{2}  \ge \frac{R}{4} \ge \kappa\xi_0,
\end{align} 
where the second and third inequalities are derived from the condition~\eqref{condition_for_m_ast_1} with $m=2$. 
From the above inequality, the definition of~$g(R)$ in \eqref{def:f_R_LR_Long} implies
 \begin{align} 
\min[ g(\dist_{L_s,L_{s_1}}), g(\dist_{L_{s_1},L_{s'}}) ] \le g(R/4) =\frac{4^\alpha}{\tilde{g}_R/\tilde{g}_{R/4}} g(R)  \le 4^\alpha g(R) ,
\end{align} 
where the last inequality is given by $\tilde{g}_R\ge\tilde{g}_{R/4}$.
Hence, we obtain
 \begin{align} 
\sum_{s_1=1}^{\tilde{n}}  g(\dist_{L_s,L_{s_1}}) g(\dist_{L_{s_1},L_{s'}})\le \sum_{s_1=1}^{\tilde{n}}  \min[ g(\dist_{L_s,L_{s_1}}), g(\dist_{L_{s_1},L_{s'}}) ] 
\cdot \Bigl [ g(\dist_{L_s,L_{s_1}}) +   g(\dist_{L_{s_1},L_{s'}}) \Bigr] \le 2\tilde{\lambda} \cdot 4^\alpha  g(R), \label{correpondence_lambda_tilde} 
\end{align} 
where in the last inequality we use 
\begin{align}
\max_{i_{s} \in \Lambda^{(\xi_0)}} \sum_{i_{s'} \in \Lambda^{(\xi_0)}} g(\dist_{L_{s'},L_{s}})  \le \tilde{\lambda}  
\label{upp_bound_lambda_ineq}
\end{align}  
with $\tilde{\lambda}$ defined in Eq.~\eqref{def_para_constant_f_proof}.
The proof of the inequality~\eqref{upp_bound_lambda_ineq} is given in Sec.~\ref{proof:upp_bound_lambda_ineq}.
We thus upper-bound Eq.~\eqref{mathdal_L_m_def_finite_second} by
 \begin{align} 
\mathcal{L}_2\le \frac{2(2\gamma^29^D|t| )^2}{2!} \abb X^{(\xi_0)} \abb \cdot \abb Y^{(\xi_0)}\abb  2\tilde{\lambda} \cdot 4^\alpha  g(R).
\end{align}

In the same way, we consider the third term $\mathcal{L}_3$:
 \begin{align} 
\mathcal{L}_3= \frac{2(2|t|)^3}{3!} \sum_{Z_1 \cap X \neq \emptyset} \|\tilde{h}_{Z_1}\|  \sum_{Z_{2} \cap Z_{1} \neq \emptyset} \|\tilde{h}_{Z_2}\|  
 \sum_{Z_{3} \cap Z_{2} \neq \emptyset, Z_{3}\cap Y\neq \emptyset} \|\tilde{h}_{Z_3}\|  .
\label{mathdal_L_m_def_finite_third}
\end{align} 
In order to estimate the summation, we first consider
 \begin{align} 
&\sum_{Z_1 \cap X \neq \emptyset} \|\tilde{h}_{Z_1}\|  \sum_{Z_{2} \cap Z_{1} \neq \emptyset} \|\tilde{h}_{Z_2}\|  
 \sum_{Z_{3} \cap Z_{2} \neq \emptyset, Z_{3}\cap Y\neq \emptyset} \|\tilde{h}_{Z_3}\|   \notag \\
 &\le 
 \sum_{i_{s} \in X^{(\xi_0)}} \sum_{i_{s'} \in Y^{(\xi_0)}}  \sum_{Z_1 \cap L_{s} \neq \emptyset} \|\tilde{h}_{Z_1}\| 
\sum_{Z_{2} \cap Z_{1} \neq \emptyset} \|\tilde{h}_{Z_2}\|   
 \sum_{Z_{3} \cap Z_{2} \neq \emptyset, Z_{3}\cap  L_{s'}\neq \emptyset} \|\tilde{h}_{Z_3}\| \notag\\
  &\le 
 \sum_{i_{s} \in X^{(\xi_0)}} \sum_{i_{s'} \in Y^{(\xi_0)}}  \sum_{s_1=1}^{\tilde{n}}   \sum_{Z_1 \cap L_{s} \neq \emptyset, Z_1 \cap L_{s_1} \neq \emptyset} \|\tilde{h}_{Z_1}\| 
\sum_{Z_{2} \cap L_{s_1} \neq \emptyset} \|\tilde{h}_{Z_2}\|   
 \sum_{Z_{3} \cap Z_{2} \neq \emptyset, Z_{3}\cap  L_{s'}\neq \emptyset} \|\tilde{h}_{Z_3}\| \notag\\
 &\le 
 \sum_{i_{s} \in X^{(\xi_0)}} \sum_{i_{s'} \in Y^{(\xi_0)}} \sum_{s_1=1}^{\tilde{n}}   \sum_{Z_1 \cap L_{s} \neq \emptyset, Z_1 \cap L_{s_1} \neq \emptyset} \|\tilde{h}_{Z_1}\| 
\sum_{s_2=1}^{\tilde{n}}  \sum_{Z_{2} \cap  L_{s_1}  \neq \emptyset, Z_{2} \cap  L_{s_2}  \neq \emptyset} \|\tilde{h}_{Z_2}\|   
 \sum_{Z_{3} \cap L_{s_2} \neq \emptyset, Z_{3}\cap  L_{s'}\neq \emptyset} \|\tilde{h}_{Z_3}\|  \notag \\
 &\le (\gamma^29^D )^3\sum_{i_{s} \in X^{(\xi_0)}} \sum_{i_{s'} \in Y^{(\xi_0)}} \sum_{s_1=1}^{\tilde{n}} \sum_{s_2=1}^{\tilde{n}}   
g(\dist_{L_s,L_{s_1}}) g(\dist_{L_{s_1},L_{s_2 }})g(\dist_{L_{s_2},L_{s'}}) .
\end{align} 
In the similar manner to the inequality~\eqref{R_le_distance_meq2_6xi0}, we obtain
 \begin{align} 
R &\le \dist_{L_s,L_{s_1}}+\dist_{L_{s_1},L_{s_2 }} + \dist_{L_{s_2},L_{s'}} + \diam(L_s)-1+ \diam(L_{s'})-1+ \diam(L_{s_1})-1 + \diam(L_{s_2}) -1\notag \\
& \le \dist_{L_s,L_{s_1}}+\dist_{L_{s_1},L_{s_2 }} + \dist_{L_{s_2},L_{s'}}+8\xi_0,
\end{align} 
which yields
 \begin{align} 
\max(\dist_{L_s,L_{s_1}},\dist_{L_{s_1},L_{s_2 }}, \dist_{L_{s_2},L_{s'}}) \ge \frac{R- 8\xi_0}{3}  \ge \frac{R}{6} \ge \kappa\xi_0
\end{align} 
from the condition~\eqref{condition_for_m_ast_1} with $m=3$. 
By combining the above inequality and the definition of~$g(R)$ in \eqref{def:f_R_LR_Long}, we also obtain 
 \begin{align} 
\min[ g(\dist_{L_s,L_{s_1}}), g(\dist_{L_{s_1},L_{s_2}}), g(\dist_{L_{s_2},L_{s'}}) ] \le g(R/6) =\frac{6^\alpha}{\tilde{g}_R/\tilde{g}_{R/6}} g(R)  \le 6^\alpha g(R) ,
\end{align} 
which yields 
 \begin{align} 
\sum_{s_1=1}^{\tilde{n}} \sum_{s_2=1}^{\tilde{n}}   
g(\dist_{L_s,L_{s_1}}) g(\dist_{L_{s_1},L_{s_2 }})g(\dist_{L_{s_2},L_{s'}}) \le 3 \tilde{\lambda}^2 \cdot 6^\alpha g(R) ,
\end{align} 
where we use the inequality~\eqref{upp_bound_lambda_ineq}.
We thus obtain
 \begin{align} 
\mathcal{L}_3\le  \frac{2(2\gamma^29^D |t|)^3}{3!} \abb X^{(\xi_0)} \abb \cdot \abb Y^{(\xi_0)}\abb3 \tilde{\lambda}^2 \cdot 6^\alpha g(R) .
\end{align} 

By repeating this process, we obtain 
 \begin{align} 
\mathcal{L}_m \le    \frac{2(2\gamma^29^D |t|)^m}{m!}  \abb X^{(\xi_0)} \abb \cdot \abb Y^{(\xi_0)} \abb m  \tilde{\lambda}^{m-1} \cdot (2m)^\alpha g(R) 
\end{align} 
for $m\le m^\ast -1$.
From the above upper bounds for $\{\mathcal{L}_m\}_{m=1}^{m^\ast-1}$, we have
 \begin{align} 
\sum_{m=1}^{m^\ast-1} \mathcal{L}_m 
&\le \frac{2^{\alpha+1} }{\tilde{\lambda}} \abb X^{(\xi_0)} \abb \cdot \abb Y^{(\xi_0)} \abb g(R)   \sum_{m=1}^{\infty} \frac{(2 \tilde{\lambda }\gamma^29^D|t| )^m }{m!} m^{\alpha+1} \notag \\
&\le \frac{2^{\alpha+1} e\Gamma(\alpha+2) }{\tilde{\lambda}} \abb X^{(\xi_0)} \abb \cdot \abb Y^{(\xi_0)} \abb \left( e^{2 e\tilde{\lambda }\gamma^29^D|t|  }-1 \right)  g(R)  ,
\label{summation_mathcal_L_m_mast-1}
\end{align} 
where the upper bound of the summation is derived as follows:
 \begin{align} 
\sum_{m=1}^{\infty} \frac{(2 \tilde{\lambda }\gamma^29^D|t| )^m }{m!}  m^{\alpha+1} 
&\le \sum_{m=1}^{\infty}  \frac{(2 \tilde{\lambda }\gamma^29^D|t| )^m }{m!} e^m \sum_{\tilde{m}=1}^{\infty} e^{-\tilde{m}} \tilde{m}^{\alpha+1} \notag \\
&\le \left( e^{2e \tilde{\lambda }\gamma^29^D|t|  }-1 \right ) \int_0^\infty (x+1)^{\alpha+1} e^{-x} \le e\Gamma(\alpha+2)\left( e^{2e \tilde{\lambda }\gamma^29^D|t|  }-1 \right) .
\end{align} 
Note that $\Gamma(x)$ is the gamma function.

Finally, for $m=m^\ast$, by regarding as $Y=\Lambda$, we can apply the same analyses as in the case of $m\le m^\ast -1$.
We obtain 
 \begin{align} 
 &\sum_{Z_1 \cap X \neq \emptyset} \|\tilde{h}_{Z_1}\|  
\sum_{Z_{2} \cap Z_{1} \neq \emptyset}  \|\tilde{h}_{Z_{2}} \|\cdots    \sum_{Z_{m^\ast} \cap Z_{m^\ast-1} \neq \emptyset}  \| \tilde{h}_{Z_{m^\ast}} \|  \notag \\
 &\le 
 \sum_{i_{s} \in X^{(\xi_0)}}  \sum_{i_{s'} \in \Lambda^{(\xi_0)}}  \sum_{s_1,s_2,\ldots,s_{m^\ast-1}=1}^{\tilde{n}}  \sum_{Z_1 \cap L_{s} \neq \emptyset, Z_1 \cap L_{s_1} \neq \emptyset} \|\tilde{h}_{Z_1}\|  \sum_{Z_{2} \cap  L_{s_1}  \neq \emptyset, Z_{2} \cap  L_{s_2}  \neq \emptyset} \|\tilde{h}_{Z_2}\| \sum_{Z_{3} \cap L_{s_2} \neq \emptyset, Z_{3}\cap  L_{s'}\neq \emptyset} \|\tilde{h}_{Z_3}\|    \notag \\
&\quad \quad \quad  \cdots    \sum_{Z_{m^{\ast}-1} \cap L_{s_{m^\ast-2}} \neq \emptyset, Z_{m^{\ast}-1}\cap  L_{m^{\ast}-1}\neq \emptyset} \|\tilde{h}_{ Z_{m^{\ast}-1}}\|      \sum_{Z_{m^{\ast}} \cap L_{s_{m^\ast-1}} \neq \emptyset, Z_{m^{\ast}}\cap  L_{s'}\neq \emptyset} \|\tilde{h}_{ Z_{m^{\ast}}}\|    \notag \\
 &\le (\gamma^2 9^D )^{m^\ast}  \sum_{i_{s} \in X^{(\xi_0)}} \sum_{s' , s_1,s_2,\ldots,s_{m^\ast-1}=1}^{\tilde{n}} 
g(\dist_{L_s,L_{s_1}}) g(\dist_{L_{s_1},L_{s_2 }})g(\dist_{L_{s_2},L_{s_3}}) \cdots g(\dist_{L_{s_{m^\ast-2}},L_{s_{m^\ast-1}}}) g(\dist_{L_{s_{m^\ast-1}},L_{s'}})  \notag \\
 &\le  (\gamma^2 9^D )^{m^\ast} \abb X^{(\xi_0)} \abb \tilde{\lambda}^{m^\ast}.
\end{align} 
From the definition of $\mathcal{L}_{m^\ast}$ as in~\eqref{mathdal_L_m_def_finite_m_ast}, the above inequality yields
\begin{align} 
\mathcal{L}_{m^\ast}= 2\abb X^{(\xi_0)} \abb  \frac{(2\tilde{\lambda}\gamma^2 9^D |t|)^{m^\ast}}{m^\ast!} \le  2\abb X^{(\xi_0)} \abb \left( \frac{2e \tilde{\lambda} \gamma^2 9^D |t|}{m^\ast} \right)^{m^\ast},
\label{mathcal_mast_upp_bound}
\end{align} 
where we use $m^\ast! \ge (m^\ast/e)^{m^\ast}$.

By applying the inequalities~\eqref{summation_mathcal_L_m_mast-1} and \eqref{mathcal_mast_upp_bound} to \eqref{Lieb-Robinson_expansion_0_finite} with Eq.~\eqref{choice_of_mast_def}, we 
obtain the main inequality~\eqref{main_ineq_long_range_LR_proof}.
This completes the proof. $\square$


\subsubsection{Derivation of the inequality~\eqref{upp_bound_lambda_ineq}} \label{proof:upp_bound_lambda_ineq}
We here derive the upper bound~\eqref{upp_bound_lambda_ineq} for 
\begin{align}
\max_{i_{s} \in \Lambda^{(\xi_0)}} \sum_{i_{s'} \in \Lambda^{(\xi_0)}}  g(\dist_{L_{s'},L_{s}}) .
\end{align}  
Note that we have defined $L_s=\bal{i_s}{\xi_0}$ and $L_{s'}=\bal{i_{s'}}{\xi_0}$.
First, by using $\dist_{L_{s'},L_{s}} \ge \dist_{i_{s'},i_{s}}-2\xi_0$ and the definition~\eqref{def:f_R_LR_Long}, we obtain
\begin{align}
\sum_{i_{s'} \in \Lambda^{(\xi_0)}} g(\dist_{L_{s'},L_{s}}) &\le 
\sum_{i_{s'} \in \Lambda^{(\xi_0)}} g(\dist_{i_{s'},i_{s}} -2\xi_0)
\le 
\sum_{\substack{i_{s'} \in \Lambda^{(\xi_0)} \\ \dist_{i_{s'},i_{s}} <(\kappa+ 2)\xi_0}}  g(\dist_{i_{s'},i_{s}} -2\xi_0)  + 
\sum_{\substack{i_{s'} \in \Lambda^{(\xi_0)} \\ \dist_{i_{s'},i_{s}} \ge (\kappa+ 2)\xi_0}} g(\dist_{i_{s'},i_{s}} -2\xi_0)  ,
\label{upp_bound_lambda_ineq_proof0}
\end{align}  
where in the first inequality we use the condition that $g(R)$ monotonically decreases with $R$.
The first term is bounded from above by
\begin{align}
\sum_{\substack{i_{s'} \in \Lambda^{(\xi_0)} \\ \dist_{i_{s'},i_{s}} <(\kappa+ 2)\xi_0}}  g(\dist_{i_{s'},i_{s}} -2\xi_0)   
&\le 
\sum_{ i \in  (i_{s}[(\kappa+2)\xi_0])^{(\xi_0)}} \tilde{g}
\le \tilde{g} \abb (i_{s}[(\kappa+2)\xi_0])^{(\xi_0)} \abb 
\le\tilde{g} \gamma [2(\kappa+3)]^D,\label{upp_bound_lambda_ineq_proof2}
\end{align}  
where the last inequality is derived by using~\eqref{geometric_parameter_gamma1} as follows:
 \begin{align}
\abb (i_{s}[(\kappa+2)\xi_0])^{(\xi_0)} \abb \le \gamma \left[ \frac{\diam\bigl(\bal{i_{s}}{(\kappa+2)\xi_0}\bigr)}{\xi_0}\right]^D \le 
\gamma \left[ \frac{2(\kappa+2)\xi_0 +1}{\xi_0}\right]^D\le \gamma [2(\kappa+3)]^D.
\end{align}

The second term in \eqref{upp_bound_lambda_ineq_proof0} is bounded by using Lemma~\ref{summation_discrete_f}.
By choosing
 \begin{align}
f(z) = g(z-2\xi_0),\quad x_0=(\kappa+2)\xi_0,
\end{align}
we obtain from Ineq.~\eqref{summation_discrete_f_main_ineq} 
\begin{align}
\sum_{\substack{i_{s'} \in \Lambda^{(\xi_0)} \\ \dist_{i_{s'},i_{s}} \ge (\kappa+ 2)\xi_0}} g(\dist_{i_{s'},i_{s}} -2\xi_0)  \le 
2^{D+1} C_{g,(\kappa+2)\xi_0} \gamma D \xi_0^{-D} [(\kappa+2)\xi_0]^{D} g(\kappa\xi_0) ,
\label{upp_bound_lambda_ineq_proof3}
\end{align}  
where $C_{g,(\kappa+2)\xi_0}$ is defined by Eq.~\eqref{definition_C_f_x_0}:
 \begin{align}
C_{g,(\kappa+2)\xi_0}= \sup_{z\in \mathbb{R}| z\ge (\kappa+2)\xi_0} \left(\frac{z^{D+1} g(z-2\xi_0)}{[(\kappa+2)\xi_0]^{D+1} g(\kappa\xi_0)} \right).
\end{align}
From the definition~\eqref{def:f_R_LR_Long}, for $z\ge (\kappa+2)\xi_0$ we have $g(z-2\xi_0)=\tilde{g}_0 \log^p(z-2\xi_0+1)  (z-2\xi_0)^{-\alpha}$ and hence 
 \begin{align}
C_{g,(\kappa+2)\xi_0}&=\frac{\kappa^\alpha}{(\kappa+2)^{D+1}} \sup_{z\in \mathbb{R}| z
\ge (\kappa+2)\xi_0} \left[ (z/\xi_0)^{D+1} [(z-2\xi_0)/\xi_0]^{-\alpha}  \frac{\log^p(z-2\xi_0+1)}{\log^p(\kappa\xi_0+1)}\right]  \notag\\
&\le \frac{\kappa^\alpha}{(\kappa+2)^{D+1}} \sup_{z\in \mathbb{R}| z\ge \kappa+2} \left[\frac{ (\log (z)+1)^p}{(z-2)^{\alpha-D-1}}\right] =:\frac{\kappa^\alpha}{(\kappa+2)^{D+1}} c_{p,\kappa} ,
\label{upp_bound_lambda_ineq_proof4}
\end{align}
where we use the upper bound of
 \begin{align}
\frac{\log (z-2\xi_0+1)}{\log (\kappa\xi_0+1)}\le \frac{\log (z/\xi_0) + \log (\xi_0)}{\log (2\xi_0+1)}\le \frac{\log (z/\xi_0) }{\log (3)}+1 \le \log (z/\xi_0)+1.
\end{align}
Note that $\kappa\ge 2$ and $\xi_0\ge1$. 
By combining the inequalities \eqref{upp_bound_lambda_ineq_proof3} and \eqref{upp_bound_lambda_ineq_proof4}, we have
\begin{align}
\sum_{\substack{i_{s'} \in \Lambda^{(\xi_0)} \\ \dist_{i_{s'},i_{s}} \ge (\kappa+ 2)\xi_0}} g(\dist_{i_{s'},i_{s}} -2\xi_0) & \le 
2^{D+1} \gamma D \xi_0^{-D} [(\kappa+2)\xi_0]^{D} g(\kappa\xi_0) \cdot \frac{\kappa^\alpha}{(\kappa+2)^{D+1}} c_{p,\kappa} \notag \\
&\le \frac{c_{p,\kappa}  \tilde{g}_0 \gamma D 2^{D+1} }{\kappa+2}  \frac{\log^p(\kappa\xi_0+1)}{\xi_0^{\alpha} } .
\label{upp_bound_lambda_ineq_proof5}
\end{align}  

By applying the inequalities~\eqref{upp_bound_lambda_ineq_proof2} and \eqref{upp_bound_lambda_ineq_proof5} to \eqref{upp_bound_lambda_ineq_proof0}, we have
\begin{align}
\sum_{i_{s'} \in \Lambda^{(\xi_0)}} g(\dist_{L_{s'},L_{s}})  &\le 
\tilde{g}\gamma [2(\kappa+3)]^D + \frac{c_{p,\kappa}  \tilde{g}_0 \gamma D 2^{D+1} }{\kappa+2}  \frac{\log^p(\kappa\xi_0+1)}{\xi_0^{\alpha} }  = \tilde{\lambda},
\end{align}  
where we have defined $\tilde{\lambda}$ in Eq.~\eqref{def_para_constant_f_proof}.
We thus obtain the inequality~\eqref{upp_bound_lambda_ineq}. This completes the proof. $\square$


%
%
%




\end{widetext}

%
%
%
%
%
%
%
%
%
%


\end{document}